\documentclass[a4paper,11pt]{article}
\usepackage{imakeidx}
\makeindex[columns=2, title=Alphabetical Index, intoc]
\usepackage{jheppub} 
\usepackage{lineno}
\usepackage{amsmath}
\usepackage{dsfont} 
\usepackage{nicefrac}
\usepackage{braket}
\usepackage{enumitem}
\usepackage{mathtools}
\usepackage{cleveref}
\usepackage{bbold}

\usepackage{rotating}

\makeatletter
\newcommand\ps@indexpagestyle{
  \renewcommand\@oddfoot{\hfill-- \thepage\ --\hfill}
  \renewcommand\@oddhead{}
}
\makeatother
\indexsetup{firstpagestyle=indexpagestyle}

\newcommand{\be}{\begin{equation}}
\newcommand{\ee}{\end{equation}}

\newcommand{\lint}{\big\langle}
\newcommand{\rint}{\big\rangle}
\newcommand{\llint}{\left\langle}
\newcommand{\rrint}{\right\rangle}

\newcommand{\sepc}{|}
\newcommand{\lec}{{_c\langle}}
\newcommand{\ric}{\rangle_c}

\DeclarePairedDelimiter\myket{\lvert}{\rangle}                                      
\DeclarePairedDelimiterX\mybraket[2]{\langle}{\rangle}{#1 \delimsize\vert #2}       
\DeclarePairedDelimiterX\mybraketOp[3]{\langle}{\rangle}%
{#1\,\delimsize\vert\,\mathopen{}#2\,\delimsize\vert\,\mathopen{}#3} 

\newcommand{\scprod}[2]{{{#1}\!\cdot\!{#2}}}

\newcommand{\Soft}[1]{S_{#1}}
\newcommand{\Coll}[1]{C_{#1}}
\newcommand{\DoubSoft}[1]{S_{#1}}
\newcommand{\iden}{\mathbb{1}}

\newcommand{\ISoft}{I_{\rm S}}
\newcommand{\ISoftONLO}{I_{\rm S}^{\Np+1}}

\newcommand{\IVirt}{I_{\rm V}}
\newcommand{\IVirtONLO}{I_{\rm V}^{\Np+1}}
\newcommand{\IVirtSoft}{I_{\rm V+S}}
\newcommand{\IColl}{I_{\rm C}}
\newcommand{\ICollONLO}{I_{\rm C}^{\Np+1}}
\newcommand{\IColltilde}{\widetilde{I}_{\rm C}}

\newcommand{\ITot}{I_{\rm T}}
\newcommand{\ITotONLO}{I_{\rm T}^{\Np+1}}
\newcommand{\ICat}{I_1}
\newcommand{\ICatbar}{\overline{I}_1}
\newcommand{\ICatTwo}{I_2}
\newcommand{\ColT}[1]{\boldsymbol{T}_{#1}}

\newcommand{\ISofttilde}{\widetilde{I}_{\rm S}}
\newcommand{\ICollFour}{I_{\rm C}^{(4)}}

\newcommand{\Itricc}{I_{\mathrm{tri}}^{\cc}}

\newcommand{\ItriRV}{I_{\mathrm{tri}}^{\mathrm{RV}}}

\newcommand{\Iccfin}{I_{\rm cc}^{\rm fin}}

\newcommand{\SmnFin}{\lint \DoubSoft{\Fp \Sp} \THmn \FLM(\Fp,\Sp)\rint_{T^2}^\text{fin}}

\newcommand{\rmd}{\mathrm{d}}
\newcommand{\sigmahat}{\hat{\sigma}_{ab}}
\newcommand{\obs}{O}

\newcommand{\SigmaRR}{\Sigma_{\rm RR}}
\newcommand{\SigmaRRc}{\Sigma_{\rm RR, 1c}}
\newcommand{\SigmaRRcc}{\Sigma_{\rm RR, 2c}}
\newcommand{\SigmaNdoubboost}{\Sigma_N^{\mathrm{db}}}
\newcommand{\SigmaNleft}{\Sigma_N^{\mathrm{lb}}}
\newcommand{\SigmaNright}{\Sigma_N^{\mathrm{rb}}}
\newcommand{\SigmaNunboost}{\Sigma_N^{\mathrm{c, el}}}
\newcommand{\pdftermNNLO}{\rmd \hat \sigma_{ab}^{\mathrm{pdf}}}

\newcommand{\SigmaRRcACDC}{\Sigma_{\rm RR,1c}^{(a,c, \rm dc)}}
\newcommand{\SigmaRRcBD}{\Sigma_{\rm RR,1c}^{(b,d)}}

\newcommand{\SigmaRRcBDiSC}{\Sigma_{{\rm RR,1c},i }^{(b,d), \rm sc}}
\newcommand{\SigmaRRcBi}{\Sigma_{{\rm RR,1c},i}^{(b)}}
\newcommand{\SigmaRRcBiSA}{\Sigma_{{\rm RR,1c},i }^{(b), \rm sa}}
\newcommand{\SigmaRRcBiSC}{\Sigma_{{\rm RR,1c},i }^{(b), \rm sc}}

\newcommand{\SigmaRRcBDiSCI}{\Sigma_{{\rm RR,1c},i }^{(b,d), {\rm sc}, I}}
\newcommand{\SigmaRRcBDiSCII}{\Sigma_{{\rm RR,1c},i }^{(b,d), {\rm sc}, II}}

\newcommand{\SigmaRRcBDiSCIone}{\Sigma_{{\rm RR,1c},i }^{(b,d), {\rm sc}, I, 1}}
\newcommand{\SigmaRRcBDiSCItwo}{\Sigma_{{\rm RR,1c},i }^{(b,d), {\rm sc}, I, 2}}
\newcommand{\SigmaRRcBDSCIone}{\Sigma_{{\rm RR,1c}}^{(b,d), {\rm sc}, I, 1}}

\newcommand{\SigmaRRcACDCone}{\Sigma_{\rm RR,1c}^{(a,c, \rm dc),  1}}
\newcommand{\SigmaRRcACDCtwo}{\Sigma_{\rm RR,1c}^{(a,c, \rm dc),  2}}
\newcommand{\SigmaRRcACDConefin}{\Sigma_{N}^{\rm fin, (2)}}
\newcommand{\SigmaRRcACDCtwofin}{\Sigma_N^{\rm fin,  (3)}}
\newcommand{\SigmaRRcBDsa}{\Sigma_{\rm RR,1c}^{(b,d), \rm sa}}
\newcommand{\SigmaRRcBDsc}{\Sigma_{\rm RR,1c}^{(b,d), \rm sc}}
\newcommand{\SigmaRRcBDsaI}{\Sigma_{\rm RR,1c}^{(b,d), {\rm sa}, I}}
\newcommand{\SigmaRRcBDsaII}{\Sigma_{\rm RR,1c}^{(b,d), {\rm sa}, II}}
\newcommand{\Sigmadivfin}{\Sigma_{N+1}^{\rm fin, (3)}}
\newcommand{\SigmaFullyRes}{\Sigma_{N+2}^{\rm fin}}
\newcommand{\SigmaRRccfinONE}{\Sigma_N^{\rm fin,(1) }}
\newcommand{\SigmaSingUnrFinONE}{\Sigma_{N+1}^{\rm fin, (1)}}
\newcommand{\SigmaSingUnrFinTWO}{\Sigma_{N+1}^{\rm fin, (2)}}
\newcommand{\SigmaNSpinAv}{\Sigma_N^{\rm sa}}
\newcommand{\SigmaNSpinAvFin}{\Sigma_N^{\rm sa, fin}}
\newcommand{\SigmaNSpinCorr}{\Sigma_N^{\rm sc}}
\newcommand{\SigmaNSpinCorrFin}{\Sigma_N^{\rm sc, fin}}

\newcommand{\SigmaNtri}{\Sigma_N^{\rm tri}}
\newcommand{\SigmaNccElbetafin}{\Sigma_{N}^{\rm fin, (6)}}
\newcommand{\SigmaNccElbeta}{\Sigma_N^{({\rm V+S}), {\rm el}, \beta_0}}
\newcommand{\SigmaNccEl}{\Sigma_N^{({\rm V+S}), {\rm el}}}
\newcommand{\SigmaNccBoost}{\Sigma_N^{({\rm V+S}), {\rm boost}}}

\newcommand{\ww}[1]{\omega^{#1}}

\newcommand{\LO}{\mathrm{LO}}
\newcommand{\NLO}{\mathrm{NLO}}
\newcommand{\NNLO}{\mathrm {NNLO}}
\newcommand{\order}[1]{\mathcal{O}(#1)}
\newcommand{\ONLO}{\mathcal{O}_\text{NLO}}

\newcommand{\Pqq}{P_{qq}}
\newcommand{\Pgg}{P_{gg}}
\newcommand{\Paa}{P_{aa}}
\newcommand{\PAP}{\hat P^{(0)}}
\newcommand{\PAPone}{\hat P^{(1)}}
\newcommand{\hatP}{\hat{P}}
\newcommand{\CalP}{\mathcal{P}}

\newcommand{\CalPgen}{\mathcal{P}^{\mathrm{gen}}}
\newcommand{\PqqGen}{\CalPgen}
\newcommand{\convPgenPgen}[2]{\big[\PqqGen_{#1} \, \bar{\otimes} \, \PqqGen_{#2}\big]}
\newcommand{\CalPgenfin}[1]{\mathcal{P}^{(#1),\rm fin}}
\newcommand{\PNLO}{\mathcal{P}^{\mathrm{NLO}}}
\newcommand{\PNNLO}{\mathcal{P}^{\mathrm{NNLO}}}
\newcommand{\PaaNLO}{\PNLO_{aa}}
\newcommand{\PbbNLO}{\PNLO_{bb}}
\newcommand{\GammaLoop}[1]{\Gamma_{#1}^{\text{1L}}}
\newcommand{\CalPLoopgen}{\mathcal{P}^{\mathrm{1L, gen}}}
\newcommand{\CalPGenFour}{\mathcal{P}^{(4),\text{gen}}}
\newcommand{\CalPfin}{\mathcal{P}^{\mathrm{fin}}}
\newcommand{\CalPkoneLgen}{\CalP^{(k),\mathrm{1L,gen}}}
\newcommand{\CalPoneLgen}{\CalP^{\mathrm{1L,gen}}}
\newcommand{\CalPkoneLfin}{\CalP^{(k),\mathrm{1L,fin}}}
\newcommand{\Ppdf}[1]{\hat{\mathbb{P}}_{#1}}

\newcommand{\amp}{\mathcal{M}}
\newcommand{\FLM}{F_{\mathrm{LM}}}
\newcommand{\FLMmunu}{F_{\mathrm{LM},\mu\nu}}
\newcommand{\FLV}{F_{\mathrm{LV}}}
\newcommand{\FLMcol}{\widetilde{F}_{\mathrm{LM}}}
\newcommand{\FLRV}{F_{\rm RV}}
\newcommand{\FLRVfin}{F_{\rm RV}^{\rm fin}}
\newcommand{\FLVfin}{F_{\mathrm{LV}}^{\mathrm{fin}}}

\newcommand{\FLVV}{F_{\rm VV}}
\newcommand{\FLVfinsq}{F^{\mathrm{fin}}_{{\rm LV}^2}}
\newcommand{\FLVVfin}{F^{\mathrm{fin}}_{\rm VV}}

\newcommand{\ampM}[1]{\mathcal{M}_{#1}}

\newcommand{\Emax}{E_{\rm max}}
\newcommand{\EulerGamma}{\gamma_{\mathrm{E}}}
\newcommand{\Sep}{S_\ep}
\newcommand{\eps}{\epsilon}
\newcommand{\ep}{\epsilon}
\newcommand{\musq}{\mu^2}
\newcommand{\Np}{N_p}

\newcommand{\colsing}{X}

\newcommand{\D}{\mathcal{D}}
\newcommand{\fl}[1]{f_{#1}}

\newcommand{\TR}{T_R}
\newcommand{\nf}{n_f}
\newcommand{\Ca}{C_A}
\newcommand{\Cf}{C_F}
\newcommand{\Nc}{N_c}

\newcommand{\gsb}{g_{s,b}}
\newcommand{\gs}{g_s}
\newcommand{\as}{\alpha_s}
\newcommand{\asbr}{[\alpha_s]}
\newcommand{\amu}{\frac{\alpha_s(\mu)}{2\pi}}

\newcommand{\inotj}{(ij)}

\newcommand{\knotl}{(kl)}
\newcommand{\doublesum}[3]{\sum_{\substack{#1 \\ #2}}^{#3}}

\newcommand{\conv}{\otimes}

\newcommand{\myRe}{\mathrm{Re}}
\newcommand{\Li}{\mathrm{Li}}
\newcommand{\Ltildei}{\widetilde{L}_i}
\newcommand{\Lmax}{L_{\rm max}}
\newcommand{\hypF}{{_2F_1}}
\newcommand{\THmn}{\Theta_{\Fp \Sp}}   
\newcommand{\THnm}{\Theta_{\Sp \Fp}}   
\newcommand{\T}{\boldsymbol{T}}
\newcommand{\Htwotc}{\mathcal{H}_{2, \rm tc}}
\newcommand{\Htwocd}{\mathcal{H}_{2, \rm cd}}
\newcommand{\HtwotcOL}{\mathcal{H}_{2, \rm tc}}
\newcommand{\HtwocdOL}{\mathcal{H}_{2, \rm cd}}

\newcommand{\Wac}[1]{\mathcal{W}_{#1}^{#1 \parallel \Sp}}
\newcommand{\Wacfin}[1]{\mathcal{W}_{#1}^{#1 \parallel \Sp,\text{fin}}}
\newcommand{\Wbd}[1]{\mathcal{W}_{#1}^{\Fp \parallel \Sp}}
\newcommand{\Wbdfin}[1]{\mathcal{W}_{#1}^{\Fp \parallel \Sp, \text{fin}}}

\newcommand{\cc}{\mathrm{(cc)}}
\newcommand{\colorprod}{\cdot}

\newcommand{\Kijtwo}{K_{ij}^{(2)}}
\newcommand{\Iplus}{I_+}
\newcommand{\Iminus}{I_-}
\newcommand{\Iplusminus}{I_{\pm}}
\newcommand{\ICatbarcc}{\ICatbar^{\cc}}
\newcommand{\Ipluscc}{I_+^{\cc}}
\newcommand{\Iminuscc}{I_-^{\cc}}
\newcommand{\Iplusminuscc}{I_{\pm}^{\cc}}
\newcommand{\ISoftcc}{\ISoft^{\cc}}

\newcommand{\fn}[1]{f_{#1}}

\newcommand{\Fp}{\mathfrak{m}}
\newcommand{\Sp}{\mathfrak{n}}

\newcommand{\oS}{\overline S}
\newcommand{\oC}{\overline C}

\newcommand{\tensprod}{\cdot}

\newcommand{\hc}{h_c}
\newcommand{\myset}{\psi}

\def\OXF{Rudolf Peierls Centre for Theoretical Physics, University of Oxford,
Clarendon Laboratory, Parks Road, Oxford OX1 3PU, UK}

\def\KITA{Institute for Theoretical Particle Physics, KIT, Wolfgang-Gaede-Straße 1, 76131, Karlsruhe, Germany}
\def\KITB{Institute for Astroparticle Physics, KIT, 
Hermann-von-Helmholtz-Platz 1, 76344 Eggenstein-Leopoldshafen,  Germany}
\def\TIF{Tif Lab, Dipartimento di Fisica, Universit\'{a} di Milano and
INFN, Sezione di Milano, Via Celoria 16, I-20133 Milano, Italy}
\def\MP{Max-Planck-Institut für Physik, Föhringer Ring 6, 80805 München, Germany}

\preprint{
\begin{flushright}
TIF-UNIMI-2023-29,
TTP23-050,
P3H-23-077, 
OUTP-23-11P,
MPP-2023-250
\end{flushright}
}

\title{\boldmath A fresh look at the nested soft-collinear subtraction scheme: NNLO QCD corrections to $N$-gluon final states in $q\bar{q}$ annihilation}

\author[a]{Federica Devoto,}
\author[b]{Kirill Melnikov,}
\author[c]{Raoul R{\"o}ntsch,}
\author[b,d,e]{Chiara Signorile-Signorile}
\author[c]{Davide Maria Tagliabue}

\emailAdd{federica.devoto@physics.ox.ac.uk}
\emailAdd{kirill.melnikov@kit.edu}
\emailAdd{raoul.rontsch@unimi.it}
\emailAdd{signoril@mpp.mpg.de}
\emailAdd{davide.tagliabue@unimi.it}
\affiliation[a]{\OXF}
\affiliation[b]{\KITA}
\affiliation[c]{\TIF}
\affiliation[d]{\KITB}
\affiliation[e]{\MP}

\abstract{We describe  how the nested soft-collinear subtraction scheme \cite{Caola:2017dug} 
can be used to compute the  
 next-to-next-to-leading order (NNLO) QCD corrections to the production of \emph{an arbitrary number of gluonic jets} in hadron collisions. 
We show that   the infrared subtraction terms  can be combined into  recurring structures that in many cases are simple iterations of those terms known from 
next-to-leading order. 
The way that these recurring structures are identified and computed is fairly general, and can be applied to any partonic process. 
 As an example, we  explicitly demonstrate  the cancellation of \emph{all} singularities in the fully-differential cross section 
for the $q\bar{q} \to X + N g $ process
 at NNLO in QCD.
 The  finite remainder of the NNLO QCD contribution, which  arises upon 
 cancellation of all $\ep$-poles, is expressed via relatively simple formulas, 
which can be implemented in a numerical code in a straightforward way.  
Our approach can be extended  to describe arbitrary 
processes at NNLO in QCD; the largest remaining challenge at this point is the combinatorics of quark and gluon collinear limits.  
}


\begin{document}
\maketitle
\flushbottom

\allowdisplaybreaks
\section{Introduction}

The theoretical description of hard scattering processes at the LHC
is based, almost entirely, on perturbative QCD.  Because of this, the development of theoretical methods that can be used to provide predictions at progressively higher orders of perturbation theory  has been one of the most active and exciting topics in theoretical particle physics in the past decade 
(see Refs~\cite{Heinrich:2020ybq,Weinzierl:2022eaz,Agarwal:2021ais,Badger:2023eqz}
for the recent 
reviews).

An important part of the theoretical toolbox that allows the description of infrared-safe observables at high orders of perturbative QCD 
is the treatment of infrared singularities. It is well-known that these singularities cancel upon combining virtual corrections,  unresolved real-radiation contributions, and the collinear renormalization of 
parton distribution functions (PDFs). An important question is then how to organize this cancellation in a process-independent way and how to arrive at finite remainders that are suitable for numerical evaluations.  

This problem was fully solved at next-to-leading order (NLO) in perturbative QCD  many years ago  \cite{Giele:1993dj, Giele:1994xd, Frixione:1995ms, Catani:1996vz, Nagy:2003qn, Bevilacqua:2013iha}
(see also Ref.~\cite{Prisco:2020kyb} for more recent work),
but its extension to next-to-next-to-leading order (NNLO) and beyond has proved to be difficult. 
In fact, there are many NNLO subtraction and slicing schemes~\cite{Frixione:2004is, Gehrmann-DeRidder:2005btv, Currie:2013vh,Somogyi:2005xz,Somogyi:2006db,DelDuca:2016csb,DelDuca:2016ily,Czakon:2010td,Czakon:2011ve,Czakon:2014oma,Anastasiou:2003gr, Caola:2017dug,Catani:2007vq,Grazzini:2017mhc, Boughezal:2011jf,Gaunt:2015pea,Boughezal:2015dva,Sborlini:2016hat,Herzog:2018ily,Magnea:2018hab,Bertolotti:2022aih,Capatti:2019ypt,TorresBobadilla:2020ekr} that have been  used to perform the many impressive computations at this perturbative order,\footnote{See Refs.~\cite{Chen:2014gva,Boughezal:2015dra,Caola:2015wna,Chen:2016zka,Campbell:2019gmd,Cacciari:2015jma,Cruz-Martinez:2018rod,Gauld:2021ule,Catani:2022mfv,Chawdhry:2019bji,Chawdhry:2021hkp,Czakon:2020coa,Gauld:2023zlv,Currie:2017eqf,Chen:2022tpk,Badger:2023mgf,Czakon:2021mjy,Czakon:2015owf,Catani:2019hip,Buonocore:2023ljm,Brucherseifer:2014ama,Berger:2016oht,Campbell:2020fhf,Alvarez:2023fhi} for a representative list of NNLO calculations by different collaborations.}
but it is fair to say that  
the complete generality achieved at NLO is still elusive at NNLO. 

A peculiar  illustration  of this 
statement  is the fact that the  cancellation of $1/\ep^n$ infrared poles\footnote{Throughout the paper we use dimensional regularization and work in $d=4-2\ep$ dimensions.} for a \emph{generic}  hadron collider process has not 
been demonstrated  in any NNLO slicing or subtraction scheme up to now, although important work in this direction, focusing on gluonic states, has recently been presented in Ref.~\cite{Chen:2022ktf}, and including other partonic channels in Ref.~\cite{Gehrmann:2023dxm}. For $e^+e^-$ collisions such a cancellation for arbitrary final states has been  shown  only  in the 
context of the so-called  local analytic sector subtraction  scheme~\cite{Magnea:2018hab, Bertolotti:2022aih}. 

The goal of this  paper is to partially address this issue in the context of the  nested soft-collinear subtraction scheme~\cite{Caola:2017dug}.
This scheme has already been successfully applied to compute the NNLO QCD corrections to a variety of processes such as color singlet production~\cite{Caola:2019nzf} and decay~\cite{Caola:2019pfz}, deep inelastic scattering~\cite{Asteriadis:2019dte}, Higgs production in WBF~\cite{Asteriadis:2021gpd}, non-factorizable corrections to $t$-channel single-top production~\cite{Bronnum-Hansen:2022tmr} as well 
as mixed QCD-electroweak corrections to the production of  electroweak gauge bosons
and dilepton pairs
~\cite{Buccioni:2020cfi, Behring:2020cqi,Buccioni:2022kgy}.
This 
suggests that the nested soft-collinear subtraction scheme  
possesses the 
flexibility and the simplicity that is needed for  studies of  multi-particle final states. 

Moreover, the computations of double-unresolved soft and collinear contributions for arbitrary 
kinematics are usually  considered to be some of the most challenging calculations  required to develop a particular subtraction scheme.  Interestingly, in the case of the nested 
soft-collinear subtraction scheme, such computations were completed several years ago~\cite{Delto:2019asp, Caola:2018pxp},
but this has not led  to immediate applications  of  this scheme to  high-multiplicity  QCD processes.
Understanding the reasons for that 
is essential 
for further developing the nested soft-collinear subtraction scheme  and for making it  applicable to the description of  arbitrary 
collider processes. 

In this paper, we take a step in that direction by  describing the application of 
this scheme to the study of NNLO QCD corrections to 
the production of an \emph{arbitrary} number of gluons 
and a colorless final state $X$ 
in $q \bar q$ annihilation. We emphasize that we restrict ourselves to the case where all resolved and unresolved final-state partons are gluons, i.e.\ splittings of the form $g^* \to q\bar{q}$ are not considered in this paper. Practically, this can easily be achieved by setting the number of light quark flavors $n_f=0$ in e.g.\ the QCD $\beta$ function. Nevertheless, we will report all such formulas with their $n_f$ contributions, with an eye on a future extension to unresolved quark final states, and with the understanding that we take $n_f=0$ throughout the paper. 

Therefore we are interested the process
$1_a + 2_{b} \to X + N \, g$
with 
$a,b \in \{q, \bar{q}\}$, i.e.  the process
$1_a + 2_{b} \to X + N \, g$
with 
$a,b \in \{q, \bar{q}\}$.\footnote{ A prototypical physical process is the gluonic contribution to  $q \bar q$ annihilation into an electroweak vector boson and a large number of jets.}  However,  we will keep the generic notation of $a$ and $b$ for the initial-state partons, in order to make a future generalization easier. 
In particular, we stress that 
the extension of this result 
to $gg$ annihilation  into $X + N\,g$
is straightforward
since
many of our arguments apply \emph{verbatim} to this case as well, and the problem reduces  to repeating certain steps  of the calculation using different splitting functions  and replacing a few  color factors. 

Moreover,
although our results are currently restricted to gluonic 
final states, they require the analysis of matrix elements containing the richest singularity structures that can possibly arise, and we are confident  that the new insights into the 
mechanisms of infrared cancellations at NNLO in QCD that we obtain in this paper are useful for generic final states. In fact, the outstanding challenge in generalizing the results from all-gluonic to arbitrary final states is the combinatorics of various collinear limits. This aspect of the problem does not show up prominently  for all-gluonic final states because 
of the symmetry of the relevant matrix elements under permutations of final-state gluons.

There is multiple evidence suggesting  that infrared subtraction terms can be organized into clear structures that iterate from NLO to NNLO and possibly, beyond. This is rather obvious in case of leading collinear singularities where the highest collinear poles at each perturbative order are described by convolutions of leading-order splitting functions.  The fact that a similar iterative description  should  hold for soft emissions as well  follows from Catani's formula for $\ep$-poles of one- and two-loop amplitudes~\cite{Catani:1998bh}. However, the iterative nature of the subtraction terms  is  not manifest in many NNLO subtraction schemes because, following the idea of FKS subtraction at NLO, one often splits  real-emission phase spaces into partitions and sectors to project matrix elements onto the minimal number of singular kinematic configurations that one has to deal with at any  point in the calculation. 

In this paper we show how these iterative structures can be recognized and constructed in the context of the nested soft-collinear subtraction scheme. We also demonstrate  that  the existence of these iterative structures provides a strong guide for organizing NNLO QCD computations and leads  to the reduction of the computational complexity, allowing  us to deal with final states of \emph{arbitrary} multiplicity.

The main  result of this paper is a  formula 
that allows the computation of NNLO QCD corrections to a process where a 
 $q \bar q$ initial state annihilates 
into $N$ final-state hard gluons and an arbitrary number of colorless particles, through a fully local subtraction procedure. 
This formula can be implemented in a computer code in a straightforward way;  
it requires finite remainders of two- and 
one-loop scattering amplitudes for a particular process and 
the corresponding Born amplitudes. Since the 
cancellation of all $1/\ep^n$ singularities is proved analytically, all 
required numerical integrations can be  performed in four-dimensional space-time. 

The rest of the paper is organized as follows. After preliminary 
 remarks in the next section, we 
 present the computation of NLO QCD corrections to the process $1_a + 2_b\to X + N \, g$ with $a,b \in \{q, \bar{q}\}$ in Section~\ref{sect:nlo}. The reader might also find it useful to refer to Appendix~\ref{sec:AppNLOdetails}, where we elaborate on the cancellation of poles at NLO. This discussion  allows us to introduce 
 the iterative structures that are crucial for  the  subsequent analyses of the NNLO QCD corrections in Section~\ref{sec:nnlo}.
 There we show how to rewrite  the double-real  contribution as a sum of terms with well-defined partonic multiplicities, and how to express these through operators corresponding to soft or collinear limits or virtual corrections. The reader who is more interested in 
 the mechanism of the pole cancellation at NNLO can skim over this section 
 and focus instead on Section~\ref{sect5new}.
The final results for the finite remainders of the NNLO QCD corrections are presented in  Section~\ref{sec:final_res}. This section  is quite self-contained so that the reader who is only interested in these results can skip to this section right away. We conclude in Section~\ref{sec:conclusions}.

Finally, we note that  the discussion of many technical details  is relegated to  multiple appendices. 
In  particular, we collect the definitions of the various constants, splitting functions and fundamental operators used throughout the manuscript in Appendix~\ref{sec:Splitting}. For the readers' convenience, the many different notations that we use in the paper are summarized in an alphabetic index that can be found at the 
end of the paper and used   to identify the place in the paper where a particular notation has been  introduced for the first time. 

\label{sec:introduction}
\section{Preliminary  considerations}

Subtraction schemes should enable  calculations of hard processes at lepton and hadron colliders at higher orders in QCD perturbation theory. In this paper, we will consider the process where  
 $N$ jets and a color-singlet system $\colsing$ are produced in hadron collisions, $pp \to X + N~{\rm jets}$. The cross section 
of this process 
 is given by the following formula 
\be
\rmd \sigma = \sum_{a,b} \int \rmd x_1  \rmd x_2 \; f_a(x_1,\mu_F) f_b(x_2,\mu_F) \, \rmd \sigmahat(x_1, x_2, \mu_R, \mu_F; \obs) \; .
\ee
Here $\rmd \sigmahat$ is the  cross section in the $ab$ partonic channel, $f_{a,b}$ are the parton distribution functions (PDFs),   $\mu_R$ and $\mu_F$ are the renormalization and factorization scales, respectively, and $\obs$ is an observable, which provides (among other things) an infrared-safe definition of  the 
$N$-jet final state. 

The partonic cross section can be expanded in the strong coupling $\alpha_s$. We write 
\index{S!$\rmd \sigmahat$}
\be
\rmd \sigmahat = \rmd \sigmahat^{\LO}+\rmd \sigmahat^{\NLO}+\rmd \sigmahat^{\NNLO}+\mathcal{O}(\alpha_s^{q+3}) \; ,
\ee
where the LO  term is proportional to $\alpha_s^q$, and we have suppressed the arguments of all the functions for brevity.

The computation of partonic cross sections and kinematic distributions requires integrating matrix elements squared 
over phase spaces of relevant final states. For a generic process, we
find it convenient to treat matrix elements 
as vectors in color space \cite{Catani:1996vz}.  A matrix element where $\Np$ partons\footnote{In the case of $pp \to X+ N$ jets we have $\Np = N+2$.}
are assigned definite color indices is then written 
as a projection on a particular color-space basis vector 
\index{Color space and algebra}
\be
    \mathcal{M}^{c_1, ...\, , c_{\Np}}(p_1, ...\, , p_{\Np}) = \lec c_1, ...\, , c_{\Np} \sepc \mathcal{M}(p_1, ...\, , p_{\Np})\ric \; .
\ee
The square of the amplitude summed over all possible color assignments is then
\be
    |\ampM{}(p_1, ...\, , p_{\Np})|^2 = \lec \ampM{}(p_1, ... \, , p_{\Np}) \sepc \ampM{}(p_1, ... \, , p_{\Np}) \ric \; .
\ee

Although it is sufficient to use the   
summed-over-colors 
amplitude squared 
to compute  leading-order cross sections, in higher orders of  QCD perturbation theory color-correlated matrix elements appear. For example, at NLO, one encounters $\langle {\cal M}| \scprod{\ColT{i}}{ \ColT{j}} | {\cal M} \rangle $,
where $\ColT{i(j)}$ 
is the color charge operator  of 
parton $i(j) \in \{1,..,N_p\}$.
To address  this possibility, it is convenient to  introduce a tensor product of leading-order matrix elements $| \ampM{0} \rangle$ in color space. We therefore define the function
\index{L!${\rm dLips}_{\rm \colsing}$}
\be
\begin{split}
\FLMcol(1_a,2_b;3, ... \,  \Np; \colsing) = & \; \sepc \ampM{0}(1_a,2_b; 3, ... \, , \Np ; \colsing) \ric \otimes \lec  \ampM {0}(1_a,2_b; 3, ... \, ,\Np ; \colsing) \sepc \\
&\times {\rm dLips}_{\rm \colsing} \; \obs(p_3, ... \, , p_{\Np}; p_\colsing) \; ,
\label{eq:defnFLMcol}
\end{split}
\ee
to describe the partonic process {$1_a+2_b\to \colsing + N$ jets at leading order. In Eq.~\eqref{eq:defnFLMcol}, $\Np=N+2$ is the number of initial- and final-state partons, the symbol $\otimes$ indicates a tensor product in color space, and ${\rm dLips}_{\rm \colsing}$ is the Lorentz-invariant phase space for the colorless system $\colsing$, including the momentum-conserving delta function. Furthermore, we always assume $f_i=g$ for $i=3, ..., \Np$,  where $f_i$ is the flavor of parton $i$, and hence we do not show a flavor index for the final state partons. 

The matrix element squared is obtained by taking the trace in color space 
\index{F!$\FLM$}
\be
\begin{split}
\mathrm{Tr} \left[ \FLMcol  \right]_c &= {\rm dLips}_{\rm \colsing} \; 
| {\ampM{0}} |^2 \; 
\obs \equiv \FLM \; ,
\label{eq:defnFLM2}
\end{split}
\ee
where the arguments of all functions 
have been suppressed. As we already mentioned, in the course of NLO and NNLO calculations we will need to act on $\FLMcol$ with a function of operators in color space, and take the trace in color space after that. Denoting such  a function as $A$, we introduce  the notation
\be
\begin{split}
    A \colorprod \FLM &\equiv \mathrm{Tr} \left[ A \, \FLMcol \right]_c  = \lec \ampM{0} \sepc A \sepc \ampM{0} \ric \; {\rm dLips}_{\rm \colsing} \, \obs \; .
    \label{eq:colorproddefn}
    \end{split}
\ee

The LO partonic cross section can be obtained by integrating $\FLM(1_a,2_b; 3, \ldots ,N_p;X)$ over the phase space of the final-state partons. We write 
\index{S!$\rmd \sigmahat^{\LO}$}
\be 
2s \;
\rmd \sigmahat^{\LO} = \mathcal{N}\int  \prod_{i=3}^{N_p} \; [\rmd p_i] \; \FLM(1_a,2_b;3, \ldots \, , \Np ;\colsing)
= \lint \FLM \rint \; ,
\label{eq:defnFLM1}
\ee
where $s=2p_1\cdot p_2$ is the partonic center-of-mass energy squared, and the angular brackets $\lint \ldots \rint$ indicate the integration over the final-state phase space.
In Eq.~\eqref{eq:defnFLM1}  $\mathcal{N}$ is a normalization factor that takes into account color and spin averages as well as symmetry factors, and  $[\rmd p_i]$ is the phase-space element of a final-state parton $i$
 \be
 [\rmd p_i] = \frac{\rmd^3 p_i}{(2\pi)^3 2 E_i} \; .
 \ee

\label{sec:gencons}
\section{Calculations at next-to-leading order}
\label{sect:nlo}

In this section we discuss  the calculation of the partonic 
cross section  of the process 
$1_a+2_b\to X + N \, g $
at next-to-leading 
order in perturbative QCD. Our main goal is to introduce an infrared finite-operator  $\ITot$, see Eq.~(\ref{eq3.2}), 
that describes the sum of virtual, soft and certain collinear contributions and, as we explain later, is important  for simplifying NNLO QCD calculations.

Computation of NLO corrections requires 
 the one-loop (virtual)  contribution, the real-emission  contribution and the contribution of the collinear renormalization of parton distribution functions\footnote{Throughout this paper we work with UV-renormalized matrix elements.}
\be
\rmd \sigmahat^{\NLO}= \rmd \sigmahat^{\rm V}+ \rmd \sigmahat^{\rm R}+ \rmd \sigmahat^{\rm pdf} \; .
\ee

It is well-known that the virtual contribution contains explicit poles in $\eps$ that arise from the integration over the loop momentum.  For a generic  process, these  poles can be written in a closed form  using Catani's function $\ICat(\eps)$ \cite{Catani:1998bh}.  On the other hand, the real-emission 
contributions do not contain explicit poles in $\ep$ 
until the integration  over the phase space of final-state  partons is performed.  Such an integration 
extends over singular 
kinematic regions that correspond to soft and/or collinear emissions and generates the $1/\ep^n$ poles.
Eventually, many of these poles will cancel with poles 
in the one-loop  contribution; therefore, we 
 would  like to parametrize them in a manner similar 
 to Catani's function for the virtual corrections. 
 Hence, we define  soft and hard-collinear analogs of Catani's function, which we call $\ISoft(\eps)$ and $\IColl(\eps)$, respectively, as well as a function $\IVirt(\eps)$ which is related to $\ICat(\eps)$.
These functions will multiply terms with leading order kinematics, 
such that the sum
\index{I!$\ITot$}
\be
    \ITot(\eps) = \IVirt(\eps) + \ISoft(\eps) + \IColl(\eps) \; ,
    \label{eq3.2}
\ee
is $\ep$-finite.%
\\

To define all the $I$-operators in Eq.~(\ref{eq3.2}) and 
to explain 
how their  combination  arises, we  begin by considering  the real-emission  contribution to the NLO cross section. 
This contribution refers to the  process $1_a+2_b\to X+ (N+1)\, g $.
We write 
\index{S!$\rmd \sigmahat^{\rm R}$}
\index{F!$\FLM(\Fp)$}
\be
2 s \;
\rmd \sigmahat^{\rm R} =   \lint  \FLM(1_a,2_b;3, ... \, , N_p+1;\colsing) \rint 
\; .
\ee
Since the observable $\obs$ in the definition of $\FLM$
requires at least $N$ resolved partons, one and only one parton among the $N+1$ final-state ones in the above equation can become unresolved, i.e. soft and/or collinear to another parton. 
To  identify the unresolved parton,   we introduce damping factors $\Delta^{(i)}$
such that they provide  a partition of unity, 
\index{Damping factor!NLO!Properties}
\be
\label{eq:damping_NLO}
\sum_{i=3}^{\Np+1} \Delta^{(i)} = 1 \; .
\ee
The explicit form of the damping factors can be found in Appendix~\ref{sect:partitions}. They are constructed 
in such a way that a damping factor $\Delta^{(i)}$
vanishes when any parton, with the exception of parton $i$,
becomes either soft or collinear to any other parton, including the incoming ones. This implies that in the 
combination $\Delta^{(i)} \FLM$,  only soft and collinear limits of parton $i$ can lead to non-integrable singularities and, eventually, to  the appearance of $1/\ep^n$ poles. 

We then write
\index{Damping factor!NLO!Properties}
\be
\lint \FLM(1_a,2_b;3, ... \, , \Np + 1;\colsing) \rint = \sum_{i=3}^{\Np+1} \lint \Delta^{(i)} \FLM(1_a,2_b;3, ... \, , \Np + 1;\colsing) \rint \; .
\ee
Since we focus on the all-gluon final state,  $\FLM$ is unchanged under any permutation of the final-state partons. Then we obtain
\be
  \sum_{i=3}^{\Np+1} \lint\Delta^{(i)} \FLM(1_a,2_b;3, ... \, , \Np + 1;\colsing) \rint = \lint (\Np-1) \Delta^{(\Fp)}  \FLM(\Fp) \rint \; . 
  \label{eq3.6}
\ee
In the above result, 
we have relabelled  the arguments  of $\FLM$ in such  a way that the damping factors become identical for 
each term in  the sum and  we denote 
the potentially-unresolved  gluon 
as $\Fp$. The remaining $N = N_p -2 $ final-state gluons are resolved. For simplicity, we do not show the dependence 
of $\FLM$ 
on their momenta and  polarizations. 
We also omit the dependence of $\FLM$ on the  kinematics of   color-singlet 
final-state particles. 

 We note that in Eq.~(\ref{eq3.6}) 
the functions $\FLM$ include $1/(\Np-1)!$ symmetry factors 
for the all-gluon final state. 
The factor $(\Np-1)$ on the right hand side 
of that equation combines with $1/(\Np-1)!$ and turns 
into $1/(\Np-2)! = 1/N!$ 
where $N$ is the minimal required 
number of \emph{resolved}  jets. 
This is the same symmetry factor as 
in e.g.~the virtual contribution 
 and we will  simply not write 
it explicitly in what follows.  Thus, by an abuse 
of notation, we will write the right-hand side of 
Eq.~(\ref{eq3.6}) as $\lint \Delta^{(\Fp)}  \FLM(\Fp) \rint $, with the understanding that symmetry factors 
in $\FLM$ refer  to \emph{resolved final-state 
gluons only}.

\vspace*{0.5cm}
To deal with matrix elements and phase spaces in 
soft and collinear limits we need the corresponding
operators. These operators were 
introduced earlier \cite{Caola:2017dug} and 
we repeat their definitions here for 
completeness. The actions of soft $S_i$ and collinear 
$C_{ij}$ operators on a function $A$ are described by 
the following formulas 
\index{S!$S_i$}
\index{C!$C_{ij}$}
\index{R!$\rho_{ij}$}
\be
    S_i A = \lim_{E_i \to 0} A \;, \qquad C_{ij} A = \lim_{\rho_{ij} \to 0} A \; ,
\ee
where $E_i$ is the energy of parton $i$ and $\rho_{ij} = 1-\cos\theta_{ij}$, with $\theta_{ij}$ is the angle between the three-momenta of partons $i$ and $j$.\footnote{Since our primary variables are energies and angles, we need to fix a reference frame at the beginning of the calculation.} 
When these operators appear in the formulas for cross sections,  it is understood that they act on all quantities to the right of them; when limits in the conventional sense do not exist,  they  extract the most singular contributions.
 
The soft and collinear 
  operators acting  on the damping 
  factors lead to  the following results\footnote{Derivation of these results can be found in Appendix~\ref{sect:partitions}.}
\index{Damping factor!NLO!Properties}
\begin{equation}
    S_\Fp \Delta^{(\Fp)} = 1 \; , \qquad C_{a \Fp} \Delta^{(\Fp)} = 1 \; , \qquad C_{i \Fp} \Delta^{(\Fp)} = \frac{E_\Fp}{E_i + E_\Fp} \equiv z_{\Fp,i} \; ,
    \label{eq:NNLOdampinglimits}
\end{equation}
for $a=1, 2$ and $i \ge 3$.

\vspace*{0.5cm}
We will now use these operators to isolate and subtract the singular contributions, starting with the soft one. We write 
\be
    \lint \Delta^{(\Fp)} \FLM(\Fp) \rint =
    \lint S_\Fp \FLM(\Fp) \rint
    + \lint \oS_\Fp \Delta^{(\Fp)} \FLM(\Fp ) \rint \; , 
\label{eq:softlimNLO0}
\ee
where we introduced the  handy notation 
\index{S!$\oS$}
\begin{equation}
    \oS_\Fp \equiv \iden - S_\Fp \; . 
\end{equation}
The soft limit of the matrix element squared reads
\index{Soft limit}
\index{S!$S_i$}
\index{G!$g_{s,b}$}
\be
    S_{\Fp} \FLM(\Fp) = -g_{s,b}^2\sum \limits_{\inotj}^{\Np} \frac{p_i  \cdot p_j}{(p_i \cdot  p_\Fp)(p_j \cdot  p_\Fp)} \left(\scprod{\ColT{i}}{ \ColT{j}} \right)\colorprod \FLM \; ,
\label{eq:softNLO0}
\ee
where $g_{s,b}$ is the bare coupling constant, and 
we have used Eq.~\eqref{eq:colorproddefn} 
to write the color-correlated matrix element squared 
in a convenient way. In Eq.~\eqref{eq:softNLO0}, the sum runs over distinct indices $i$ and $j$.
We remind the reader that the color-charge operators of different particles $\ColT{i}$ commute with each  other. Furthermore,  we use the Casimir operators to compute squares of color-charge operators
with
$\ColT{q}^2 =\ColT{\bar q}^2= C_F$ and $\ColT{g}^2 = C_A$.
\index{Color space and algebra}

Since the unresolved gluon $\Fp$  decouples from $\FLM$, we can integrate Eq.~\eqref{eq:softNLO0} over its $d$-dimensional phase space. To do so, we introduce an upper bound on the soft gluon energy,  $E_\Fp \le \Emax$.\footnote{ 
$\Emax$ is an arbitrary quantity that should be larger than the largest energy that a particle  in a particular process can have.  For additional information, see Ref.~\cite{Caola:2017dug}.} 
Performing this integration, we find
\index{I!$\ISoft$}
\be
\begin{split}
    \lint S_{\Fp}  \FLM( \Fp ) \rint = & - [\alpha_s] \frac{(2\Emax/\mu)^{-2\eps}}{\ep^2}  \sum_{\inotj}^{\Np} \lint \eta_{ij}^{-\eps} K_{ij} \left(\scprod{\ColT{i}}{\ColT{j}} \right)\colorprod \FLM \rint \\
    \equiv & ~ [\alpha_s] \lint \ISoft (\eps)\colorprod \FLM \rint \; ,
\label{eq:NLO_soft0}
\end{split}
\end{equation}
where 
\index{A!$\asbr$}
\index{A!$\alpha_s$}
\be
    [\alpha_s] = \amu \frac{e^{\eps \gamma_\text{E}}}{\Gamma(1-\eps)} \; ,
\ee
and 
\index{K!$K_{ij}$}
\index{E!$\eta_{ij}$}
\be
    K_{ij} = \frac{\Gamma^2(1-\eps)}{\Gamma(1-2\eps)} \eta_{ij}^{1+\epsilon} \hypF(1,1,1-\eps,1-\eta_{ij}) \; ,
    \qquad 
    \eta_{ij} = \rho_{ij}/2 \; .
    \label{eq:Kij_defn0}
    \ee

We now return to Eq.~\eqref{eq:softlimNLO0} and focus on the second term on the right-hand side. This term is soft-regulated, but contains collinear singularities. In order to remove them, we introduce angular partitions of unity $\ww{\Fp i}$, which satisfy the 
following equations
\index{Partition functions!Properties}
\be
\label{eq:part_NLO_prop}
\sum_{i=1}^{\Np} \ww{\Fp i}=1 ;
\qquad
        \Coll{j \Fp} \ww{\Fp  i} = \delta^{ij} \; .
\ee
Generic expressions which satisfy these constraints are presented in Eq.~\eqref{eq:sector_NLO}. 
We thus write
\be
\begin{split}
    \lint \Delta^{(\Fp)} \FLM(\Fp) \rint = & ~
    \lint S_\Fp \FLM(\Fp)  \rint
    +  \sum_{i=1}^{\Np} \lint \oS_\Fp C_{i \Fp } \Delta^{(\Fp)} \FLM(\Fp)  \rint
    \\
    &  + \sum_{i=1}^{\Np} \lint \oS_\Fp \oC_{i \Fp} \, \ww{\Fp i}   \Delta^{(\Fp)} \FLM(\Fp) \rint \; ,
\end{split}
\label{eq:NLOregulated0}
\ee
where  
\index{C!$\oC_{ij}$}
\begin{equation}
    \oC_{i\Fp} \equiv \iden - C_{i\Fp} \; . 
\end{equation}
The last  term  on the right-hand side of Eq.~\eqref{eq:NLOregulated0}
is fully regulated and can be integrated in four dimensions.
In the   hard-collinear limits that appear in the second term on the right-hand side in  Eq.~\eqref{eq:NLOregulated0},
the  gluon decouples from $\FLM$ either partially or fully, allowing  us to integrate over its phase space  in $d$ dimensions.
 
We continue with the second term on the right-hand side of Eq.~\eqref{eq:NLOregulated0}, and consider the situation where the gluon $\Fp$ becomes  collinear to the  final-state gluon $i$ 
and produces a single final-state gluon that we label as $[i\Fp]$. Integrating over the phase space of gluon $\Fp$ and renaming $[i\Fp] \mapsto i$, we find 
\index{Hard-collinear limit}
\be
   \lint \oS_\Fp C_{i \Fp }\Delta^{(\Fp)} \FLM(\Fp)  \rint = [\alpha_s] 
   \llint
   \frac{\Gamma_{i,g}}{\eps} \,   \FLM\rrint \; .
   \label{eq3.21}
 \ee
In Eq.~\eqref{eq3.21}
we have introduced the \emph{generalized energy-dependent final-state gluon anomalous dimension} 
\index{G!$\Gamma_{i,g}$}
\be
    \Gamma_{i,g} = \left(\frac{2E_i}{\mu}\right)^{-2\eps} \frac{\Gamma^2(1-\eps)}{\Gamma(1-2\eps)} \,  \; \gamma_{z,g \to gg}^{22}(\eps,L_i)  \; , \qquad i=3, ...\, , \Np \; ,
    \label{Eq:Gamma_i_Leg3_to_N_definition}
\ee
where, for any  function 
$f(z)$ regular at $z = 1$, we define
\index{G!$\gamma_{f(z),g \to g g}^{nk}(\eps,L_i)$}
\be
\begin{split}
    \gamma_{f(z),g \to g g}^{nk}(\eps,L_i) = & - \int\limits_{0}^{1} \rmd z\, (1-S_z) \left[z^{-n \eps} (1-z)^{-k\eps} \; f(z) \, \Pgg(z)\right] 
    \\
    & + 2\ColT{g}^2 \frac{1 - e^{ -k\eps L_i}}{k \eps} f(1) \; ,
    \label{eq:defn_gamma_nk0}
    \end{split}
\ee
\index{L!$L_i$}
and $L_i = \log(\Emax/E_i)$. In Eq.~\eqref{eq:defn_gamma_nk0},  
we  introduced an operator $S_z$ which extracts  the (soft)  $z \to 1$ limit of the expression it acts upon, and used $\Pgg$ to denote the spin-averaged gluon splitting function defined in Eq.~\eqref{def:Pgg_av}.  
We emphasize that 
$\Gamma_{i,g}$ depends on the energy of the 
hard-collinear parton and on $E_{\rm max}$,  but 
we do not show these dependencies in what follows. \\

We continue with the case 
when the gluon $\Fp$ becomes  collinear to one of the initial-state partons, say $1_a$. 
\index{Z!$z$}
The matrix element squared that enters the definition of the function  $\FLM$ depends on the energy fraction $z=1-E_{\Fp}/E_1$, which implies that  one cannot integrate over the energy of the collinear gluon. 
However, integrating over the relative 
angle between $\Fp$ and $a$ 
is possible; performing this integration, we find
\index{Hard-collinear limit}
\index{P!$\CalPgen$}
\be
   \lint \oS_\Fp  C_{a\Fp}  \Delta^{(\Fp)} \FLM(\Fp ) \rint  =  [\alpha_s]  \llint \frac{\Gamma_{1,a}}{\epsilon} \, \FLM\rrint  +  \frac{[\alpha_s]}{\epsilon} \llint \CalPgen_{aa} (\eps)\conv \FLM \rrint \; .
   \label{eq:NLO_hard_coll_final0}
\ee
In Eq.~(\ref{eq:NLO_hard_coll_final0})
$\Gamma_{1,a}$ is the \emph{generalized 
initial-state anomalous dimension} which reads 
\index{G!$\Gamma_{1,a}$}
\be
    \Gamma_{1,a} = \left(\frac{2E_1}{\mu}\right)^{-2\eps} \frac{\Gamma^2(1-\eps)}{\Gamma(1-2\eps)} \left( \gamma_{a} + \ColT{a}^2 \frac{1-e^{-2\eps L_1}}{\epsilon}  \right) \; , 
    \label{eq:gamma_expansion_is0}
\ee
where $\gamma_a$ is  the anomalous dimensions of the initial-state parton $a$.\footnote{ 
We remind the reader that the 
quark and gluon anomalous dimensions read  $\gamma_q = 3/2 \, C_F$ and $\gamma_g = \beta_0= 11/6 \, \Ca - 2/3 \, \TR \nf$. Since in this paper we only deal with gluon final states, 
we systematically set $n_f$ to zero
in what follows.
}
When  writing Eq.~\eqref{eq:NLO_hard_coll_final0} we have 
used the fact that we only consider final-state gluons; 
because of that the parton type does not change after the collinear splitting. 
The function  
$\CalPgen_{aa}$ in Eq.~\eqref{eq:NLO_hard_coll_final0}
is the \emph{generalized splitting function} 
\index{P!$\CalPgen_{ij}$}
\be
    \CalPgen_{aa}(z,E_1) =   \left(\frac{2E_1}{\mu}\right)^{-2\eps} \frac{\Gamma^2(1-\eps)}{\Gamma(1-2\eps)} \left[- \PAP_{aa}(z) + \epsilon \, \CalPfin_{aa}(z)\right] \; .
\label{eq:calpgen_exp}
\ee
where $\PAP_{aa}$ are the Altarelli-Parisi splitting functions which can be found in Appendix~\ref{sec:Splitting}, together with the definition of the function $\CalPfin_{aa}$.\footnote{We note that $\CalPfin_{aa}$ is a function of  $\eps$; for brevity, we do not show this dependence.} 
Furthermore,
in Eq.~\eqref{eq:NLO_hard_coll_final0}
we  also used the shorthand notation 
\be 
    \CalPgen_{aa} \conv \FLM \equiv  \int\limits_{0}^{1} \rmd  z\; \CalPgen_{aa}(z) \; \frac{\FLM(z\cdot 1_a, 2_b; \dots)}{z} \; .
    \label{eq3.26}
\ee

The case when the gluon $\Fp$
becomes collinear to the initial-state parton $2_b$
is described by an equation which is 
analogous to 
Eq.~\eqref{eq:NLO_hard_coll_final0} but contains $\Gamma_{2,b}$ instead of $\Gamma_{1,a}$, and the ``right'' convolution 
\be 
    \FLM \conv \CalPgen_{bb} \equiv  \int\limits_{0}^{1} \rmd  z \;  \CalPgen_{bb}(z) 
    \;
    \frac{\FLM(1_a, z\cdot 2_b; \dots )}{z} \; .
    \label{Eq:convolution_leg_2_def}
\ee

We can now combine the various contributions  and  write the real-emission part of  the  NLO 
cross section. We find\footnote{We note that, since we consider gluonic final states,  $\CalPgen_{aa}$ is the same as  $  \CalPgen_{bb}$. Nevertheless, we find it convenient to distinguish between these two.} 
\index{S!$\rmd \sigmahat^{\rm R}$}
\be
\begin{split}
    2 s \; \rmd \sigmahat^{\rm R} = & ~ [\alpha_s] \lint \left ( \ISoft 
    (\eps)  
    + \IColl(\eps)
    \right )
    \colorprod \FLM \rint 
    +\frac{[\alpha_s]}{\eps} \Big[\llint \CalPgen_{aa} \conv \FLM\rrint + \llint \FLM \conv \CalPgen_{bb} \rrint\Big] 
    \\
    & +  \sum_{i=1}^{\Np} \lint \oS_\Fp \oC_{i \Fp} \, \ww{\Fp i} \Delta^{(\Fp)} \FLM(\Fp) \rint \; ,
    \label{eq:NLO_real_poles0}
\end{split}
\ee
where we introduced the hard-collinear operator 
\index{I!$\IColl$}
\be
    \IColl(\eps) = \sum_{i=1}^{\Np}  \frac{\Gamma_{i,f_i}}{\epsilon} \; ,
\label{eq:IColl_definition0}
\ee
with $f_1 = a$ and $f_2 = b$.

\vspace*{0.3cm}
The infrared poles in Eq.~\eqref{eq:NLO_real_poles0}  cancel against those  in  the virtual contribution and the collinear renormalization of the PDFs,  producing a finite remainder proportional to terms with lower parton multiplicities.
To show this, we note that the infrared poles of the one-loop amplitude $\ampM{1}$ can be written using Catani's formula~\cite{Catani:1998bh} 
\index{Virtual amplitudes}
\be
\begin{split}
    \ampM{1}(1_a,2_b;3, ... \, , \Np;X) = & ~ \amu\; \ICat(\eps) \; \ampM{0}(1_a,2_b;3, ... \, , \Np;X) \\ 
    &+ \ampM{1}^{\rm fin}(1_a,2_b;3, ... \, , \Np;X) \; , 
\end{split}
\label{eq:oneloop0}
\ee
where $\ampM{1}^{\rm fin}$ is the infrared finite one-loop amplitude and 
\index{I!$\ICat$}
\index{V!$\mathcal{V}_i^\text{sing}$}
\index{L!$\lambda_{ij}$}
\be
    \ICat(\eps) = \frac{1}{2}\frac{e^{\eps\gamma_E}}{\Gamma(1-\eps)} \sum_{\inotj}^{\Np} \frac{\mathcal{V}_i^\text{sing}(\eps)}{\ColT{i}^2}    \, (\scprod{\ColT{i}}{\ColT{j}} )\left(\frac{\musq} {2p_i\cdot p_j}\right)^\epsilon e^{i \pi\lambda_{ij} \ep} \; ,
    \quad
    \mathcal{V}_i^\text{sing}(\eps) = \frac{\ColT{i}^2}{\epsilon^2} + \frac{\gamma_i}{\epsilon} \; .
 \label{eq:I1Cat0}
\ee
 The parameters 
$\lambda_{ij}$ in Eq.~\eqref{eq:I1Cat0} are 1 if $i$ and $j$ are both incoming or outgoing partons  and zero otherwise. Therefore, we can write
\index{F!$\FLVfin$}
\be
    2s \; \rmd \sigmahat^V = \lint \FLV \rint = [\alpha_s] \lint \IVirt(\eps) \colorprod \FLM \rint +  \lint \FLVfin \rint \; ,
    \label{eq3.32}
\ee
where 
\index{I!$\IVirt$}
\begin{equation}
 \IVirt(\eps) =  \ICatbar(\ep) + \ICatbar^\dagger(\ep) \; .
    \label{eq:IVirt_defn0}
\end{equation}
In the equation above we have introduced
the operator $\ICatbar$ in place 
of Catani's original operator to factor
out the same  strong coupling $[\alpha_s]$
that  appears in the real-emission contribution.
It is defined by the following equation
\index{I!$\ICatbar$}
\index{Virtual operators!Relations}
\begin{equation}
    \ICatbar(\ep) = \frac{\Gamma(1-\ep)}{e^{\ep \gamma_E}}\, \ICat(\ep)\; ,
\end{equation}
such that 
\be
[\alpha_s] \, \ICatbar(\ep) = \amu \, \ICat(\ep) \; .
\label{eq:ICatbar_defn}
\ee
Furthermore, $\FLVfin$ in Eq.~\eqref{eq3.32} is analogous to $\FLM$ in Eq.~\eqref{eq:defnFLM2} but 
with  
$2\,\myRe \left[ \ampM{1}^{\mathrm{fin}} \;\ampM{0}^* \right] $ 
instead of 
$|\ampM{0}|^2$.

The collinear renormalization of parton distribution functions is standard. The NLO  contribution to the cross section reads
\index{PDFs' collinear renormalization!NLO}
\begin{equation}
    2 s \; \rmd  \sigmahat^{\rm pdf} =
    \frac{\alpha_s(\mu)}{2\pi \ep} \left[\lint \PAP_{aa} \conv \FLM \rint + \lint \FLM \conv \PAP_{bb} \rint \right] \; .
\label{eq3.36a}
\end{equation}

 Finally, combining 
 virtual (see Eq.~\eqref{eq3.32}), real-emission  (see Eq.~\eqref{eq:NLO_real_poles0})
and PDF-renormalization
(see Eq.~\eqref{eq3.36a})
contributions,
we derive  the following \emph{finite} formula
for the NLO cross section
\index{S!$\rmd \sigmahat^{\NLO}$}
\begin{equation}
\begin{split}
   2 s \; \rmd \sigmahat^{\NLO} = & ~ \rmd \sigmahat^{\rm V}+ \rmd \sigmahat^{\rm R}+ \rmd \sigmahat^{\rm pdf}
    =~ \frac{\alpha_s(\mu)}{2\pi}\lint \ITot^{(0)} \colorprod \FLM \rint 
    + \langle\FLV^\text{fin}\rangle 
    \\
   & + \frac{\alpha_s(\mu)}{2\pi} \Big[\llint \PaaNLO \conv \FLM \rrint + \llint \FLM \conv \PbbNLO \rrint\Big]
   + \lint \ONLO \, \Delta^{(\Fp)} \FLM(\Fp) \rint \; ,
\label{Eq:final_expression_NLO}
\end{split}
\end{equation}
where $\ITot^{(0)}$ is the $\order{\ep^0}$ coefficient in the expansion of $\ITot(\ep)$, displayed in Eq.~\eqref{eq3.2}.

A few comments about this result are in order. 
First, as we have anticipated at the beginning of this section,  we have defined an infrared-finite sum\footnote{We show that  this sum is $\ep$-finite in Appendix~\ref{sec:AppNLOdetails}.} of the virtual, soft, and collinear $I$-operators that appears 
in the fully-unresolved part of $\rmd \hat{\sigma}^{\rm NLO}_{ab}$
\index{I!$\ITot$}
\be
    \lint \ITot(\ep)\colorprod \FLM \rint =  \llint \big[\IVirt(\eps) + \ISoft(\eps) + \IColl(\eps)\big] \colorprod  \FLM\rrint = \order{\eps^0} \; .
    \label{Eq:ITot_def}
\ee
As we show in the next section,  \emph{iterations} of this  operator will appear in the result for the NNLO contribution to the cross section; this fact will play 
an important role in proving the cancellation of  poles at NNLO as well. 
Second, we  have denoted  the subtraction operator for the fully-regulated real-emission  contribution as
\index{O!$\ONLO$}
\be
\ONLO = \sum_{i=1}^{\Np} \oS_\Fp \oC_{i \Fp } \, \omega^{\Fp i} \; .
\label{eq:ONLO0}
\ee
Finally we have exploited  the expansion of $\CalPgen_{aa}$
\index{Splitting functions!Expansion}
\index{P!$\CalPgen_{ij}$}
\be
    \CalPgen_{aa}(z,E_i) = - \PAP_{aa}(z) + \ep \, \PNLO_{aa}(z,E_i) + \order{\ep^2} \; ,
\label{Eq:PqqGEN_expansion_order}
\ee
to obtain a manifestly finite quantity once we combine the hard-collinear 
subtraction terms with the PDF-renormalization contributions. The function $\PaaNLO$ is defined in Eq.~\eqref{eq:PNLO_app_final_res}.
When using this function 
it is understood that 
$E_i$ should be set to  $E_1$  
in $\llint\PaaNLO \conv \FLM\rrint$ and to $E_2$ in  $\llint\FLM \conv \PbbNLO\rrint$. 

For the reader's convenience, the definitions introduced in this section are repeated in  Appendix~\ref{sec:Splitting}. A more detailed discussion of the NLO calculation, including expansions of the various functions in powers of $\ep$  and a demonstration of the  cancellation of the $\ep$-poles, is presented in Appendix~\ref{sec:AppNLOdetails}.

\label{sec:nlo}
\section{Calculations at next-to-next-to-leading order}
\label{sec:nnlo}

In this section  we extend the NLO QCD analysis described in the previous section to NNLO. 
At this order of perturbation theory 
we  have to combine the double-virtual, the real-virtual, the double-real and the PDF renormalization contributions to compute the 
differential cross section. Hence, we write
\index{S!$\rmd \sigmahat^{\NNLO}$}
\index{S!$\rmd \sigmahat^{\rm VV}$}
\index{S!$\rmd \sigmahat^{\rm RV}$}
\index{S!$\rmd \sigmahat^{\rm RR}$}
\index{S!$\rmd \sigmahat^{\rm pdf}$}
\be
\label{eq:partoni_sigma_NNLO}
\rmd \sigmahat^{\NNLO} = \rmd \sigmahat^{\rm VV}+ \rmd \sigmahat^{\rm RV} + \rmd \sigmahat^{\rm RR} + \rmd \sigmahat^{\rm pdf} \; .
\ee
Although the NNLO computation is significantly 
more involved  than 
the NLO one,  our aim is to replicate the latter  as much as possible. 
In doing so, we face the following dilemma. 
On the one hand, the double-real
contributions need to 
be split into partitions 
and sectors in order to define the approach to collinear 
singular limits in a unique way. On the other hand, this ``sectoring'' destroys the emergence of  
structures that can be combined in a natural 
way with the double-virtual and real-virtual corrections.  Hence, finding   an optimal  balance between splitting the real-emission contributions into many  well-defined pieces and identifying
proper structures early in the calculation  is the central challenge  to organizing the NNLO  computation  in an efficient way.  We explain 
how we address this challenge 
 in this section. 

\vspace*{0.3cm}
Similar to the NLO case, we distinguish between resolved and potentially
unresolved partons with the help of  the partitions $\Delta^{(i)}$ and $\Delta^{(ij)}$ defined in Appendix~\ref{sect:partitions}. We use 
symmetries of the 
final-state gluons
to define the NNLO contribution to the cross section without the PDFs renormalization  
in the following way
\index{S!$\rmd  \bar{\sigma}^{\rm NNLO}$}
\index{F!$\FLRV$}
\index{F!$\FLVV$}
\begin{equation}
\label{eq:sigma_NNLO}
\begin{split}
    2s \; \rmd  \bar{\sigma}^{\rm NNLO} = & ~ \lint \FLVV(1_a,2_b;3, ... , N_p)\rint
    + \lint \Delta^{(\Fp)} \FLRV(1_a,2_b; 3, ... , N_p, \Fp_g)\rint \\
    & + \frac{1}{2!} \lint \Delta^{(\Fp \Sp)} \FLM(1_a,2_b; 3, ... , N_p, \Fp_g, \Sp_g)\rint \; .
\end{split}
\end{equation}
Here, $\FLVV$ and $\FLRV$ are defined analogously to Eq.~\eqref{eq:defnFLM2}, but using double-virtual and real-virtual matrix elements, while $\Fp$ and $\Sp$ are potentially-unresolved partons. Furthermore,
all the functions $F_{\rm VV}$, $F_{\rm RV}$ and $F_{\rm RR}$ include 
the symmetry factor $1/(N_p-2)!$ arising from the  $N=N_p-2$ identical \emph{resolved} 
gluons in the final state. The dependence of 
the matrix elements and phase spaces
on colorless final-state particles is not shown.

It is convenient to remove the (remaining) symmetry factor $1/2!$ from the double-real  contribution
by introducing the energy ordering of the unresolved gluons $\Fp$ and $\Sp$ 
\index{T!$\THmn$}
\begin{equation}
    \frac{1}{2!} \lint \Delta^{(\Fp \Sp)} \FLM(... , \Fp_g, \Sp_g)\rint = \lint \Delta^{(\Fp \Sp)} \THmn \FLM(..., \Fp_g, \Sp_g)\rint \; ,
\end{equation}
where $\THmn = \Theta(E_\Fp - E_\Sp)$. We obtain
\begin{equation}
\begin{split}
    2s \; \rmd  \bar{\sigma}^{\rm NNLO} = & ~ \lint \FLVV(1_a,2_b;3, ... , \Np)\rint
    + \lint \Delta^{(\Fp)} \FLRV(1_a,2_b; 3, ..., \Np, \Fp_g)\rint \\
    & + \lint \Delta^{(\Fp \Sp)} \THmn \FLM(1_a,2_b; 3, ..., \Np, \Fp_g, \Sp_g)\rint \; .
    \label{eq4.4}
\end{split}
\end{equation}

The above  equation provides the starting point for our calculation. It 
follows  that the NNLO QCD corrections
to the cross section contain contributions that exist in three distinct phase spaces.
These phase spaces  overlap in  configurations where the gluons
labelled as $\Fp$ and $\Sp$ become unresolved.
When this happens, the corresponding amplitudes become singular and 
integrating over unresolved phase
spaces leads  to the appearance of $1/\ep^n$  poles, similar to the NLO case. Our goal is to isolate 
and remove these  singularities locally in the phase space, demonstrate the cancellation of poles between the different 
contributions in Eq.~\eqref{eq4.4},
and determine the finite remainder. 

\vspace*{0.3cm}
We begin by isolating the soft limits of the real-emission  contributions. As already discussed 
in Ref.~\cite{Caola:2017dug}, two soft limits are 
needed: one to describe the double-soft limit 
$E_{\Fp} \sim E_{\Sp} \to 0$, which we denote as 
$S_{\Fp \Sp}$, and one for the single-soft limit 
$E_{\Sp} \to 0$ at fixed $E_{\Fp}$, which we denote 
as $S_{\Sp}$.  We write 
\index{S!$\oS$}
\begin{equation}
\begin{split}
    2s \; \rmd  \bar{\sigma}^{\rm NNLO} = & ~ \lint\FLVV\rint + \lint S_{\Fp \Sp} \Delta^{(\Fp \Sp)} \THmn \FLM(\Fp,\Sp)\rint \\
    & + \lint \Delta^{(\Fp)} \FLRV(\Fp)\rint + \lint \oS_{\Fp \Sp} S_\Sp \, \Delta^{(\Fp \Sp)} \THmn \FLM(\Fp,\Sp)\rint \\
    & + \lint \oS_{\Fp \Sp} \oS_\Sp \, \Delta^{(\Fp \Sp)} \THmn \FLM(\Fp,\Sp)\rint \; ,
\end{split}
\label{eq4.5}
\end{equation}
where the operator $\oS_x = I - S_x$ has already 
been introduced in the context of the NLO QCD computation. Furthermore, when 
writing Eq.~(\ref{eq4.5}),  we have dropped the arguments related to the resolved partons, i.e. 
\index{F!$\FLM(\Fp, \Sp)$}
\begin{equation}
  \FLM(\Fp, \Sp) \equiv  \FLM(1_a, 2_b, 3, ... \, , N_p, \Fp_g, \Sp_g) \; .
\end{equation}

Next, we take the fourth term on the right-hand side 
of Eq.~(\ref{eq4.5}) 
\be
    \lint \oS_{\Fp \Sp} S_\Sp \, \Delta^{(\Fp \Sp)} \THmn \FLM(\Fp,\Sp)\rint \; ,
\ee
make use of the fact that
\be
    \lint \oS_{\Fp \Sp} S_\Sp \, \Delta^{(\Fp \Sp)} \THmn \FLM(\Fp,\Sp)\rint = \lint \oS_\Fp S_\Sp \, \Delta^{(\Fp \Sp)} \THmn \FLM(\Fp,\Sp)\rint \; ,
\ee
and add collinear subtractions for the gluon $\Fp$. We find
\begin{equation}
\begin{split}
    \lint \oS_{\Fp \Sp} S_\Sp \, \Delta^{(\Fp \Sp)} \THmn \FLM(\Fp,\Sp)\rint = & ~ \lint \ONLO \Delta^{(\Fp)} S_\Sp  \THmn \FLM(\Fp,\Sp)\rint \\
    & + \sum_{i=1}^{\Np} \lint \oS_\Fp C_{i\Fp} \, \Delta^{(\Fp)} S_\Sp \THmn \FLM(\Fp,\Sp)\rint \; .
    \label{eq:4.9A}
\end{split}
\end{equation}
We remind the reader that the operator $\ONLO$, defined  in Eq.~\eqref{eq:ONLO0}, subtracts singularities associated with parton $\Fp$, and we have 
used Eq.~\eqref{eq:SoftonDeltaij} to simplify Eq.~\eqref{eq:4.9A}. To obtain a similar structure for the real-virtual contribution, we rewrite $\FLRV$ as 
\begin{equation}
    \lint \Delta^{(\Fp)} \FLRV(\Fp) \rint = \lint S_\Fp \FLRV(\Fp) \rint 
    + \sum_{i=1}^{\Np} \lint \oS_\Fp C_{i\Fp} \Delta^{(\Fp)} \FLRV(\Fp) \rint 
    + \lint \ONLO \, \Delta^{(\Fp)} \FLRV(\Fp) \rint \; . 
    \label{eq4.8}
\end{equation}

\vspace*{0.3cm}
Since  the cancellation of infrared singularities can only occur among  terms with similar kinematics of the hard final-state partons, we would like to write the NNLO QCD cross section  in  such a way that
contributions with the same   number of \emph{resolved} final-state partons are combined. 
At NNLO this number  varies 
between $N$ and $N+2$, so there are three terms that need to be considered. Hence, we aim to 
write the cross section in the following way
\index{S!$\Sigma_N$} 
\index{S!$\Sigma_{N+1}$} 
\index{S!$\Sigma_{N+2}$}
\be
  2 s \;  \rmd   \bar{\sigma}^{\rm NNLO} =  \Sigma_{N} + \Sigma_{N+1} + \Sigma_{N+2} \; .
  \label{eq:cross_sec_resolved_partons}
\ee
Most of the contributions to  the above equation are yet to be determined. However, as a first step, we can use Eq.~\eqref{eq4.5}
and the rearrangement of terms that led to 
Eqs.~(\ref{eq:4.9A}) and (\ref{eq4.8}) to write\footnote{
All the steps that are needed for the rearrangements can be found in Fig.~\ref{fig:graph}.
} 
\be
\label{eq:split_sigma_NNLO}
   2 s \; \rmd  \bar \sigma^{\rm NNLO} = \Sigma_{N}^{(1)} +   
\Sigma^{(1)}_{N+1}  + \SigmaRR \; ,
\ee
where 
\index{S!$\Sigma_N^{(1)}$}
\index{S!$\Sigma_{N+1}^{(1)}$}
\index{S!$\Sigma_{\rm RR}$}
\begin{equation}
\begin{split}
   & \Sigma_{N}^{(1)} =  ~ \lint\FLVV\rint + \lint S_{\Fp \Sp} \THmn \FLM(\Fp,\Sp)\rint + \lint S_\Fp \FLRV(\Fp)\rint \\
    & \;\;\;\;\;\;\;\;\;\;+ \sum_{i=1}^{\Np} \lint \oS_\Fp C_{i\Fp} \Delta^{(\Fp)} \big[\FLRV(\Fp) + S_\Sp \THmn \FLM(\Fp,\Sp)\big] \rint \; , \\
   &  \Sigma_{N+1}^{(1)} =  ~ \lint \ONLO \Delta^{(\Fp)} \big[\FLRV(\Fp) + S_\Sp \THmn \FLM(\Fp,\Sp)\big]\rint \; , \\
   &  \SigmaRR =  ~ \lint \oS_{\Fp \Sp} \oS_\Sp \, \Delta^{(\Fp \Sp)} \THmn \FLM(\Fp,\Sp)\rint \; .
    \label{Eq:Sigma_N_1_and_Sigma_N+1_1_and_Sigma_RR_definitions}
\end{split}
\end{equation}
The quantity $\Sigma_{N}^{(1)}$
is double-unresolved, in the sense that both gluons $\Fp$ and $\Sp$ are either
soft or collinear. The superscript indicates that 
this is the \emph{first} of several contributions  to $\Sigma_N$ that has been identified.
Similarly, the quantity $\Sigma_{N+1}^{(1)}$  is the first single-unresolved term contributing to $\Sigma_{N+1}$ that we identify.
On the contrary,  $\SigmaRR$ is a mix of various contributions as it contains
unregulated collinear singularities.  As we will see, upon extracting 
these singularities,
some parts of $\SigmaRR$  will contribute  to $\Sigma_N$ and $\Sigma_{N+1}$  
and will play an important role in the cancellation 
of infrared poles.

\vspace*{0.3cm}
It is well-known that extracting  all singularities  from the double-real contribution  is a complicated problem as many of them overlap.
To disentangle them, we partition the angular phase space~\cite{Caola:2017dug,Czakon:2010td,Czakon:2011ve,Caola:2019nzf}. Further details are given in Appendices~\ref{appendix:b} and~\ref{sect:collinearseparation}.
Using these  results,  we split $\SigmaRR$ into four distinct
terms. We write 
\index{S!$\Sigma_{N+2}^{\rm fin}$}
\index{S!$\Sigma_N^{(2)}$}
\index{S!$\Sigma_{\rm RR, 1c}$}
\index{S!$\Sigma_{\rm RR, 2c}$}
\begin{equation}
    \SigmaRR = \SigmaFullyRes + \Sigma_{N}^{(2)} + \SigmaRRcc + \SigmaRRc \; ,
    \label{eq4.13}
\end{equation}
where, as we already mentioned, the subscripts of the first two  terms on the right-hand side  indicate the number 
of resolved partons. 
In brief, the first term on the right-hand side in Eq.~\eqref{eq4.13} is fully resolved, the second is the triple-collinear subtraction term, the third is the double-collinear term and the last term is the single-collinear contribution. To elaborate further,  the first term
$\SigmaFullyRes$ is the fully-regulated  contribution given by
\be
   \SigmaFullyRes = \lint \oS_{\Fp \Sp} \oS_{\Sp} \, \Omega_1 \, \Delta^{(\Fp \Sp)} \, \THmn \FLM(\Fp,\Sp) \rint \; , 
   \label{Eq:Sigma_N+2^fin_def}
\ee
where   $\Omega_1$ 
is a function of collinear-subtraction operators and partition functions 
defined in  Eq.~(\ref{eq:Omega1}).\footnote{We note that this contribution will contribute to final states with $N+2$, $N+1$ and $N$ gluons, as a result of the subtracted limits contained in the definition of $\Omega_1$.} The quantity $\SigmaFullyRes$ is the only contribution to 
the NNLO cross section with $N+2$ resolved final-state partons and it can be  implemented in 
a numerical code without further ado.

The second term
$\Sigma_{N}^{(2)}$ is the triple-collinear contribution. It reads 
\begin{equation}
    \Sigma_{N}^{(2)} = \lint \oS_{\Fp \Sp} \oS_{\Sp} \, \Omega_2 \, \Delta^{(\Fp \Sp)} \, \THmn \FLM(\Fp,\Sp) \rint \; , 
   \label{Eq:Sigma_N_2}
\end{equation}
where  $\Omega_2$ is a triple-collinear projection operator that  can be found  in  Eq.~(\ref{eq:Omega2}).
We note that $\Sigma_{N}^{(2)}$ 
was computed in
Ref.~\cite{Delto:2019asp} and can be immediately borrowed from
there. It represents the second contribution to the fully-unresolved term $\Sigma_N$ that we have identified. 

The third term $\SigmaRRcc$ is  the double-collinear contribution where  
gluons are emitted from different legs
\begin{equation}
\begin{split}
    \SigmaRRcc = & ~ \lint \oS_{\Fp \Sp} \oS_{\Sp} \, \Omega_3 \, \Delta^{(\Fp \Sp)} \, \THmn \FLM(\Fp,\Sp) \rint \\
    = & - \sum_{\inotj}^{\Np} \lint \oS_{\Fp \Sp} \oS_{\Sp} C_{j \Sp} C_{i \Fp} [\rmd p_\Fp] [\rmd p_\Sp] \, \omega^{\Fp i, \Sp j} \Delta^{(\Fp \Sp)} \, \THmn \FLM(\Fp,\Sp) \rint \; ,
\end{split}
\label{eq4.16}
\end{equation}
where the angular partition functions $\omega^{\Fp i, \Sp j}$ are defined in Eq.~\eqref{eq:DC_partitions}.
Although this contribution is fairly simple, it is useful to rewrite it before proceeding further. 
According to Eq.~\eqref{eq4.16}
  both collinear operators $C_{i\Fp}$ and $C_{j\Sp}$  act  on the 
  phase space of partons $\Fp$ and $\Sp$. This is necessary to be able to use the results for $\Sigma_N^{(2)}$ from Ref.~\cite{Delto:2019asp}. Eventually, we will have to combine these double-collinear 
  contributions with  collinear limits  
of the single-soft, the real-virtual and other 
terms, where by definition the collinear operators
do not act on the potentially unresolved 
phase spaces.  Hence, it is convenient to rewrite 
Eq.~(\ref{eq4.16})  in the same way, ensuring   
that $C_{i \Fp}$ and $C_{j \Sp}$  
do not act on the phase space of the unresolved 
partons. 
We explain how to do this in Appendix~\ref{sect:pppandcl}. Here, 
we just state  the result and 
write $\SigmaRRcc$ as follows 
\index{S!$\Sigma_N^{(3)}$}
\index{S!$\Sigma_N^{\rm fin, (1)}$}
\begin{equation}
\label{eq:SigmaRR2c_split}
    \SigmaRRcc = \Sigma_{N}^{(3)} + \SigmaRRccfinONE \; ,
\end{equation}
where $\Sigma_{N}^{(3)}$ is the third (divergent) double-unresolved contribution that we have extracted. Likewise,  $\SigmaRRccfinONE$ is the first $\eps$-finite contribution to $\Sigma_N$ that we have encountered. We stress that this is \emph{not} the same as the finite part of $\Sigma_N^{(1)}$ defined previously.
The two terms read
\be
\begin{split}
    \Sigma_{N}^{(3)} = & - \sum_{\inotj}^{\Np} \lint \oS_{\Sp} C_{j \Sp} C_{i \Fp} \Delta^{(\Fp \Sp)} \, \THmn \FLM(\Fp,\Sp) \rint \; ,  \\
\SigmaRRccfinONE = & - \bigg[\left(\frac{\Gamma(1-2\epsilon)}{\Gamma^2(1-\epsilon)}\right)^2 - 1\bigg] \sum_{\inotj}^{\Np} \lint \oS_{\Sp} C_{j \Sp} C_{i \Fp} \, \Delta^{(\Fp \Sp)} \, \THmn \FLM(\Fp,\Sp) \rint \; .
\end{split}
\label{Eq:Sigma_N_3}
\ee
We note  that the unresolved phase space $[\rmd p_\Fp] [\rmd p_\Sp]$  does not appear in the 
above formulas, indicating 
that  collinear operators  do not 
act on it anymore. In addition, we have used $C_{j \Sp} C_{i \Fp} \, \omega^{\Fp i, \Sp j} = 1$ to remove the partition functions.  Furthermore, if the  gluons are emitted off  
 different external legs (which is ensured by the two collinear operators), we have 
 \be
    S_{\Fp \Sp} \oS_\Sp \, [...]= 0 \; ,
    \label{eq:SbarS}
 \ee
 allowing us to write $\oS_{\Fp \Sp} \oS_{\Sp} = \oS_{\Sp}$.
Finally, to see that $\SigmaRRccfinONE$ is 
finite, we observe that 
Eq.~\eqref{Eq:Sigma_N_3} is completely soft-regulated, while the two collinear operators $C_{i \Fp}$ and $C_{j \Sp}$ each produce an  $\order{\ep^{-1}}$ singularity  upon integrating over the phase space of gluons $\Fp$ and $\Sp$. This is compensated by the prefactor $\left( \Gamma(1-2\epsilon)/\Gamma^2(1-\ep)\right)^2 - 1 \sim \order{\ep^2}$, 
leading to an infrared finite quantity. To summarize, we have written $\SigmaRRcc$ as the sum of two double-unresolved contributions, one of which contains poles and one of which is $\ep$-finite. \\

We are left with $\SigmaRRc$, which is the  double-real single-collinear contribution. It reads 
\index{T!$\theta^{(a)}$}
\index{T!$\theta^{(b)}$}
\index{T!$\theta^{(c)}$}
\index{T!$\theta^{(d)}$}
\begin{equation}
\label{eq4.21REF}
\begin{split}
    \SigmaRRc = & ~ \lint \oS_{\Fp \Sp} \oS_{\Sp} \, \Omega_4 \, \Delta^{(\Fp \Sp)} \, \THmn \FLM(\Fp,\Sp) \rint \\
    = & \sum_{\inotj}^{\Np} \llint \oS_{\Fp \Sp} \oS_{\Sp} \Big[C_{i \Fp}[\rmd p_\Fp] + C_{j \Sp}[\rmd p_\Sp] \Big] \, \omega^{\Fp i, \Sp j} \Delta^{(\Fp \Sp)} \, \THmn \FLM(\Fp,\Sp) \rrint  \\
    & + \sum_{i=1}^{\Np} \Big\langle \oS_{\Fp \Sp} \oS_{\Sp} \Big[C_{i \Sp} \theta^{(a)} + C_{\Fp \Sp} \theta^{(b)} + C_{i \Fp} \theta^{(c)} + C_{\Fp \Sp} \theta^{(d)}\Big] \\
    & \times [\rmd p_\Fp] [\rmd p_\Sp] \, \omega^{\Fp i, \Sp i} \Delta^{(\Fp \Sp)} \, \THmn \FLM(\Fp,\Sp) \Big\rangle \; ,
\end{split}
\end{equation}
where the partitions $\omega^{\Fp i,\Sp j}$ and $\omega^{\Fp i,\Sp i}$ can be found in Eq.~\eqref{eq:DC_partitions} and Eq.~\eqref{eq:TC_partitions}, respectively.
the functions $\theta^{(\alpha)}$ with  $\alpha = a,b,c,d$ indicate that 
a particular contribution is confined to a certain 
phase-space \emph{sector}. These sectors, together with the 
corresponding phase-space parameterizations, are defined in Appendices~\ref{sect:collinearseparation} and~\ref{sect:phasespace}, respectively. 
The  challenge therefore is to write $\SigmaRRc$
as a sum of terms with a well-defined number of resolved partons. To do this, we need 
to 
extract the remaining collinear singularities from 
$\SigmaRRc$.\footnote{The label ``single-collinear'' for this contribution refers 
to the one collinear limit appearing in Eq.~\eqref{eq4.21REF}.
Such collinear limit is 
relevant for only one potentially unresolved parton, but the remaining one is still unregulated.
For this reason we need 
to further extract the singularities associated to the second extra emission. This procedure will lead to double-unresolved terms.} We do so in the next section. We do so in the next section.

\subsection{Analyzing single-collinear contributions} \label{Sec:Extracting_singularities_from_Sigma_RR_1c}

The $1/\ep^n$ singularities in  $\SigmaRRc$ simplify if the contributions of   different partitions and sectors are \emph{combined}.  To appreciate why
doing so  is non-trivial,  we need to remind ourselves why partitions and sectors were 
introduced in the first place.  The reason
was to disentangle  overlapping 
singular
limits, making them uniquely defined.  However, it also
complicates the identification of   physical quantities such as e.g.~collinear  anomalous dimensions
and splitting functions. We emphasize that the ability to recognize these universal structures in the early stages of the 
calculation is very useful for canceling the  infrared  divergences in an efficient and 
transparent manner. 
Hence,  our strategy will be   to remove sectors  in a  \emph{controlled}  way, eventually 
getting to the point where
various contributions can be rearranged  into  recognizable universal structures.

As a result of this   analysis we are able to represent  $\SigmaRRc$ by a sum 
of five  divergent
$\Sigma^{(4,\dots,8)}_N$ 
and four finite  $(\Sigma^{{\rm fin}, (2, \dots ,5)}_N)$ double-unresolved quantities, and two divergent 
$\Sigma^{(2,3)}_{N+1}$ and two 
finite $\Sigma^{{\rm fin} (1,2)}_{N+1}$
 single-unresolved quantities, see  Fig.~\ref{fig:graph}.
 These quantities 
 are  used 
in Eq.~(\ref{eq4.65a}) 
and Eq.~(\ref{Eq:Sigma_N+1_div}), 
respectively, 
to construct relevant contributions to the NNLO cross section. 
The remainder of this section describes manipulations of 
$\SigmaRRc$ that lead to such a representation.

\vspace*{0.3cm}
We begin by  separating sectors $\theta^{(b)}$ and $\theta^{(d)}$ from the remaining contributions 
to  $\SigmaRRc$.
We write
\index{S!$\Sigma_{\rm RR,1c}^{(a,c, \rm dc)}$}
\index{S!$\Sigma_{\rm RR,1c}^{(b,d)}$}
\be
\label{eq:SigmaRRc_split}
\SigmaRRc = \SigmaRRcACDC +\SigmaRRcBD \; ,
\ee
where\footnote{The superscript $(a,c,{\rm dc})$ reminds us that $\SigmaRRcACDC$ includes contributions of sectors $a$ and $c$ and of the double-collinear partitions.}
\be
\begin{split}
    \SigmaRRcACDC = & ~ \bigg\langle \oS_{\Fp \Sp} \oS_\Sp \bigg[\sum_{\inotj}^{\Np}(C_{i\Fp} + C_{j\Sp})\omega^{\Fp i,\Sp j} + \sum_{i =1}^{\Np} (C_{i\Sp} \theta^{(a)} + C_{i\Fp} \theta^{(c)})\omega^{\Fp i,\Sp i}\bigg] \\
    & \times [\rmd p_{\Fp}][\rmd p_{\Sp}] \Delta^{(\Fp \Sp)} \THmn \FLM(\Fp,\Sp)\bigg\rangle \; ,
\label{Eq:Sigma_RR1c_a_definition} 
\end{split}
\ee
and
\begin{equation}
    \SigmaRRcBD = \sum_{i =1}^{\Np} \llint \oS_{\Fp \Sp} \oS_\Sp C_{\Fp \Sp}(\theta^{(b)} + \theta^{(d)})[\rmd p_{\Fp}][\rmd p_{\Sp}] \omega^{\Fp i,\Sp i} \Delta^{(\Fp \Sp)} \THmn \FLM(\Fp,\Sp)\rrint \; .
    \label{Eq:Sigma_RR1c_b_definition}
\end{equation}

We first consider   $\SigmaRRcACDC$. In this case Eq.~\eqref{eq:SbarS} holds, so that  $ \oS_{\Fp \Sp} \oS_{\Sp}$ can be replaced by $\oS_{\Sp}$.
We then write 
\be
\begin{split}
    \SigmaRRcACDC = & ~ \bigg\langle \oS_\Sp \bigg[\sum_{\inotj}^{\Np}(C_{i\Fp} + C_{j\Sp})\omega^{\Fp i,\Sp j}  \\
    & + \sum_{i =1}^{\Np} \big(C_{i\Sp} \theta^{(a)} + C_{i\Fp} \theta^{(c)}\big)\omega^{\Fp i,\Sp i}\bigg] [\rmd p_{\Fp}][\rmd p_{\Sp}] \Delta^{(\Fp \Sp)} \THmn \FLM(\Fp,\Sp)\bigg\rangle \; .
\end{split}
\ee
We can simplify this expression  by renaming 
gluons $\Fp$ and  $\Sp$ in such a way that the collinear operators always refer to the gluon $\Fp$.  We also exploit the fact that under such a relabelling sector $\theta^{(a)}$ becomes sector $\theta^{(c)}$, see Eq.~\eqref{eq:theta}. Hence, we obtain 
\be
\begin{split}
    \SigmaRRcACDC = &~ \bigg\langle \mathcal{S}(\Fp,\Sp) \bigg[\sum_{\inotj}^{\Np} C_{i\Fp} \, \omega^{\Fp i,\Sp j}  \\
    & + \sum_{i =1}^{\Np} C_{i\Fp} \theta^{(c)} \omega^{\Fp i,\Sp i}
    \bigg] [\rmd p_{\Fp}][\rmd p_{\Sp}] \Delta^{(\Fp \Sp)} \FLM(\Fp,\Sp)\bigg\rangle \; ,
    \label{Eq:Sigma_RR1c^a}
\end{split}
\ee
where the soft-regulating operator ${\cal S}(\Fp,\Sp)$ reads
\index{S!${\cal S}(\Fp,\Sp)$}
\be
    {\cal S}(\Fp,\Sp)  =  \oS_\Sp \THmn + \oS_\Fp \THnm \; .
\ee
We note that we can rewrite the operator ${\cal S}(\Fp,\Sp)$ in several equivalent 
ways
\be
\begin{split}
   {\cal S}(\Fp,\Sp) = & ~ \oS_\Sp \THmn + \oS_\Fp \THnm  = \iden -  S_\Sp \THmn - S_\Fp \THnm \\
   = & ~ \oS_\Fp \oS_\Sp + S_\Fp \oS_\Sp \THmn  + S_\Sp \oS_\Fp \THnm
   = \oS_\Sp (\iden  - S_\Fp \THnm)  + S_\Sp \oS_\Fp \THnm \; ,
\label{eq:Soft_comb}
\end{split}
  \ee
and we will use the different  representations displayed above in what follows. \\

To simplify $\SigmaRRcACDC$ we need to extract the remaining 
collinear singularities. Since we relabeled gluons so 
that the collinear operators refer to the gluon $\Fp$, the unregulated  
singularities affect gluon $\Sp$ only. However, there is an additional  technical detail that should be highlighted before proceeding. 

As we already mentioned, the many  single-collinear contributions will have to be combined with collinear limits from single-soft, real-virtual and other terms   where  the collinear operators  do not act on the phase space. Therefore, it is useful to rewrite $\SigmaRRcACDC$ in such 
a way that: {\it i}) $C_{i\Fp}$ does not act on the phase space and 
{\it ii}) restrictions imposed by the presence of  sector $\theta^{(c)}$  are lifted. 
We explain how to do this in Appendix~\ref{sect:pppandcl}. Here, we just report the final result,
which is obtained once we insert 
  $\iden = \oC_{i \Fp} + C_{i \Fp}$ 
in the equation for $\SigmaRRcACDC$. We find 
\index{S!$\Sigma_{\rm RR,1c}^{(a,c,{\rm dc}),1}$}
\index{S!$\Sigma_{\rm RR,1c}^{(a,c,{\rm dc}),2}$}
\be
\label{eq:Sigma_ACDC_split}
   \SigmaRRcACDC = 
   \SigmaRRcACDCone
   +
   \SigmaRRcACDCtwo \; ,
\ee
where 
\be
\begin{split}
    \SigmaRRcACDCone = &~ \bigg\langle \mathcal{S}(\Fp,\Sp) \bigg[\sum_{\inotj}^{\Np} \oC_{j\Sp} C_{i\Fp} \, \omega^{\Fp i,\Sp j} \\
    & + \sum_{i =1}^{\Np} (\eta_{i\Sp}/2)^{-\ep} \oC_{i\Sp} C_{i\Fp} \, \omega^{\Fp i,\Sp i}\bigg] \Delta^{(\Fp \Sp)} \FLM(\Fp,\Sp)\bigg\rangle \; ,
\label{eq4.33}
\end{split}
\ee
and
\begin{equation}
\begin{split}
    \SigmaRRcACDCtwo = &~ \frac{\Gamma(1-2\epsilon)}{\Gamma^2(1-\epsilon)} \,  \bigg\langle \mathcal{S}(\Fp,\Sp) \bigg[\sum_{\inotj}^{\Np} C_{j\Sp} C_{i\Fp} \\
    & + \sum_{i =1}^{\Np} (\eta_{i\Sp}/2)^{-\ep} C_{i\Sp} C_{i\Fp} \bigg] \Delta^{(\Fp \Sp)}\FLM(\Fp,\Sp)\bigg\rangle \; .
    \label{eq4.34a}
\end{split}
\end{equation}
As was the case in Eq.~\eqref{Eq:Sigma_N_3},  the phase space $[\rmd p_\Fp] [\rmd p_\Sp]$ does not appear in these formulas anymore, indicating  that  collinear operators there do not  act on it.  We also note that the sector function  $\theta^{(c)}$  disappeared from  Eq.~(\ref{eq4.33}), leaving as a remnant the factor of $(\eta_{i \Sp}/2)^{-\ep}$. Furthermore, Eq.~\eqref{eq4.34a} becomes \emph{potentially ambiguous} because the collinear operators 
$C_{i \Sp}$ and $ C_{i \Fp}$   do not commute in general. Therefore the order 
in which they appear in the above formula (and in  similar formulas) is important.\footnote{For the all-gluonic final states that we consider, these limits will always commute} On the other hand, since the operator $\mathcal{S}(\Fp,\Sp)$ represents a soft subtraction, it commutes with the collinear operators. Finally, we have omitted an overall 
factor $\Gamma(1-2\ep)/\Gamma^2(1-\ep)$ 
in $\SigmaRRcACDCone$
because it would only generate ${\cal O}(\ep)$
terms in the result.

\vspace*{0.3cm}
We will continue  with the discussion of the
 contribution $\SigmaRRcACDCone$.  It is convenient 
to rewrite the factor  $(\eta_{i\Sp}/2)^{-\ep}$ in Eq.~(\ref{eq4.33}) as follows
\begin{equation}
    (\eta_{i\Sp}/2)^{-\ep} = \big[(\eta_{i\Sp}/2)^{-\ep} -1\big] + 1 \; ,
\end{equation}
and combine the second term with the $i \ne j$ sum in that equation.   We find
\index{S!$\Sigma_{\rm RR,1c}^{(a,c,{\rm dc}),1}$}
%
\begin{equation}
\begin{split}
    \SigmaRRcACDCone ={}& \bigg\langle \mathcal{S}(\Fp,\Sp) \bigg[\sum_{i,j = 1}^{\Np} \oC_{j\Sp} C_{i\Fp} \, \omega^{\Fp i,\Sp j}
    \\
    & + \sum_{i =1}^{\Np} \big[(\eta_{i\Sp}/2)^{-\ep} -1\big] \oC_{i\Sp} C_{i\Fp} \, \omega^{\Fp i,\Sp i}\bigg] \Delta^{(\Fp \Sp)} \FLM(\Fp,\Sp)\bigg\rangle \; ,
\end{split}    
\label{Eq:Sigma_RR1c_a_intermediate_step} 
\end{equation}
where we emphasize that the first sum includes terms with $i=j$.
We also note that the comment concerning the non-commutativity of operators $C_{i \Sp}$ and $ C_{i \Fp}$ that we
just made  applies to 
Eq.~\eqref{Eq:Sigma_RR1c_a_intermediate_step} 
as well. 

Another important point is that the second term in  Eq.~\eqref{Eq:Sigma_RR1c_a_intermediate_step} is 
\emph{finite} in the limit
$\ep \to 0$. The reason for this is that the  only singularity 
present in this term comes from the collinear limit
 $i || \Fp$, which gives an $\order{\ep^{-1}}$ contribution once integrated over the phase space of gluon $\Fp$. On the other hand, the presence of $\oC_{i\Sp}$ allows us to expand the difference $\big[(\eta_{i\Sp}/2)^{-\ep} -1\big]$, giving an ${\cal O}(\ep)$ quantity. 

Furthermore, we note that,  in the first term on the  right-hand side 
of Eq.~(\ref{Eq:Sigma_RR1c_a_intermediate_step}), 
the partitioning can be replaced with another, more suitable one. Indeed, since by construction 
\index{Partition functions!Properties}
\begin{equation}
     \sum_{j=1}^{\Np} C_{i\Fp} \, \omega^{\Fp i, \Sp j} \equiv \sum_{j=1}^{\Np} \omega_{i || \Fp}^{\Fp i, \Sp j} \, C_{i\Fp} = C_{i\Fp} \; , \qquad C_{j\Sp} C_{i\Fp} \,  \omega^{\Fp i, \Sp j} = C_{j\Sp} C_{i\Fp}\; , 
\end{equation}
one finds
\begin{equation}
    \sum_{i,j=1}^{\Np} \oC_{j\Sp} C_{i\Fp} \, \omega^{\Fp i,\Sp j}
    = \sum_{i=1}^{\Np} C_{i\Fp} - \sum_{i,j=1}^{\Np} C_{j\Sp} C_{i\Fp} \\
    \equiv \sum_{i,j=1}^{\Np} \oC_{j\Sp} \,  \omega^{\Sp j} \;  C_{i\Fp} \; ,
    \label{Eq:sector_ac_change_partition_functions}
\end{equation}
where $\omega^{\Sp j}$ is, e.g., a NLO partition where the unresolved gluon is $\Sp$.

 Finally, it is convenient to split the soft subtraction operator $\mathcal{S}(\Fp,\Sp)$  acting  on the first term in 
Eq.~\eqref{Eq:Sigma_RR1c_a_intermediate_step} in 
a particular way. 
Employing  the following representation (cf. Eq.~\eqref{eq:Soft_comb}) 
\index{S!$\mathcal{S}(\Fp,\Sp)$}
\be
    \mathcal{S}(\Fp,\Sp) = \oS_\Sp \THmn + \oS_\Fp \THnm = \oS_\Sp (\iden  - S_\Fp \THnm) + S_\Sp \oS_\Fp \THnm \; ,
\ee
we rewrite the formula for $\SigmaRRcACDCone$ in such a way that partonic multiplicities are clearly separated
\index{S!$\Sigma_{N+1}^{(2)}$}
\index{S!$\Sigma_N^{(4)}$}
\index{S!$\Sigma_{N+1}^{\rm fin, (1)}$}
\index{S!$\Sigma_N^{\rm fin, (2)}$}
\be
   \SigmaRRcACDCone   =\Sigma_{N+1}^{(2)} + \Sigma_{N}^{(4)} + 
   \SigmaSingUnrFinONE +\SigmaRRcACDConefin \; .
   \label{eq:4.37A}
\ee
We note that in Eq.~\eqref{eq:4.37A}, the first $\eps$-finite contribution to the single-unresolved cross section is denoted as $\SigmaSingUnrFinONE$. We emphasize again that this does not correspond to the finite part of $\Sigma_{N+1}^{(1)}$. 
The individual contributions read
\be
\begin{split}
    \Sigma_{N+1}^{(2)} = & \sum_{i,j = 1}^{\Np} \llint \oS_\Sp (\iden - S_\Fp \THnm) \oC_{j\Sp} \, \omega^{\Sp j} C_{i\Fp} \Delta^{(\Fp \Sp)} \FLM(\Fp,\Sp) \rrint  \\
    = & \sum_{i = 1}^{\Np} \llint \ONLO (\iden - S_\Sp \THmn) C_{i\Sp} \Delta^{(\Fp \Sp)} \FLM(\Fp,\Sp)\rrint \; ,   \\
    \Sigma_{N}^{(4)}\;\;\; = & \sum_{i,j = 1}^{\Np} \llint S_\Sp \oS_\Fp \oC_{j \Sp} \, \omega^{\Sp j} C_{i \Fp} \Delta^{(\Fp\Sp)} \THnm \FLM(\Fp,\Sp) \rrint \; ,   \\
    %
    \SigmaSingUnrFinONE = & \sum_{i =1}^{\Np} \llint \big[(\eta_{i\Sp}/2)^{-\ep} -1\big] \oS_\Sp (\iden - S_\Fp \THnm)  \oC_{i\Sp} C_{i\Fp} \, \omega^{\Fp i,\Sp i} \Delta^{(\Fp \Sp)} \FLM(\Fp,\Sp)\rrint \\
    = & \sum_{i =1}^{\Np} \llint \ONLO^{(i)} \, \omega_{i \parallel \Fp}^{\Fp i,\Sp i} \, \big[(\eta_{i\Fp}/2)^{-\ep} -1\big] (\iden - S_\Sp \THmn) C_{i\Sp} \Delta^{(\Fp \Sp)} \FLM(\Fp,\Sp)\rrint \; ,\\
    \SigmaRRcACDConefin = & \sum_{i =1}^{\Np} \llint \big[(\eta_{i\Sp}/2)^{-\ep} -1\big] S_\Sp \oS_\Fp  \oC_{i\Sp} C_{i\Fp} \, \omega^{\Fp i,\Sp i} \Delta^{(\Fp \Sp)} \THnm \FLM(\Fp,\Sp)\rrint \; ,
\end{split}
\label{Eq:Sigma_N_5}
\ee
\index{O!$\ONLO^{(i)}$}
where we define $\ONLO^{(i)} = \oS_\Fp \oC_{i\Fp}$ so that $\ONLO = \sum_{i=1}^{\Np} \ONLO^{(i)} \omega^{\Fp i}$. 
We note that when moving  from the first to the second line in $\Sigma_{N+1}^{(2)}$ and $\SigmaSingUnrFinONE$ we have relabelled $\Fp$ to $\Sp$ and \emph{vice versa}.

\vspace{0.3cm}
We now return to $\SigmaRRcACDCtwo$ (see Eq.~\eqref{eq4.34a}) and rewrite it as follows 
\index{S!$\Sigma_N^{(5)}$}
\index{S!$\Sigma_N^{\rm fin, (3)}$}
\begin{equation}
\label{eq:SigmaACDC2_split}
    \SigmaRRcACDCtwo = \Sigma_{N}^{(5)} + \SigmaRRcACDCtwofin \; ,
\end{equation}
where
\be
\label{eq4.48}
\begin{split}
    \Sigma_{N}^{(5)} = & ~ \bigg\langle \mathcal{S}(\Fp,\Sp) \bigg[\sum_{\inotj}^{\Np} C_{j\Sp} C_{i\Fp} + \sum_{i =1}^{\Np} (\eta_{i\Sp}/2)^{-\ep} C_{i\Sp} C_{i\Fp} \bigg] \Delta^{(\Fp \Sp)}\FLM(\Fp,\Sp)\bigg\rangle \; , \\
    \SigmaRRcACDCtwofin = & \left[\frac{\Gamma(1-2\epsilon)}{\Gamma^2(1-\epsilon)} - 1\right] \bigg\langle \mathcal{S}(\Fp,\Sp) \bigg[\sum_{\inotj}^{\Np} C_{j\Sp} C_{i\Fp} + \sum_{i =1}^{\Np} (\eta_{i\Sp}/2)^{-\ep} C_{i\Sp} C_{i\Fp} \bigg]
\\
& \times \Delta^{(\Fp \Sp)}\FLM(\Fp,\Sp)\bigg\rangle \; . 
\end{split}
\ee
Again, we note that $\SigmaRRcACDCtwofin$ is finite because the soft-regulated  collinear limits $C_{j\Sp} C_{i\Fp}$ produce an $\order{\ep^{-2}}$ pole when integrated 
over the angles of $\Fp$ and $\Sp$, 
and the prefactor $\Gamma(1-2\ep)/\Gamma^2(1-\ep)-1$
is ${\cal O}(\ep^2)$. This concludes our discussion of all single-collinear limits, except for those in triple-collinear sectors $(b)$ and $(d)$.

\vspace{0.3cm}
We now turn to $\SigmaRRcBD$, defined in  Eq.~\eqref{Eq:Sigma_RR1c_b_definition}.
We start by mapping sector $\theta^{(d)}$ onto sector $\theta^{(b)}$
by renaming   gluons $\Fp$ to  $\Sp$  and \emph{vice versa} where appropriate.\footnote{We note that this exchange of sectors $b$ and $d$ is only possible at the level of integrated subtraction terms, and is not possible for the fully-regulated term $\Sigma_{N+2}^{\rm fin}$.}   We find
\begin{equation}
\begin{split}
    & \SigmaRRcBD = \sum_{i =1}^{\Np} \llint \oS_{\Fp \Sp} (\oS_\Sp \THmn + \oS_\Fp \THnm) C_{\Fp \Sp} \, \theta^{(b)}[dp_{\Fp}][dp_{\Sp}] \omega^{\Fp i,\Sp i} \Delta^{(\Fp \Sp)} \FLM(\Fp,\Sp)\rrint \\
    & = \sum_{i =1}^{\Np} \llint \oS_{\Fp \Sp} (\iden - S_\Sp \THmn - S_\Fp \THnm) C_{\Fp \Sp} \, \theta^{(b)}[\rmd p_{\Fp}][\rmd p_{\Sp}] \, \omega^{\Fp i,\Sp i} \Delta^{(\Fp \Sp)} \FLM(\Fp,\Sp)\rrint \; . 
\end{split}
\label{eq4.42a}
\end{equation}
Making use of the fact that the action of the collinear 
operator $C_{\Fp\Sp}$ on  the function $\FLM(\Fp,\Sp)$ is  symmetric in $\Fp$ and $\Sp$, 
   we can exchange $\Fp \leftrightarrow \Sp$  in the term 
   with $\THnm$ in 
Eq.~(\ref{eq4.42a}). 
We obtain
%
\begin{equation}
    \SigmaRRcBD = \sum_{i =1}^{\Np} \llint \oS_{\Fp \Sp} (\iden - 2S_\Sp \THmn) C_{\Fp \Sp} \, \theta^{(b)}[\rmd p_{\Fp}][\rmd p_{\Sp}] \, \ww{\Fp i,\Sp i} \Delta^{(\Fp \Sp)} \FLM(\Fp,\Sp)\rrint \; .
    \label{Eq:Sigma_RR_1c_b_def}
\end{equation}

The action of the collinear operator 
$C_{\Fp \Sp}$ on the phase space of two 
unresolved partons leads to a non-trivial result.
To derive it, we consider the specific phase-space 
parametrization 
described in Appendix \ref{sect:phasespace} and find
\index{Phase-space parametrization}
\index{L!$\Lambda$}
\index{C!$C_{\Fp \Sp}$}
\begin{equation}
\begin{split}
    &\; C_{\Fp \Sp} [\rmd \Omega_{\Fp}^{(d-1)}] [\rmd \Omega_{\Sp}^{(d-1)}] \theta^{(b)} 
    \ww{\Fp i,\Sp i}\FLM(\Fp,\Sp)
    = \bigg[\frac{1}{8\pi^2} \frac{(4\pi)^\ep}{\Gamma(1-\ep)}\bigg] N_\epsilon^{(b,d)} 
    \ww{\Fp i,\Sp i}_{\Fp \parallel \Sp}
    \\
    & \times \eta_{i[\Fp \Sp]}^{-\epsilon} (1-\eta_{i[\Fp \Sp]})^\epsilon [\rmd \Omega_{[\Fp \Sp]}^{(d-1)}] \left[\rho_{\Fp \Sp}\frac{\rmd x_4}{x_4^{1+2\epsilon}} \frac{[\rmd \Omega_a^{(d-3)}]}{[\Omega^{(d-3)}]} \rmd \Lambda\right] C_{\Fp \Sp} \FLM(\Fp,\Sp) \; .
\end{split}
\end{equation}
 Here $[\Fp \Sp]$ labels a clustered gluon whose momentum is $p_{[\Fp \Sp]} = p_\Fp + p_\Sp$ calculated \emph{in the strict collinear
 limit} and  the expression for $C_{\Fp \Sp} \FLM(\Fp,\Sp)$ is reported in Eq.~\eqref{eq:NNLO_C45}. From this equation, it follows that $C_{\Fp \Sp} \FLM(\Fp,\Sp) \sim \FLM([\Fp\Sp])$. Since it depends on the kinematics of the clustered parton $[\Fp \Sp]$ only, we can integrate  over $\rmd x_4$, $ \rmd\Omega_a^{(d-3)}$ and $\rmd \Lambda$. We find (see Appendix
 \ref{sec:splin_correlation} for details)
\begin{equation}
\begin{split}
    \SigmaRRcBD = & - \sum_{i=1}^{\Np} \frac{[\alpha_s]}{2\ep} N_\ep^{(b,d)}\, 
    \bigg\langle
    \int \limits_{0}^{\Emax} \frac{\rmd E_\Fp}{E_\Fp^{2\ep-1}} \frac{\rmd E_\Sp}{E_\Sp^{2\ep-1}} \int [ \rmd\Omega_{[\Fp \Sp]}^{(d-1)} ] \; \sigma_{i[\Fp\Sp]}^{-\ep} \, \omega_{\Fp || \Sp}^{\Fp i, \Sp i} \; \\
    & \times
    \oS_{\Fp \Sp} (\iden - 2S_\Sp \THmn) \Delta^{([\Fp \Sp])}
    \frac{1}{E_\Fp E_\Sp} \bigg[\Pgg(z) \FLM([\Fp \Sp])
    \\
   &  + \epsilon \Big[ \Pgg^\perp(z) (r_{i,(b)}^\mu r_{i,(b)}^\nu + g^{\mu\nu}) -  \Pgg^{\perp,r}(z) g^{\mu\nu}\Big]  F_{\text{LM},\mu\nu}([\Fp \Sp])\bigg]
   \bigg\rangle \; .
    \label{Eq:Sigma_RR_1c_b_spin_uncorrelation_plus_spin_correlations_writing}
\end{split}
\end{equation}
In Eq.~\eqref{Eq:Sigma_RR_1c_b_spin_uncorrelation_plus_spin_correlations_writing} we use $z = E_\Fp/(E_\Fp + E_\Sp)$ and $\Pgg^\perp$ and $\Pgg^{\perp,r}$ are splitting functions defined  in Eqs.~\eqref{def:Pgg_bot} and \eqref{def:Pgg_bot_r}, respectively. Furthermore, we have introduced 
\index{S!$\sigma_{ij}$}
\begin{equation}
\sigma_{ij} = \frac{\eta_{ij}}{1 - \eta_{ij}}\; .
\end{equation}

The four-vector  $r_{i,(b)}$ describes spin correlations that arise in the collinear  limit, see Appendix~\ref{sect:pppandcl} for further details. In particular, we note that $r_{i,(b)}$
is \emph{partition-dependent} as indicated by the subscript $i$ (cf. Eq.~\eqref{Eq:r_i_b_definition}). \\

Following the discussion in Ref.~\cite{Caola:2017dug}, it is convenient to split  Eq.~\eqref{Eq:Sigma_RR_1c_b_spin_uncorrelation_plus_spin_correlations_writing} into two terms 
\index{S!$\Sigma_{\rm RR,1c}^{(b,d),\mathrm{sa}}$}
\index{S!$\Sigma_{\rm RR,1c}^{(b,d),\mathrm{sc}}$}
\begin{equation}
\label{eq:SigmaBD_split}
\SigmaRRcBD=\SigmaRRcBDsa + \SigmaRRcBDsc   \; , 
\end{equation}
where the first term on the right-hand side is \emph{spin-averaged}, while the second is  \emph{spin-correlated}.
The spin-averaged contribution depends on  the spin-averaged splitting function $\Pgg$. It provides  the most divergent part of $\SigmaRRcBD$, with its Laurent expansion  starting  at $\order{\ep^{-2}}$.
   The spin-correlated contribution 
   $\SigmaRRcBDsc$ refers to   all terms in Eq.~\eqref{Eq:Sigma_RR_1c_b_spin_uncorrelation_plus_spin_correlations_writing}
   that are proportional to $\FLMmunu([\Fp \Sp])$. 
   Since such terms are multiplied by $\ep$, the spin-correlated 
part is  less divergent than the spin-averaged one; its Laurent expansion  starts at  $\order{\ep^{-1}}$. For this reason, in the following paragraphs we focus on the spin-averaged contribution $\SigmaRRcBDsa$  and 
relegate a detailed discussion of $\SigmaRRcBDsc$ to Appendix \ref{sec:splin_correlation}.

\vspace*{0.3cm}
Our starting point is the following expression 
for the  spin-averaged contribution
\begin{equation}
\begin{split}
    \SigmaRRcBDsa = & - \sum_{i=1}^{\Np} \frac{[\alpha_s]}{2\ep} N_\ep^{(b,d)} \, 
    \bigg\langle
    \int \limits_{0}^{\Emax} \frac{\rmd E_\Fp}{E_\Fp^{2\ep-1}} \frac{\rmd E_\Sp}{E_\Sp^{2\ep-1}} \int [\rmd \Omega_{[\Fp \Sp]}^{(d-1)}] \, \sigma_{i[\Fp\Sp]}^{-\ep} \, \omega_{\Fp || \Sp}^{\Fp i, \Sp i} \; \\
    & \times \oS_{\Fp \Sp} (\iden - 2S_\Sp \THmn) \Delta^{([\Fp \Sp])} \frac{1}{E_\Fp E_\Sp} \Pgg(z) \FLM([\Fp \Sp])\bigg\rangle \; .
    \label{eq:SigmaRR1cbdsa}
\end{split}
\end{equation}
To rewrite it,  
it is convenient to ``undo'' the collinear limit.  We find  
\begin{equation}
    - \frac{[\alpha_s]}{2\ep} N_\ep^{(b,d)} \frac{1}{E_\Fp E_\Sp} \Pgg(z) \FLM([\Fp \Sp]) \equiv \frac{N_{\Fp || \Sp}(\eps)}{2} \int [\rmd \Omega_\Sp^{(d-1)}] \, C_{\Fp \Sp} \FLM(\Fp,\Sp) \; ,
    \label{eq:4.46}
\end{equation}
where 
\index{N!$N_{\Fp || \Sp}$}
\begin{equation}
    N_{\Fp || \Sp}(\eps) = 2^{2\epsilon} \frac{\Gamma(1+2\epsilon) \Gamma(1-2\epsilon)}{\Gamma(1+\epsilon) \Gamma(1-\epsilon)} \; .
\end{equation}
Note that the integration on the right-hand side of Eq.~\eqref{eq:4.46} is performed over  the angular phase space of the unresolved parton $\Sp$ only. 
As a result,  $\SigmaRRcBDsa$ becomes
\begin{equation}
\begin{split}
   \SigmaRRcBDsa = &~ \frac{N_{\Fp || \Sp}(\eps)}{2}  \sum_{i=1}^{\Np}
    \bigg\langle
     \int \limits_{0}^{\Emax} \frac{\rmd E_\Fp}{E_\Fp^{2\ep-1}} \frac{\rmd E_\Sp}{E_\Sp^{2\ep-1}} \int [\rmd \Omega_{[\Fp \Sp]}^{(d-1)}][\rmd \Omega_\Sp^{(d-1)}] \, \sigma_{i[\Fp\Sp]}^{-\ep} \, \omega_{\Fp || \Sp}^{\Fp i, \Sp i} \; \\
    & \times \oS_{\Fp \Sp} (\iden - 2S_\Sp \THmn) \Delta^{([\Fp \Sp])} C_{\Fp \Sp} 
    \FLM(\Fp,\Sp) \bigg\rangle \; .
\end{split}
\end{equation}
Next, we note that the action of $S_{\Fp \Sp}$ on $C_{\Fp \Sp} \FLM(\Fp,\Sp)$ is equivalent to 
the action of a soft operator  $S_{[\Fp\Sp]}$, which refers 
to the zero-energy limit of a  clustered parton $[\Fp \Sp]$. 
We also note that  the joint action of 
$S_\Sp$ and  $S_{\Fp \Sp}$ can also be described as 
$S_{\Fp \Sp} S_\Sp \equiv S_\Fp S_\Sp$, and that the action of $S_{\Sp}$ on the clustered parton $[\Fp \Sp]$ gives $\Fp$.
Following these observations, 
we find
\begin{equation}
\begin{split}
    \SigmaRRcBDsa = &~ \frac{N_{\Fp || \Sp}(\eps)}{2}\sum_{i=1}^{\Np}  
    \bigg[\bigg\langle\int \limits_{0}^{\Emax} \frac{\rmd E_\Fp}{E_\Fp^{2\ep-1}} \frac{\rmd E_\Sp}{E_\Sp^{2\ep-1}} \int [\rmd \Omega_{[\Fp \Sp]}^{(d-1)}][\rmd \Omega_\Sp^{(d-1)}] \, \sigma_{i[\Fp\Sp]}^{-\ep} \, \omega_{\Fp || \Sp}^{\Fp i, \Sp i} \; \\
    & \times \oS_{[\Fp \Sp]} \Delta^{([\Fp \Sp])} C_{\Fp\Sp} \FLM(\Fp,\Sp) \bigg\rangle  \\
    & - 
    \llint 2 \THmn \oS_{\Fp} S_\Sp \; \sigma_{i\Fp}^{-\epsilon} \Delta^{(\Fp)} \omega_{\Fp\parallel \Sp}^{\Fp i,\Sp i} C_{\Fp \Sp} \FLM(\Fp,\Sp) \rrint
    \bigg]
    \; .
\label{Eq:Sigma_RR_1c_b_sa_def}
\end{split}
\end{equation}

\vspace*{0.3cm}
We focus on the first term on the right-hand side in Eq.~\eqref{Eq:Sigma_RR_1c_b_sa_def}.   Thanks 
to the constraints on the energies of $\Fp$ and $\Sp$, 
the energy of the clustered parton  $E_{[\Fp \Sp]}$ may exceed $\Emax$ and go all the way up to $2\Emax$. The two regions for the energy of the clustered particle, namely $E_{[\Fp \Sp]} \in [0,\Emax]$ and $E_{[\Fp \Sp]} \in [\Emax, 2\Emax]$, are very different: the first one is physical whereas the second one is not. 
By this we mean that $
\FLM([\Fp \Sp]) = 0$ for   $E_{[\Fp \Sp]} > \Emax$, since  $\Emax$ is chosen to exceed the maximal energy that a parton can have in a physical process. On the other hand, this unphysical region gives a non-zero contribution in the soft limit because the parton $[\Fp \Sp]$ does not appear in the matrix element.\footnote{We note that this term combines with soft subtractions in other sectors such that the final result is not affected by unphysical contributions. We refer the reader to Ref.~\cite{Caola:2017dug} for a full discussion of this issue.}
Following this discussion, we write $\SigmaRRcBDsa$ as the sum of two terms 
\index{S!$\Sigma_{\rm RR,1c}^{(b,d),{\rm sa},I }$}
\index{S!$\Sigma_{\rm RR,1c}^{(b,d),{\rm sa},II}$}
\be
\SigmaRRcBDsa = 
\SigmaRRcBDsaI
+
\SigmaRRcBDsaII\; .
\label{eq:4.52A}
\ee
The first term $\SigmaRRcBDsaI$ includes the contribution 
where the energy of the clustered particle $[\Fp \Sp]$ does not 
exceed $E_{\rm max}$ as well as the last term on the right-hand side 
of Eq.~(\ref{Eq:Sigma_RR_1c_b_sa_def}), while  $\SigmaRRcBDsaII$ accommodates the  contribution with 
the energy of the clustered particle exceeding  
$ E_{\rm max}$.\\
The term $\SigmaRRcBDsaI$ can be written in the following way
\begin{equation}
\begin{split}
    \SigmaRRcBDsaI = & \frac{N_{\Fp || \Sp}(\eps)}{2} \sum_{i=1}^{\Np} \llint \oS_{\Fp}(\iden - 2 \THmn S_\Sp) \sigma_{i\Fp}^{-\epsilon} \Delta^{(\Fp)} \omega_{\Fp\parallel \Sp}^{\Fp i,\Sp i} C_{\Fp \Sp} \FLM(\Fp,\Sp)\rrint \; ,
    \label{Eq:Sigma_RR_1c_b_sa_1_def}
\end{split}
\end{equation}
where in the first ($\THmn$-independent) term we renamed $[\Fp \Sp] \to [\Fp]$.
The above expression 
contains  divergences which arise 
when gluon $\Fp$ becomes  collinear to parton $i$. We extract these divergences by introducing 
collinear operators and write 
\index{S!$\Sigma_{N+1}^{\rm fin, (2)}$}
\index{S!$\Sigma_N^{(6)}$}
\index{S!$\Sigma_{N+1}^{(3)}$}
\begin{equation}
\label{eq:SigmaBDsaI_split}
   \SigmaRRcBDsaI = \SigmaSingUnrFinTWO + \Sigma_{N}^{(6)} + \Sigma_{N+1}^{(3)} \; ,
\end{equation}
where
\begin{equation}
\begin{split}
    & \SigmaSingUnrFinTWO =  \sum_{i=1}^{\Np} \frac{1}{2} \llint \ONLO^{(i)} \omega_{\Fp\parallel \Sp}^{\Fp i,\Sp i} (\iden - 2 \THmn S_\Sp) \big[N_{\Fp || \Sp}(\eps) \sigma_{i\Fp}^{-\epsilon} - 1\big] \Delta^{(\Fp)} C_{\Fp \Sp} \FLM(\Fp,\Sp)\rrint \; ,\\
   &  \Sigma_{N}^{(6)} =  \sum_{i=1}^{\Np} \frac{N_{\Fp || \Sp}(\eps)}{2} \llint \oS_{\Fp} C_{i\Fp} \sigma_{i\Fp}^{-\ep} \,  (\iden - 2 \THmn S_\Sp) \Delta^{(\Fp)} C_{\Fp \Sp} \FLM(\Fp,\Sp) \rrint \; , \\
   &  \Sigma_{N+1}^{(3)} = \frac{1}{2} \lint \ONLO (\iden - 2 \THmn S_\Sp) \Delta^{(\Fp)} C_{\Fp \Sp} \FLM(\Fp,\Sp)\rint \; .
    \label{Eq:Sigma_N+1_2fin_and_Simga_N_7_and_Sigma_N+1_3_defs}
\end{split}
\end{equation}
The first term in the above formula is finite in the limit $\ep \to 0$, the second term is double-unresolved, and the last one is 
single-unresolved.  We remind the reader that  $\ONLO^{(i)} = \oS_\Fp \oC_{i\Fp}$ and $\ONLO = \sum_{i} \ONLO^{(i)} \omega^{\Fp i}$. 
We also note  that in  $\Sigma_{N+1}^{(3)}$ we  replaced the  NNLO partition functions $\omega_{\Fp || \Sp}^{\Fp i, \Sp i}$ with 
NLO partion functions  $\omega^{\Fp i}$, c.f.~Eq.~\eqref{Eq:sector_ac_change_partition_functions}.

\vspace{0.3cm}
We continue with the 
discussion of  $\SigmaRRcBDsaII$. It can be 
obtained  from Eq.~(\ref{Eq:Sigma_RR_1c_b_sa_def}) 
upon neglecting the last term on the right-hand side 
and restricting the integration over energies 
to the region $E_{[\Fp \Sp]} > E_{\rm max}$. We find 
\begin{equation}
\begin{split}
    \SigmaRRcBDsaII = &~ \frac{N_{\Fp || \Sp}(\ep)}{2} \sum_{i=1}^{\Np} 
    \bigg\langle
    \int \limits_{0}^{\Emax} \frac{\rmd E_\Fp}{E_\Fp^{2\ep-1}} \frac{\rmd E_\Sp}{E_\Sp^{2\ep-1}} \int [\rmd \Omega_{[\Fp \Sp]}^{(d-1)}][\rmd \Omega_\Sp^{(d-1)}] \, \sigma_{i[\Fp\Sp]}^{-\ep} \, \omega_{\Fp || \Sp}^{\Fp i, \Sp i} \; \\
    & \times \; \Theta(E_{\Fp} + E_{\Sp} - E_{\rm max} ) \oS_{[\Fp \Sp]} \Delta^{([\Fp \Sp])} C_{\Fp \Sp} \FLM(\Fp,\Sp) 
    \bigg\rangle \; .
    \label{Eq:Sigma_RR_1c_b_sa_def2}
\end{split}
\end{equation}

We can also replace 
$\oS_{[\Fp \Sp]} $  with $-S_{[\Fp \Sp]}$ in the above equation  as $\FLM([\Fp \Sp])$ has zero support if the energy of the clustered parton  exceeds $\Emax$.  Finally, changing the integration variables 
to $E_{[\Fp \Sp]}=E_{\Fp} + E_{\Sp}$ and $z = E_{\Fp}/(E_{\Fp} + E_{\Sp})$,
computing the collinear $[\Fp \Sp] || \Sp$ limit of $\FLM$ and 
integrating over the angular phase space of 
the gluon $\Sp$, we obtain 
\begin{equation}
\begin{split}
    \SigmaRRcBDsaII = 
    &~ \frac{N_\ep^{(b,d)}}{2\ep} \sum_{i=1}^{\Np} 
    \bigg\langle
    \int \limits_{\Emax}^{2\Emax} \frac{\rmd E_{[\Fp \Sp]}}{E_{[\Fp \Sp]}^{4\ep - 1}} \int \limits_{1-\frac{\Emax}{E_{[\Fp \Sp]}}}^{\frac{\Emax}{E_{[\Fp \Sp]}}} \rmd z \, [z(1-z)]^{-2\ep} \Pgg(z) \\
    & \times \int [\rmd \Omega_{[\Fp \Sp]}^{(d-1)}] \, \sigma_{i[\Fp\Sp]}^{-\ep} \, \omega_{\Fp || \Sp}^{\Fp i, \Sp i} S_{[\Fp \Sp]}  \FLM([\Fp \Sp]) \bigg\rangle \; .
    \label{eq:4.55A}
\end{split}
\end{equation}
Using the standard result for the remaining 
soft limit $S_{[\Fp \Sp]}  \FLM([\Fp \Sp])$ in Eq.~\eqref{eq:4.55A}, we find
\begin{equation}
\begin{split}
    \SigmaRRcBDsaII = & ~ -\frac{[\alpha_s]^2 \, \delta_g^{\rm sa}(\ep) (\Emax/\mu)^{-4\ep}}{\ep} \\
    & \times   \; \sum_{i=1}^{\Np} \sum_{\knotl}^{\Np} \int \frac{[\rmd \Omega_{[\Fp \Sp]}^{(d-1)}]}{[\Omega^{(d-2)}]} \llint \sigma_{i[\Fp\Sp]}^{-\ep} \, \omega_{\Fp || \Sp}^{\Fp i, \Sp i} \frac{\rho_{kl}}{\rho_{k [\Fp \Sp]} \, \rho_{l [\Fp \Sp]}} 
   \left(\scprod{\ColT{k}}{ \ColT{l}} \right)
    \FLM \rrint \; ,
\end{split}
\label{eq4.58a}
\end{equation}
where 
\index{D!$\delta_g^{\rm sa}$}
\begin{equation}
    \delta_g^{\rm sa}(\epsilon) =
    \frac{N_\epsilon^{(b,d)} \Emax^{4\epsilon}}{2} 
    \int \limits_{\Emax}^{2\Emax} \frac{\rmd E_{[\Fp \Sp]}}{E_{[\Fp \Sp]}^{1+4\ep }} \int\limits_{1-\frac{\Emax}{E_{[\Fp \Sp]}}}^{\frac{\Emax}{E_{[\Fp \Sp]}}} \rmd z\, [z(1-z)]^{-2\epsilon} \Pgg(z) \; .
\end{equation}

The integration over the angle of the clustered gluon $[\Fp \Sp]$ in Eq.~(\ref{eq4.58a}) 
is described in 
Appendix~\ref{sec:part_dep}.  The result reads 
\index{S!$\Sigma_N^{\rm sa}$}
\index{S!$\Sigma_N^{\rm sa, fin}$}
\begin{equation}
    \SigmaRRcBDsaII  = \SigmaNSpinAv+\SigmaNSpinAvFin \; ,
\end{equation}
where $\SigmaNSpinAv$ is given by 
\begin{equation}
\begin{split}
    \SigmaNSpinAv = & ~ 2 [\alpha_s]^2 \delta_g^{\rm sa}(\epsilon) \left(\frac{\Emax}{\mu} \right)^{-2\epsilon} \\
    & \times \left[-\lint \ISoft(\epsilon) \colorprod \FLM \rint + \frac{(2\Emax/\mu)^{-2\epsilon}}{2\epsilon^2} N_c(\epsilon) \sum_{i=1}^{N} \T_i^2 \; \lint\FLM\rint \right] \; , 
\end{split}
\end{equation}
with $N_c(\epsilon)$ reported in Eq.~\eqref{eq:normalisation}. The quantity  $\SigmaNSpinAvFin$ is finite and reads
\begin{equation}
    \SigmaNSpinAvFin = [\alpha_s]^2 \, 2^{-2\ep} \, \delta_g^{\rm sa}(\ep) \left(\frac{\Emax}{\mu}\right)^{-4\ep} \sum_{i=1}^{\Np} \lint \Wbdfin{i} \colorprod \FLM \rint \; ,
\end{equation}
where $\Wbdfin{a}$ 
is computed in 
Appendix~\ref{sec:part_dep} with the result given in
Eq.~\eqref{Eq:Wbd_fin_def}.

The final contribution  to consider is the spin-correlated term $\SigmaRRcBDsc$ in Eq.~\eqref{Eq:Sigma_RR_1c_b_spin_uncorrelation_plus_spin_correlations_writing}. In Appendix \ref{sec:splin_correlation}, we show (see Eq.~\eqref{eq:SigmabdscII_AppF}) that among the contributions that  the spin-correlated term of Eq.~\eqref{Eq:Sigma_RR_1c_b_spin_uncorrelation_plus_spin_correlations_writing}  can produce, 
there are two that are \emph{identical} to $\SigmaNSpinAv$ and $\SigmaNSpinAvFin$, provided we substitute $\delta_g^{\rm sa} \mapsto \delta_g^{\perp,r}$, where $\delta_g^{\perp,r}$ is defined in Eq.~\eqref{Eq:delta_g_def}. We call these 
contributions $\SigmaNSpinCorr$ and $\SigmaNSpinCorrFin$. Combining them with $\SigmaNSpinAv$ and $\SigmaNSpinAvFin$, 
respectively, we define  the following quantities 
\index{S!$\Sigma_N^{(7)}$}
\index{S!$\Sigma_N^{\rm sc}$}
\begin{equation}
\begin{split}
    \Sigma_{N}^{(7)} = \SigmaNSpinAv + \SigmaNSpinCorr 
    = & ~ 2 [\alpha_s]^2 \delta_g(\epsilon) \left(\frac{\Emax}{\mu} \right)^{-2\epsilon} \\
    & \times \left[-\lint \ISoft(\epsilon) \colorprod \FLM \rint + \frac{(2\Emax/\mu)^{-2\epsilon}}{2\epsilon^2} N_c(\epsilon) \sum_{i=1}^{N} \T_i^2 \; \lint\FLM\rint \right] \; , 
    \label{eq4.49}
\end{split}
\end{equation}
and
\index{S!$\Sigma_N^{\rm fin, (4)}$}
\index{S!$\Sigma_N^{\rm sc, fin}$}
\begin{equation}
\begin{split}
    \Sigma_{N}^{\rm fin, (4)} = \SigmaNSpinAvFin + \SigmaNSpinCorrFin
    = [\alpha_s]^2 \, 2^{-2\ep} \, 
    \delta_g(\ep) \left(\frac{\Emax}{\mu}\right)^{-4\ep} \sum_{i=1}^{\Np} \lint \Wbdfin{i} \colorprod \FLM \rint \; ,
    \label{eq:SigmafinFour}
\end{split}
\end{equation}
\index{D!$\delta_g$}
\index{D!$\delta_g$}
\index{D!$\delta_g^{\perp, r}$}
with $\delta_g(\ep) = \delta_g^{\rm sa}(\ep) + \delta_g^{\perp, r}(\ep)$, see  Eq.~\eqref{Eq:delta_g_def}.
We denote the remaining spin-correlated terms as 
\index{S!$\Sigma_N^{(8)}$}
\be
\Sigma_N^{(8)} = 
\SigmaRRcBDSCIone \; ,
\label{eq:SigmaEight}
\ee
and 
\index{S!$\Sigma_N^{\rm fin, (5)}$}
\index{W!${\cal W}_r^{(i)}$}
\be
\Sigma_{N}^{\rm fin, (5)} = \asbr^2 \, 
\delta_{g}^{\bot} \, 
\left(\frac{E_{\rm max}}{\mu}\right)^{-4\ep} 
\sum_{i=1}^{\Np}
\lint
{\cal W}_r^{(i)}  \tensprod \FLM \rint \; ,
\label{eq:SigmafinFive}
\ee
where $\SigmaRRcBDSCIone$ is given in Eq.~\eqref{eq.h45}. \\

\clearpage

\begin{sidewaysfigure}
\setlength{\unitlength}{\columnwidth}
\begin{picture}(1,0.50)
  \put(0,0){\includegraphics[width=1.1\linewidth]{graph.pdf}}
  \put(0.335,0.547){ \footnotesize{\ref{eq:split_sigma_NNLO}}}
  \put(0.64,0.48){ \footnotesize{\ref{eq4.13}}}
\put(0.695,0.42){\footnotesize{\ref{eq:SigmaRRc_split}}}
\put(0.515,0.349){\footnotesize{\ref{eq:Sigma_ACDC_split}}}
\put(0.312,0.42){\footnotesize{\ref{eq:SigmaRR2c_split}}}
\put(0.455,0.283){\footnotesize{\ref{eq:4.37A}}}
\put(0.581,0.283){\footnotesize{\ref{eq:SigmaACDC2_split}}}
\put(0.77,0.283){\footnotesize{\ref{eq:4.52A}}}
\put(0.874,0.349){\footnotesize{\ref{eq:SigmaBD_split}}}
\put(0.825,0.22){\footnotesize{\ref{eq:4.52A}}}
\put(0.71,0.22){\footnotesize{\ref{eq:SigmaBDsaI_split}}}
  \end{picture}
  \caption{A chart illustrating  rearrangement of the subtraction terms according to 
  final-state multiplicities. 
   Clickable references to relevant equations  are provided. 
   Terms with same parton multiplicities are shown at the same heights.
    }
\index{Graphical overview of the calculation}
\index{S!$\rmd  \bar{\sigma}^{\rm NNLO}$}
\index{S!$\Sigma_{\rm RR}$}
\index{S!$\Sigma_{\rm RR, 1c}$}
\index{S!$\Sigma_{\rm RR, 2c}$}
\index{S!$\Sigma_{\rm RR,1c}^{(a,c, \rm dc)}$}
\index{S!$\Sigma_{\rm RR,1c}^{(b,d)}$}
\index{S!$\Sigma_{\rm RR,1c}^{(a,c,{\rm dc}),1}$}
\index{S!$\Sigma_{\rm RR,1c}^{(a,c,{\rm dc}),2}$}
\index{S!$\Sigma_{\rm RR,1c}^{(b,d),\mathrm{sa}}$}
\index{S!$\Sigma_{\rm RR,1c}^{(b,d),\mathrm{sc}}$}
\index{S!$\Sigma_{\rm RR,1c}^{(b,d),{\rm sa},I }$}
\index{S!$\Sigma_{\rm RR,1c}^{(b,d),{\rm sa},II}$}
\index{S!$\Sigma_N^{(1)}$}
\index{S!$\Sigma_N^{(2)}$}
\index{S!$\Sigma_N^{(3)}$}
\index{S!$\Sigma_N^{(4)}$}
\index{S!$\Sigma_N^{(5)}$}
\index{S!$\Sigma_N^{(6)}$}
\index{S!$\Sigma_N^{(8)}$}
\index{S!$\Sigma_{N+1}^{(1)}$}
\index{S!$\Sigma_{N+1}^{(2)}$}
\index{S!$\Sigma_{N+1}^{(3)}$}
\index{S!$\Sigma_{N+1}^{\rm fin, (1)}$}
\index{S!$\Sigma_{N+1}^{\rm fin, (2)}$}
\index{S!$\Sigma_N^{\rm fin, (1)}$}
\index{S!$\Sigma_N^{\rm fin, (2)}$}
\index{S!$\Sigma_N^{\rm fin, (3)}$}
\index{S!$\Sigma_N^{\rm fin, (5)}$}
\index{S!$\Sigma_N^{\rm sa}$}
\index{S!$\Sigma_N^{\rm sa, fin}$}
\index{S!$\Sigma_N^{\rm sc}$}
\index{S!$\Sigma_N^{\rm sc, fin}$}
\index{S!$\Sigma_{N+1}^{\rm sp}$}
\index{S!$\Sigma_{N+2}^{\rm fin}$}
  \label{fig:graph}
\end{sidewaysfigure}

\clearpage

To recapitulate, we have succeeded in writing  $\SigmaRRc$ as a sum of  contributions to the single- and double-unresolved terms $\Sigma_{N+1}$ and $\Sigma_N$. We can combine them with the corresponding contributions of $\SigmaRRcc$ as well as those of Eq.~\eqref{Eq:Sigma_N_1_and_Sigma_N+1_1_and_Sigma_RR_definitions}, and explore  the cancellation of the $\ep$-poles in $\Sigma_{N+1}$ and $\Sigma_N$. We study    such cancellations in Section~\ref{sect5new} but before diving  into this discussion we need to rearrange double-unresolved terms to make the investigation  of the pole cancellation easier.   We discuss a suitable rearrangement  in the next subsections. 

\subsection{Rearranging double-unresolved terms}
\label{sec:double_unr_terms}

We now turn our attention to the question of how the double-unresolved terms can be rearranged. Once this is accomplished, the preparatory
work will be complete and the cancellation of singularities between the different contributions can be explored.  

We have seen that the contributions with two unresolved partons 
can be written as a sum of eight divergent   and five finite terms, i.e. 
%
\be
\Sigma_N = \sum \limits_{i=1}^{8} \Sigma_N^{(i)} + \sum \limits_{i=1}^{5} \Sigma_{N}^{{\rm fin}, (i)} \; .
\label{eq4.65a}
\ee
The contributions 
are in  Eqs~(\ref{Eq:Sigma_N_1_and_Sigma_N+1_1_and_Sigma_RR_definitions},~\ref{Eq:Sigma_N_2},~\ref{Eq:Sigma_N_3},~\ref{Eq:Sigma_N_5},~\ref{eq4.48},~\ref{Eq:Sigma_N+1_2fin_and_Simga_N_7_and_Sigma_N+1_3_defs},~\ref{eq4.49},~\ref{eq:SigmafinFour},~\ref{eq:SigmaEight},~\ref{eq:SigmafinFive}). Three of the divergent contributions, namely $\Sigma_N^{(3,4,5)}$,
contain various collinear limits and we find that combining and rearranging them is helpful for understanding the cancellation of poles.

To make the required manipulations  more transparent, in $\Sigma_N^{(4)}$  we write $\oC_{j \Sp}$ as $(\iden-C_{j \Sp})$, 
use the fact that
$\sum \limits_{j=1} \omega^{j\Sp} = 1$ and separate the $i \ne j$ and $i=j$ sums. We find
\begin{equation}
\begin{split}
    & \Sigma_{N}^{(3)} + \Sigma_{N}^{(4)} + \Sigma_{N}^{(5)} 
    \\
    = &
    -\sum_{\inotj}^{\Np} \lint \left ( \oS_{\Sp} C_{j \Sp} C_{i \Fp} 
    \THmn
    - (\oS_\Sp \THmn + \oS_\Fp \THnm) C_{j \Sp} C_{i \Fp} 
    \right ) 
    \Delta^{(\Fp \Sp)} \FLM(\Fp,\Sp) \rint \\
    & + \sum_{i = 1}^{\Np} \lint S_\Sp \oS_\Fp C_{i \Fp} \Delta^{(\Fp\Sp)} \THnm \FLM(\Fp,\Sp) \rint - \sum_{\inotj}^{\Np} \lint S_\Sp \oS_\Fp C_{j \Sp} C_{i \Fp} \Delta^{(\Fp\Sp)} \THnm \FLM(\Fp,\Sp) \rint \\
    &
    - \sum_{i = 1}^{\Np} \lint \left ( S_\Sp \oS_\Fp C_{i \Sp} C_{i \Fp}  \THnm 
    - (\eta_{i\Sp}/2)^{-\ep} \mathcal{S}(\Fp,\Sp) 
    C_{i\Sp} C_{i\Fp} 
    \right ) 
    \Delta^{(\Fp \Sp)} \FLM(\Fp,\Sp)\rint \; .
\end{split}
\end{equation}
Combining terms with  $i \ne j$ sums in the above equation, we obtain  
\begin{equation}
\begin{split}
    & \Sigma_{N}^{(3)} + \Sigma_{N}^{(4)} + \Sigma_{N}^{(5)} = \\
    = & \sum_{i = 1}^{\Np} \lint S_\Sp \oS_\Fp C_{i \Fp} \Delta^{(\Fp\Sp)} \THnm \FLM(\Fp,\Sp) \rint + \sum_{\inotj}^{\Np} \lint \oS_\Sp \oS_\Fp C_{j \Sp} C_{i \Fp} \Delta^{(\Fp\Sp)} \THnm \FLM(\Fp,\Sp) \rint \\
    & - \sum_{i = 1}^{\Np} \lint \left (  S_\Sp \oS_\Fp C_{i \Sp} C_{i \Fp} \THmn  - (\eta_{i\Sp}/2)^{-\ep} \mathcal{S}(\Fp,\Sp) \, C_{i\Sp} C_{i\Fp} \right ) \Delta^{(\Fp \Sp)} \FLM(\Fp,\Sp)\rint \; . 
    \label{Eq:Sigma_N_3+4+5_intermediate_step}
\end{split}
\end{equation}
We can further simplify the above equation  if we rewrite the $i \ne j$ sum as follows 
\begin{equation}
\begin{split}
   &  \sum_{\inotj}^{\Np} \lint \oS_\Sp \oS_\Fp C_{j \Sp} C_{i \Fp} \Delta^{(\Fp\Sp)} \THnm \FLM(\Fp,\Sp) \rint  \\
   = &  
    ~ \frac{1}{2} \sum_{i,j=1}^{\Np} \lint \oS_\Sp \oS_\Fp C_{j \Sp} C_{i \Fp} \Delta^{(\Fp\Sp)} \FLM(\Fp,\Sp) \rint 
       - \frac{1}{2} \sum_{i=1}^{\Np} \lint \oS_\Sp \oS_\Fp C_{i \Sp} C_{i \Fp} \Delta^{(\Fp\Sp)} \FLM(\Fp,\Sp) \rint \; . 
\end{split}
\end{equation}
We now take the last term on  the right-hand side of the above equation and combine it with the next-to-last term in Eq.~\eqref{Eq:Sigma_N_3+4+5_intermediate_step}.
We find
\begin{equation}
    - \frac{1}{2} \sum_{i=1}^{\Np} \lint (\oS_\Sp \oS_\Fp + 2 S_\Sp \oS_\Fp \THnm) \, C_{i \Sp} C_{i \Fp} \Delta^{(\Fp\Sp)} \FLM(\Fp,\Sp) \rint \;
    .\label{eq4.77a}
\end{equation}
We split the second  term under the sum sign in 
Eq.~(\ref{eq4.77a}) into two identical ones, and change $\Fp \leftrightarrow  \Sp$ in one of them. We obtain
\begin{equation}
\begin{split}
    & - \frac{1}{2} \sum_{i=1}^{\Np} \lint (\oS_\Sp \oS_\Fp + S_\Sp \oS_\Fp \THnm + S_\Fp \oS_\Sp \THmn) \, C_{i \Sp} C_{i \Fp} \Delta^{(\Fp\Sp)} \FLM(\Fp,\Sp) \rint \\
    & + \frac{1}{2} \sum_{i=1}^{\Np} \lint S_\Fp \oS_\Sp \THmn \, [C_{i \Sp}, C_{i \Fp}] \Delta^{(\Fp\Sp)} \FLM(\Fp,\Sp) \rint \\
    = & - \frac{1}{2} \sum_{i=1}^{\Np} \lint \left (\mathcal{S}(\Fp,\Sp) \, C_{i \Sp} C_{i \Fp} - S_\Fp \oS_\Sp \THmn \, [C_{i \Sp}, C_{i \Fp}] 
    \right )\Delta^{(\Fp\Sp)} \FLM(\Fp,\Sp)\rint \; . 
\end{split}
\end{equation}
Putting everything together, we find 
\begin{equation}
\begin{split}
    & \Sigma_{N}^{(3)} + \Sigma_{N}^{(4)} + \Sigma_{N}^{(5)} = \\
    = & \sum_{i = 1}^{\Np} \lint S_\Sp \oS_\Fp C_{i \Fp} \Delta^{(\Fp\Sp)} \THnm \FLM(\Fp,\Sp) \rint + \frac{1}{2} \sum_{i,j=1}^{\Np} \lint \oS_\Sp \oS_\Fp C_{j \Sp} C_{i \Fp} \Delta^{(\Fp\Sp)} \FLM(\Fp,\Sp) \rint  \\
    & + \frac{1}{2} \sum_{i=1}^{\Np} \lint \big[2(\eta_{i\Sp}/2)^{-\ep} -1\big] \mathcal{S}(\Fp,\Sp) \, C_{i \Sp} C_{i \Fp} \Delta^{(\Fp\Sp)} \FLM(\Fp,\Sp)\rint \\
    & + \frac{1}{2} \sum_{i=1}^{\Np} \lint S_\Fp \oS_\Sp \THmn \, [C_{i \Sp}, C_{i \Fp}] \Delta^{(\Fp\Sp)} \FLM(\Fp,\Sp)\rint \; .
\end{split}
\end{equation}

We can now combine this result with the remaining double-unresolved contributions. We find
%
\begin{align}
    &   \Sigma_{N} =  ~  \lint\FLVV\rint + \lint S_{\Fp \Sp} \THmn \FLM(\Fp,\Sp)\rint + \lint S_\Fp \FLRV(\Fp)\rint  \allowdisplaybreaks
    \nonumber 
    \\
    & + \sum_{i=1}^{\Np} \lint \oS_\Fp C_{i\Fp} \Delta^{(\Fp)} \big[\FLRV(\Fp) + S_\Sp \THmn \FLM(\Fp,\Sp)\big] \rint \;  \allowdisplaybreaks
    \nonumber 
    \\
    & + \sum_{i = 1}^{\Np} \lint S_\Sp \oS_\Fp C_{i \Fp} \Delta^{(\Fp\Sp)} \THnm \FLM(\Fp,\Sp) \rint
    + \frac{1}{2} \sum_{i,j=1}^{\Np} \lint \oS_\Sp \oS_\Fp C_{j \Sp} C_{i \Fp} \Delta^{(\Fp\Sp)} \FLM(\Fp,\Sp) \rint   \allowdisplaybreaks
    \nonumber 
    \\
    & + \frac{1}{2} \sum_{i=1}^{\Np} \lint \big[2(\eta_{i\Sp}/2)^{-\ep} -1\big] \mathcal{S}(\Fp,\Sp) \, C_{i \Sp} C_{i \Fp} \Delta^{(\Fp\Sp)} \FLM(\Fp,\Sp)\rint \allowdisplaybreaks \label{eq:NNLO_double_unresolved_unsimplified}\\
    & + \frac{1}{2} \sum_{i=1}^{\Np} \lint S_\Fp \oS_\Sp \THmn \, [C_{i \Sp}, C_{i \Fp}] \Delta^{(\Fp\Sp)} \FLM(\Fp,\Sp)\rint  \allowdisplaybreaks
    \nonumber 
    \\
    & - 2  [\alpha_s]^2 \; \delta_g(\epsilon) \left ( \frac{ \Emax}{\mu }  \right )^{-2\epsilon} \Bigg [\lint \ISoft(\epsilon) \colorprod \FLM \rint  
    - \frac{(2\Emax/\mu)^{-2\epsilon}}{2\epsilon^2} N_c(\epsilon) \sum_{i=1}^{\Np} \T_i^2 \; \lint\FLM\rint \Bigg ]  \allowdisplaybreaks
    \nonumber 
    \\ 
    & + \sum_{i=1}^{\Np} \frac{N_{\Fp || \Sp}(\ep)}{2} \lint \oS_{\Fp} C_{i\Fp} \, \sigma_{i\Fp}^{-\ep} \;(\iden - 2 \THmn S_\Sp) \Delta^{(\Fp)} C_{\Fp \Sp} \FLM(\Fp,\Sp)  \rint \; 
    \nonumber 
    \\
    & +  \Sigma_{N}^{(2)} +  \Sigma_{N}^{(8)} + \sum\limits_{i=1}^5 \Sigma_N^{{\rm fin},(i)} \; . 
    \nonumber 
\end{align}

It is clear from the above formula 
that $\Sigma_N$ contains a large number of terms of different physical origin that exhibit infrared and collinear singularities, which will cancel  when combined with the PDFs renormalization contributions.
To simplify the  discussion of how this happens,   we will identify groups of terms 
which exhibit shared features. These features  include quartic, triple and  quadratic correlations of color-charge operators, which  originate from exchanges of soft real and virtual gluons,  as well as  double- and single-boosted kinematics that are generated by hard-collinear initial-state
emissions.  We will focus on these different categories in turn, since the cancellation of $\ep$-poles has to occur independently for each of them.

In Subsection~\ref{sec:Soft_and_Virtual}
we describe some manipulations of the virtual and soft contributions 
to Eq.~(\ref{eq:NNLO_double_unresolved_unsimplified}), which  
set the stage for the discussion of the cancellation 
of poles in color-correlated contributions 
that can be found in 
Subsections~\ref{sec:trip_color_corr} 
and \ref{sec:color_correlations}. 
With color-correlated infrared singularities out of the way, we are left with terms  that are proportional to squares of color charges of the resolved partons, 
which include both boosted and unboosted contributions.  Such terms primarily come from 
collinear emissions.   We discuss such  contributions and the cancellation of the corresponding singularities in Subsections~\ref{sec:Collinear}
and~\ref{sect:collinearcombine}.

\subsection{Simplifying virtual and soft corrections}
\label{sec:Soft_and_Virtual}
In this subsection we focus on the  color-correlated contributions to the fully-unresolved quantity $\Sigma_N$. To this end, we will examine those terms in Eq.~\eqref{eq:NNLO_double_unresolved_unsimplified} that contain soft limits and/or loop amplitudes. 
Similar to what will be done in Section~\ref{SubSec:X_fin}, we will write the results in terms of generalizations of the operators $\ISoft$, $\IVirt$ and $\IColl$, with an eye on combining these into manifestly-finite $\ITot$ structures. Furthermore, we will observe the appearance of terms involving {triple} correlators of color 
charges, which we will discuss separately in Section~\ref{sec:trip_color_corr}.

\vspace*{0.3cm}
We begin by considering the double-virtual contribution  $ \langle \FLVV \rangle $ to  Eq.~\eqref{eq:NNLO_double_unresolved_unsimplified}.
We write the loop expansion of the amplitude 
of the 
$1_a+2_b\to X+ N \, g$ process 
to $\order{\alpha_s^2}$ with respect to the LO as
\index{Color space and algebra}
\be
    \myket{\ampM{}}_c = \myket{\ampM{0}}_c + \left[\amu\right] \myket{\ampM{1}}_c + \left[\amu\right]^2 \myket{\ampM{2}}_c + \order{\alpha_s^3} \; .
\ee
The double-virtual contribution to the cross section is obtained by squaring the amplitude and retaining the $\order{\alpha_s^2}$ terms. The result   reads\footnote{We drop the subscript ``c'' in the notation for the color vector of a  matrix element.}
\index{Virtual amplitudes}
\be
 \mybraket{\ampM{}}{\ampM{}}_{\alpha_s^2} = \mybraket{\ampM{0}}{\ampM{2}} +\mybraket{\ampM{2}}{\ampM{0}}+ \mybraket{\ampM{1}}{\ampM{1}} \; .
\ee
Following Ref.~\cite{Catani:1998bh}, we extract the infrared poles of $\myket{\ampM{1}}$ 
and $\myket{\ampM{2}}$
and write them as 
\index{M!$\ket{\ampM{1}}$}
\index{M!$\ket{\ampM{2}}$}
\index{M!$\ket{\ampM{2}^{\rm fin}}$}
\be
\begin{split}
    \ket{\ampM{1}} = ~ & \ICat(\ep)\ket{\ampM{0}} + \ket{\ampM{1}^{\rm fin}}\; , \\
    \ket{\ampM{2}} = ~ & \ICat(\ep)\ket{\ampM{1}} + I_2(\ep) \ket{\amp_0} + \ket{\ampM{2}^{\rm fin}} \; ,
\end{split}
\end{equation}
where $\ket{\ampM{1}^{\rm fin}}$ and $\ket{\ampM{2}^{\rm fin}}$ are infrared-finite. The operator $\ICat(\ep)$ was introduced in the 
context of the NLO calculation  and is given in  Eq.~\eqref{eq:I1Cat0}. The operator 
$\ICatTwo(\ep)$ reads
\index{I!$\ICatTwo$}
\begin{equation}
    \ICatTwo(\ep) = - \frac{1}{2} \ICat(\ep) \left(\ICat(\ep) + \frac{2\beta_0}{\ep}\right) + c_\ep \left(\frac{\beta_0}{\ep} + K\right) \ICat(2\ep) + \mathcal{H}_2 \; ,
    \label{Eq:I2_definition_maintext}
\end{equation}
with\footnote{We remind the reader that in this paper we are accounting for gluonic final states only. For this reason $\nf$ should be set to $0$, and $\beta_0$ to $11/6\, \Ca$.}
\index{K!$K$}
\index{C!$c_\ep$}
\begin{equation}
    K = \left( \frac{67}{18} - \frac{\pi^2}{6} \right) C_A  - \frac{10}{9} \TR \nf \; , \qquad c_\ep = \frac{e^{-\ep \gamma_E} \Gamma(1-2\ep)}{\Gamma(1-\ep)} \; .
\end{equation}
The operator $\mathcal{H}_2$ contains $\order{\ep^{-1}}$ poles only. We split this function into a term containing triple color correlations and a color-diagonal term
\index{H!$\mathcal{H}_2$}
\index{H!$\Htwotc$}
\index{H!$\Htwocd$}
\be
\label{eq:H2_decomp}
\mathcal{H}_2(\eps) = \Htwotc(\ep) + \Htwocd(\ep) \; .
\ee
The two quantities $\Htwotc$ and $\Htwocd$ were explicitly  computed in Refs.~\cite{Becher:2009qa, Becher:2013vva}. The triple color-correlated term  $\Htwotc$ is given in Eq.~\eqref{eq:5.27A}.  The color-diagonal piece reads 
\begin{equation}
 \label{eq:H2cd_def_becher}
    \Htwocd(\ep) = \frac{1}{2\eps}\sum_{i=1}^{N_p} H_{f_i}
 \; ,
\end{equation}
where $f_i$ denotes the flavor of parton $i$. Explicitly one has
\begin{equation}
\begin{split}
    H_g = C_A^2 \left(\frac{5}{12}+\frac{11}{144}\pi^2+\frac{\zeta_3}{2}\right) +C_A n_f \left(-\frac{29}{27}-\frac{\pi^2}{72}\right) + \frac{C_F n_f}{2}+\frac{5}{27}n_f^2 \; ,
 \label{eq:H2cd_def_g}
\end{split}
\end{equation}
and
\begin{equation}
    H_q = C_F^2 \left(\frac{\pi^2}{2}-6\zeta_3-\frac{3}{8}\right) + C_A C_F \left(\frac{245}{216}-\frac{23}{48}\pi^2+\frac{13}{2}\zeta_3\right) +C_F n_f\left(\frac{\pi^2}{24}-\frac{25}{108}\right) \; .
\label{eq:H2cd_def_q}
\end{equation}

The matrix element  squared 
that appears  in the double-virtual term 
$\FLVV$ is then
\begin{equation}
\begin{split}
      & \mybraket{\ampM{}}{\ampM{}}_{\alpha_s^2}
     =  \mybraketOp*{\ampM{0}}{\frac{1}{2}\ICat^2(\ep) + \frac{1}{2} \big(\ICat^\dagger(\ep)\big)^2 + \ICat^\dagger(\ep) \ICat(\ep) + \left(\mathcal{H}_2 + \mathcal{H}_2^\dagger\right)}{\ampM{0}} \\
    {}& 
    + \mybraketOp*{\ampM{0}}{- \frac{\beta_0}{\ep} \Big(\ICat(\ep) + \ICat^\dagger(\ep)\Big) + c_\ep \left(\frac{\beta_0}{\ep} + K\right)\Big(\ICat(2\ep) + \ICat^\dagger(2\ep)\Big)} {\ampM{0}} \\
    {}& 
    + 2\myRe{\left[\mybraketOp{\ampM{0}}{\ICat(\ep) + \ICat^\dagger(\ep)}{\ampM{1}^{\rm fin}}\right]} + 2\myRe{\big[\mybraket{\ampM{0}}{\ampM{2}^{\rm fin}}\big]} + \mybraket{\ampM{1}^{\rm fin}}{\amp_1^{\rm fin}}  \; .
    \label{Eq:double_virtual_me}
\end{split}
\end{equation}
The one-loop operators $\ICat$ in the second and third lines appear as the sum of $I_1$ and $I_1^\dag$; for 
this reason, 
they can immediately be written using the function $\IVirt$ defined in Eq.~\eqref{eq:IVirt_defn0}. 
However, this does not happen automatically 
for entries in  the first line in Eq.~\eqref{Eq:double_virtual_me}. 
To force the appearance of $\IVirt$, we write
\be
\begin{split}
\frac{1}{2}\ICatbar^2(\ep) + \frac{1}{2} \big(\ICatbar^\dagger(\ep)\big)^2 + \ICatbar^\dagger(\ep) \ICatbar(\ep) =  
\frac{1}{2}\IVirt^2(\eps)
- \frac{1}{2} \left[\ICatbar, \ICatbar^\dagger\right] \; .
\label{eq4.100}
\end{split}
\ee
As we will see, in the general case 
the commutator in the 
above equation contains triple color-correlated poles.
We will study them in detail in Section~\ref{sec:trip_color_corr}. 
For now,  we use 
Eq.~(\ref{eq4.100})
and write the double-virtual contribution as follows
\index{F!$\FLVV$}
\index{F!$\FLVfinsq$}
\index{F!$\FLVVfin$}
\begin{equation}
\begin{split}
    \lint \FLVV \rint & = 
     [\alpha_s]^2 
     \llint
     \left[\frac{1}{2} \IVirt^2(\eps) -
     \frac{\Gamma(1-\ep)}{e^{\ep \EulerGamma}} \left (
     \frac{\beta_0}{\epsilon}  \IVirt(\epsilon) -  \left(\frac{\beta_0}{\epsilon} + K\right) \IVirt(2\epsilon) 
     \right )
     \right] \tensprod \FLM \rrint \\
     & + \asbr^2 
     \llint \left[ -\frac{1}{2}\left[\ICatbar(\ep), \ICatbar^\dagger(\ep)\right] + \HtwotcOL +\HtwotcOL ^\dag 
     + \HtwocdOL +\HtwocdOL^\dag 
          \right] \colorprod \FLM  \rrint 
       \\
     &
     + \left[\alpha_s\right] \lint  \IVirt(\eps) \tensprod \FLVfin \rint
     + \lint \FLVfinsq \rint 
     + \lint \FLVVfin \rint \; .
\end{split}
\label{eq:DoubVirt_exp}
\end{equation}
In Eq.~(\ref{eq:DoubVirt_exp}) 
 $\FLVfinsq$ and $\FLVVfin$ contain the finite remainders of the one-loop squared and two-loop amplitudes interfered with the tree level, respectively. Furthermore, we have made use of the fact that $\mathcal{H}_2 \sim \order{\ep^{-1}}$ to replace the coupling $\alpha_s(\mu)/(2\pi)$ with $\asbr$ in front of it. This concludes our discussion of the double-virtual contribution, and we will make use of Eq.~\eqref{eq:DoubVirt_exp} in Section~\ref{sect5new} to discuss the cancellation of poles.
 \\

  Next, we  consider the double-soft term $\lint  \DoubSoft{\Fp \Sp } \THmn  \FLM \rint $ in Eq.~\eqref{eq:NNLO_double_unresolved_unsimplified}. As was mentioned earlier, it was computed in Ref.~\cite{Caola:2018pxp} for an arbitrary opening angle between the hard radiators. We can write the result in terms of a double color-correlated  
  and a quartic color-correlated 
  component
\index{Soft limit}
\index{S!$\lint \DoubSoft{\Fp \Sp} \THmn \FLM(\Fp,\Sp)\rint_{T^2}$}
\index{S!$\lint \DoubSoft{\Fp \Sp}\THmn \FLM(\Fp,\Sp)\rint_{T^4}$}
   \be
    \lint \DoubSoft{\Fp \Sp} \THmn \FLM(\Fp,\Sp)\rint = \lint \DoubSoft{\Fp \Sp} \THmn \FLM(\Fp,\Sp)\rint_{T^2} + \lint \DoubSoft{\Fp \Sp}\THmn \FLM(\Fp,\Sp)\rint_{T^4} \; .
\ee
The quartic color-correlated component has a simple
(factorized) form 
\index{I!$\ISoft^2$}
\be
\begin{split}
    \llint \DoubSoft{\Fp \Sp} \THmn \FLM(\Fp,\Sp)\rrint_{T^4} 
    &=  
    2 \gsb^4 \sum_{
    (i j), (kl)}^{\Np}
    \Big\langle
    \int [\rmd p_{\Fp}] [\rmd p_{\Sp}] \Theta(E_{\Fp} - E_{\Sp}) 
    S_{ij}(p_{\Fp})  
    S_{kl}(p_{\Sp}) \\
    & \;\;\; \times
    \{\scprod{\ColT{i}}{ \ColT{j}}, \scprod{\ColT{k}}{ \ColT{l}}\} \cdot \FLM
    \Big\rangle
    \\ 
    & =  \asbr^2 \frac{1}{2}
    \llint
    \ISoft^2(\ep)  \cdot \FLM
    \rrint \; .
    \end{split}
    \label{eq:double_soft_T4}
\ee
In the above, we have introduced the short-hand notation $S_{ij}(p_{\Fp})$ for the eikonal function
\index{S!$S_{ij}$}
\be
S_{ij}(p_\Fp) = \frac{p_i  \cdot p_j}{2(p_i \cdot  p_\Fp)(p_j \cdot  p_\Fp)} \; .
\label{eq:defn_eikonal}
\ee 
The (double) color-correlated term appears to be 
significantly more  complex   \cite{Caola:2018pxp}.
However, upon careful inspection, we find 
that its \emph{poles} 
can be written in a reasonably simple manner. 
We obtain
\index{S!$\widetilde{S}_{ij}$}
\index{C!$c_1$}
\index{C!$c_2$}
\index{C!$c_3$}
\index{I!$\ISofttilde$}
\be
\begin{split}
    & \lint\DoubSoft{\Fp \Sp} \THmn \FLM(\Fp,\Sp)\rint_{T^2} \\
    = &~ \gsb^4      \sum_{i<j}^{\Np}
    \int [\rmd p_{\Fp}] [\rmd p_{\Sp}] \Theta(E_{\Fp} - E_{\Sp})  \lint \widetilde{S}_{ij}(p_{\Fp},p_{  \Sp}) \,   \left(\scprod{\ColT{i}}{ \ColT{j}} \right)\cdot 
    \FLM
    \rint \\
    = &~  
    \asbr^2 \left[\frac{\Ca}{\ep^2} c_1(\ep) + \frac{\beta_0}{\ep}  c_2(\ep)
    +
    {\beta_0} \,c_3(\ep)
    \right] 
    \lint 
   \ISofttilde(2\epsilon) \cdot \FLM
    \rint + \big\langle 
     \DoubSoft{\Fp \Sp} \THmn \FLM(\Fp,\Sp)\big\rangle_{T^2}^\text{fin} \; ,
\end{split}
\label{eq4.105}
\ee
where $\widetilde{S}_{ij}$ is the double-soft current defined in Ref.~\cite{Catani:1999ss}. We note that the last term in Eq.~\eqref{eq4.105} is $\ep$-finite and can be found in Eq.~\eqref{eq_SmnFin}. Furthermore,  the quantities $c_{1,2,3}$ are polynomials in $\ep$ and are given in Eq.~\eqref{eq:constants_SS}.
Additionally, we have introduced
\be
\ISofttilde(2\epsilon) = - \frac{(2\Emax/\mu)^{-4\epsilon}}{(2\epsilon)^2} \sum_{\substack{i,j=1 \\i \not= j}}^{\Np}  \eta_{ij}^{-2\epsilon} 
\widetilde{K}_{ij}(\eps) \, \left(\scprod{\ColT{i}}{ \ColT{j}} \right) \; ,
\ee
where
\index{K!$\widetilde{K}_{ij}$}
\be
\label{eq:Ktilde_def}
\widetilde{K}_{ij}(\eps) 
    =
    \frac{\Gamma^2(1-2\epsilon)}{\Gamma(1-4\epsilon)} \,  \eta_{ij}^{1+3\epsilon} {}_2F_{1}(1+\epsilon, 1+\epsilon, 1-\epsilon, 1-\eta_{ij}) \; .
\ee

We note apparent similarities between $\ISofttilde$ 
and $\widetilde K_{ij}$ and $\ISoft$ and $K_{ij}$ defined in Eqs.~\eqref{eq:NLO_soft0} and~\eqref{eq:Kij_defn0}. In fact, one can use the following property  of the hypergeometric functions
\begin{equation}
    \hypF(a,b,c,z) = (1-z)^{c-a-b} \hypF(c-a,c-b,c,z) \; ,
\end{equation}
to show that 
\begin{align}
\label{eq:Ktilde_vs_K}
\widetilde{K}_{ij}(\ep)
={}&
{K}_{ij}(2\ep) \, 
\frac{{}_2F_{1}(-2\ep, -2\ep; 1-\ep, 1-\eta_{ij})}{{}_{2}F_{1}(-2\ep,-2\ep,1-2\ep,1-\eta_{ij})}
=
{K}_{ij}(2\ep) 
+
\mathcal{O}(\ep^3) \; .
\end{align}
It follows that 
\be
\ISofttilde(2\ep) = \ISoft(2\ep) + \order{\eps} \; .
\label{eq:ISofttilde_and_ISoft}
\ee
This relation will be very helpful for  demonstrating   the cancellation of poles in color-correlated 
terms. Following this discussion, 
we write the double-soft term as
\index{S!$\SmnFin$}
\be
\begin{split}
    & \lint \DoubSoft{\Fp \Sp} 
    \THmn \FLM(\Fp,\Sp)\rint
    \\
    = &~ \asbr^2 
    \bigg\langle    
    \bigg[ \frac{1}{2}\ISoft^2(\ep)    
    +\left(\frac{\Ca}{\ep^2} c_1(\ep) + \frac{\beta_0}{\ep}  c_2(\ep)
    +
    {\beta_0} \,c_3(\ep)
    \right) \ISofttilde(2\eps) \bigg] \cdot \FLM 
    \bigg\rangle 
    \\
    & + \SmnFin \; .
    \end{split}
\label{eq:DoubSoft_exp}
    \ee
This concludes our discussion of the double-soft limits.\\

We now move on to the third term on the right-hand side of Eq.~\eqref{eq:NNLO_double_unresolved_unsimplified}, which involves the soft limit of the real-virtual
contribution. This limit reads~\cite{Catani:2000pi, Bern:1999ry}
\index{Soft limit}
\be
\begin{split}
    & \Soft{\Fp} \, \FLRV({\Fp}) \\
    = & - \gsb^2  \, \sum_{\inotj}^{\Np}
    \bigg\{2\, S_{ij}(p_{\Fp}) \, \left(\scprod{\ColT{i}}{ \ColT{j}} \right)\colorprod \FLV - \amu\,  \frac{\beta_0}{\ep} \,2\, S_{ij}(p_{\Fp}) \, 
    \left(\scprod{\ColT{i}}{ \ColT{j}} \right)\colorprod \FLM \\
    & - 2\, \frac{\asbr}{\ep^2} \, 
    \Ca \,
    A_K(\ep) \, 
     \Big( S_{ij}(p_{\Fp}) \Big)^{1+\ep} \, 
    \left(\scprod{\ColT{i}}{ \ColT{j}} \right) \colorprod \FLM
     \\
&
    - \asbr \, 
 \frac{4\pi \,\Gamma(1+\ep) \Gamma^3(1-\ep)}{\ep \, \Gamma(1-2\ep)}  \sum_{\substack{k=1 \\ k \neq i,j}}^{\Np} \kappa_{ij} \, 
 S_{ki}(p_{\Fp}) \,
\Big(
S_{ij}(p_{\Fp})\Big)^\ep f_{abc} \,  T_k^a \, T_i^b \, T_j ^c \,  \FLM
 \bigg\} \; 
 ,
 \label{eq:SoftLim_RV}
\end{split}
\ee
\index{K!$\kappa_{ij}$}
where $\kappa_{ij} \equiv \big( \lambda_{ij}-\lambda_{i \Fp}-\lambda_{j \Fp}\big)=+1$ when both $i$ and $j$ are incoming momenta
and $\kappa_{ij}=-1$ otherwise. We point out that $\kappa_{ij}$ is 
 symmetric under the exchange $i \leftrightarrow j$. Moreover, we have introduced 
the constant (cf. Eq.~\eqref{eq:constants_RV})
\index{A!$A_K$}
\be
A_K(\ep) = \frac{\Gamma^3(1+\ep) \, \Gamma^5(1-\ep)}{\Gamma(1+2\ep) \, \Gamma^2(1-2\ep)} = 1 + \order{\eps^2} \; .
\ee

The terms in Eq.~\eqref{eq:SoftLim_RV} that include $S_{ij}(p_\Fp)$ can be integrated over the unresolved phase space along the same lines as the soft subtraction term at NLO (see Eq.~\eqref{eq:NLO_soft0}), giving rise to the operator $\ISoft$.   The  term  with $F_{\rm LV}$  
in Eq.~\eqref{eq:SoftLim_RV} can be further  simplified 
using  Catani's formula (Eq.~\eqref{eq:oneloop0}) to extract divergences from  the loop amplitude. 
However, care is needed since the operators $I_1$ and $\ISoft$ do not commute in general. Hence, upon  integrating the first term on the right-hand side 
of Eq.~\eqref{eq:SoftLim_RV}
over the phase space of gluon $\Fp$, we find the following expression for the 
combination of divergent loop and soft-emission 
contributions 
\be
[\alpha_s]^2  \llint\left[\scprod{\ISoft(\ep)}{\ICatbar(\ep)}+ \scprod{\ICatbar^\dagger(\ep)}{ \ISoft(\ep)} \right] \colorprod \FLM \rrint \; .
\ee
We can rewrite the above quantity 
using the identity
\be   
\ISoft \ICatbar + \ICatbar^\dagger \ISoft =  \frac{1}{2}\left( \Big(\ICatbar + \ICatbar^\dagger \Big) \ISoft + \ISoft \Big(\ICatbar + \ICatbar^\dagger\Big) + \left[\ISoft \, , \, \ICatbar - \ICatbar^\dagger\right] \right) \; ,
\ee
where the first and second terms can be expressed through  $\IVirt$ and $\ISoft$, and the third term contains triple color correlations  and will be discussed in  detail in Section~\ref{sec:trip_color_corr}.

The integration  of the third term on the right-hand side of Eq.~\eqref{eq:SoftLim_RV}, which includes the factor $\left(S_{ij}(p_\Fp)\right)^{1+\eps}$,  leads to 
\be
\begin{split}
   - 
       2 \gsb^2 \, 
       {}&
       \sum_{\inotj}^{\Np}
        \llint
       \big( S_{ij}(p_{\Fp}) \big)^{1+\ep}
        \, 
\left(\scprod{\ColT{i}}{ \ColT{j}} \right) \colorprod \FLM 
       \rrint 
       \\
       ={} &
       -
       \frac{\asbr}{4\ep^2} \,  
 \left(\frac{2E_{\rm max}}{\mu}\right)^{-4\ep} 
 \sum_{\inotj}^{\Np} \llint   
 \eta_{ij}^{-2\ep}   \widetilde{K}_{ij}(\eps)  \, 
\left(\scprod{\ColT{i}}{ \ColT{j}} \right) \colorprod \FLM
 \rrint \\
  ={}& [\alpha_s] 
  \llint \ISofttilde(2\eps) \cdot \FLM \rrint \; .
  \end{split}
\ee

The last  term  on the right-hand side of Eq.~\eqref{eq:SoftLim_RV} contains explicit triple color correlators. Integrating this term over the phase space of gluon $\Fp$ is non-trivial and is discussed at length in Appendix \ref{sec:trip_color_corr_RV}.
In what follows we will refer to 
it  as the triple color-correlated real-virtual subtraction term, $\ItriRV$. Putting everything together, we find that the soft limit
of the real-virtual correction can be written in the 
following way 
\be
\begin{split}
 \lint \Soft{\Fp} \, \FLRV(\Fp) \rint
=&~
 \asbr^2  \llint \frac{1}{2} \Big[ \scprod{\ISoft(\eps)}{\IVirt(\eps)}+\scprod{\IVirt(\eps)}{\ISoft(\eps)}  \Big] \colorprod \FLM \rrint \\ 
 & + \asbr \llint \ISoft(\eps) \colorprod \FLVfin \rrint
  - 
\asbr^2 \frac{\Gamma(1-\eps)}{e^{\eps \gamma_E}}  
\frac{\beta_0}{\ep} \,  
\Big\langle
\ISoft(\eps)
\FLM
\Big\rangle \\
& -
\frac{\asbr^2}{\ep^2} \, \Ca \,
A_K(\ep) \,
\llint
 \ISofttilde(2\eps) \colorprod
\FLM
\rrint
\\
&
+
\asbr^2 \,
\llint
\left( \frac{1}{2}\left[\ISoft(\eps) \, , \, \ICatbar(\eps) - \ICatbar^\dagger(\eps)\right] + \ItriRV(\eps)\right) \colorprod \FLM
\rrint \; .
\end{split}
\label{eq:SoftRV_exp}
\ee

We have now analyzed all terms with quartic and triple-color correlators. These arose due to soft limits of real emission amplitudes and virtual corrections; because of that, they are associated with unboosted kinematics. We have also found a number of terms with double-color correlations. Further terms of this kind emerge when a soft or virtual operator appears in conjunction with a collinear limit, and such terms can also lead to unboosted kinematic configurations. Our next goal 
 is to identify such contributions 
 in Eq.~\eqref{eq:NNLO_double_unresolved_unsimplified}.\\

We begin with the term that describes 
the hard-collinear limits of the real-virtual amplitude 
squared $ \lint \oS_\Fp \Coll{i\Fp} \omega^{\Fp i} \Delta^{(\Fp)} \FLRV (\Fp)\rint$.
 These limits were studied in Refs.~\cite{Bern:1999ry,Kosower:1999rx}. They 
 involve both  the tree-level splitting function $P_{ii}$ as well as the the one-loop splitting function $P^{\rm 1L}_{ii}$, 
 whose explicit form can be found in Appendix~\ref{sec:Splitting}. Even though $P_{ii}^{1 \rm L}$ is more complicated than the corresponding tree-level splitting function, the  integration over unresolved 
phase space of the gluon $\Fp$ 
proceeds in exactly the same 
way as in the NLO computation. 

Similar to the NLO case, it is useful to distinguish 
between the initial-state and the final-state splittings. 
When the unresolved parton $\Fp$ becomes collinear to a final-state parton $i$ we find 
\begin{equation}
\begin{split}
    & \lint \oS_\Fp \Coll{i\Fp} \omega^{\Fp i} \Delta^{(\Fp)} \FLRV (\Fp)\rint
    =  [\alpha_s]^2 \llint \frac{\Gamma_{i,g}}{\epsilon} \IVirt(\eps)  \tensprod \FLM\rrint \\
    & - [\alpha_s]^2 \frac{\beta_0}{\epsilon} \frac{\Gamma(1-\eps)}{e^{\eps \EulerGamma}} \llint \frac{\Gamma_{i,g}}{\epsilon} \FLM \rrint - \frac{[\alpha_s]^2}{\epsilon^2} C_A \hc(\epsilon)  \llint \frac{\GammaLoop{i,g}}{2\epsilon} \FLM\rrint + [\alpha_s]  \llint\frac{\Gamma_{i,g}}{\epsilon} \FLVfin\rrint \; ,
\end{split}
\end{equation}
where 
\index{H!$\hc$}
\be
      \hc(\eps) = \frac{\Gamma^2(1-2\eps) \Gamma(1+\eps)}{\Gamma(1-3\eps)} = 1 + \order{\eps^3} \; .
\ee
Furthermore, the one-loop generalized anomalous dimension for the final-state splitting  reads
\index{G!$\GammaLoop{i,\fl{i}}$}
\index{G!$\gamma_{z,g \to gg}^{33,\mathrm{1L}}$}
\begin{equation}
    \GammaLoop{i,g} = - \left[\left(\frac{2E_i}{\mu}\right)^{-2\epsilon} \frac{\Gamma^2(1-\epsilon)}{\Gamma(1-2\epsilon)}\right]^{2} \frac{\epsilon^2 \cos(\pi\epsilon)}{C_A} \gamma_{z,g \to gg}^{33,\mathrm{1L}}(\epsilon,L_i) \, , \hspace{1cm} i = 3 , ...,  \Np  \; ,
\end{equation}
where $\gamma_{z,g \to gg}^{33,\mathrm{1L}}$ is defined analogously to Eq.~\eqref{eq:defn_gamma_nk0}, but with  the splitting function 
$P^{\rm 1L}_{gg}$ instead of  $P_{gg}$.
The $\ep$-expansion of the one-loop generalized anomalous dimension reads
\be
\GammaLoop{i,g} = \gamma_i + 2\ColT{i}^2 L_i + \order{\eps} \, , \hspace{1cm} i = 3, ...,  \Np \; .
\ee

We continue with the case where 
the unresolved parton $\Fp$ becomes  collinear to an initial state parton, say $1_a$. In this case we find 
\begin{equation}
\begin{split}
      \lint \oS_\Fp \Coll{1\Fp} \omega^{\Fp 1} \Delta^{(\Fp)} \FLRV (\Fp)\rint
    = &~   [\alpha_s]^2  \llint \frac{\Gamma_{1,\fl{1}}}{\epsilon} \IVirt(\eps)  \tensprod \FLM\rrint + [\alpha_s] \llint \frac{\Gamma_{1,\fl{1}}}{\epsilon}  \FLVfin\rrint \\
    &+  \frac{[\alpha_s]^2}{\epsilon} \lint \CalPgen_{aa} \conv   \left( \IVirt(\eps) \colorprod \FLM  \right) \rint + \frac{[\alpha_s]}{\epsilon} \lint \CalPgen_{aa} \conv   \FLVfin \rint \\
    &  - [\alpha_s]^2 \frac{\Gamma(1-\eps)}{e^{\eps \gamma_E}}  \frac{\beta_0}{\epsilon} \left[ \llint \frac{\Gamma_{1,\fl{1}}}{\epsilon}\FLM\rrint + \frac{1}{\epsilon} \lint \CalPgen_{aa} \conv  \FLM\rangle\right] \\
    &  - \frac{[\alpha_s]^2}{\epsilon^2} C_A \hc(\epsilon) \bigg\langle \CalPgen_{aa}  \otimes \bigg(\frac{\GammaLoop{1,\fl{1}}(\eps)}{2\epsilon} \FLM
    \bigg)\bigg\rangle \\ 
    &- \frac{[\alpha_s]^2}{2\epsilon^3} C_A \hc(\epsilon) \lint\CalPoneLgen_{aa} \conv \FLM\rint \; ,
    \label{eq:RV_hard_coll}
\end{split}
\end{equation}
where the one-loop initial-state generalized anomalous dimension is 
\index{G!$\GammaLoop{i,\fl{i}}$}
\index{P!$\CalPoneLgen_{ij}$}
\index{Generalized anomalous dimension!Expansion}
\be
\begin{split}
   \GammaLoop{1,\fl{1}}(\eps) =  & \left[\left(\frac{2E_1}{\mu}\right)^{-2\epsilon} \frac{\Gamma^2(1-\epsilon)}{\Gamma(1-2\epsilon)}\right]^{2} \left[\gamma_{\fl{1}} + 2 \ColT{\fl{1}}^2 \frac{1- e^{-4 \epsilon L_1}}{4} \pi \cot(\pi \epsilon)\right]\\ 
   =& ~ \gamma_{\fl{1}} + 2\ColT{\fl{1}}^2 L_1 + \order{\eps} \; ,
   \end{split}
   \ee
and we have also introduced a generalized splitting function at one-loop 
\index{Splitting functions!Expansion}
\begin{equation}
    \CalPoneLgen_{aa}(z,E_1) = \left[\left(\frac{2E_1}{\mu}\right)^{-2\epsilon} \frac{\Gamma^2(1-\epsilon)}{\Gamma(1-2\epsilon)}\right]^{2} \Big[-\PAP_{aa}(z) + \eps \hatP_{aa}^{1\text{L,fin}}(z)\Big] \; .
\label{eq:CalPoneLgen}
\end{equation}
We observe that the one-loop generalized anomalous dimension $\GammaLoop{i,g}$ coincides with its tree-level counterpart  $\Gamma_{i,g}$ at $\order{\ep^0}$, cf.~Eq.~\eqref{eq:gamma_expansion_fsis}. Similarly, the one-loop and tree-level generalized splitting functions $\CalPoneLgen_{aa}$ and $\CalPgen_{aa}$ have the same expansion at $\order{\ep^0}$. 
Further details concerning these one-loop generalized anomalous dimensions and splitting functions can be found in Appendix~\ref{sec:Splitting}. Finally, we note that in Eq.~\eqref{eq:RV_hard_coll} some terms involve the convolution of a splitting function with the product of  $\IVirt$ or the anomalous dimensions and $\FLM$. In these cases, the relevant energy in $\IVirt$ or  $\GammaLoop{1,\fl{1}}$ is also multiplied by a factor of $z$.

Summing the initial and final state collinear limits we find
\index{Soft limit}
\be
\begin{split}
 \sum \limits_{i=1}^{\Np} \lint \oS_{\Fp}  \Coll{i \Fp } \omega^{ \Fp i} \Delta^{(\Fp)} &  F_{\rm RV}  \rint =   [\alpha_s]^2 \llint \IColl(\eps) \IVirt(\eps)  \tensprod \FLM\rrint + [\alpha_s] \lint \IColl(\eps) \tensprod \FLVfin\rint \\
    & +\frac{[\alpha_s]^2}{\epsilon} \lint \CalPgen_{aa} \otimes  \left( \IVirt(\eps) \colorprod\FLM \right) \rint + \frac{[\alpha_s]}{\epsilon} \lint \CalPgen_{aa} \otimes  \FLVfin \rint  \\
    &+\frac{[\alpha_s]^2}{\epsilon} \llint  \left( \IVirt(\eps) \colorprod\FLM \right) \otimes \CalPgen_{bb} \rrint + \frac{[\alpha_s]}{\epsilon} \lint   \FLVfin \otimes \CalPgen_{bb} \rint \\
     &   - [\alpha_s]^2  \frac{\beta_0}{\epsilon}\frac{\Gamma(1-\eps)}{e^{\ep \EulerGamma}} \bigg[ \frac{1}{\eps} \llint \CalPgen_{aa} \otimes  \FLM\rrint +  \frac{1}{\eps} \llint \FLM \otimes \CalPgen_{bb} \rrint  \\
     &+ \llint \IColl(\epsilon) \tensprod \FLM\rrint \bigg] - \frac{[\alpha_s]^2}{\epsilon^2} C_A \hc(\epsilon) \lint \IColltilde(2\epsilon) \colorprod \FLM\rint \\
     & - \frac{[\alpha_s]^2}{2\epsilon^3} C_A \hc(\epsilon) \lint \CalPoneLgen_{aa} \otimes \FLM + \FLM \otimes \CalPoneLgen_{bb} \rint \; ,
     \label{eq:CollRV}
\end{split}
\ee
with
\index{I!$\IColltilde$}
\begin{equation}
    \IColltilde(2\epsilon)= \sum_{i=1}^{\Np} \frac{\GammaLoop{i,\fl{i}}(\eps)}{2\epsilon} \; .
\end{equation}
 We point out that the relation between the one-loop and tree-level hard-collinear operators 
 \index{Collinear operators!Relations}
 \begin{equation}
\IColltilde(2\epsilon) \equiv \IColl(2\epsilon) + \order{\ep} \; ,
    \label{Eq:relation_between_IColltilde_and_IColl_main}
\end{equation}
  is analogous to that of the soft operators, see  Eq.~\eqref{eq:ISofttilde_and_ISoft}.
\\

We now consider the fifth and sixth terms in Eq.~\eqref{eq:NNLO_double_unresolved_unsimplified} 
\be
\lint \oS_{\Fp}  \Coll{i \Fp}  \Delta^{(\Fp)}  
\big(
\Soft{\Sp} \THmn F_{\rm LM} (\Fp, \Sp)
\big)
\rint 
+ \lint \Soft{\Sp} 
\big(
\oS_{\Fp}        \Coll{i \Fp}  \;  \Delta^{(\Fp )}  \THnm  F_{\rm LM}(\Fp, \Sp)
\big)
\rint \; ,
\label{eq:fifth_and_sixth_terms}
\ee
where we have used $\Soft{\Sp} \Delta^{(\Fp \Sp )} = \Delta^{(\Fp)}$.
At first glance, it may seem that the two  terms 
in Eq.~(\ref{eq:fifth_and_sixth_terms})
can be trivially combined, since the first  contains an energy-ordering theta-function which
enforces $E_{\Fp} > E_{\Sp}$, while the second requires $E_{\Sp} > E_{\Fp}$. However, one should be careful about the order  in which 
the various operators act on $\FLM$. In the first term, one should compute the soft limit $\Soft{\Sp}$ of $\FLM$ first, then integrate over the unresolved phase space of $\Sp$, and then compute the hard-collinear limit $\oS_{\Fp} \Coll{i \Fp}$ and integrate over the phase space of $\Fp$. In the second term, the hard-collinear limit $\oS_{\Fp} \Coll{i \Fp}$ is evaluated first, followed by 
the integration over the phase space of $\Fp$.  Then we take the  soft limit $\Soft{\Sp}$ and integrate  over the phase space of $\Sp$. We emphasize that these operations do not commute. Indeed, one can show by explicit calculation that the following holds
true
\be
\begin{split}
    & \lint 
    \Soft\Sp
    \big(\oS_\Fp \Coll{i\Fp}  \Delta^{(\Fp)}  \THnm \FLM(\Fp,\Sp) 
    \big)\rint \\ 
    = &~ \lint \oS_\Fp \Coll{i\Fp} \Delta^{(\Fp)} 
    \big(\Soft{\Sp} \THnm \FLM(\Fp,\Sp) \big)\rint  - \frac{[\alpha_s]}{\epsilon^2} \Ca \frac{\Gamma^3(1-\epsilon) \Gamma(1+\epsilon)}{\Gamma(1-2\epsilon)} \\
    & \times \bigg\langle \eta_{i\Fp}^{-\epsilon} \, \oS_\Fp \Coll{i\Fp} 
    \bigg[\left(\frac{2\Emax}{\mu}\right)^{-2\epsilon} - \left(\frac{2E_{\Fp}}{\mu}\right)^{-2\epsilon}\bigg] \Delta^{(\Fp)} \FLM(\Fp)\bigg\rangle \; .
\end{split}
    \end{equation}
 Thus we can rewrite  Eq.~\eqref{eq:fifth_and_sixth_terms} as follows
\be
\begin{split}
&
\llint \oS_{\Fp} 
\Coll{i \Fp}  \Delta^{(\Fp)}   \Soft{\Sp} \THmn \FLM (\Fp, \Sp) \rrint 
 + \llint \Soft{\Sp} \oS_\Fp       \Coll{i \Fp}  \;  \Delta^{(\Fp )}  \THnm  \FLM(\Fp, \Sp) \rrint \\
 = & \llint \oS_{\Fp} \Coll{i\Fp} \Delta^{(\Fp)}  \Soft{\Sp}  \FLM(\Fp,\Sp)  \rrint 
  - \frac{[\alpha_s]}{\epsilon^2} \Ca \frac{\Gamma^3(1-\epsilon) \Gamma(1+\epsilon)}{\Gamma(1-2\epsilon)}  \\ &
 \times \llint \eta_{i\Fp}^{-\epsilon} \, \oS_\Fp \Coll{i\Fp} \bigg[\left(\frac{2\Emax}{\mu}\right)^{-2\epsilon} - \left(\frac{2E_{\Fp}}{\mu}\right)^{-2\epsilon}\bigg] \Delta^{(\Fp)} \FLM(\Fp)\rrint \; .
\label{eq:soft_unordered}
\end{split}
\ee

It is straightforward to integrate the 
second term on the right-hand side of  Eq.~\eqref{eq:soft_unordered} over the  phase space 
of parton $\Fp$ since the required  calculation is NLO-like. 
On the contrary, the first term on the right-hand side in  Eq.~\eqref{eq:soft_unordered} requires some discussion. We  begin by acting with the soft operator  $\Soft{\Sp}$  
on $\FLM(\Fp, \Sp)$ and integrating 
over the phase space of $\Sp$. We find 
\be
\begin{split}
    & \llint \oS_\Fp \Coll{i\Fp} \Delta^{(\Fp)}  \Soft\Sp  \FLM(\Fp,\Sp)  \rrint \\
    = & - \frac{[\alpha_s]}{\epsilon^2} 
    \left (\frac{2\Emax}{\mu } \right )^{-2\epsilon} \sum_{\knotl}^{\Np+1} \llint \oS_\Fp \Coll{i\Fp} \eta_{kl}^{-\epsilon} \,  K_{kl} \, \Delta^{(\Fp)} \, 
    \left(\scprod{\ColT{k}}{\ColT{l}} \right)\colorprod \FLM(\Fp)\rrint \; .
\end{split}
\ee
The important point is that the sum in the above 
expression runs over $N_p+1$ partons which 
includes the parton $\Fp$. To simplify such an  
expression, we split the sum into the following 
contributions 
\be
\begin{split}
\sum_{\knotl}^{\Np+1} A_{kl} \scprod{\ColT{k}}{\ColT{l}} = & \sum_{\substack{k,l \neq i \\ k \neq l}}^{\Np} A_{kl}\scprod{\ColT{k}}{\ColT{l}} +  \sum_{k\neq i}^{\Np} \left(A_{ik}\ColT{i}+A_{\Fp k}\ColT{\Fp} \right) \cdot  \ColT{k} \\
& + \sum_{k\neq i}^{\Np} \ColT{k} \cdot \left( A_{ki} \ColT{i} + A_{k\Fp} \ColT{\Fp} \right)  + 2 A_{i\Fp} \scprod{\ColT{i}}{\ColT{\Fp}} \; ,
 \label{eq4.119}
\end{split}
\ee
for an arbitrary symmetric $A_{ij}$.
We consider the action of the operator $\oS_{\Fp} C_{i \Fp}$ in each of the terms in 
Eq.~(\ref{eq4.119}).  In the first term,  these 
operators act directly on $F_{\rm LM}(\Fp)$. 
In the second term, the factor
$A_{\Fp k}$ becomes $A_{i k}$ because of the collinear 
$i || \Fp$ limit. Thus the corresponding color factors combine 
into $( \ColT{i} + \ColT{\Fp} ) \cdot \ColT{k}  = \ColT{[i \Fp ]} \cdot \ColT{k}$. The same occurs in the 
third term, leading to  $\ColT{k} \cdot \ColT{[i \Fp ]}$.
Finally, in the last term, the product of the color charges is   
$2 \ColT{i} \cdot  \ColT{\Fp} 
 = -C_A$, because 
 the parton $\Fp$ is a gluon.
 Using the limit
 \be
 \lim_{\eta_{ij} \rightarrow 0} K_{ij} = \frac{\Gamma^3(1-\eps) \Gamma(1+\eps)}{\Gamma(1-2\eps)} \; ,
 \ee
 we find 
 \be
 \begin{split}
     & \sum_{\knotl}^{\Np+1} \oS_{\Fp} C_{i\Fp} \eta_{kl}^{-\epsilon} K_{kl} \Big[ \left(\scprod{\ColT{k}}{\ColT{l}} 
     \right)\colorprod \FLM(\Fp) \Big]
     =
     \sum_{\knotl}^{\Np} \eta_{kl}^{-\epsilon} K_{kl}  \left(
     \scprod{\ColT{k}}{\ColT{l}} 
     \right) \, 
      \oS_{\Fp} C_{i \Fp} \colorprod \FLM(\Fp) \\
      & - \Ca \frac{\Gamma^3(1-\eps) \Gamma(1+\eps)}{\Gamma(1-2\eps)} \eta_{i\Fp}^{-\epsilon} \; \oS_{\Fp}C_{i\Fp}  \FLM(\Fp) \; ,
\end{split}
\ee
where in the first term on the right-hand side the sum over partons $k$ and $l$ includes a clustered 
parton $[i \Fp]$ in place  of  parton $i$.

 Putting everything together 
 and  including the sum over all unresolved partons, we find 
\be
\begin{split}
   & \sum_{i=1}^{\Np}  \Big[  \lint \oS_{\Fp}  \Coll{i \Fp}  \Delta^{(\Fp)}   \Soft{\Sp} \THmn F_{\rm LM}  \rint + \lint \Soft{\Sp}  \oS_{\Fp}       \Coll{i \Fp}  \;  \Delta^{(\Fp )}  \THnm  F_{\rm LM} \rint \Big] \\
    = & ~ [\alpha_s]^2 \lint \scprod{\ISoft(\eps)}{\IColl(\eps)} \colorprod \FLM \rint +  \frac{[\alpha_s]^2}{\epsilon^2} \hc(\epsilon) \Ca \lint\ICollFour(\epsilon) \colorprod  \FLM \rint \\
    &+  \frac{[\alpha_s]^2}{\epsilon} \lint \CalPgen_{aa} \otimes   \ISoft(\epsilon) \colorprod \FLM + \ISoft(\epsilon) \colorprod \FLM \otimes \CalPgen_{bb} \rint \\
 & + \frac{[\alpha_s]^2}{2\epsilon^3} \Ca \hc(\epsilon) \lint \CalPGenFour_{aa} \otimes \FLM + \FLM \otimes \CalPGenFour_{bb} \rint \; .
    \label{eq:DoubSoft_reg_SoftColl}
    \end{split}
\ee
In the above formula, we have employed generalizations of $\IColl$ and $\CalPgen_{ab}$. They are defined in Appendix~\ref{sec:Splitting}. For the specific case that we are interested in here, we have
\index{I!$\ICollFour$}
\index{G!$\Gamma_{i,\fl{i}}^{(4)}$}
\index{G!$\Gamma_{i,g}^{(4)}$}
\be
\ICollFour(\eps) = \sum_{i=1}^{\Np}\frac{\Gamma^{(4)}_{i,\fl{i}}(\eps)}{2\eps} \; ,
\label{eq:ICollFour_defn}
\ee
where
   \begin{equation}
   \begin{split}
    \Gamma_{i,\fl{i}}^{(4)} = &  \left(\frac{2E_i}{\mu}\right)^{-4\epsilon} \frac{\Gamma^4(1-\epsilon)}{\Gamma^2(1-2\epsilon)} \left[\gamma_{\fl{i}} + \ColT{\fl{i}}^2 \frac{1 - e^{-4\epsilon L_i}}{2 \epsilon}\right] \; , \hspace{1cm} i = 1,2 \, , \\
        \Gamma_{i,g}^{(4)} =& \left(\frac{2E_i}{\mu}\right)^{-4\eps} \frac{\Gamma^4(1-\eps)}{\Gamma^2(1-2\eps)} \gamma_{z,g\to gg}^{24}(\eps,L_i) \; ,  \hspace{2.6cm} i=3, ..., \Np \; ,
           \end{split}
    \end{equation}
and 
\index{P!$\CalPGenFour_{ij}$}
\index{Splitting functions!Expansion}
\be
\CalPGenFour_{ab}(z,E_a) = \left[\left(\frac{2E_a}{\mu}\right)^{-2\epsilon} \frac{\Gamma^2(1-\epsilon)}{\Gamma(1-2\epsilon)}\right]^2 \left[- \PAP_{ab}(z) + \epsilon \,  \CalPgenfin{4}_{ab}(z)\right] \; .
\label{eq:CalPGenFour}
\ee
The function $\CalPgenfin{4}_{ab}$ is given in Eq.~\eqref{eq_P_aa_k_fin_def}. It follows from the above formulas that $\Gamma_{i,f_i}^{(4)}$ and $\CalPGenFour_{ab}$ coincide with $\Gamma_{i,f_i}$ and $\CalPgen_{ab}$  to $\order{\eps^0}$. Similarly, 
$\IColl$ and $\IColl^{(4)}$
have the same pole structure
\index{Collinear operators!Relations}
\begin{equation}
    \IColl^{(4)}(\ep) \equiv \IColl(2\epsilon) + \order{\ep^0} \; .
\label{eq:IC_k_exp}
\end{equation}

Before closing this section, we make a brief comment about the term on the third-to-last line of  Eq.~\eqref{eq:NNLO_double_unresolved_unsimplified}, which is proportional to $\delta_g(\ep)$.  It turns 
out that one can rewrite it in the following way 
\begin{equation}
    \begin{split}
        & 2 \asbr^2  \delta_g(\eps) \left ( \frac{ \Emax}{\mu} \right )^{-2\eps} \bigg[-\lint \ISoft(\eps) \colorprod \FLM\rint + \frac{(2\Emax/\mu)^{-2\eps}}{2\eps^2} N_c(\eps) \sum_{i=1}^{\Np} \ColT{i}^2 \; \lint \FLM\rint \bigg] \\
        = & - \asbr^2 \, 2^{2+2\eps} \left( \Ca \delta^{\Ca}_g(\eps) + \beta_0 \delta^{\beta_0}_g(\eps) \right) \llint \ISofttilde(2\eps) \colorprod \FLM\rrint + \order{\eps^0} \; ,
    \end{split}
    \label{eq4.21}
    \end{equation}
    where 
\index{D!$\delta^{\Ca}_g$}
\index{D!$\delta_g^{\beta_0}$}
    \be
    \delta^{\Ca}_g(\eps) =  \left(-\frac{131}{72}+\frac{\pi^2}{6}\right)  + \order{\eps} \; ;  \qquad \qquad 
    \delta_g^{\beta_0}(\eps) = \log 2 + \order{\eps} \; .
    \ee
    The reason why this rewriting  is  useful will     become clear when we discuss 
    the cancellation of color-correlated contributions 
    with unboosted kinematics. \\

In summary, we have derived expressions for all the divergent  terms in Eq.~\eqref{eq:NNLO_double_unresolved_unsimplified} that involve virtual amplitudes  and the 
various soft limits. 
Such contributions involve infrared poles in color-correlated 
matrix elements that don't appear in 
other parts of the calculations. Thus, 
we anticipate that the poles of the color-correlated contributions 
cancel amongst themselves. 
We describe this cancellation, as well as the cancellation of the poles of the single-unresolved and color-uncorrelated double-unresolved contributions, in the following section.}

\section{Cancellation of poles}
\label{sect5new}

 We begin our discussion of the infrared poles by focusing on the  single-unresolved contribution. We show 
that the cancellation of poles there is \emph{equivalent} to that in the NLO QCD contribution to the 
process $q \bar q \to 
X + (N+1) g$. We then continue with the discussion of the various contributions to the double-unresolved term $\Sigma_N$, starting from the color-correlated ones.

\subsection{Single-unresolved terms}\label{SubSec:X_fin}
As explained in the 
previous section,  when extracting singularities from 
the double-real and real-virtual contributions, we find terms featuring $N+1$ resolved partons.
In this section we will show that,  once combined,  these terms exhibit significant simplifications, allowing us to cancel the poles in the same way as we did for the NLO contribution.
We consider  $\Sigma_{N+1}^{(1)}$,
$\Sigma_{N+1}^{(2)}$  and  $\Sigma_{N+1}^{(3)}$, given 
in  Eqs.~\eqref{Eq:Sigma_N_1_and_Sigma_N+1_1_and_Sigma_RR_definitions},
\eqref{Eq:Sigma_N_5}
and~\eqref{Eq:Sigma_N+1_2fin_and_Simga_N_7_and_Sigma_N+1_3_defs}, respectively. 
We will refer to the sum of these 
contributions as $\Sigma_{N+1}^{\rm div}$. It reads
\index{S!$\Sigma_{N+1}^{\rm div}$}
\begin{equation}
\begin{split}
    \Sigma_{N+1}^{\rm div} = \sum \limits_{i=1}^{3} \Sigma_{N+1}^{(i)}
    = & ~ \lint \ONLO \Delta^{(\Fp)} \big[\FLRV(\Fp) + S_\Sp \THmn \FLM(\Fp,\Sp)\big]\rint \\
    & + \sum_{i = 1}^{\Np} \lint \ONLO (\iden - S_\Sp \THmn) C_{i\Sp} \Delta^{(\Fp \Sp)} \FLM(\Fp,\Sp)\rint \\
    & + \frac{1}{2} \lint \ONLO \Delta^{(\Fp)} (\iden - 2 S_\Sp \THmn) C_{\Fp \Sp} \FLM(\Fp,\Sp) \rint \; .
    \label{Eq:Sigma_N+1_div}
\end{split}
\end{equation}   
In the equation above, gluon $\Fp$ is resolved, since all the singularities associated with its emission are regulated by the $\ONLO$ operator (see Eq.~\eqref{eq:ONLO0}). The gluon $\Sp$, on the 
other hand, plays the  
role of an unresolved parton in NLO computations.   
Such a structure suggests a close relation between $\Sigma_{N+1}^{\rm div}$ and the NLO cross section for the production of $(N+1)$ jets. In order to make this correspondence transparent, we need to rewrite Eq.~\eqref{Eq:Sigma_N+1_div} in terms of  virtual, soft and collinear operators defined in the phase space for $(N+1)$ partons.

We begin our analysis with the first
term in Eq.~\eqref{Eq:Sigma_N+1_div}. It 
contains the one-loop amplitudes
 with $(N+1)$ final-state partons and 
a contribution from the soft limit of gluon $\Sp$. 
The former term can be treated analogously to what has been done in Section~\ref{sect:nlo}; its  infrared  singularities  can be written  with the help of  Catani's formula. The latter contribution, once integrated over the $\Sp$-parton phase space, returns the same structure as in Eq.~\eqref{eq:NLO_soft0}, up to replacing $\Emax$ with $E_\Fp$. This is due to the energy-ordering factor $\THmn$ appearing in Eq.~\eqref{Eq:Sigma_N+1_div}, which forces the energy of  
gluon $\Fp$, rather than $\Emax$, to serve as the upper cut-off for the integration over the energy  of gluon $\Sp$ 
in the soft limit.
We thus find 
\begin{equation}
\begin{split}
    & \lint \ONLO \Delta^{(\Fp)} \big[\FLRV(\Fp) + S_\Sp \THmn \FLM(\Fp,\Sp)\big]\rint 
    = \\
    = & ~ [\alpha_s] \llint \ONLO \Delta^{(\Fp)}\left[\IVirtONLO + \ISoftONLO(E_{\Fp})\right] \colorprod \FLM(\Fp) \rrint  + \lint \ONLO \Delta^{(\Fp)} \FLRVfin(\Fp)\rint \; ,
\end{split} 
\label{Eq:Sigma_N+1^div_1st_lines_def}
\end{equation}
\index{I!$\IVirtONLO$}
\index{I!$\ISoftONLO$}
where $\IVirtONLO$ is constructed
in analogy with Eq.~\eqref{eq:IVirt_defn0}, but  starting from Catani's operator $\ICat$ in Eq.~\eqref{eq:I1Cat0} with $\Np \mapsto \Np + 1$. Similarly, $\ISoftONLO(E_{\Fp})$ can be obtained by 
replacing $\Np \mapsto \Np + 1$ in Eq.~\eqref{eq:NLO_soft0} and using $E_\Fp$ in place of $E_{\rm max}$. 

We then address the contributions shown in the second  and   third  lines 
in Eq.~(\ref{Eq:Sigma_N+1_div}).
Both of these contributions describe soft-subtracted collinear limits; as such they 
provide either generalized anomalous dimensions (in case of final state splittings) or generalized anomalous dimensions 
and splitting functions (in case 
of initial state splittings). It follows from Eq.~(\ref{Eq:Sigma_N+1_div}) that in both of these cases integrations over the energy of the soft-collinear  parton 
$\Sp$  extends to $E_{\Fp}$ and not to $E_{\rm max}$.  

We would like to assemble these two terms to create the collinear operator $\IColl$ for the process with $(\Np+1)$ partons, which could then be combined with the terms in Eq.~\eqref{Eq:Sigma_N+1^div_1st_lines_def} to produce an infrared-finite operator $\ITot$, similar to what we did when describing the   NLO calculation in Section~\ref{sect:nlo}.  At first glance it appears simple to do that. Indeed,   the second line of Eq.~\eqref{Eq:Sigma_N+1_div} contains terms with collinear limits of  $\Np$ (and not $\Np+1$)  partons, and the required collinear limit of one additional parton  is supplied by the third line of this equation.
However, there seems to be a mismatch between these terms because the 
final state collinear 
operators   acting 
on $\Delta^{(\Fp \Sp)}$ in the second line produce $z_{i, \Sp} \Delta^{(\Fp)} $, whereas in the 
third line the collinear operator 
does not act on $\Delta^{(\Fp)}$ 
and, therefore,   cannot produce such a  factor.  
The resolution 
of this hypothetical problem boils 
down to the 
fact that we consider a gluon-only final state, which is highly 
symmetric.  The additional factor of $z_{i,\Sp}$ effectively lowers this symmetry, and hence plays the same role as the factor $1/2$ in 
the last term in Eq.~(\ref{Eq:Sigma_N+1_div}).\footnote{More explicitly, the additional factor $z_{i,\Sp}$ produces an additional factor of $1/2$ upon integrating over the final-state  phase space.}
We can thus write the second and third lines on the right-hand side of Eq.~\eqref{Eq:Sigma_N+1_div} as 
\begin{equation}
\begin{split}
    \sum_{i = 1}^{\Np}
    {}&
    \lint \ONLO \, (\iden - S_\Sp \THmn) C_{i\Sp} \Delta^{(\Fp \Sp)} \FLM(\Fp,\Sp)\rint \\
    & + \frac{1}{2} \lint \ONLO \, \Delta^{(\Fp)} (\iden - 2 S_\Sp \THmn) C_{\Fp \Sp} \FLM(\Fp,\Sp) \rint  \\
    ={}& [\alpha_s] \Big[\llint 
    \ONLO \, \Delta^{(\Fp)}
    \big(
    \CalPgen_{aa} \conv   \FLM(\Fp) \big)
    \rrint 
    + \llint \ONLO \, \Delta^{(\Fp)} \big(
    \FLM(\Fp) \conv \CalPgen_{bb} 
    \big)
    \rrint \Big] \\
    {}& + [\alpha_s] \llint \ONLO \left[\ICollONLO(E_{\Fp}) \colorprod \Delta^{(\Fp)} \FLM(\Fp) \right] \rrint \; .
    \label{Eq:Sigma_N+1^div_2nd_3rd_lines_def}
\end{split}
\end{equation}
\index{I!$\ICollONLO$}
Similar to the ($\Np+1$) virtual and soft operators, $\ICollONLO(E_{\Fp})$ is defined as in Eq.~\eqref{eq:IColl_definition0}, but with $\Np \mapsto \Np+1$ and setting $\Emax \mapsto E_\Fp$ in the definition of $\Gamma_{i,f_i}$.
We emphasize that the $\ONLO$ operator does not commute with the collinear operator $\ICollONLO$ or the splitting function $\CalPgen_{ab}$. Indeed the latter depends on the energy of parton $\Fp$, which is sensitive to the action of the soft limit encoded in $\ONLO$.\\

The expression for $\Sigma_{N+1}^{\rm div}$  is the sum of Eqs.~\eqref{Eq:Sigma_N+1^div_1st_lines_def} and \eqref{Eq:Sigma_N+1^div_2nd_3rd_lines_def}. We note that this quantity still contains hard-collinear singularities related to initial state emissions. To remove them, we need to add the PDF renormalization contribution proportional to the $\ONLO$ operator, i.e. 
\be
\label{eq:Sigma_NplusOne_div_pdf}
    \Sigma_{N+1}^{\rm div,pdf} =
    \frac{\alpha_s(\mu)}{2\pi \ep} \Big[\lint \PAP_{aa} \conv \ONLO \Delta^{(\Fp)} \FLM(\Fp) \rint + \lint \ONLO \Delta^{(\Fp)} \FLM(\Fp) \conv \PAP_{bb}\rint\Big] \; .
\ee
In contrast with the observation made below Eq.~\eqref{Eq:Sigma_N+1^div_2nd_3rd_lines_def}, in the expression of $\Sigma_{N+1}^{\rm div,pdf}$ we can exchange the order of the Altarelli-Parisi splitting functions and the $\ONLO$ operator. In fact, $\PAP_{qq}$ is independent of any energy variables, and thus can be moved ``inside" the fully-resolved operator. 
Given this, we can write Eq.~\eqref{eq:Sigma_NplusOne_div_pdf} as
\index{PDFs' collinear renormalization!NLO}
\index{S!$\Sigma_{N+1}^{\rm div,pdf}$}
\be
\label{eq:Sigma_NplusOne_div_pdf_v2}
    \Sigma_{N+1}^{\rm div,pdf} =
    \frac{\alpha_s(\mu)}{2\pi \ep} \Big[\lint
    \ONLO \Delta^{(\Fp)} \big(\PAP_{aa} \conv \FLM(\Fp) \big)\rint 
    + \lint \ONLO \Delta^{(\Fp)}
    \big(\FLM(\Fp) \conv \PAP_{bb} \big)\rint\Big] \; ,
\ee
and combine it with $\Sigma^{\rm div}_{N+1}$. We obtain
%
\begin{equation}
\begin{split}
   &\Sigmadivfin 
    = \Sigma^{\rm div}_{N+1}
     + \Sigma^{\rm div, pdf}_{N+1} \\
    = & ~ [\alpha_s] \llint \ONLO \, 
    \Delta^{(\Fp)}
    \left(\ITotONLO(E_{\Fp} ) \colorprod  \FLM(\Fp) \right) \rrint \\
    & + [\alpha_s] \left[\llint 
    \ONLO \, \Delta^{(\Fp)} \, \left(\PNLO_{aa} \conv  \FLM(\Fp) \right)
    \rrint + \llint \ONLO \, \Delta^{(\Fp)} \, \left(
    \FLM(\Fp) \conv \PNLO_{bb}\right) \rrint \right] \\
    & + \llint \ONLO \, \Delta^{(\Fp)} \, \FLRVfin(\Fp)\rrint \; . 
\end{split}
\label{eq4.68a}
\end{equation}
As expected, Eq.~(\ref{eq4.68a}) contains  a generalized version of the $\ep$-finite operator $\ITot$ given in Eq.~\eqref{Eq:ITot_def}. It reads
\index{I!$\ITotONLO$}
\begin{equation}
    \ITotONLO(E_{\Fp}) \equiv \IVirtONLO + \ISoftONLO(E_{\Fp}) + \ICollONLO(E_{\Fp}) \; . 
\end{equation}
 Note also that, as we mentioned at the beginning 
of this section, $\Sigmadivfin$ contains almost exactly the NLO contribution 
to the $(N+1)$-jet production cross section; the only missing piece is the fully-regulated term with up to $N+2$ resolved jets.

In addition to $\Sigmadivfin$,  
there are three other contributions 
with $N+1$ resolved final state partons 
that are explicitly  $\ep$-finite; they
appeared  in the 
course of simplifying  $\SigmaRRc$, discussed in the  previous section. We combine all these contributions 
into a single quantity that we will refer to as $\rmd \hat \sigma^{\rm NNLO}_{N+1}$.  It is given by 
\index{S!$\rmd \sigma_{N+1}^\text{NNLO}$}
\index{S!$\Sigma_{N+1}^{\rm fin, (1)}$}
\index{S!$\Sigma_{N+1}^{\rm fin, (2)}$}
\index{S!$\Sigma_{N+1}^{\rm fin, (3)}$}
\begin{equation}
    2s \; \rmd\hat \sigma_{N+1}^\text{NNLO} =  \SigmaSingUnrFinONE+ \SigmaSingUnrFinTWO
    + \Sigmadivfin + \Sigma_{N+1}^{\rm sp} \;,
\label{Eq:dsigma_NLO_final_def}
 \end{equation}
where $\SigmaSingUnrFinONE$ is given in Eq.~\eqref{Eq:Sigma_N_5} and $\SigmaSingUnrFinTWO$  in Eq.~\eqref{Eq:Sigma_N+1_2fin_and_Simga_N_7_and_Sigma_N+1_3_defs}. The final term originates from the
spin-correlated contributions discussed in Appendix \ref{sec:splin_correlation}; 
in particular, it describes the $\ONLO$ piece of the expression given in Eq.~\eqref{eq.h21}. We can expand these three terms in $\ep$, leading to the following $\order{\ep^0}$ result
\index{S!$\Sigma_{N+1}^{\rm sp}$}
\begin{equation}
\begin{split}
    \SigmaSingUnrFinONE = 
    & ~ [\alpha_s] \, \llint \PAP_{qq} \conv \Big[\ONLO^{(1)} \,  \ww{\Fp 1, \Sp 1}_{1 \parallel \Sp} 
    \log\Big(\frac{\eta_{1 \Fp}}2\Big) \Delta^{(\Fp)} \FLM(\Fp) \Big] \rrint \\
    & + [\alpha_s] \llint \Big[\ONLO^{(2)} \,  \ww{\Fp 2, \Sp 2}_{2 \parallel \Sp} \log\Big(\frac{\eta_{2 \Fp}}2\Big) \Delta^{(\Fp)} \FLM(\Fp) \Big] \conv \PAP_{qq} \rrint \\
    & -\sum_{i=1}^{\Np} [\alpha_s] \llint \ONLO^{(i)} \,  \ww{\Fp i, \Sp i}_{i \parallel \Sp} \, \Gamma_{i,f_i}   \log \Big(\frac{\eta_{i\Fp}}2\Big) \,\Delta^{(\Fp)} \FLM(\Fp)\rrint_{\Emax \mapsto E_\Fp} \; ,\\
    \SigmaSingUnrFinTWO = & -\sum_{i=1}^{\Np} [\alpha_s] \; \gamma_{z,g \to gg}^{22} \; \llint \ONLO^{(i)} \, \omega_{\Fp\parallel \Sp}^{\Fp i,\Sp i} 
    \log \left ( \frac{\eta_{i \Fp}}{4 (1-\eta_{i \Fp})} \right )  \Delta^{(\Fp)} \FLM(\Fp) \rrint \;, \\
    \Sigma_{N+1}^{\rm sp} = & ~ \sum_{i=1}^{\Np} \frac{[\alpha_s]}{2} \,  \gamma_{\perp, g \to gg}^{22}  \llint \ONLO^{(i)} \, \ww{\Fp i,\Sp i}_{\Fp\parallel \Sp} \Delta^{(\Fp)} (r_i^\mu r_i^\nu + g^{\mu\nu})  \FLMmunu(\Fp) \rrint \\
    & + \sum_{i=1}^{\Np} \frac{[\alpha_s]}{2} \, \gamma_{\perp, g \to gg}^{22,r} \llint \ONLO^{(i)} \, \omega_{\Fp\parallel \Sp}^{\Fp i,\Sp i} \Delta^{(\Fp)} \FLM(\Fp)\rrint \; ,
\end{split}
\end{equation}
where $\gamma_{z,g \to gg}^{22}$ is equal to the function reported in Eq.~\eqref{eq:defn_gamma_nk0} upon setting $L_i=0$, 
and $\gamma_{\perp, g \to gg}^{22}$ and $\gamma_{\perp, g \to gg}^{22,r}$ in Eq.~\eqref{eq:gamma_tilde}. 

\subsection{Double-unresolved triple color-correlated contributions }
\label{sec:trip_color_corr}

\index{Color space and algebra}
Having demonstrated how $\ep$-poles in single-unresolved terms disappear, 
 we continue with the discussion  of poles  in the double-unresolved contribution $\Sigma_N$.  We begin with the investigation of $\ep$-poles 
 that involve matrix elements of 
triple correlators of color-charge operators $\lint\ampM{0}| 
f_{abc}\, T_{i}^a T_{j}^b T_{k}^c | \ampM{0} \rint $. Such terms
vanish for processes with three or fewer partons at tree level, but are non-zero in general. 

As we explained in the previous subsection, triple color-correlated terms 
arise in two distinct ways. First, there are two contributions that contain triple color correlators explicitly. One of these is the $\Htwotc$ term of the double-virtual contribution in Eq.~\eqref{eq:DoubVirt_exp} and the other  one was denoted by $\ItriRV$ in the integrated soft limit of the real-virtual correction in Eq.~\eqref{eq:SoftRV_exp}.

Second, triple correlators of color 
charges appear in  
\emph{commutators} of various $I$-operators. Such 
commutators are present in Eqs.~(\ref{eq:DoubVirt_exp},
\ref{eq:SoftRV_exp}); they arise because we expressed the double-virtual contribution 
and the soft limit of the real-virtual corrections  through an 
operator $\IVirt$. All in all, combining the 
relevant terms, we find\footnote{In general, 
there are triple color correlators in 
$\ep$-finite terms present in $\Sigma_N$
that are not included in Eq.~(\ref{eq4.138}).} 
\index{S!$\Sigma_N^{\rm tri}$}
\be
\begin{split}
\SigmaNtri
= & ~ \asbr^2 \,
\Big\langle
\left( \frac{1}{2}\left[\ISoft(\eps) \, , \, \ICatbar(\eps) - \ICatbar^\dagger(\eps)\right] + \ItriRV(\eps)\right) \colorprod \FLM
\Big\rangle \, ,
\\
& + \asbr^2 
     \llint \left( -\frac{1}{2}\left[\ICatbar(\ep), \ICatbar^\dagger(\ep)\right] + \HtwotcOL +\HtwotcOL^\dag \right) \colorprod \FLM  \rrint \; .
     \label{eq4.138}
     \end{split}
\ee
We find it convenient to rewrite the commutators that 
appear in Eq.~\eqref{eq4.138} as follows 
\be
 \frac{1}{2}\left[\ISoft \, , \, \ICatbar - \ICatbar^\dagger\right] 
-\frac{1}{2}\left[\ICatbar, \ICatbar^\dagger\right]
=
- \left[\Iplus,\Iminus\right] 
+ \left[2 \Iplus + \ISoft, 
\Iminus \right] \; ,
\label{eq:I_tri_defn}
 \ee
where we introduced two additional $I$-operators 
\index{I!$\Iplus$}
\index{I!$\Iminus$}
\index{I!$\IVirt$}
\be
\Iplus(\ep) = \frac{\ICatbar(\ep) + \ICatbar^\dag(\ep)}{2} \; ,
\qquad 
\Iminus(\ep) = \frac{ \ICatbar(\ep) - \ICatbar^\dag(\ep)}{2} \; ,
\ee
such that 
\index{Virtual operators!Relations}
\be
\IVirt(\ep) = \ICatbar(\ep) + \ICatbar^\dagger(\ep) \equiv 2 I_+(\ep) \; .
\ee  

We combine the commutators and the operator $\Htwotc$, and write 
\index{I!$\ItriRV$}
\index{I!$\Itricc$}
\begin{equation}
\begin{split}
    \Sigma_N^{\rm tri}
    = \asbr^2 \, \llint \left(\ItriRV + \Itricc \right) \colorprod \FLM  \rrint \; ,
     \label{eq4.141}
\end{split}
\end{equation}
where $\Itricc$ is defined as
\begin{equation}
\begin{split}
    \Itricc = - \left[\Iplus,\Iminus\right] + \left[2 \Iplus + \ISoft, \Iminus \right] + \HtwotcOL +\HtwotcOL^\dag \; .
    \label{eq:tripole_VVplusRVa}
\end{split}
\end{equation}
Eq.~\eqref{eq4.141} collects all potentially divergent terms  where the triple color-correlated contributions can appear and 
provides the starting point for their 
analysis.

To proceed, we need to compute the commutators 
of the various $I$-operators that appear in Eq.~\eqref{eq:tripole_VVplusRVa}.
To do that, we write 
$\ICatbar$ 
as (see Eqs.~\eqref{eq:I1Cat0} and~\eqref{eq:ICatbar_defn}) 
\index{I!$\ICatbar$}
\index{L!$L_{ij}$}
\index{S!$s_{ij}$}
\be
\begin{split} 
\ICatbar & = 
- \frac12 \sum_{i=1}^{\Np} \left ( \frac{\ColT{i}^2}{\ep^2} + \frac{\gamma_i}{\ep} \right )
+ \frac{1}{2} \sum_{\inotj}^{\Np} \left ( \frac{1}{\ep^2} + \frac{\gamma_i}{\ColT{i}^2\ep}
\right ) \left(\scprod{\ColT{i}}{\ColT{j}} \right)\big( e^{i \lambda_{ij} \pi \ep} e^{\ep L_{ij}}  - 1 \big) \; ,
\end{split}
\label{eq4.142}
\ee
where $L_{ij} = \log \big( \mu^2/s_{ij}\big)$ with $s_{ij} = 2 p_i \cdot p_j $,
and $\lambda_{ij} = 1$ if both $i$ and $j$ are 
either incoming or outgoing, and $\lambda_{ij} = 0$ 
otherwise. Since we are interested in commutators of $I$-operators, in general 
the only non-vanishing contributions come from  color-correlated terms.
Therefore, the first term 
on the right hand side in Eq.~(\ref{eq4.142}) 
is irrelevant, and only the term with the $\scprod{\ColT{i}}{\ColT{j}}$  product can play a role.
Hence, we define
\be
\ICatbarcc = \frac12
\sum_{\inotj}^{\Np} \left ( \frac{1}{\ep^2} + \frac{\gamma_i}{\ColT{i}^2\ep}
\right ) \; \left(\scprod{\ColT{i}}{\ColT{j}}\right) \; \big( e^{i \lambda_{ij} \pi \ep} e^{\ep L_{ij}}  - 1 \big) \; , 
\ee
and we can use this operator instead of $\ICatbar$ to compute the commutators 
in Eq.~(\ref{eq4.141}).
To this end, we compute the color-correlated versions of 
$\Iplusminus$ using $\ICatbarcc$
and find\footnote{ We defined the color-correlated 
versions of $I_\pm$ operators $
2\Iplusminuscc = \ICatbarcc \pm \big(\ICatbarcc\big)^\dag$.
}
\index{I!$\Ipluscc$}
\index{I!$\Iminuscc$}
\index{D!$\delta^+_{ij}$}
\index{D!$\delta^-_{ij}$}
\be
\begin{split} 
    & \Ipluscc =  \frac12 \sum_{\inotj}^{\Np} \left(\scprod{\ColT{i}}{\ColT{j}} \right)
     \left (
    \frac{1}{\ep} L_{ij}  + \delta^+_{ij}
     \right ) + \order{\ep} \; ,
    \\
    & \Iminuscc  = \frac{i \pi}2 \sum_{\inotj}^{\Np}
    \left(\scprod{\ColT{i}}{\ColT{j}}\right) \,
    \left ( \frac{1}{\ep} \lambda_{ij} + \delta^-_{ij} \right ) + \order{\ep} \; ,
\label{eq:Iminus_color_corr}
\end{split}
\ee
 where
 \be
\delta^+_{ij} = \frac{1}{2} L_{ij}^2
+\frac{\gamma_i}{\ColT{i}^2} \, L_{ij}
-\frac{1}{2} \pi^2 \lambda_{ij}^2 \; , 
\qquad \quad 
\delta^-_{ij}=\frac{\gamma_i}{\ColT{i}^2} \, \lambda_{ij}  + L_{ij}\,  \lambda_{ij} \; .
\ee
We note that the objects shown in Eq.~\eqref{eq:Iminus_color_corr} are sufficient to compute the poles 
in the triple color-correlated contributions to $\Sigma_N$.

We can now proceed with the 
calculation of the commutators in Eq.~\eqref{eq4.141}. Since they involve objects such as 
$\big[ \scprod{\ColT{k}}{\ColT{l}}, \scprod{\ColT{i}}{\ColT{j}} \big]$, it is convenient to report the following general relation:
given two operators $A$ and $B$  defined as 
\begin{equation}
    A = \sum_{\inotj}^{\Np} a_{ij} (\T_i \cdot \T_j) \; ,
    \qquad
    B = \sum_{\inotj}^{\Np} b_{ij} (\T_i \cdot \T_j) \; ,
    \label{eq:5.21A}
\end{equation}
where $a_{ij}$ and $b_{ij}$ are  symmetric tensors,\footnote{If one starts with non-symmetric tensors, as is the case for the $\delta^{\pm}_{ij}$ functions, then it is clear that only their symmetric components will contribute to the sums of the type shown in Eq.~\eqref{eq:5.21A}.}
their commutator reads
\index{Color space and algebra}
\index{F!$F^{(kij)}$}
\begin{equation}
    [A,B] = i  \sum \limits_{(ijk)}^{N_p}
        \big( a_{kj} + a_{jk} \big) \big(b_{ij} + b_{ji} \big) F^{(kij)} \; , 
    \qquad 
    F^{(kij)} = f_{abc} \, T_k^a \,  T_i^b \, T_j^c \; .
  \label{eq:TkTlTiTjcomm}
\end{equation}
Note that in the above equation, we  introduced the handy notation  $(ijk)$ to label triplets with different $i$, $j$ and $k$ in the sum.

  Eq.~(\ref{eq:TkTlTiTjcomm})
  can be  used to compute the commutators in Eq.~\eqref{eq4.141},  replacing $\Iplusminus$ with their color-correlated analogues $\Iplusminuscc$, as discussed above. We find 
\be
\begin{split}
    \left[\Ipluscc,\Iminuscc \right]
    = &- \frac{\pi}{2} \sum \limits_{(ijk)}^{N_p} F^{(kij)} \bigg[\frac{ 2 L_{kj} \, \lambda_{ij} }{\ep^2} + \frac{\lambda_{ij}\big(\delta^+_{kj} + \delta^+_{jk}\big) }{\ep} + \frac{L_{kj} \big(\delta^-_{ij} + \delta^-_{ji}\big) }{\ep} \\
    & + \order{\ep^0} \; .
    \label{eq:tripole_term1}     
\end{split}
\ee

\vspace*{0.3cm}
The second commutator that we need is 
$[2 I_+ + \ISoft,I_-]$. 
 To compute it, we  extract the color-correlated contributions to $\ISoft$. Proceeding 
 along the same lines as in the derivation 
 of $\Iplusminuscc$, we obtain  
\index{I!$\ISoftcc$}
\index{F!$\phi_{ij}$}
\be
    \ISoftcc = \sum_{\inotj}^{\Np} 
    \scprod{\ColT{i}}{\ColT{j}}
    \left[\frac{\log(\eta_{ij})}{\ep} + \phi_{ij}\right] + \order{\ep} \; ,
    \label{eq:ISoft_color_corr}
\end{equation}
with 
\begin{equation}
    \phi_{ij} = -2 \log\left(\frac{2\Emax}{\mu}\right) \log(\eta_{ij}) - \frac{1}{2} \log^2(\eta_{ij}) - \Li_2(1-\eta_{ij}) \; .
\end{equation}
Considering the expressions in Eq.~\eqref{eq:Iminus_color_corr} and Eq.~\eqref{eq:ISoft_color_corr}, and following the discussion in Appendix~\ref{sec:AppNLOdetails}, it is easy to show that 
the equation 
\begin{equation}
    2\Ipluscc + \ISoftcc = \sum_{\inotj}^{\Np}  \T_i \cdot \T_j \left(\delta^+_{ij} + \phi_{ij} \right) + \order{\ep} \; , 
\end{equation}
holds.
With this representation at hand, 
we can calculate the second color-correlated commutator 
required in Eq.~\eqref{eq:tripole_VVplusRVa}
\be
\begin{split}
    \left[ 2\Ipluscc + \ISoftcc,\Iminuscc\right]
    = & -\frac{\pi}{2} \sum \limits_{(ijk)}^{N_p}  F^{(kij)} \bigg[\frac{2\lambda_{ij}}{\ep} \left(\delta^+_{kj}  + {\cal \phi}_{kj} + \delta^+_{jk} + {\cal \phi}_{jk} \right) + \order{\ep^0} \; .
   \label{eq:tripole_term2}
\end{split}
\ee
\\

It remains to determine  a suitable representation for the 
triple color-correlated part of the 
operator ${\cal H}_2$, which we denote as  $\Htwotc$. According to Ref.~\cite{Becher:2013vva}, one can write 
$\Htwotc$ as a commutator
\index{H!$\Htwotc$}
\index{G!$\Gamma$}
\index{C!$C$}
\be
 \Htwotc = \frac{1}{2\ep}
 \left [ \Gamma , C \right] \; ,
 \label{eq:5.27A}
 \ee
where  the two operators $\Gamma$ and $C$ 
are  related to the $\ep$-expansion of the 
$\ICatbarcc$ operator
\be
\ICatbarcc = \frac{\Gamma}{\ep} + C+{\cal O}(\ep) \; .
\ee
Since $\ICatbarcc = I_+ + I_-$, 
we easily obtain 
\be
\Gamma = \frac12 \sum_{(ij)}
 \scprod{\ColT{i}}{\ColT{j}} \,
 \big( L_{ij} + i \pi \lambda_{ij}  \big) \; ,
 \qquad 
C  = \frac12 \sum_{(ij)}
 \scprod{\ColT{i}}{\ColT{j}} \,
 \left ( 
  \delta^+_{ij} 
  + i \pi \, \delta^-_{ij} \right ) \; .  
 \ee
Here, in analogy with Eq.~\eqref{eq:TkTlTiTjcomm}, we have used the 
 shorthand notation $(ij)$ to indicate that the sum runs over all possible pairs of distinct partons.
It is then straightforward to compute the 
commutator of these two operators following the 
preceding discussion. The result reads 
\be
 \begin{split} 
 \HtwotcOL + \HtwotcOL^\dag 
 ={}& 
-\frac{\pi}{2\ep}
\sum_{(ijk)}^{\Np} F^{(kij)} \Big[
L_{kj} 
\big( \delta^-_{ij} + \delta^-_{ji} \big)
+ \lambda_{kj} \big(\delta^+_{ij} + \delta^+_{ji} \big)
\Big] \; .
  \end{split} 
  \label{eq:H2_tripole}
   \ee

We can now combine the three triple color-correlated terms in Eqs.~\eqref{eq:tripole_term1}, \eqref{eq:tripole_term2} and \eqref{eq:H2_tripole} to obtain the final expression the operator $\Itricc$ of Eq.~\eqref{eq:tripole_VVplusRVa}, i.e.
\index{I!$\Itricc$}
\begin{equation}
\begin{split}
    \Itricc(\ep)
    = &~ \frac{\pi}{2} \sum \limits_{(ijk)}^{\Np} F^{(kij)} \bigg[ \frac{2 L_{kj} \, \lambda_{ij}}{\ep^2} - \frac{4 \phi_{jk} \lambda_{ij}}{\ep} \bigg]
    + \order{\ep^0}\; ,
    \label{eq:Itricc}
\end{split}
\end{equation}
where we have used $\phi_{jk} = \phi_{kj}$ and have omitted $\order{\ep^0}$ terms.
\\

The calculation of $\Sigma_N^{\rm tri}$ requires us to compute $\ItriRV$ up to finite terms in $\epsilon$. Such a calculation is non-trivial; we describe it in Appendix~\ref{sec:trip_color_corr_RV}. The  result for 
$\ItriRV$
is given in Eq.~\eqref{eq:trip_RV_final}. Once $\ItriRV$ is computed, it is then 
possible to show that 
$\Sigma_N^{\rm tri}$
 is 
free of $\ep$-poles. 
To do that, we need to  rewrite 
Eq.~(\ref{eq:Itricc}) 
to make the role of the factors $\lambda_{ij}$ clear.
We recall that $\lambda_{ij}$ 
are phase factors that distinguish between 
time-like and space-like processes. In fact, 
$\lambda_{ij} =1$ if 
 partons $i$ and $j$ are both either incoming or outgoing, and zero otherwise. Furthermore, important 
 simplifications  in Eq.~(\ref{eq:Itricc}) 
 occur because 
 $\phi_{jk}$ and $L_{kj} = \log(\mu^2/s_{kj})$ are symmetric and $\FLM^{(kij)}$ is antisymmetric 
 with respect to 
$k \leftrightarrow j$ exchange. Thus, for a process with only outgoing (or only incoming) partons, 
we have $\lambda_{ij}=1$ for all $i,j$ and hence the triple 
color-correlated poles in Eq.~\eqref{eq:Itricc} vanish. A similar analysis shows that $\ep$-poles in $\ItriRV$ also vanish if all resolved partons are in the final state. 

To understand what happens in processes where both 
incoming and outgoing partons are present, 
it is convenient to write  $\lambda_{ij}$ in the 
following way 
\index{L!$\lambda_{ij}$}
 \be
 \label{eq:delta_dec_sec45}
  \lambda_{ij} = 1 - \delta_{i1} 
  -\delta_{i2} 
  - \delta_{j 1} 
  - \delta_{j2}
  + 2 \delta_{i1} \, 
  \delta_{j 2} 
  + 2 \delta_{i2} \, \delta_{j1} \; ,
  \ee
  where $1$ and $2$ label the initial state partons. We have already argued that the first term on the right-hand side provides a vanishing contribution  to Eq.~\eqref{eq:Itricc}. Terms in Eq.~\eqref{eq:delta_dec_sec45} that depend on the index $i$ only also do not contribute since 
  they do not break the $k \leftrightarrow j$  (anti)symmetry.   
  The terms that  depend on the index $j$ also vanish. 
  To see this, we write 
  \be
  \begin{split}
\sum_{(ijk)}
\langle {\cal M}| 
 F^{(kij)} A_{kj} C_j 
| {\cal M} \rangle
={}&
\sum_{(jk) }^{\Np}
\sum_{i \ne j,k }^{\Np}
\langle {\cal M}| 
 f_{abc} \, A_{kj} \,  C_j  \,  T^{a}_k  T^c_j T^b_i
| {\cal M}\rangle
\\
 ={}&  - \sum_{(jk) }^{\Np}
 \langle {\cal M}| 
  f_{abc} \, A_{kj} \, C_j \, 
 T^{a}_k T^c_j (T^b_j + T^b_k) | {\cal M} \rangle 
 \\
 ={}& 
\frac{i C_A}{2} 
\sum_{(jk) }^{\Np}
\langle {\cal M}| 
  A_{kj} C_j \left ( 
 \ColT{k} \cdot \ColT{j}  - \ColT{j} \cdot \ColT{k}  
 \right ) | {\cal M} \rangle =0 \; ,
 \label{eq:5.33A}
 \end{split}
 \ee
where $A_{kj}$ stands for $L_{kj}$ or $\phi_{kj}+\phi_{jk}$. Furthermore,  we have used color conservation 
\be
\sum \limits_{i \ne j,k}^{\Np} 
 T_i^b | {\cal M} \rangle = -(T^b_j + T^b_k) | {\cal M}  
 \rangle \;, 
 \ee
 to go from the first line to the second in Eq.~\eqref{eq:5.33A}, 
 and the $\rm SU(3)$ commutation relations for color charges 
 in the next step. 
 
  Finally, we write $L_{jk} = \log(\mu/(2E_j)) + \log(\mu/(2E_k)) - \log (\eta_{jk})$.  Using the same 
  arguments as above, it is easy to show that
    the first two of these terms  do not contribute to $\Itricc$. The only terms that remain include $\log(\eta_{jk})$ and the final two terms of Eq.~\eqref{eq:delta_dec_sec45}. Combining 
    all these results, we finally arrive at an expression for the triple color-correlated poles
\index{I!$\Itricc$}
 \be
 \begin{split}
\Itricc   =  &
   \sum \limits_{k \neq 1, 2}^{\Np} F^{(k 1 2)} \bigg[  -\frac{2\pi}{\ep^2}  
    \log \left ( \frac{\eta_{2k}}{\eta_{1 k} } \right )  - \frac{2 \pi}{\ep}\bigg(
2 \log\left(\frac{4 \Emax^2}{\mu^2}\right)
\log \left( \frac{\eta_{1 k}}{\eta_{2 k}}\right) \\
&{} +\log^2 \eta_{1k}
-\log^2 \eta_{2k}
  + 2 \Li(1-\eta_{1k})
- 2 \Li(1-\eta_{2k})
\bigg) \bigg] + \order{\ep^0}
  \; .
  \label{eq:Icctri}
  \end{split}
  \ee
 Comparing this result 
 with the expression for $\ItriRV$ in Eq.~\eqref{eq:trip_RV_final}, we find that 
 their  poles are equal and opposite in sign.\footnote{One could equally well understand this as the $\order{\ep^{-2}}$ poles cancelling between the commutator terms in $\Itricc$ (see Eq.~\eqref{eq:tripole_VVplusRVa}) and $\ItriRV$, leaving a simple pole which cancels against the contribution originating from the double-virtual   amplitude, cf. Eq.~\eqref{eq:H2_tripole}.} 
 This establishes the  cancellation of  
 $\ep$-poles in   triple color-correlated contributions  for a generic $1_a+2_b\to X+N \, g $ process.

\subsection{Other color-correlated double-unresolved contributions}
\label{sec:color_correlations}
We continue with the discussion of 
divergent contributions to $\Sigma_{N}$
that contain 
double and quartic color-correlated matrix elements 
 squared 
 with double-unresolved kinematics. As these contributions must involve either a loop amplitude or a soft limit, we are interested in those terms in 
 Eqs.~(\ref{eq:DoubVirt_exp},
 \ref{eq:DoubSoft_exp},
 \ref{eq:SoftRV_exp},
 \ref{eq:CollRV},
 \ref{eq:DoubSoft_reg_SoftColl}, \ref{eq4.21}) that contain
 either $\IVirt$ or  $\ISoft$ or both. 

The sum of the elastic (i.e. unboosted) terms involving color correlations, which  we denote as  $\SigmaNccEl$, 
reads 
\index{S!$\Sigma_N^{({\rm V+S}), {\rm el}}$}
\index{C!$\tilde c$}
\be
\begin{split}
    \SigmaNccEl = &~ \asbr^2 \frac{1}{2} \lint 
    \left[ 
    \IVirt^2 
    + \IVirt \ISoft
    + \ISoft \IVirt
    + \ISoft^2
    + 2 \IColl \IVirt 
    + 2 \IColl \ISoft 
    \right] \colorprod \FLM  
    \rint \\
    & + \asbr^2 \frac{\beta_0}{\ep} 
    \frac{\Gamma(1-\eps)}{e^{\eps \EulerGamma}} 
    \llint \Bigl[ -\big[\ISoft(\eps)+\IVirt(\eps)\big]
    + \IVirt(2\ep) + \tilde c(\ep)  \, \ISofttilde(2\eps)    \Bigr]\colorprod  \FLM \rrint \\ 
    & + \asbr^2 
    \bigg\langle \bigg[ K \frac{\Gamma(1-\ep)}{e^{\ep \EulerGamma}} \; \IVirt(2\eps) +  \Ca \bigg(\frac{c_1(\eps)}{\eps^2} - \frac{A_K(\ep)}{\eps^2} - 2^{2+2\ep}  \delta_g^{\Ca}(\eps) \bigg) \\
    &  \times \ISofttilde(2\eps) \bigg] \colorprod \FLM \bigg\rangle + \asbr \lint \big[ \IVirt(\eps) + \ISoft(\eps) \big] \colorprod \FLVfin
    \rint \; , 
    \label{eq:NNLO_col_corr_unboosted}
\end{split}
\ee
where 
\be
\tilde c(\ep) = \frac{e^{\eps \EulerGamma}}{\Gamma(1-\eps)} 
\left (c_2(\ep) + \eps \, c_3(\eps) - 2^{2+2\ep} \ep \, \delta_g^{\beta_0}(\ep) \right ) \; .
\ee
Before continuing, we recall that the soft and virtual operators $\ISoft$ and $\IVirt$ have \emph{color-correlated} poles starting at $\order{\ep^{-1}}$, while $\IColl$ does not contain any color-correlated terms and $\ITot$ is finite. It follows  that the combination $\IVirtSoft=\IVirt+\ISoft=\ITot-\IColl$ contains color-correlated contributions starting at $\order{\ep^0}$. 

Using these properties, it is easy to see that the first and last lines of Eq.~\eqref{eq:NNLO_col_corr_unboosted} do not contain divergent color-correlated contributions. Indeed, the sum of $I$-operators in the first line gives  $\ITot^2 - \IColl^2 $, while the final line  yields $\IVirtSoft$.
Further details about this  rearrangement and the origin of each term can be found in Ref.~\cite{Signorile-Signorile:2023blp}.
We note that all  quartic color correlations $\sim \left( \ColT{i} \cdot \ColT{j} \right) \left( \ColT{k} \cdot \ColT{l} \right)$ appear in the first line, so this demonstrates the complete cancellation of infrared singularities associated with this color structure.

We  continue with 
the discussion of terms proportional 
to $\beta_0$ that appear in the 
second line of 
Eq.~\eqref{eq:NNLO_col_corr_unboosted}. 
Here we can reconstruct two different versions of $\IVirtSoft$. Indeed,  the first two terms in square brackets return $\IVirtSoft(\ep)$, while the 
third and fourth terms suggest that the combination 
$\IVirtSoft(2\ep)$ can be assembled. To do so, we add and subtract the  soft operators $\ISoft(2\ep)$ and $\ISofttilde(2\ep)$ such that 
\index{S!$\Sigma_N^{({\rm V+S}), {\rm el}, \beta_0}$}
\be
\begin{split} 
\SigmaNccElbeta = &~
\asbr^2 \frac{\beta_0}{\ep} 
\frac{\Gamma(1-\eps)}{e^{\eps \EulerGamma}}
\llint\left [ 
-\big[\ISoft(\eps)+\IVirt(\eps)\big]
 + \IVirt(2\ep) + \tilde c(\ep) \ISofttilde(2\eps)   
 \right ] \cdot \FLM \rrint 
 \\
= &~ \asbr^2 \frac{\beta_0}{\ep} 
\frac{\Gamma(1-\eps)}{e^{\eps \EulerGamma}}
\Big \langle \Big[ 
-\IVirtSoft(\ep)
+\IVirtSoft(2\ep)
+ \big( \tilde c(\ep) -1 \big) \; \ISofttilde(2\ep) \\
& + \ISofttilde(2\ep) - \ISoft(2\ep)
\Big ] \cdot \FLM \Big\rangle \; .
\end{split} 
\label{eq4.165}
\ee

We now argue that this contribution does not 
contain divergent color-correlated 
terms. 
First, since $\IVirtSoft(2\ep)$ and $\IVirtSoft(\ep)$ must coincide at $\order{\ep^0}$, the difference $\IVirtSoft(2\ep)-\IVirtSoft(\ep)$ contains color-correlated terms at $\order{\ep}$ only. 
Second, it is easy to check that 
\be
\tilde c(\ep) - 1 = {\cal O}(\ep^2) \; ,
\label{eq:cep_expansion}
\ee
and since color-correlated terms in $\ISofttilde(2 \ep)$
appear for the first time  at  order ${\cal O}(\ep^{-1})$, the  third 
term in Eq.~(\ref{eq4.165}) also does not 
give rise to color-correlated poles. Finally, as we have mentioned previously (cf. Eq.~\eqref{eq:ISofttilde_and_ISoft}), the difference
\index{Soft operators!Relations}
\be
\ISofttilde(2 \ep) - \ISoft(2\ep) = {\cal O}(\ep) \; , 
\label{eq:ItildeSoftISoft}
\ee
which implies that the combination of 
the fourth and the 
fifth term in Eq.~(\ref{eq4.165}) 
is also finite.  Hence, we have proved 
that all terms proportional to $\beta_0$
in Eq.~(\ref{eq:NNLO_col_corr_unboosted}) 
are free of  divergent color-correlated 
contributions.
Finally, for future purposes, it is convenient to introduce the following decomposition
\index{!S${\Sigma_N^{({\rm V+S}), {\rm el}, \beta_0}}$}
\index{S!$\Sigma_N^{\rm fin, (6)}$}
\be
\begin{split} 
\SigmaNccElbeta={}&
\asbr^2\frac{\beta_0}{\ep} 
\frac{\Gamma(1-\eps)}{e^{\eps \EulerGamma}}
\llint \left [ 
-\IVirtSoft(\ep)
+\IVirtSoft(2\ep)
+ \big( \tilde c(\ep) -1 \big) \; \ISofttilde(2\ep)
\right ] \cdot \FLM \rrint \\
{} &+ \SigmaNccElbetafin \; , 
\end{split} 
\label{eq:Sigma_N_cc_el_beta0}
\ee
where
\be
\SigmaNccElbetafin = 
\asbr^2 \frac{\beta_0}{\ep} 
\frac{\Gamma(1-\eps)}{e^{\eps \EulerGamma}}
\llint \left [ 
\ISofttilde(2\ep) - \ISoft(2\ep)
\right ] \cdot \FLM \rrint \; .
\label{eq:SigmaNfinSix}
\ee

The term in the third line in Eq.~\eqref{eq:NNLO_col_corr_unboosted} can 
be analyzed in a similar manner. We write 
\be
\begin{split}
&  K \; \IVirt(2\ep) +  \Ca \left(\frac{c_1(\eps)}{\eps^2} - \frac{A_K(\ep)}{\eps^2} - 2^{2+2\ep}  \delta_g^{\Ca}(\eps)   \right) \ISofttilde(2\eps) 
=
 K \; \IVirtSoft(2\ep) 
\\
& 
+ 
\left [ 
\Ca \left(\frac{c_1(\eps)}{\eps^2} - \frac{A_K(\ep)}{\eps^2} - 2^{2+2\ep}  \delta_g^{\Ca}(\eps)   \right) -K 
\right ]\ISofttilde(2\eps) 
+ K  \left(\ISofttilde(2\eps) - \ISoft(2\eps) \right) \; ,
\end{split} 
\label{eq4.168}
\ee
where we dropped the factor $\Gamma(1-\ep)/e^{\ep \EulerGamma}$ as it contributes at $\order{\ep^0}$ only. We observe that the 
first and the third terms on the 
right-hand side of the above equation 
do not contain singular 
 color-correlated terms for the reasons 
 discussed above. The second term 
 on the right-hand side in 
 Eq.~(\ref{eq4.168}) 
 also does not 
 contain  divergent color-correlated
 contributions 
 because 
\be
C_A 
\left ( 
\frac{c_1(\eps)}{\eps^2} - \frac{A_K(\ep)}{\eps^2} - 2^{2+2\eps} \delta_g^{\Ca}(\eps) 
\right ) - K 
=  \order{\eps} \; . 
\label{eq:4.166}
\ee
This completes the analysis of the unboosted color-correlated contributions. 

Additionally, there are boosted terms with color correlations in Eqs.~\eqref{eq:CollRV} and~\eqref{eq:DoubSoft_reg_SoftColl}. It is straightforward to show that the sum of  these terms assumes a particularly simple form 
\index{S!$\Sigma_N^{({\rm V+S}), {\rm boost}}$}
\be
\label{eq:SigmaNccBoost}
\SigmaNccBoost = \frac{\asbr^2}{\ep} \lint \CalPgen_{aa} \otimes   \big[ \IVirtSoft(\eps) \colorprod \FLM \big]+ \big[ \IVirtSoft(\epsilon) \colorprod \FLM \big] \otimes \CalPgen_{bb} \rint \; .
\ee
Given the properties of $\IVirtSoft(\eps)$ stated above, it is clear that $\SigmaNccBoost$ contains color-correlated divergences at  $\order{\ep^{-1}}$. These divergences get canceled upon combining Eq.~\eqref{eq:SigmaNccBoost}  with similar contributions that arise as the result  of the  collinear renormalization of parton distribution functions.  We briefly discuss this point at  the end of Sec.~\ref{sec:Collinear}, 
after Eq.~(\ref{eq4.183}).

Hence,  the analysis performed 
in the current and 
previous  sections proves  the cancellation of  \emph{all} color-correlated divergent terms
in a generic process 
$1_a+2_b\to X+N \, g $.  The remaining divergences in the double-unresolved contribution $\Sigma_N$ are \emph{not} color-correlated and, instead, are proportional to the squares of color charges of the external partons. These are 
related 
to collinear emissions and we 
continue with their analysis in the 
next section. 

\subsection{Collinear double-unresolved contributions}
\label{sec:Collinear}

Having demonstrated the cancellation of poles in the color-correlated contributions  to $\Sigma_N$ in the previous two sections,  we need to discuss the remaining terms in this quantity. Such terms are related to collinear emissions and, therefore, are proportional to 
the squares of color charges of the external hard partons.  In this subsection we  manipulate the corresponding 
contributions to Eq.~\eqref{eq:NNLO_double_unresolved_unsimplified}
in order to write them in terms of collinear operators $\IColl$ and splitting functions $\CalPgen_{ab}$. This will pave the way for  demonstrating the cancellation of the poles, which we undertake in Subsections~\ref{sect:collinearcombine}.

\vspace*{0.3cm}
The first term that we have yet to discuss
is the last one in the third line of Eq.~\eqref{eq:NNLO_double_unresolved_unsimplified}. We find it convenient to split it into two pieces 
\be
\begin{split}
& \frac{1}{2} \sum_{i,j=1}^{\Np}
\llint \oS_\Sp \oS_\Fp    C_{j \Sp } C_{i \Fp }   \Delta^{(\Fp \Sp)}  \FLM(\Fp, \Sp) \rrint \\
= &~
\sum_{\substack{i,j=1\\i < j}}^{\Np}
\llint \oS_\Sp \oS_\Fp    C_{ j \Sp} C_{i \Fp }  \Delta^{(\Fp \Sp)}  \FLM(\Fp, \Sp) \rrint 
+ \frac{1}{2} 
\sum_{i=1}^{\Np} \llint \oS_\Sp \oS_\Fp  C_{i \Sp} C_{ i \Fp}    \Delta^{(\Fp \Sp)}  \FLM(\Fp, \Sp) \rrint \; .
\end{split}
\label{eq4.157}
\ee
In the first term on the right-hand side of Eq.~(\ref{eq4.157}) the unresolved partons $\Fp$ and $\Sp$ become collinear to two different resolved partons $i$ and $j$,  and we have used the symmetry of the limits to remove the factor  $1/2$. In the second term 
in Eq.~(\ref{eq4.157})
both $\Fp$ and $\Sp$ become collinear to the same parton $i$. It is straightforward to evaluate the first term since all we need to do 
is perform the NLO-like computation twice. The result reads
\begin{equation}
\begin{split}
    &  \sum_{\substack{i,j=1 \\i<j}}^{\Np} \llint \oS_\Sp \oS_\Fp \Coll{j \Sp } \Coll{i \Fp} \Delta^{(\Fp \Sp)} \FLM(\Fp,\Sp)\rrint \\
    = & ~  
    \frac{\asbr^2}{\ep^2}
    \bigg\{\frac12
    \sum_{\inotj}^{\Np}  \lint \Gamma_{i,f_i} \Gamma_{j,f_j}  \cdot  \FLM\rint 
    + \llint 
    \PqqGen_{aa} \conv  \FLM \conv \PqqGen_{bb} \rrint 
    \\
    & +  \sum_{\substack{i=1 \\i\not= 1}}^{\Np} \lint \PqqGen_{aa} \conv     \big[\Gamma_{i,f_i} \cdot \FLM \big] \rint  
    +  \sum_{\substack{i=1 \\ i\not= 2}}^{\Np} \lint \big[\Gamma_{i,f_i} \cdot \FLM \big]  \conv \PqqGen_{bb} \rint
    \bigg\}\;  .
\label{Eq:NNLO_generalized_double_collinear_i_not_j}
\end{split}
\end{equation}

The last term in Eq.~(\ref{eq4.157}) requires more care, as it involves a product of two  operators that 
describe the soft-subtracted collinear limits of gluons $\Fp$ and $\Sp$ relative to the 
same hard parton.  We would like to relate 
this contribution to the iteration of two collinear 
emissions and write it 
in terms of the functions $\CalPgen_{ab}$ and $\Gamma_{i,f_i}$, as done in Eq.~\eqref{Eq:NNLO_generalized_double_collinear_i_not_j}.
It turns out that this is \emph{nearly} possible but 
that the intertwined phase space of the two collinear gluons leads to one additional term when such a rewriting is performed. Indeed, for  $i=1$  we find
\be
\begin{split}
       & \frac{1}{2} \llint \oS_\Fp \oS_\Sp C_{ 1 \Fp } C_{1 \Sp } \Delta^{(\Fp \Sp)}  \FLM(\Fp,\Sp) \rrint 
        =
   \frac{\asbr^2}{2\ep^2} \llint \Gamma_{1,a}^2 \cdot \FLM \rrint 
      \\
        & +
    \frac{\asbr^2}{\eps^2} \llint \PqqGen_{aa} \conv  \big[ \Gamma_{1,a} \cdot \FLM  \big] \rrint  +\frac{\asbr^2}{2\eps^2}\llint \convPgenPgen{aa}{aa}\conv  \FLM\rrint
    \\
         & + \frac{\asbr^2}{2\eps^2} 
        \llint G_1(z) \conv \FLM\rrint \; .
    \end{split}
    \label{eq4.159}
\ee
The ``bar''-convolution $[f \bar{\otimes} g]$ is defined as 
\index{Convolution!$[f \bar{\otimes} g]$}
\be
\begin{split} 
\label{eq_bar_conv}
\big[f(z_1,E_i) \, \bar{\otimes} \, g(z_2, E_i)\big] (z,E_i) = \int\limits_0^1 \rmd z_1 \, \rmd z_2 \, f(z_1,E_i) g(z_2, z_1 E_i) \delta(z-z_1 z_2) \; .
\end{split}
\ee
The first three terms on the right-hand side of Eq.~\eqref{eq4.159}
represent  the ``naive'' product of two soft-subtracted 
collinear limits and the function $G_1$ incorporates
the modifications required by the non-trivial dependence of the double-collinear phase space of two
unresolved gluons on their energies. To obtain the results for  $i=2$ we can use 
Eq.~(\ref{eq4.159}) and replace $i=1$ with $i=2$ and exchange  ``left'' and ``right'' convolutions.  The functions $G_i$ read 
\index{G!$G_i(z,E_i)$}
\index{G!$G_i$}
\index{G!$\Gamma_{i,f_i}$}
\be
   G_i(z,E_i) =  \big[\Gamma_{i,f_i} - \Gamma_{i,f_i}(z) \big] \; \PqqGen_{f_i f_i}(z,E_i) \; , 
   \qquad i=1,2 \; ,
   \label{eq:Gi_IS}
   \ee
   with
\be
 \Gamma_{i,f_i}(z) = \left(\frac{2 z E_i}{\mu}\right)^{-2\eps} \frac{\Gamma^2(1-\eps)}{\Gamma(1-2\eps)} \left[\gamma_{f_i} + \ColT{f_i}^2 \frac{1-e^{-2\eps L_{z \cdot i}}}{\eps}\right] \; .
 \ee
In the above equation $L_{z \cdot i} = 
\log E_{{\rm max}}/(z E_i)$. A similar computation for the final-state parton $i$ yields 
 \be
         \frac{1}{2} \lint \oS_\Fp \oS_\Sp \Coll{i \Fp } 
         \Coll{i \Sp} \Delta^{(\Fp \Sp)} \FLM(\Fp , \Sp) \rint = \frac{\asbr^2}{2 \ep^2} \lint \Gamma_{i,f_i}^2  \FLM \rint +  \frac{\asbr^2}{2 \eps^2} \lint G_i  \FLM \rint \;  , 
         \label{eq4.163}
         \ee
         where 
         \be
         \label{eq:Gi_FS}
         \begin{split}
           G_i =& \bigg[\left(\frac{2E_i}{\mu}\right)^{-2\eps} \frac{\Gamma^2(1-\eps)}{\Gamma(1-2\eps)}\bigg]^2 \left[\gamma_{z,g\to gg}^{22}(\eps, L_i) + \frac{\T_g^2}{\epsilon} e^{-2\eps L_i}\right] \\
           &\times \Big[\gamma_{z,g\to gg}^{42}(\epsilon, L_i) - \gamma_{z,g\to gg}^{22}(\eps, L_i) \Big] \; ,  
           \end{split}
         \ee
            and $i=3, ... \, , \Np$.
         Combining Eqs.~\eqref{Eq:NNLO_generalized_double_collinear_i_not_j},~\eqref{eq4.159} and \eqref{eq4.163} and summing over the final-state partons, we find the following result for 
         the last term in the third line of Eq.~\eqref{eq:NNLO_double_unresolved_unsimplified}
\be
\begin{split}
    & \frac{1}{2} \sum_{i,j=1}^{\Np} \llint \oS_\Fp \oS_\Sp \, C_{i \Fp } C_{j \Sp}   \Delta^{(\Fp \Sp)}  \FLM(\Fp, \Sp) \rrint
    = \asbr^2 \bigg\{\frac{1}{2} \llint \IColl^2(\eps) \colorprod \FLM\rrint   \\
    & + \frac{1}{2\eps^2} \sum_{i=3}^{\Np} \llint G_i  \,  \FLM\rrint
    + \frac{1}{\eps} \Big[ \lint \PqqGen_{aa} \conv \big[ \IColl(\eps) \colorprod \FLM\big] \rint +  \lint \big[ \IColl(\eps) \colorprod \FLM \big] \conv \PqqGen_{bb} \rint\Big]  \\
    & + \frac{1}{2\eps^2} \Big( \lint  \convPgenPgen{aa}{aa} \conv  \FLM\rint  + \lint   \FLM \conv \convPgenPgen{bb}{bb}\rint \Big) \\
    &  + \frac{1}{2\eps^2} \Big[\lint G_{1} \conv \FLM\rint + \lint \FLM \conv G_{2}  \rint \Big] 
    + \frac{1}{\eps^2} \lint \PqqGen_{aa} \conv  \FLM \conv \PqqGen_{bb} \rint 
    \bigg\}\; . 
\label{eq:Hard_DoubColl_ac}
\end{split}
\ee
As we will show in the next subsection, the above 
equation is already in a suitable form to discuss the 
cancellation of some $1/\ep$ collinear contributions 
to $\Sigma_N$. 

\vspace*{0.3cm}
We now briefly discuss  the terms in the fourth and  fifth  lines of Eq.~\eqref{eq:NNLO_double_unresolved_unsimplified}. The term in the fourth line contains two soft-subtracted collinear operators $C_{i \Sp} C_{i \Fp}$ and a factor $[2 (\eta_{i
\Sp}/2)^{- \ep} - 1]$. The two soft-subtracted collinear limits  produce an ${\cal O}(\ep^{-2})$ term but the prefactor is arranged in such a way that the actual singularity is just ${\cal O}(\ep^{-1})$. In what follows we will mostly focus on  the cancellation of $1/\ep^2$ collinear singularities and 
for this reason we do not need to discuss how this term can be rewritten. 
Furthermore,  the term on the fifth line includes a commutator of the limits $\Coll{i\Fp}$ and $\Coll{i\Sp}$. Since  we consider final-state gluons only, this contribution is identically zero for the purposes of this paper. However, we note that this term would no longer vanish when one considers processes with both quarks and gluons in the final state.
The only term in Eq.~\eqref{eq:NNLO_double_unresolved_unsimplified} that we have yet to consider is the one on the penultimate line, which originates from the soft-regulated double collinear limits in sectors $(b)$ and $(d)$. The first part of the computation proceeds similarly to the NLO case, and results in
 \be
 \begin{split}
     & \llint C_{\Fp \Sp} \Big[\FLM(\Fp,\Sp) - 2S_\Sp \THmn \FLM (\Fp,\Sp)\Big]\rrint \\
     = &~ \frac{\asbr}{\eps} \left(\frac{2E_\Fp}{\mu}\right)^{-2\eps} \frac{\Gamma^2(1-\eps)}{\Gamma(1-2\eps)}  \llint 2 \, \gamma_{z,g \to gg}^{22}(\eps)  \FLM(\Fp)\rrint \;,
     \end{split}
\ee
\index{G!$\gamma_{z,g \to gg}^{22}(\ep)$}
where $\gamma_{z,g \to gg}^{22}(\eps) = \gamma_{z,g \to gg}^{22}(\eps,L_i = 0)$ and we have renamed the clustered parton $[\Fp \Sp] \to \Fp$.\footnote{We are free to do so because both $\Fp$ and $\Sp$ are gluons, and hence the clustered parton $[\Fp \Sp]$ is also a gluon.} To complete the calculation,  we  need to evaluate the soft-regulated collinear limit $\oS_\Fp C_{i\Fp}$. Recalling that $\sigma_{ij} = \eta_{ij}/(1-\eta_{ij})$, we  find    $C_{i \Fp} 
\sigma_{i\Fp}^{-\ep} =  \eta_{i\Fp}^{-\eps}$  and obtain 
 \begin{equation}
 \begin{split}
     & \sum_{i = 1}^{\Np}  \frac{N_{\Fp\parallel \Sp}(\ep)}{2} \llint \oS_\Fp C_{i\Fp} \sigma_{i\Fp}^{-\eps} \Delta^{(\Fp)} C_{\Fp \Sp} \Big[\FLM(\Fp,\Sp) - 2S_\Sp \THmn \FLM(\Fp,\Sp) \Big] \rrint \\
     = &~ \frac{\asbr^2}{\eps} N_{sc}^{(b,d)} \lint \gamma_{z,g \to gg}^{22}(\eps)   \big[\ICollFour(\eps) \colorprod \FLM \big]\rint \\ 
     & + \frac{\asbr^2}{2\eps^2} N_{sc}^{(b,d)} \Big\langle \gamma_{z,g \to gg}^{22}(\eps) 
     \Big[  
     \CalPGenFour_{aa} \conv \FLM 
     +  \FLM \conv \CalPGenFour_{bb} \Big]
     \Big\rangle \; ,
     \label{eq:Hard_DoubColl_bd}
 \end{split}
 \end{equation}
where $\ICollFour$ and $\CalPGenFour_{ab}$ are defined in Eqs.~\eqref{eq:ICollFour_defn} and~\eqref{eq:CalPGenFour}, respectively, and  the normalization constants are collected in Appendix~\ref{sebsec:useful_constants}.

 This concludes the  discussion  of the  collinear 
 contributions to $\Sigma_N$; through ${\cal O}(\ep^{-2})$ they are given by the sum of  
 Eqs.~\eqref{eq:Hard_DoubColl_ac} and~\eqref{eq:Hard_DoubColl_bd}. In addition to these terms, there are also remnants of virtual 
 and soft contributions that are not color correlated.  All these terms will have to be  combined together with the collinear renormalizations of parton distribution functions to demonstrate the cancellation of  singularities. 
 
 Before discussing the details of this cancellation, we  will write down the  term in Eq.~\eqref{eq:partoni_sigma_NNLO} that arises from the collinear renormalization of the parton distribution functions  at $\order{\alpha_s^2}$. It reads
\index{PDFs' collinear renormalization!NNLO}
 \index{S!$\rmd \sigmahat^{\rm pdf}$}
 \begin{equation}
\begin{split}
    \pdftermNNLO = & \left[\amu\right] \sum_x \Big[
 \Ppdf{1, xa} \conv \rmd\bar{\sigma}_{xb}^\text{NLO} 
 + \rmd\bar{\sigma}_{ax}^\text{NLO} \conv \Ppdf{1, xb}\Big] \\
	& + \left[\amu\right]^2 \sum_{x,y} \Big[\Ppdf{1, xa} \conv \rmd\bar{\sigma}_{xy}^\text{LO} \conv \Ppdf{1, yb} + \Ppdf{2, xa} \conv \rmd\bar{\sigma}_{xb}^\text{LO} + \rmd\bar{\sigma}_{ax}^\text{LO} \conv\Ppdf{2, xb}\Big] \; ,	\label{Eq:PDFs_renormalization_general_expression_NNLO}
\end{split}
\end{equation}
where we have used 
the following short-hand notation 
\index{P!$\Ppdf{1, ij}$}
\index{P!$\Ppdf{2, ij}$}
\begin{equation}
    \Ppdf{1, ab}(z) = \frac{\hat{P}_{ab}^{(0)}(z)}{\epsilon} \; , 
    \qquad
    \Ppdf{2, ab}(z) =  \frac{\big[\hat{P}_{ax}^{(0)} \conv \hat{P}_{xb}^{(0)}\big](z) - \beta_0 \hat{P}_{ab}^{(0)}(z)}{2\epsilon^2} + \frac{\hat{P}_{ab}^{(1)}(z)}{2\epsilon} \; .
\end{equation}
We note that at  variance with Eq.~\eqref{Eq:final_expression_NLO},  $\rmd \bar{\sigma}^{\rm NLO}$ does not include the PDFs renormalization. 
Furthermore, the summation in Eq.~\eqref{Eq:PDFs_renormalization_general_expression_NNLO} is performed over all initial-state parton flavors. However, since we consider processes with $q\bar q$ initial states and  gluonic final states, the Altarelli-Parisi splitting functions always have identical indices. We can therefore write the contribution from the PDFs renormalization as follows
\be
\begin{split}
    \pdftermNNLO = &~ \left[\amu\right] \bigg[ \frac{\PAP_{aa} \conv \rmd\bar{\sigma}_{ab}^\text{NLO}}{\ep}+
    \frac{\rmd\bar{\sigma}_{ab}^\text{NLO} \conv \PAP_{bb}}{\ep}\bigg] \\
    & +\left[\amu\right]^2 \Bigg\{  \left[ \frac{\PAP_{aa} \conv \PAP_{aa} - \beta_0 \PAP_{aa}}{2\ep^2} - \frac{\PAPone_{aa}}{2\ep} \right] \conv \lint\FLM\rint \\ 
    & + \lint\FLM\rint \conv \left[\frac{\PAP_{bb} \conv \PAP_{bb} - \beta_0 \PAP_{bb}}{2\ep^2} - \frac{\PAPone_{bb}}{2\ep} \right]  
    + \frac{\PAP_{aa} \conv \lint\FLM\rint \conv \PAP_{bb}}{\ep^2} \Bigg \} \; .
\end{split}
\label{eq:pdfrenorm_NNLO}
\ee
The NLO cross section  $\rmd\bar{\sigma}_{ab}^\text{NLO}$ can be obtained from the results of Section~\ref{sec:nlo}  and reads 
\index{S!$\rmd\bar{\sigma}_{ab}^\text{NLO}$}
\begin{equation}
\begin{split}
    \rmd\bar{\sigma}_{ab}^\text{NLO} = &~ [\alpha_s]
    \lint \ITot(\ep) \colorprod \FLM \rint 
    + \lint\FLV^\text{fin}\rint
    + \frac{[\alpha_s]}{\ep} \Big[\lint \CalPgen_{aa} \conv \FLM \rint + \lint \FLM \conv \CalPgen_{bb} \rint\Big] \\
   &+ \llint \ONLO \, \Delta^{(\Fp)} \FLM(\Fp) \rrint \;  . 
\end{split}
\label{eq4.183}
\end{equation}
As mentioned earlier, we do not include the $\order{\alpha_s}$ contribution of the collinear renormalization of PDFs in the definition of $\rmd\bar{\sigma}_{ab}^\text{NLO}$, and therefore 
this quantity still contains unsubtracted hard-collinear poles.  We also note that we already used the convolution
 of the Altarelli-Parisi splitting function with the $\ONLO$ term 
 in Eq.~(\ref{eq4.183})
 to cancel $\ep$-poles 
 in single-unresolved contributions to $\rmd\hat \sigma_{N+1}^\text{NNLO}$ shown in Eq.~\eqref{Eq:dsigma_NLO_final_def}.

Before continuing with the discussion of the double-unresolved collinear contributions, we can use 
Eq.~\eqref{eq:pdfrenorm_NNLO} to complete the  demonstration of the 
cancellation of the color-correlated divergences, 
see the discussion after Eq.~\eqref{eq:SigmaNccBoost}.
We note that 
terms in the 
first line of 
Eq.~(\ref{eq:pdfrenorm_NNLO}) contain 
divergent contributions that involve a convolution of a splitting function and a next-to-leading order cross section. 
The latter contains the $\ITot$
operator which 
has color-correlated terms at ${\cal O}(\ep^0)$. These 
terms are 
identical to 
those that 
appear in the operator $\IVirtSoft$ in Eq.~\eqref{eq:SigmaNccBoost}. Using 
the relation between $\CalPgen_{ab}$
and $\PAP_{ab}$
shown in  Eq.~\eqref{Eq:PqqGEN_expansion_order},
it is easy to check that the color-correlated contribution to the
$\ep$-poles cancel when Eq.~\eqref{eq:SigmaNccBoost} and 
the first  line of 
Eq.~\eqref{eq:pdfrenorm_NNLO} are combined.

\subsection{Pole cancellation in double-unresolved color-uncorrelated contributions}
\label{sect:collinearcombine}

We are now in the position to discuss the double-unresolved terms that
are free of color correlations. These terms must be collected  from 
 Eqs.~(\ref{eq:DoubVirt_exp},
 \ref{eq:DoubSoft_exp},
 \ref{eq:SoftRV_exp},
 \ref{eq:CollRV},
 \ref{eq:DoubSoft_reg_SoftColl}, 
 \ref{eq4.21}, 
 \ref{eq:Hard_DoubColl_ac}, 
 \ref{eq:Hard_DoubColl_bd})
 and Eq.~\eqref{eq:pdfrenorm_NNLO}.  They include terms with double-boosted kinematics (db), terms with a single boost from either the right (rb) or the left (lb), as well as elastic  terms (el).  We discuss these contributions separately.
We write
\index{S!$\Sigma_N^{\mathrm{coll}}$}
\index{S!$\Sigma_N^{\mathrm{c, el}}$}
\index{S!$\Sigma_N^{\mathrm{lb}}$}
\index{S!$\Sigma_N^{\mathrm{rb}}$}
\be
\Sigma_N^{\mathrm{coll}} = \SigmaNunboost + \SigmaNleft + \SigmaNright + \SigmaNdoubboost \; ,
\label{eq:SigmaNboosts}
\ee
where the superscript $``c "$ emphasizes that the first term on the right-hand side originates from collinear limits.
We begin by 
considering 
the double-boosted term, which only receives contributions from the double-collinear limits in Eq.~\eqref{eq:Hard_DoubColl_ac} and the PDFs renormalization in Eq.~\eqref{eq:pdfrenorm_NNLO}. Their sum reads
\begin{equation}
\begin{split}
    \SigmaNdoubboost = & ~ \frac{[\alpha_s]^2}{\epsilon^2} \lint \CalPgen_{aa} \conv  \FLM \conv \CalPgen_{bb} \rint + \left[\amu\right]^2 \frac{1}{\epsilon^2} \lint \PAP_{aa} \conv  \FLM \conv \PAP_{bb} \rint \\
    & + \frac{[\alpha_s]}{\epsilon^2} \left[\amu\right] \Big[\lint \PAP_{aa} \conv \FLM \conv \CalPgen_{bb}\rint + \lint \CalPgen_{aa} \conv \FLM \conv \PAP_{bb} \rint \Big] \;.
\end{split}
\label{eq4.173}
\end{equation}
Using the expansion
\index{Splitting functions!Expansion}
\be
    \CalPgen_{aa} = - \PAP_{aa} + \ep \, \PNLO_{aa} + \order{\ep^2} \; ,
\label{Eq:PqqGEN_expansion_order_v2}
\ee
and the fact that $\alpha_s(\mu)/(2\pi)=\asbr +\order{\ep^2} $, we can simplify the expression of
$\SigmaNdoubboost$ and find
\begin{equation}
    \SigmaNdoubboost = [\alpha_s]^2 \, \lint \PNLO_{aa} \conv \FLM\conv \PNLO_{bb} \rint \; , 
    \label{eq:double_boost_finite}
\end{equation}
which is finite in $\ep$.

We continue with the single-boosted terms, and demonstrate the pole
cancellation up to $\order{\ep^{-1}}$. Focusing on the left boost, i.e.~the boost applied to the 
 initial-state parton with momentum $p_1$, 
and combining selected  
contributions from Eqs.~(\ref{eq:CollRV}, \ref{eq:DoubSoft_reg_SoftColl}, \ref{eq:Hard_DoubColl_ac}, \ref{eq:Hard_DoubColl_bd}) and Eq.~\eqref{eq:pdfrenorm_NNLO}, 
we obtain the following result
\be
\begin{split}
    \SigmaNleft =&~ 
    \asbr^2 \llint \PNLO_{aa} \conv \big[ \ITot(\ep) \colorprod \FLM \big] \rrint
    + \asbr \lint \PNLO_{aa} \conv \FLVfin \rint \\
    & + \frac{1}{2\ep^2} \bigg\langle \bigg\{ \asbr^2  \convPgenPgen{aa}{aa} + \left[\amu\right]^2   \big[ \PAP_{aa} \conv \PAP_{aa} \big] 
    \\
     & + 2\asbr \left[\amu\right] \big[ \PAP_{aa} \, \bar{\conv} \,  \CalPgen_{aa} \big] \bigg\} \conv  \FLM \bigg\rangle  \\
    & + \frac{\asbr^2}{2\epsilon^3} \llint \left[ \Ca \hc(\eps)   \left(\CalPGenFour_{aa} - \CalPLoopgen_{aa}\right) + \epsilon \, G_1 \right]  \conv \FLM\rrint \\
    & - \left[\amu\right]^2 \frac{\beta_0}{2\eps^2} \lint \PAP_{aa}  \conv \FLM\rint- \asbr \left[\amu\right] \frac{\beta_0}{\eps^2} \lint\CalPgen_{aa} \conv \FLM\rint  \\
  &  + \frac{\asbr^2}{2\eps^2} N_{sc}^{(b,d)} \llint \gamma_{z,g \to gg}^{22}(\eps) \CalPGenFour_{aa} \conv \FLM \rrint  \; ,
  \label{eq:leftboosted_epm2}
\end{split}
\ee
where we have dropped irrelevant  $\order{\ep}$
terms in the first line, and we used  the bar-convolution, defined in Eq.~\eqref{eq_bar_conv}.

The two terms on the first line of Eq.~\eqref{eq:leftboosted_epm2} are clearly finite in $\eps$. As for the sum of the second and third lines, we recall that 
\index{Splitting functions!Expansion}
\be
\CalP_{aa}^{(k),\rm gen} = -\PAP_{aa} + \order{\ep} \; ,
    \label{Eq_Pqq_kGEN_expansion_order_ep}
\end{equation}
and hence
\begin{equation}
\begin{split}
    \asbr^2  \convPgenPgen{aa}{aa} = &~ \asbr^2 \big[\PAP_{aa} \bar{\conv} \PAP_{aa} \big] + \order{\ep} \; , \\
    2\asbr \left[\amu\right] \big[ \PAP_{aa} \, \bar{\conv} \,  \CalPgen_{aa} \big] = &~ - 2 \asbr^2 \big[\PAP_{aa} \bar{\conv} \PAP_{aa} \big] + \order{\ep} \; .
\end{split}
\end{equation}
The two convolutions of Altarelli-Parisi splitting functions are related by
\index{Splitting functions!Relations}
\begin{equation}
\begin{split}
    & \left[\amu\right]^2 \big[\PAP_{aa} \conv \PAP_{aa} \big](z) - \asbr^2 \big[\PAP_{aa} \bar{\conv} \PAP_{aa} \big](z)\\
    = &~ \asbr^2 \int \limits_{0}^{1} \rmd z_1 \, \rmd z_2 \,  \big(1 - z_1^{-2\ep}\big) \, \PAP_{aa}(z_1) \, \PAP_{aa}(z_2) \, \delta(z-z_1z_2) + \order{\ep} \equiv \order{\ep} \; .
\end{split}
\end{equation}
It follows that the $\order{\ep^{-2}}$ poles on the second and third lines of Eq.~\eqref{eq:leftboosted_epm2} vanish.
 
To discuss the cancellation of the poles in the fourth line on the right-hand side of Eq.~(\ref{eq:leftboosted_epm2}) 
we require the functions $P_{aa}^{\rm 1L, gen}$, $P_{aa}^{(4),\rm gen}$, and  $G_1$. These quantities are defined in Eqs.~\eqref{eq:CalPoneLgen}, \eqref{eq:CalPGenFour} and (\ref{eq:Gi_IS}), respectively. Using both Eq.~\eqref{Eq_Pqq_kGEN_expansion_order_ep} and the relation
\index{Splitting functions!Expansion}
\begin{equation}
    \CalPoneLgen_{aa} = -\PAP_{aa}+\order{\ep} \; ,
\end{equation}
we see that $\order{\ep^{-3}}$ poles disappears. Furthermore, using
\index{Splitting functions!Relations}
\begin{equation}
\begin{split}
    &
    \hc(\ep) \Big(\CalPGenFour_{aa}(z,E_1) - \CalPLoopgen_{aa}(z,E_1)\Big) =   2 \ep \log(z) \PAP_{aa}(z) + \order{\ep^2}\; , \\
    &\ep \, G_i(z,E_1) =  - 2  \ep \log(z) \PAP_{aa}(z) + \order{\ep^2} \; ,
\end{split}
\end{equation}
we observe the cancellation of $\order{\ep^{-2}}$ poles in the fourth line of Eq.~(\ref{eq:leftboosted_epm2}). The cancellation of the $\order{\ep^{-2}}$ poles in the last two lines of Eq.~\eqref{eq:leftboosted_epm2}
follows from the expansions  $\gamma_{z,g \to gg}^{22}(\eps) =11/6\, \Ca +\order{\ep}$  and  $N_{sc}^{(b,d)} = 1 + \order{\ep}$, and recalling that $\beta_0 = 11/6\, \Ca$ in our setup. Demonstrating the complete cancellation of the single poles takes more effort. We comment on this point at the end of this section.

Finally, we discuss the
pole cancellation 
in elastic terms. We begin by summing terms that arise 
from hard-collinear limits and that do not involve contributions from virtual loops.
These terms can be found in  Eqs~(\ref{eq:CollRV}, \ref{eq:DoubSoft_reg_SoftColl}, \ref{eq:Hard_DoubColl_ac}) and \eqref{eq:Hard_DoubColl_bd}. The result reads
\be
\begin{split}
\SigmaNunboost ={}&
 \asbr^2
 \bigg\{
 \llint  \left[ \frac{\IColl^2(\ep)}{2}- \frac{\beta_0}{\eps} \frac{\Gamma(1-\ep)}{e^{\ep \EulerGamma}}  \IColl(\ep)   \right] \colorprod \FLM \rrint \\
& + \frac{1}{\ep^2} \llint \bigg[ \Ca \hc(\eps) \left(  \ICollFour(\eps)   
- \IColltilde(2\eps)\right) + \frac{1}{2} \sum_{i=3}^{\Np} G_i  \bigg] \colorprod \FLM\rrint \\
&+\frac{1}{\eps} \llint N_{sc}^{(b,d)}  \left[ \gamma_{z,g \to gg}^{22}(\eps)   \ICollFour(\ep) \right] \colorprod \FLM \rrint 
\bigg\}\\
&+ \asbr \lint \IColl(\ep) \colorprod \FLVfin \rint \; . 
\label{eq:unboosted_epm2}
\end{split}
\ee
In Eq.~\eqref{eq:NNLO_col_corr_unboosted} we defined the color-correlated component of the elastic term $\SigmaNccEl$, and in the discussion that followed we demonstrated that the color-correlated poles vanish. However, this still left color-uncorrelated poles in $\SigmaNccEl$, starting at $\order{\ep^{-2}}$. Combining this term with $\SigmaNunboost$ we find
\be
\begin{split}
    \SigmaNccEl & + \SigmaNunboost =
    \asbr \lint
 \ITot(\eps) \colorprod \FLVfin \rint
 + \asbr^2 
 \bigg\{
 \frac{1}{2} \lint \ITot^2(\eps) \colorprod \FLM \rint  + K \lint \ITot(2\ep) \colorprod \FLM \rint \\
 & +  \frac{\beta_0}{\ep} 
\frac{\Gamma(1-\eps)}{e^{\eps \EulerGamma}} 
 \llint \Big( 
 \ITot(2\ep) -\ITot(\eps) \Big) \colorprod  \FLM 
 \rrint  \\
& +   \frac{\beta_0}{\ep} 
\frac{\Gamma(1-\eps)}{e^{\eps \EulerGamma}} \llint \big( \tilde c(\ep) -1 \big) \; \ISofttilde(2\ep)
\colorprod  \FLM \rrint  +  \SigmaNccElbetafin \\ 
 &+  
 \llint  \left [ 
\Ca \left(\frac{c_1(\eps)}{\eps^2} - \frac{A_K(\ep)}{\eps} - 2^{2+2\ep}  \delta_g^{\Ca}(\eps)   \right) -K 
\right ]\ISofttilde(2\eps) \cdot \FLM \rrint \\
& + 
 \Big\langle  K\left(\ISofttilde(2\eps) - \ISoft(2\eps) \right)  \colorprod \FLM \Big\rangle \\
 & + \frac{1}{\ep^2} \llint \bigg[ \Ca \hc(\eps) \left(  \ICollFour(\eps)   - \IColltilde(2\eps)\right) - K \ep^2 \IColl(2 \ep) + \frac{1}{2} \sum_{i=3}^{\Np} G_i  \bigg] \colorprod \FLM\rrint \\
 &+\frac{1}{\eps} \llint \left[ N_{sc}^{(b,d)}   \gamma_{z,g \to gg}^{22}(\eps)   \ICollFour(\ep) 
 - \beta_0 \frac{\Gamma(1-\ep)}{e^{\ep \EulerGamma}} \IColl(2\ep) \right] \colorprod \FLM \rrint \bigg\} \; .
 \label{eq:unboosted_all}
\end{split}
\ee
The terms in the first line are manifestly finite. We explained in  Section~\ref{sec:color_correlations} that $\ITot(2\ep) - \ITot(\ep) = \order{\ep}$; thus the second line is finite as well. The third, fourth, and fifth lines give rise to $\order{\ep^{-1}}$ poles only; this follows from the fact that the 
highest pole in $\ISofttilde$ is $\order{\ep^{-2}}$,  but the  coefficients
of $\ISofttilde$
suppress this singularity as can be seen by using Eqs.~\eqref{eq:cep_expansion},~\eqref{eq:ItildeSoftISoft} and~\eqref{eq:4.166}. Likewise, the fifth line contains poles of $\order{\ep^{-1}}$, since 
\index{Collinear operators!Relations}
\begin{equation}
    \Ca \hc(\eps) \left(\ICollFour(\eps) - \IColltilde(2\eps)\right) = \sum_{i=3}^{\Np} \Ca \, \ColT{i}^2 \left(-\frac{65}{72}+\frac{\pi^2}{3}\right) + \order{\ep}
    \; , 
\end{equation}
and
\index{Splitting functions!Expansion}
\begin{equation}
    G_i  = 2 \Ca \, \ColT{i}^2 \left(\frac{65}{72}-\frac{\pi^2}{3}\right) + \order{\ep} \; ,
\end{equation}
while $\ep^2  \IColl(2 \ep) = \order{\ep}$.
Finally, using Eq.~\eqref{eq:IC_k_exp} and expansions already employed in this section, we can easily check  that the last line of Eq.~\eqref{eq:unboosted_all} contains $\order{\ep^{-1}}$ poles only.\\

At this point, it is useful to 
review what we have accomplished  regarding the double-unresolved contributions. In Sections~\ref{sec:trip_color_corr}  and~\ref{sec:color_correlations}, we have combined contributions of  soft limits of real-emission amplitudes and contributions of loop amplitudes to demonstrate the cancellation of all  $\ep$-poles that contain correlators of color-charge operators. 
We are then left with $\ep$-poles proportional to squares of the color charges of the external partons. In this section, we combined these remaining divergences with the ones from hard-collinear limits and showed that all  poles multiplying double-boosted matrix elements vanish, and that poles multiplying single-boosted and elastic contributions vanish up to $\order{\ep^{-1}}$. 

We have done this by combining structures  that emerge from virtual, soft, and collinear singularities into finite operators such as $\ITot$, or, where this has not been possible, we have used simple relationships between the $\ep$-expansions of the various operators. This  dramatically simplifies the cancellation of the singularities. As a result we are able to demonstrate the cancellation of poles without resorting to excessive evaluations of multiple singular terms,  which would have been 
 needed had we followed the approach of Refs~\cite{Caola:2017dug, Caola:2019nzf}.

In order to investigate how the remaining $\order{\ep^{-1}}$ color-uncorrelated poles cancel, we need to consider the $\order{\ep^{-1}}$ terms from Eqs.~\eqref{eq:leftboosted_epm2} and~\eqref{eq:unboosted_all}, the triple-collinear and spin-correlated terms $\Sigma_N^{(2)}$ and $\Sigma_N^{(8)}$ in Eq.~\eqref{eq:NNLO_double_unresolved_unsimplified}, the term in the fourth line in Eq.~\eqref{eq:NNLO_double_unresolved_unsimplified}, and the contribution from the NLO Altarelli-Parisi kernel $\hat{P}_{qq}^{(1)}$ in the  collinear renormalization
of parton distribution functions.\footnote{Since we consider gluonic final states only, we need to remove the contribution of final state quarks from $P_{qq}^{(1)}$. The resulting expression $\PAPone_{qq,\widetilde{\mathrm{NS}}}$ is shown in Eq.~\eqref{Eq:P_AP_1_def}.}
Although it should be possible to organize the cancellation of the remaining $1/\ep$ terms following  what has been done for higher poles, it becomes much more cumbersome to do so. 
For this reason, we simply note that the cancellation of the remaining $\order{\ep^{-1}}$ poles has been checked by means of a straightforward, 
but tedious, 
algebraic computation. We emphasize that this computation is done with $N$ as a parameter, and thus holds for an \emph{arbitrary} number of gluons. Everything that is needed to confirm this cancellation  is provided in the main body of this paper and the relevant appendices. 

Having cancelled all the poles, we can take the $\ep \to 0$ limit and obtain a finite result for the NNLO contribution to the cross section $\rmd \hat{\sigma}^{\rm NNLO}_{q\bar{q}}$ for the process $1_q + 2_{\bar q} \to X + N g $. We present this result in the following section.

\section{Final result}
\label{sec:final_res}
\index{Final result for the NNLO cross section}
In this section we present a formula  for the \emph{finite}  NNLO QCD contribution
 $\rmd \hat{\sigma}_{q\bar q}^{\rm NNLO}$
to the partonic cross section of the process $1_q + 2_{\bar q} \to X + N g $.  This formula is the
main result of this paper.   As explained in the preceding sections, we arrive at this result by
considering double-real, double-virtual, real-virtual and PDF-renormalization contributions to
$\rmd \hat{\sigma}_{q\bar q}^{\rm NNLO}$ and manipulating them to remove all singularities without
impacting the fully-differential nature of the result. 
An important feature of our approach is the organization of the subtraction terms into iterations of NLO-like structures, which allows us to ameliorate the proliferation of subtraction terms that plagues NNLO calculations. As a result, 
the NNLO remainder can be written
in a very compact form.

We
split $\rmd \hat{\sigma}_{q\bar q}^{\rm NNLO}$ into contributions with $N+2$, $N+1$ and $N$
resolved final-state partons
 (c.f.~Eq.~\eqref{eq:cross_sec_resolved_partons}) and write
 \index{S!$\rmd \hat{\sigma}_{q\bar q}^{\rm NNLO}$}
  \index{S!$\rmd \hat{\sigma}_{N+2}^{\rm NNLO}$}
   \index{S!$\rmd \hat{\sigma}_{N+1}^{\rm NNLO}$}
    \index{S!$\rmd \hat{\sigma}_{N}^{\rm NNLO}$}
\begin{equation}
    \rmd \hat{\sigma}_{q\bar q}^{\rm NNLO} =    
    \rmd \hat{\sigma}_{N+2}^{\rm NNLO} + \,
    \rmd \hat{\sigma}_{N+1}^{\rm NNLO} + \,
    \rmd \hat{\sigma}_{N}^{\rm NNLO}.
\label{Eq:NNLO_partonic_cross_section_mult}
\end{equation}
The first term on the right-hand side is the finite, fully-regulated contribution  given in Eq.~
\eqref{Eq:Sigma_N+2^fin_def}.
The single-unresolved cross section  $\rmd \hat{\sigma}_{N+1}^{\rm NNLO}$  can be found  in Eq.~\eqref{Eq:dsigma_NLO_final_def}.  The double-unresolved contribution $\rmd \hat{\sigma}_{N}^{\rm NNLO}$
is obtained by combining the many different terms calculated in the previous sections.   
As was  explained there, it is convenient 
to write $\rmd \hat{\sigma}_{N}^{\rm NNLO}$
as the sum  of double-boosted, single-boosted 
and elastic terms 
 \index{S!$\rmd \hat{\sigma}_{\rm db}^{\rm NNLO}$}
  \index{S!$\rmd \hat{\sigma}_{\rm sb}^{\rm NNLO}$}
   \index{S!$\rmd\hat{\sigma}_{\rm el}^{\rm NNLO}$}
\begin{equation}
    \rmd \hat{\sigma}_N^{\rm NNLO} = 
     \rmd \hat{\sigma}_{\rm db}^{\rm NNLO} + \rmd \hat{\sigma}_{\rm sb}^{\rm NNLO} + \rmd\hat{\sigma}_{\rm el}^{\rm NNLO} \; .\label{Eq:NNLO_partonic_cross_section_final_result}
\end{equation}
We now present each contribution separately, 
using several functions that we collect in 
Appendix~\ref{app:final_res_fun}.
The double-boosted contribution is described by the very simple expression
\begin{equation}
\label{eq:finite_double_boosted}
    2s\;\rmd \hat{\sigma}_{\rm db}^{\rm NNLO} = \left[\amu\right]^2 \lint \PNLO_{qq} \conv \FLM \conv \PNLO_{qq} \rint \; ,
\end{equation}
where $\PNLO_{qq}$ is the finite remainder of NLO splitting functions, and can be found in Eq.~\eqref{eq:PNLO_app_final_res}. As expected, this contribution is independent of the multiplicity of the final state. 

The expression for the single-boosted contribution is slightly more complex and corresponds to
\begin{equation}
\label{eq:finite_reminder_SB}
\begin{split}
    2s\; \rmd \hat{\sigma}_{\rm sb}^{\rm NNLO}  = 
    & \left[\amu\right]^2 
    \bigg\{
    \lint \PNLO_{qq} \conv \big[\ITot^{(0)} \colorprod \FLM \big] \rint + \lint \big[\ITot^{(0)} \colorprod \FLM \big] \conv \PNLO_{qq} \rint \\
    & +  \lint \CalP_{qq}^{\cal W} \conv \big[\Wacfin{1} \colorprod \FLM \big] \rint + \lint \big[ \Wacfin{2} \colorprod \FLM \big] \conv \CalP_{qq}^{\cal W} \rint  \\
    & +  \lint \PNNLO_{qq} \conv \FLM \rint + \lint \FLM \conv \PNNLO_{qq} \rint  \\
    & +  \lint \PNLO_{qq} \conv \FLVfin \rint + \lint \FLVfin \conv \PNLO_{qq} \rint 
    \bigg\}\; ,
\end{split}
\end{equation}
Here, we remind the reader that $I_{\rm T}^{(0)}$ is the $\ep \to 0$ 
limit of the finite operator $I_{\rm T}(\ep)$. Its explicit expression is reported in Eq.~\eqref{eq:IT_fin}. The function $\Wacfin{i}$, appearing in the second line of Eq.~\eqref{eq:finite_reminder_SB}, is given in Eq.~\eqref{Eq:Appendix_Wacfin_i_def}, while the NNLO splitting function $\PNNLO_{qq}$ is reported in Eq.~\eqref{eq:PNNLO_app}. 

Finally, the elastic contribution reads
\begin{equation} 
\label{sigma_elastic_def}
\begin{split}
    2s\; \rmd \hat{\sigma}_{\rm el}^{\rm NNLO} = 
    & ~ \left[\amu\right]^2 \bigg\{\lint\big[ \Iccfin + I_{\rm tri}^{\rm fin} +I_{\rm unc}^{\rm fin} \big] \colorprod \FLM \rint 
    \\
     & 
    + \sum_{i=1}^{\Np}  \llint \Big[\gamma^{\cal W}(L_i) \, \theta_{i2} \, \Wacfin{i} + \delta_g^{(0)} \, \Wbdfin{i} + \delta_g^{\perp} {\cal W}_r^{(i)} \Big] \colorprod \FLM \rrint \bigg\} \\
    & + \left[\amu\right] \lint \ITot^{(0)} \colorprod \FLVfin\rint 
    + \SmnFin + \lint\FLVfinsq\rint + \lint\FLVVfin\rint \, .
\end{split}
\end{equation}
In this equation 
$\theta_{i2} = 1$ if $i$ is the final-state parton ($i > 2$) and $0$ otherwise.  
In the first line we have the combination of a double color-correlated contribution, a triple color-correlated component, and a
color-uncorrelated part. 
They are presented in Eq.~\eqref{eq:Iccfin_defn}, \eqref{eq:Itrifin_defn}, and \eqref{eq:cunc_fin_app} respectively. 
In the second line of Eq.~\eqref{sigma_elastic_def}, the functions $\gamma^{\cal W}$, $\Wacfin{i}$, $\Wbdfin{i}$ and ${\cal W}_r^{(i)}$ appear. They are
given in Eqs~\eqref{Eq:Sec5_gamma_calW_def}, \eqref{Eq:Appendix_Wacfin_i_def}, \eqref{Eq:Wbd_fin_def} and \eqref{eq:part_dep_contr}. The constants $\delta_g^{(0)}$ and $\delta_g^{\perp}$ are reported in Eq.~\eqref{eq:delta_fin_app}.
The term $\SmnFin$ in Eq.~\eqref{sigma_elastic_def} refers to the finite 
remainder of the double-soft integrated subtraction term.
It can be extracted from Ref.~\cite{Caola:2018pxp}, and its explicit expression is reported in Eq.~\eqref{eq_SmnFin}. Finally, $\FLVfinsq$ and $\FLVVfin$ are the process-dependent finite remainders of virtual amplitudes.
\\

We claim that the above result for  $\rmd \hat{\sigma}_{q\bar q}^{\rm NNLO}$ can be 
used, without further ado,  to implement the finite remainder of NNLO QCD corrections to a 
process $q \bar q \to X + Ng$ in a computer code.  In theory, this can be  done 
for \emph{arbitrary} $N$, but the practical realization of this idea will have to wait until   finite remainders for two-loop amplitudes for  such processes become available.

Nevertheless, it is important to emphasize that the form of the final results is well-suited for numerical implementation, in the sense that the parameter $N$ that controls the final state multiplicity only appears in relatively few places. Indeed, the splitting functions that appear in the boosted contributions are universal and are determined only by the flavor of the external partons and their energies. In the elastic contribution, the final state multiplicity only affects the upper limit in the sum over partons, see e.g.~Eqs~\eqref{sigma_elastic_def}, \eqref{eq:Iccfin_defn}, \eqref{eq:Itrifin_defn} and \eqref{eq:cunc_fin_app}. It follows that implementing the color-uncorrelated elastic terms in a numerical code is also quite simple for any $N$. 
It could be less trivial to implement contributions containing color correlations (e.g.\ $\ITot^{(0)}$), as these require one to evaluate color-correlated matrix elements for high-multiplicity processes.
However, even in this case a numerical implementation for a given $N$ should be straightforward, using e.g.~the ideas of color ordering.

Results of the general computation reported here can be compared 
with those obtained for specific values of $N$. The $N=0$ case corresponds to the Drell-Yan process, and the $N=1$ case to the gluonic contribution to the $V+{\rm jet}$ production. 
It is well-known that, in both  cases, the  correlators of color-charge operators can be expressed through Casimir operators. 
For example, in the case of $q_1 \bar q_2 \to V+g_3$, 
we find 
\index{Color space and algebra}
\be
\ColT{1} \cdot \ColT{2} 
= \frac{C_A}{2} - C_F \; , 
\qquad \quad
\ColT{1} \cdot \ColT{3} 
=
\ColT{2} \cdot \ColT{3} 
= -\frac{C_A}{2} \; . 
\ee
Using such expressions it is straightforward to replace  
all products of color-charge 
operators in Eqs.~(\ref{eq:finite_double_boosted},
\ref{eq:finite_reminder_SB},
\ref{sigma_elastic_def}) with the 
corresponding Casimir operators.
One can also easily check that the partition functions defined 
for generic $N$ turn into structures already used in earlier  computations.  It follows from the 
definition in Eq.~\eqref{eq:Delta_NNLO_def} that $\Delta^{(ij)} =1$
for  the Drell-Yan process and  
\be
\Delta^{(\Fp \Sp)}
= \frac{p_{\perp,3}}{
p_{\perp,3} + p_{\perp,\Fp}
+p_{\perp,\Sp}} \; ,
\ee
for $V+{\rm jet}$ partitioning.   
Similarly, it is easy to see 
that $\omega$-partitions are the 
same as those used in Ref.~\cite{Caola:2017dug, Caola:2019nzf} for $N=0$ and Ref.~\cite{Boughezal:2013uia} for $N=1$.

We have reproduced the analytic results for the finite NNLO remainders for Drell-Yan production that were reported in Ref.~\cite{Caola:2019nzf} starting from Eqs~(\ref{eq:finite_double_boosted},
\ref{eq:finite_reminder_SB},
\ref{sigma_elastic_def}), and setting $N=0$.
We have also checked 
 that, upon setting $N=1$,  the  general formulas 
 reproduce the results of a dedicated  computation of the  NNLO QCD corrections to the process $q \bar q \to 
 V+g$ that we performed  earlier. 
 Although this computation  was also based on the nested soft-collinear subtraction scheme, it was organized very differently,  with an emphasis on separately  integrating all the different  subtraction terms over unresolved phase spaces  \emph{before} combining and simplifying them.  The two approaches are sufficiently independent to provide an important check of the general-$N$ formula that we reported in this section.
\section{Conclusions}

In this paper, we have shown how to use  the nested soft-collinear subtraction scheme to describe the production of a 
generic  color-singlet state  accompanied by  an \emph{arbitrary} number of gluons in quark-antiquark annihilation at NNLO QCD. 
We have identified recurring structures associated with the sums of single-soft, single-collinear  and one-loop virtual corrections. We have also shown that by organizing the calculation  in such a way that the iterative nature of these finite contributions is fully exposed, 
much of the complexity of NNLO computations related to an interplay of soft and collinear singularities can be ameliorated. This has allowed us to demonstrate the cancellation of all color-correlated poles, as well as color-uncorrelated poles through $\order{\ep^{-2}}$, in a  straightforward manner. We have also confirmed the cancellation of the remaining $\eps$-poles, and obtained compact expressions for the finite subtraction terms, which we have checked, where possible, against previous results and independent calculations. 
To the best of our knowledge, it is the first time that such expressions have been presented for the production of an arbitrary  number of gluons at a hadron collider.\footnote{See Ref.~\cite{Chen:2022ktf} for a related analysis in the context of the antenna subtraction scheme.} 

Although we considered a $q\bar{q}$ initial state in this paper, many of our arguments apply to $gg$ annihilation as well; the only modifications required for this channel would be the use of gluon  splitting functions in place of the quark  ones as well as the necessary changes in the color charges where appropriate. These modifications are clearly minor and do not impact the logic of the  computation that we report in this paper. 

 The results of this study provide a necessary step towards the complete generalization of the nested soft-collinear subtraction scheme to arbitrary initial and final states. 
 Indeed, on the one hand, the gluonic final state ensures that the maximal number of infrared and collinear singularities are present, so processes with final state quarks should have a simpler singularity structure. On the other hand, we relied on the symmetries of the final state and particular 
 features of the initial state, 
 and this will not be possible if generic processes are considered. Although nothing will change as a matter of principle, the 
 combinatorics of 
 collinear limits will become   more complicated. We look forward to addressing these issues in future studies.

\label{sec:conclusions}

\section*{Note added}
We thank Matteo Tresoldi for carefully cross-checking our calculations and identifying minor errors, which have now been corrected.

\acknowledgments
We would like to thank Konstantin Asteriadis, Federico Buccioni, Fabrizio Caola, Lorenzo Magnea and Paolo Torrielli for their helpful comments on the manuscript. 
We would also like to thank Federico Buccioni for providing an independent calculation of the phase space 
integration of the soft real-virtual subtraction term.
RR and DMT are thankful to the Institute for Theoretical Particle Physics at KIT for hospitality 
extended to them during the course of the work on this paper.
KM would like to thank the Galileo Galilei Institute for Theoretical Physics in Florence for the hospitality, 
and the INFN for the partial support during the
completion of this work. 
The research of KM and CSS was partially supported by the Deutsche
Forschungsgemeinschaft (DFG, German Research Foundation) under grant
396021762-TRR 257. The research of FD was supported by the ERC Starting Grant 804394 \textsc{hipQCD}.


\appendix

\section{Constants, angular integrals, splitting functions, anomalous dimensions and fundamental operators}
\label{sec:Splitting}
In this section 
we provide a collection of formulas that 
are used throughout the main text of this paper. They include:
\begin{enumerate}[label=(\roman*)]
    \item various constants in Appendix~\ref{sebsec:useful_constants};
    \item angular integrals in Appendix~\ref{app:useful_integral};
    \item the relevant Altarelli-Parisi splitting functions in Appendix~\ref{subsec:ap_splitting};
    \item generalized splitting functions and anomalous dimensions in Appendix~\ref{subsec:split_anomdim};
    \item operators arising from soft and collinear limits as well as from virtual corrections, and useful relations between them in Appendix~\ref{sec:subsec_operator_defn};
\end{enumerate}

\subsection{Useful constants}
\label{sebsec:useful_constants}

\index{Color space and algebra}
Here we summarize the various constants that we have introduced throughout the manuscript.
First we discuss the notations related to color.  Following Ref.~\cite{Catani:1996jh}, we denote the color-charge operators  with $\ColT{i}$; squares 
of color-charge operators are the 
Casimir operators of the corresponding representations of SU(3). They read 
$\ColT{q}^2 =\ColT{\bar q}^2= C_F$, $\ColT{g}^2 = C_A$,  where $\Cf=(\Nc^2-1)/(2\Nc)$, $\Ca=\Nc$, and $N_c= 3$ is the number of colors. Quark and gluon anomalous dimensions read  $\gamma_q = 3/2 \, \Cf$ and $\gamma_g = 11/6 \, \Ca - 2/3 \, \TR \nf$, where $T_R = 1/2$ and $n_f$ is the number of massless quark flavors.  

We renormalize the strong coupling in the $\overline{\rm MS}$ scheme, i.e.
\index{G!$\gsb$} 
\begin{align}
  \label{eq:as-renorm}
  \gsb^2
  ={}&
  \gs^2
  \Sep
  \mu^{2\ep}
  \left[
    1
    -
    \amu
    \frac{\beta_0}{\ep}
    +
    \left(
      \amu
    \right)^2
    \left(
      \frac{\beta_0^2}{\ep^2}
      -
      \frac{\beta_1}{2\ep}
    \right)
    +
    \mathcal{O}(\as^3)
  \right]
  \;,
\end{align}
where $\Sep = (4\pi)^{-\ep} e^{\ep\EulerGamma}$ and 
\index{B!$\beta_0$} 
\index{B!$\beta_1$} 
\be
\begin{split}
  \beta_0
  =
  \frac{11}{6}\Ca
  - \frac{2}{3}\TR\nf
  = \gamma_g
  \;,
  \qquad
  \beta_1
  ={}&
  \frac{ 17}{6} \Ca^2 - \frac{5}{3}\Ca\TR\nf - \Cf\TR\nf \; .
\end{split}
\ee
We note that we only consider gluons in the final state, so that $n_f$ is set to zero throughout this paper. 
Furthermore,  it is convenient 
to define the following coupling
\index{A!$\asbr$} 
\index{A!$\alpha_s$} 
\begin{align}
  \label{eq:asbr-def}
  \asbr
  \equiv{}&
  \frac{ \as(\mu) }{ 2\pi }
  \frac{e^{\ep\EulerGamma} }{ \Gamma(1-\ep) }
  \;.
\end{align}

Then, combining Eqs.~\eqref{eq:as-renorm}
and~\eqref{eq:asbr-def}, we find 
\begin{align}
  \gsb^2
  ={}&
  8\pi^2
   \asbr \,
   \frac{ \Gamma(1-\ep) }{ (4\pi)^\ep } \,
   \Big[ 1 + \mathcal{O}(\as) \Big]
  \;.
\end{align}
\\

In the main text of this paper we encounter a number of angular integrals, for which we introduce the following normalization constants:
\index{N!$N_\ep^{(b,d)}$}
\index{N!$N_{\Fp \parallel \Sp}$}
\index{N!$N_c$}
\be
\label{eq:normalisation}
\begin{split}
N_\ep^{(b,d)} ={}&  \frac{\Gamma(1-\ep) \, \Gamma(1+2\ep)}{\Gamma(1+\ep)} 
= 1+\frac{\pi^2}3\ep^2 + \mathcal{O}(\ep^3)\; ,
\\
N_{\Fp || \Sp}(\ep) ={}& 2^{2\epsilon} \frac{\Gamma(1+2\epsilon) \Gamma(1-2\epsilon)}{\Gamma(1+\epsilon) \Gamma(1-\epsilon)}
=
1
+2 \epsilon  \log 2
+\frac{1}{2} \ep^2 \big(\pi ^2
+4 \log^2 2\big)
+\mathcal{O}(\ep^3)\; ,
\\
N_c(\ep) ={}& -
\frac{\Gamma(1-\ep) \Gamma(1-2\ep)}{\Gamma(1-3\ep)}
+ 
\frac{2\Gamma^2(1-\ep)}{\Gamma(1-2\ep)}
= 1 + \mathcal{O}(\ep^3) \; .
\end{split}
\ee 
We note that all the above normalization constants are equal to one to zeroth order in $\eps$.\\

\noindent To describe virtual corrections we have used the convention of Ref.~\cite{Catani:1998bh,Catani:2000pi}
\index{L!$\lambda_{ij}$}
\index{K!$\kappa_{ij}$}
\be
\label{eq:phase_factors}
\begin{split}
&
\lambda_{ij}=
\begin{cases}
+ 1 & i \text{ and } j \text{ are both incoming or outgoing} \; ,
\\
0  & \text{otherwise} \; ,
\end{cases} 
\\
&
\kappa_{ij} = \big( \lambda_{ij}-\lambda_{i \Fp}-\lambda_{j \Fp}\big)
= 
\begin{cases}
+ 1 & i \text{ and } j \text{ are both incoming or outgoing} \; ,
\\
-1 & \text{otherwise} \; .
\end{cases}
\end{split}
\ee
For double-virtual
amplitudes we have 
used the following constants~\cite{Catani:1998bh} 
\index{K!$K$}
\index{C!$c_\ep$}
\be
\label{eq:constants_VV}
\begin{split}
    K ={}& \left( \frac{67}{18} - \frac{\pi^2}{6} \right) C_A  - \frac{10}{9} \TR \nf \; , \\
    c_\ep ={}& \frac{e^{-\ep \gamma_E} \Gamma(1-2\ep)}{\Gamma(1-\ep)} = 1+ \frac{\pi^2}{4} \, \ep^2 + \frac73 \zeta_3 \, \ep^3 + \mathcal{O}(\ep^4)  \; .
\end{split}
\ee
To describe integrated 
 double-soft limits  (see Eq.~\eqref{eq4.105}),  we have introduced
\index{C!$c_1$}
\index{C!$c_2$}
\index{C!$c_3$}
\be
\label{eq:constants_SS}
\begin{split}
    c_1(\epsilon) 
    ={}&
    1 + \left(\frac{\pi^2}{6} - \frac{32}{9}\right)\epsilon^2 + \left(\frac{217}{27} - \frac{137}{9}\log2 - 22 \log^2 2 + \frac{11 \zeta_3}{2}\right)\epsilon^3 \; , \\
    c_2(\epsilon) ={}&
    1 + \frac{\pi^2}{3} \epsilon^2  \; ,
    \\
    c_3(\epsilon) ={}& 4 \log 2+ 8 \ep \log^2 2 \; .
\end{split}
\ee
We emphasize that $c_{1,2,3}$ do not contain powers of $\eps$ beyond those shown above. \\

\noindent To compute soft and collinear  limits 
of the real-virtual contribution 
$\FLRV$, we used 
\index{A!$A_K$}
\index{H!$\hc$}
\be
\label{eq:constants_RV}
\begin{split}
A_K(\ep) ={}& \frac{\Gamma^3(1+\ep) \, \Gamma^5(1-\ep)}{\Gamma(1+2\ep) \, \Gamma^2(1-2\ep)} = 1 - \frac{\pi^2}{3} \eps^2 + \order{\eps^3} \; ,
\\
\hc(\ep) ={}&
\frac{\Gamma^2(1-2\eps) \Gamma(1+\eps)}{\Gamma(1-3\eps)} = 1 + \order{\eps^3} \; .
\end{split}
\ee
We also have defined (see Eq.~\eqref{eq4.21})
\index{D!$\delta^{\Ca}_g(\eps)$}
\index{D!$\delta_g^{\beta0}(\eps)$}
\be
\label{eq.constants_spin_av}
    \delta^{\Ca}_g(\eps) =  \left(-\frac{131}{72}+\frac{\pi^2}{6}\right)  + \order{\eps} \; ,  
    \qquad 
    \delta_g^{\beta0}(\eps) = \log 2 + \order{\eps} \; .
    \ee
When combining the 
unboosted terms involving color correlations (see Section~\ref{sec:color_correlations}), we require the following combinations of some of the above constants
\index{C!$\tilde c$}
\be
\begin{split}
& \tilde c(\ep) = \frac{e^{\eps \EulerGamma}} {\Gamma(1-\eps)}
\left (c_2(\ep) + \eps c_3(\eps) - 2^{2+2\ep} \ep \, \delta_g^{\beta_0}(\ep) \right )
= 1
+ \order{\ep^2} \; ,\\ 
&\Ca
\left ( 
\frac{c_1(\eps)}{\eps^2} - \frac{A_K(\ep)}{\eps^2} - 2^{2+2\eps} \delta_g^{\Ca}(\eps) 
\right ) - K 
=  \order{\eps} \; . 
\end{split}
\ee

\subsection{A collection of simple angular integrals}
\label{app:useful_integral} 

Throughout the manuscript we make use 
of various integrals over the angles of 
unresolved gluons. We summarize some of the useful formulas here. 
First, we define the normalized element 
of the solid angle in $(d-1)$- and 
$(d-2)$-dimensions
\index{O!$[\rmd \Omega_i^{(d-1)}]$}
\index{O!$[\rmd \Omega_i^{(d-2)}]$}
\index{O!$[\Omega^{(d-2)}]$}
\be
[\rmd \Omega_i^{(d-1)}]
\equiv
\frac{\rmd \Omega_i^{(d-1)}}{2 (2\pi)^{d-1}} \;, 
\qquad
[\rmd \Omega_i^{(d-2)}]
\equiv
\frac{\rmd \Omega_i^{(d-2)}}{2 (2\pi)^{d-1}} \; .
\ee
Then, we find 
\be
[\Omega^{(d-2)}]
\equiv
\int [\rmd  \Omega^{(d-2)}]
= \frac{1}{8\pi^2} \, 
\frac{(4\pi)^\ep}{\Gamma(1-\ep)} \; .
\ee
Furthermore,  we use 
\be
\begin{split}
\int \frac{[\rmd\Omega_a^{(d-1)}]}{[\Omega^{(d-2)}]} \;  \frac{ \rho_{ij} }{ \rho_{ia} \rho_{ja} }
={}&
-\, \frac{2^{1-2\ep}}{\ep} \, \eta_{ij}^{-\ep} \, K_{ij} \; , 
 \\
\int \frac{[\rmd\Omega_a^{(d-1)}]}{[\Omega^{(d-2)}]} \;  \frac{ 1 }{  \rho_{ia} }  
={}&
-\frac{2^{-2\ep}}{\ep} \, \frac{\Gamma^2(1-\ep)}{\Gamma(1-2\ep)}  \; , 
 \\
\int \frac{[\rmd\Omega_a^{(d-1)}]}{[\Omega^{(d-2)}]} \
\Big(\frac{\rho_{ia}}4\Big)^{-\ep} \,  \frac{1}{\rho_{ia}}
={}&
- \frac{2^{-2\ep}}{2\ep} \, 
\frac{2^{\ep} \, \Gamma(1-\ep) \, \Gamma(1-2\ep)}{\Gamma(1-3\ep)}  \; ,
\end{split}
\ee
\index{K!$K_{ij}$}
where $K_{ij}$ is given by (cf.~Eq.~\eqref{eq:Kij_defn0})
\be
    K_{ij} = \frac{\Gamma^2(1-\eps)}{\Gamma(1-2\eps)} \eta_{ij}^{1+\epsilon} \hypF(1,1,1-\eps,1-\eta_{ij}) \; .
    \label{eq:Kij_defn} 
    \ee

Other integrals that we require involve the collinear limits acting on the angular phase space measure; they can be computed using the phase space parametrization described in  Appendix~\ref{sect:phasespace}. Here we just give two examples that appear frequently  
\be
\int [\Coll{ij} \,  \rmd \Omega_{j}^{(d-1)}] \, \frac{1}{\rho_{ij}}
=
-\frac{2^{-2\ep}}{\ep} \, 
\Big[\frac{1}{8\pi^2} \, 
\frac{(4\pi)^\ep}{\Gamma(1-\ep)}\Big] \; ,
\ee
and
\be
\int [\Coll{ij} \,  \rmd \Omega_{j}^{(d-1)}] \, \frac{1}{\rho_{ij}} \, 
\Theta\Big(\eta_{ij} < \frac{\eta_{ik}}{2} \Big)
=
-\frac1{\ep} \, 
\Big[\frac{1}{8\pi^2} \, 
\frac{(4\pi)^\ep}{\Gamma(1-\ep)}\Big] \rho_{ik}^{-\ep} \;  .
\ee

\subsection{Altarelli-Parisi splitting functions}
\label{subsec:ap_splitting}
In this section we report the Altarelli-Parisi splitting functions that we use in this paper. The only leading order splitting function that we require reads 
\index{P!$\PAP_{qq}$}
\be
\label{Eq:PAP_0_definition}
\begin{split}
    \PAP_{qq}(z) = & ~ \Cf \left[2\D_0(z) - (1+z) + \frac{3}{2} \delta(1-z) \right] \; , 
\end{split}
\ee
where
\index{D!$\D_n$}
\begin{equation}
    \D_n(z) \equiv \left[\frac{\log^n(1-z)}{1-z}\right]_+ \; .
\end{equation}
At NLO, we need the non-singlet splitting
function from which the contribution of identical quarks has been subtracted, which reads 
\index{P!$\PAPone_{qq,\widetilde{\mathrm{NS}}}$}
\begin{equation}
\begin{split}
	\PAPone_{qq,\widetilde{\mathrm{NS}}}(z) = & ~ \Ca \Cf \bigg[\frac{\pi^2}{6}(1+z) - \frac{62}{9}z - \frac{19}{18} + \left(\frac{67}{9} - \frac{\pi^2}{3}\right) \D_0(z) \\  
	& + \frac{2 + 11z^2}{6(1-z)}\log z - \frac{1+z^2}{1-z} \Li_2(1-z) + \delta(1-z)\left(\frac{17}{24} + \frac{11}{18}\pi^2 - 3\zeta_3\right)\bigg] \\
	& + \Cf^2 \bigg[3 - 2z - 2\frac{1+z^2}{1-z} \log(1-z) \log z + 2\log z + \frac{1+3z^2}{2(1-z) }\log^2 z \\
	& + 2\frac{1+z^2}{1-z} \text{Li}_2(1-z) + \delta(1-z) \left(\frac{3}{8} - \frac{\pi^2}{2} + 6\zeta_3\right)\bigg] \; .
    \label{Eq:P_AP_1_def}
\end{split}
\end{equation}

\subsection{Generalized splittings and anomalous dimensions}
\label{subsec:split_anomdim}
\subsubsection{Tree-level}
\label{subsec:split_tree_level}
We start by introducing the two tree-level splitting functions needed throughout the paper
\index{P!$\Pqq$}
\index{P!$\Pgg^{\mu\nu}$}
\begin{equation}
\begin{split}
\label{eq:def_hatPqq_hatP_gg_mu_nu}
    \Pqq(z) = & ~ C_F \left[\frac{1+z^2}{1-z} - \epsilon(1-z)\right] \; , \\
	\Pgg^{\mu\nu}(z) = & ~ 2C_A \left[-g^{\mu\nu}\left(\frac{1-z}{z} + \frac{z}{1-z}\right) + 2(1-\epsilon) z(1-z) \kappa_\perp^\mu \kappa_\perp^\nu\right] \;  ,
\end{split}
\end{equation}
where $\kappa_\perp^\mu$ is a transverse momentum defined as
\index{K!$\kappa_\perp^\mu$}
\begin{equation}
	\kappa_\perp^\mu = \frac{k_\perp^\mu}{\sqrt{-k_\perp^2}} \; , \qquad \kappa_\perp^2 = -1 \;  .
\end{equation}
We also need the gluon spin-averaged splitting function
\index{P!$\Pgg(z)$}
\begin{align}
	\Pgg(z) = ~ 2C_A \left[\frac{z}{1-z} + \frac{1-z}{z} + z(1-z)\right] \; .
    \label{def:Pgg_av}
\end{align}
To describe the spin-correlated component arising from sectors $(b)$ and $(d)$ we have introduced the functions
\index{P!$\Pgg^{\bot}$}
\index{P!$\Pgg^{\bot, r}$}
\index{P!$\Pgg^{(0)}$}
\index{P!$\Pgg(z,\ep)$}
\begin{align}
\Pgg^{\bot}(z)  ={}& 4 \Ca (1-\ep) \, z \, (1-z)\; , \label{def:Pgg_bot}
\\
\Pgg^{\bot, r}(z) ={}& 2 \Ca\, z (1-z) 
 (1 - 2 \ep) \; ,
 \label{def:Pgg_bot_r}
 \\
\Pgg^{(0)}(z) ={}& 2 \Ca \left( \frac{z}{1-z} + \frac{1-z}{z}\right) \; ,
 \\
\Pgg(z,\ep) ={}& \Pgg^{(0)}(z)
+  \frac12
\Pgg^\bot(z) 
=
2 \Ca \left( \frac{1-z}{z}
+
\frac{z}{1-z}
+z (1-z) \, (1-\ep) \right) \; .
\label{def:Pgg_z_ep}
\end{align}
We also require the following integral of the soft-subtracted function $\Pgg$ over $z$ 
\index{G!$\gamma_{f(z),g \to g g}^{nk}(\eps,L_i)$}
\be
\begin{split}
    \gamma_{f(z),g \to g g}^{nk}(\eps,L_i) = & - \int\limits_{0}^{1} \rmd z\, (1-S_z) \left[z^{-n \eps} (1-z)^{-k\eps} \; f(z) \, \Pgg(z)\right] 
    \\
    & + 2\Ca \frac{1 - e^{k\eps L_i}}{k \eps} f(1) \; ,
    \label{eq:defn_gamma_nk_appA}
    \end{split}
\ee
where $S_z$ stands for the soft  $z \to 1$ limit and $L_i = \log(\Emax/E_i)$. We also define the following integrals over $z$
\index{G!${\gamma}_{\bot, g\rightarrow gg}^{22}$}
\index{G!${\gamma}_{\bot, g\rightarrow gg}^{22, r}$}
\be
\label{eq:gamma_tilde}
{\gamma}_{\bot, g\rightarrow gg}^{22}
= - \int \limits_0^1 \rmd z \; \frac{\Pgg^\bot(z)}{ [ z (1-z) ]^{2\ep}} 
 \; ,
\qquad 
{\gamma}_{\bot, g\rightarrow gg}^{22, r}
=
-\int\limits_{0}^1 \rmd z \; \frac{\Pgg^{\bot, r}(z)}{[ z (1-z) ]^{2\ep}}
 \; ,
\ee
as well as  integrals over $z$ and the  energy of the unresolved parton 
\index{D!$\delta_g^{\rm sa}$}
\index{D!$\delta_g^{\bot, r}$}
\index{D!$\delta_g^\bot$}
\index{D!$\delta_g$}
\be
\label{Eq:delta_g_def}
\begin{split}
    \delta_g^{\rm sa}(\epsilon) =&~
\frac{N_\epsilon^{(b,d)}}{2} \Emax^{4\epsilon} 
    \int \limits_{\Emax}^{2\Emax} \frac{\rmd E_{\Fp }}{E_{\Fp}^{1+4\ep }} \int\limits_{1-\xi}^{\xi} \rmd z\, [z(1-z)]^{-2\epsilon} \Pgg(z) \; ,
    \\
    \delta_g^{\bot, r}(\ep) = &~ \frac{N_\ep^{(b,d)}}2 \, E_{\rm max}^{4\ep} \, \int \limits_{E_{\rm max}}^{2E_{\rm max}} \frac{\rmd E_{\Fp }}{E_{\Fp}^{1+4\ep }}   \int \limits_{1-\xi}^{\xi}
    \rmd z \, [ z (1-z) ]^{-2\ep} \; \ep \, \Pgg^{\bot,r}(z) \; , 
    \\
    \delta_g^\bot(\ep) = &~ \frac{N_\ep^{(b,d)}}2 \, E_{\rm max}^{4\ep} \, \int \limits_{E_{\rm max}}^{2E_{\rm max}} \frac{\rmd E_{\Fp }}{E_{\Fp}^{1+4\ep }}  \int \limits_{1-\xi}^{\xi}
    \rmd z \, [ z (1-z) ]^{-2\ep} \; \Pgg^\bot(z) \; , 
    \\
    \delta_g(\ep) = &~ \frac{N_\ep^{(b,d)}}2 \, E_{\rm max}^{4\ep} \, \int \limits_{E_{\rm max}}^{2E_{\rm max}} \frac{\rmd E_{\Fp }}{E_{\Fp}^{1+4\ep }}  \int \limits_{1-\xi}^{\xi}
    \rmd z \, [ z (1-z) ]^{-2\ep} \; \Big(\Pgg(z, \ep) + \ep \, \Pgg^\bot(z) \Big) \; ,
\end{split}
\ee
where we have defined $\xi = \Emax/E_{\Fp}$.

For the configurations where a final state gluon becomes collinear to an initial state parton $1_a$, we require convolutions of the type
\index{Convolution}
\begin{equation}
\begin{split}
    \int \limits_{0}^{1} \rmd z\, \CalP_{aa}^{(k)}(z,E_1) \,  g(z) = & \int \limits_{0}^{1} \rmd z\, \bigg[(1 - S_z) \Big[(1-z)^{-k \epsilon} \Paa(z)\Big] \\
    & - 2\T_{a}^2 \, \frac{1 - e^{-k \epsilon L_1}}{k \epsilon}\delta(1-z)\bigg] \, g(z) \; ,
    \label{Eq:P_aa_k_definition}
\end{split}
\end{equation}
where $k=2$ at NLO and $k=4$ at NNLO.
It is worth rewriting the above splitting function as
\begin{equation}
   - \left[\left(\frac{2E_1}{\mu}\right)^{-2\epsilon} \frac{\Gamma^2(1-\epsilon)}{\Gamma(1-2\epsilon)}\right]^{\frac{k}{2}} \CalP_{aa}^{(k)}(z,E_1) = \Gamma_{1,a}^{(k)} \, \delta(1-z) + \mathcal{P}_{aa}^{(k),\text{gen}}(z,E_1) \; ,
   \label{Eq:Paa_k_relation_with_Gamma_a_and_Paa_GEN_k}
\end{equation}
with
\index{G!$\Gamma_{1,a}^{(k)}$}
\index{P!$\CalP_{ij}^{(k),\text{gen}}$}
\begin{align}
    \Gamma_{1,a}^{(k)} = & \left[\left(\frac{2E_1}{\mu}\right)^{-2\epsilon} \frac{\Gamma^2(1-\epsilon)}{\Gamma(1-2\epsilon)}\right]^{\frac{k}{2}} \left[\gamma_{a} + 2\T_{a}^2 \frac{1 - e^{-k\epsilon L_1}}{k \epsilon}\right] \; , \label{Eq:Gamma_1_2_definition_k_general}\\
    \CalP_{aa}^{(k),\text{gen}}(z,E_1) = & \left[\left(\frac{2E_1}{\mu}\right)^{-2\epsilon} \frac{\Gamma^2(1-\epsilon)}{\Gamma(1-2\epsilon)}\right]^{\frac{k}{2}} \left[- \PAP_{aa}(z) + \epsilon \,  \CalP_{aa}^{(k),\text{fin}}(z)\right] \; . \label{Eq:Paa_GEN_definition_k_general}
\end{align}
Here
$\PAP_{aa}$ is the Altarelli-Parisi splitting function given in Eq.~\eqref{Eq:PAP_0_definition},
while $\CalP_{aa}^{(k),\text{fin}}$ is an $\order{\eps^0}$ function that can be obtained by comparing Eqs.~\eqref{Eq:P_aa_k_definition} and \eqref{Eq:Paa_GEN_definition_k_general}, 
namely
\be\label{eq_P_aa_k_fin_def}
\CalP_{aa}^{(k),\text{fin}}(z)
= -\frac{\Cf}{\ep} \bigg[
2 \sum_{n=1}^\infty \frac{(-1)^n (k \ep)^n}{n!} \mathcal{D}_n(z)
+ (1-z)^{-k\ep} P_{aa}^{\rm reg}(z) + (1+z)
\bigg]\; ,
\ee
with
\be
P_{aa}^{\rm reg}(z) = -\big[(1+z)+\ep(1-z)\big] \; .
\ee
If the unresolved final state gluon goes collinear to another final state parton $i_g$, the generalized gluon final-state anomalous dimension reads
\index{G!$\Gamma_{i,g}^{(k)}$}
\begin{equation}
    \Gamma_{i,g}^{(k)} = \left[\left(\frac{2E_i}{\mu}\right)^{-2\epsilon} \frac{\Gamma^2(1-\epsilon)}{\Gamma(1-2\epsilon)}\right]^{\frac{k}{2}} \gamma_{z,g \to g g}^{2k}(\epsilon,L_i) \; , \qquad i \in [3,\Np] \; ,
    \label{Eq:Gamma_i_3...N_definition_k_general}
\end{equation}
where
$\gamma_{f(z),g \to gg}^{nk}$ is defined in Eq.~\eqref{eq:defn_gamma_nk0} and repeated in Eq.~\eqref{eq:defn_gamma_nk_appA}.
Throughout the paper, we use
\index{P!$\CalP_{ij}^{\text{gen}}$}
\index{P!$\CalP_{ij}^{\text{fin}}$}
\be
    \CalP_{aa}^{\text{gen}}  = \CalP_{aa}^{(2),\text{gen}} \;,
    \qquad
    \CalP_{aa}^{\text{fin}}  = \CalP_{aa}^{(2),\text{fin}} \;, 
    \qquad
    \Gamma_{i,f_i} = \Gamma_{i,f_i}^{(2)} \;,
\ee
to lighten the notation.

\subsubsection{One-loop}\label{SubSub:definitions_one_loop}
\par When computing the real-virtual contributions, one finds a convolution similar to the one in  Eq.~\eqref{Eq:P_aa_k_definition} for the case when a final-state gluon is collinear to initial state parton $1_a$. It reads 
\index{Convolution}
\begin{equation}
\begin{split}
    \int \limits_{0}^{1} \rmd z\, \CalP_{aa}^{(k),\mathrm{1L}}(z,E_1) g(z)
    = & \int \limits_{0}^{1} \rmd z\, \bigg[(1 - S_z) \Big[(1-z)^{-k \epsilon} P_{aa,\mathrm{i}}^{\text{1L}}(z)\Big] \\
    & + 2 \T_{a}^2 \, \frac{1- e^{-(2+k) \epsilon L_1}}{(2+k)} \pi \cot(\pi \epsilon) \delta(1-z)\bigg] g(z) \; .
    \label{Eq:P_aa_k_1_loop_definition}
\end{split}
\end{equation}
The initial-state one-loop splitting function for a $q\to q$ splitting is given by~\cite{Bern:1999ry,Kosower:1999rx,Campbell:1999ah} 
\index{P!$P_{qq,\mathrm{i}}^{\text{1L}}$}
\begin{equation}
\begin{split}
    P_{qq,\mathrm{i}}^{\text{1L}}(z) =
    & - \frac{\Ca}{\ep^2} \left[\frac{\Gamma^2(1-\ep) \Gamma^2(1+\ep)}{\Gamma(1-2\ep) \Gamma(1+2\ep)} (1-z)^{-\ep} + 2 \sum_{n=1}^{\infty} \ep^{2n} \, \Li_{2n}(1-z) \right] \\
    & \times (1-z)^{-\ep} \Pqq(z) + \frac{2\Cf}{\ep^2} (1-z)^{-\ep} \Pqq(z) \sum_{n=1}^{\infty} \ep^n \, \Li_n(1-z) \\
    & - \Cf (\Ca - \Cf)   \frac{z + \ep(1 - z)}{1-2\ep} (1-z)^{-\ep} \; .
\end{split}
\end{equation}
We rewrite $\CalP_{aa}^{(k),\mathrm{1L}}$ in analogy with Eq.~\eqref{Eq:Paa_k_relation_with_Gamma_a_and_Paa_GEN_k}, getting 
\begin{equation}
    \left[\left(\frac{2E_1}{\mu}\right)^{-2\epsilon} \frac{\Gamma^2(1-\epsilon)}{\Gamma(1-2\epsilon)}\right]^{k} \CalP_{aa}^{(k),\mathrm{1L}}(z,E_1) = \frac{\Ca}{\ep^2} \Big[\Gamma_{1,a}^{(k),\mathrm{1L}} \delta(1-z) + \mathcal{P}_{aa}^{(k),\mathrm{1L,gen}}(z,E_1)\Big] \; ,
\end{equation}
where 
\index{G!$\Gamma_{1,a}^{(k),\mathrm{1L}}$}
\index{P!$\CalPkoneLgen_{ij}$}
\begin{align}
    & \Gamma_{1,a}^{(k),\mathrm{1L}} = \left[\left(\frac{2E_1}{\mu}\right)^{-2\epsilon} \frac{\Gamma^2(1-\epsilon)}{\Gamma(1-2\epsilon)}\right]^{k} \left[\gamma_{a} + 2 \T_{a}^2 \, \frac{1- e^{-(2+k) \epsilon L_1}}{(2+k)} \pi \frac{\cos(\pi \epsilon)}{\sin(\pi \ep)} \right] \; , \label{Eq:Gamma_i_1L_1_2_definition} \\
    & \CalPkoneLgen_{aa}(z,E_1) = \left[\left(\frac{2E_1}{\mu}\right)^{-2\epsilon} \frac{\Gamma^2(1-\epsilon)}{\Gamma(1-2\epsilon)}\right]^{k} \Big[-\PAP_{aa}(z) + \ep \, \CalPkoneLfin_{aa}(z)\Big] \; \label{Eq:Paa_1L_GEN_definition} \; .
\end{align}
In Eq.~\eqref{Eq:Paa_1L_GEN_definition}, the function $\CalPkoneLfin_{aa}$ is finite in $\ep$ and can be extracted from Refs \cite{Bern:1999ry,Kosower:1999rx,Campbell:1999ah}.

For the final state collinear limits, the equivalent of Eq.~\eqref{Eq:Gamma_i_3...N_definition_k_general} is the generalized gluon one-loop, final-state anomalous dimension
\index{G!$\Gamma_{i,g}^{(k), \mathrm{1L}}$}
\begin{equation}
    \Gamma_{i,g}^{(k), \mathrm{1L}} = - \left[\left(\frac{2E_i}{\mu}\right)^{-2\epsilon} \frac{\Gamma^2(1-\epsilon)}{\Gamma(1-2\epsilon)}\right]^{k} \frac{\epsilon^2 \cos(\pi\epsilon)}{C_A} \, \gamma_{z,g \to gg}^{3(k+1),\mathrm{1L}}(\epsilon,L_i) \; , \label{Eq:Gamma_i_1L_i_3_...N_definition}
    \hspace{1cm} i \in [3,\Np] \; , 
\end{equation}
with
\index{G!$\gamma_{f(z),g \to gg}^{n(k+1), \rm 1L}$}
\begin{equation}
\begin{split}
    \gamma_{f(z),g \to gg}^{n(k+1), \rm 1L}(\eps,L_i) = & - \int\limits_{0}^{1} \rmd z\, (1-S_z) \left[z^{-n \eps} (1-z)^{1-(k+1)\eps} f(z) \, P_{gg}^{\text{1L}}(z)\right] \\
    & - 2 \Ca^2 \, \frac{1- e^{-(2+k) \epsilon L_i}}{(2+k)} \frac{\pi}{\ep^2 \sin(\pi \epsilon)} f(1) \; . 
\end{split}
\end{equation}
The above formula requires the following splitting function
\index{P!$\Pgg^{\text{1L}}$}
\begin{equation}
    \Pgg^{\text{1L}}(z) = \Ca \Pgg(z) \left[\frac{a(z) + \tilde{b}(z)}{2}\right] + \widetilde{P}_{gg}^{\rm new}(z) \left[\frac{\nf - \Ca(1-\ep)}{(1-\ep) (1-2\ep) (3-2\ep)}\right] \; ,
\end{equation}
where
\begin{equation}
\begin{split}
    a(z) = & ~ (1-z) F_1(1-z) \; , \\
    \tilde{b}(z) = & ~\frac{2}{\ep^2}  +  z F_1(z) \; , \\
\end{split}
\end{equation}
with 
\be
    F_1(z) =  ~ \frac{2}{z \ep^2} \Big[-\Gamma(1-\ep) \Gamma(1+\ep) z^{-\ep} (1-z)^\ep -1 + (1-z)^\ep \hypF(\ep, \ep, 1+\ep, z)\Big] \; ,
\ee
and
\begin{equation}
    \widetilde{P}_{gg}^{\mathrm{new}}(z) = - \Ca \left[\frac{1 - 2z(1-z)\ep}{1-\ep}\right] \; . 
\end{equation}

\subsection{Definitions of the main operators, commutators, and expansions}
\label{sec:subsec_operator_defn}
Throughout this paper we have used virtual, soft, and collinear operators to encode singularities, and have made use of various relations between them. For the reader's convenience we list these definitions and relations here. 

We begin with Catani's operator~\cite{Catani:1998bh} 
\index{I!$\ICat$}
\be
   \ICat(\ep) = \frac{1}{2}\frac{e^{\ep\gamma_E}}{\Gamma(1-\ep)} \sum_{\inotj}^{\Np} \frac{\mathcal{V}_i^\text{sing}(\ep)}{\ColT{i}^2} \,   \scprod{\ColT{i}}{\ColT{j}} \left(\frac{\musq} {2p_i\cdot p_j}\right)^\ep e^{i \pi\lambda_{ij} \ep}\; ,
\ee
where the relevant constants are defined in Subsection~\ref{sebsec:useful_constants}.
We find it convenient to modify the normalization slightly, yielding
\index{I!$\ICatbar$}
\be
    \ICatbar(\ep) =\frac{1}{2} \sum_{\inotj}^{\Np} \frac{\mathcal{V}_i^\text{sing}(\eps)}{\ColT{i}^2} \,   \scprod{\ColT{i}}{\ColT{j}} \left(\frac{\musq} {2p_i\cdot p_j}\right)^\epsilon e^{i \pi\lambda_{ij} \ep} \; ,
    \label{eq:Icatbar_app}
\end{equation}
 from which we define the operators for amplitudes-squared
\index{I!$I_{\pm}$}
\index{I!$\IVirt$}
\be
\begin{split}
I_{\pm}(\ep) = \frac{\ICatbar(\ep) \pm \ICatbar^\dagger(\ep)}{2} \; , 
\qquad 
    \IVirt(\ep) = \ICatbar(\ep) + \ICatbar^\dagger(\ep) \equiv 2 I_+(\ep) \; .
    \label{Eq:Appendix_IV_definition}
\end{split}
\ee
The Laurent expansion for  $\IVirt(\ep)$ reads 
\begin{equation}
    \IVirt(\ep) = \sum_{n=-2}^{\infty} \ep^n \IVirt^{(n)} \; ,
\end{equation}
where 
 \index{Virtual operators!Expansion}
\index{I!$\IVirt^{(-2)}$}
\index{I!$\IVirt^{(-1)}$}
\be
\begin{split}
\IVirt^{(-2)} = - \sum_{i=1}^{\Np} \ColT{i}^2 \; ,
 \qquad 
\IVirt^{(-1)} = \sum_{\substack{\inotj}}^{\Np} \scprod{\ColT{i}}{\ColT{j}} L_{ij}
- \sum_{\substack{i=1}}^{\Np} \gamma_i \; .
\end{split}
\ee
The soft operator is equal to
\index{I!$\ISoft$}
\begin{equation}
    \ISoft(\ep) = - \frac{(2\Emax/\mu)^{-2\ep}}{\ep^2} \sum_{\inotj}^{\Np} \eta_{ij}^{-\ep} K_{ij} \, (\T_i \cdot \T_j) \; ,
    \label{Eq:ISoft_definition_appendix}
\end{equation}
where $K_{ij}$ is defined in Eq.~\eqref{eq:Kij_defn0}.
The Laurent expansion of $\ISoft$ reads
\index{Soft operators!Expansion}
\begin{equation}
\label{eq:IS_exp}
    \ISoft(\ep) = \sum_{n=-2}^{\infty} \ep^n \ISoft^{(n)} \; , 
\end{equation}
and we require the following terms in the above expansion
\index{I!$\ISoft^{(-2)}$}
\index{I!$\ISoft^{(-1)}$}
\index{I!$\ISoft^{(0)}$}
\index{I!$\ISoft^{(1)}$}
\begin{align}
    \ISoft^{(-2)}={}& \sum_{i=1}^{\Np} \ColT{i}^2 \; , \notag \\
    \ISoft^{(-1)}={}& \sum_{\substack{\inotj}}^{\Np} \scprod{\ColT{i}}{\ColT{j}} \log \eta_{ij}
    - 2 \Lmax
    \sum_{\substack{i=1}}^{\Np} \ColT{i}^2 \; ,  \notag \allowdisplaybreaks \\
    \ISoft^{(0)}={}& -
    \sum_{\substack{\inotj}}^{\Np} \scprod{\ColT{i}}{\ColT{j}} \bigg[ 2 \Lmax \log \eta_{ij} 
    + \frac12 \log^2 \eta_{ij}
    + \Li_2(1-\eta_{ij})\bigg]  \notag \allowdisplaybreaks\\
    {}&
    + 
    \bigg[2 \Lmax^2
    -\frac{\pi^2}6\bigg] \sum_{\substack{i=1}}^{\Np} \ColT{i}^2 \; , \label{eq:IS_exp_coeff} \allowdisplaybreaks
    \\
    \ISoft^{(1)}={}& 
    \sum_{\substack{\inotj}}^{\Np} \scprod{\ColT{i}}{\ColT{j}} \bigg[ 2 \Lmax^2 \log \eta_{ij} 
    +  \bigg( \Lmax
    - \frac12 \log(1- \eta_{ij}) \bigg) \log^2 \eta_{ij}  \notag \allowdisplaybreaks \\
    {}&
    + \frac16 \log^3 \eta_{ij}
    + 2 \Lmax \, \Li_2(1-\eta_{ij})
    - \Li_3(1-\eta_{ij})
    - \Li_3(\eta_{ij}) \bigg]  \notag \allowdisplaybreaks \\
    {}&
    -
    \bigg[\Lmax\left(\frac43 \Lmax^2 -\frac{\pi^2}3\right) + 3\zeta_3\bigg] \sum_{\substack{i=1}}^{\Np} \ColT{i}^2 \; , \notag
\end{align}
\index{L!$\Lmax$}
where $\Lmax = \log(2\Emax/\mu)$.

The computation of the soft contributions requires a variant of the soft operator, namely
\index{I!$\ISofttilde$}
\begin{equation}
    \ISofttilde(2\epsilon) = - \frac{(2\Emax/\mu)^{-4\epsilon}}{(2\epsilon)^2} \sum_{\inotj}^{\Np}  \eta_{ij}^{-2\epsilon} \widetilde{K}_{ij} \, (\T_i \cdot \T_j) \; ,
\end{equation}
where $\widetilde{K}_{ij}$ is defined in Eq.~\eqref{eq:Ktilde_def}.
The following property relates 
$\ISoft$
and 
$\ISofttilde$ 
\begin{equation}
    \ISofttilde(2\eps) = \ISoft(2\eps) + \order{\ep} \; .
    \label{Eq:relation_between_ISofttilde_and_ISoft}
\end{equation}
We also require an $\eps$-expansion  for $\ISofttilde$. Given Eq.~\eqref{Eq:relation_between_ISofttilde_and_ISoft}, the first three coefficients $\ISofttilde^{\,(n)}$ with $n=-2, -1, 0$ can be directly obtained from those in Eq.~\eqref{eq:IS_exp_coeff}, up to a rescaling by factors of $1/4, 1/2$ and $1$ respectively.
The coefficient at  $\order{\ep}$ reads
\index{I!$\ISofttilde^{\,(1)}$}
\index{Soft operators!Expansion}
\be
\begin{split}
\ISofttilde^{\,(1)}
={}&
\sum_{\substack{\inotj}}^{\Np} \scprod{\ColT{i}}{\ColT{j}} \bigg[\bigg(2 \Lmax -\frac32 \log(1-\eta_{ij}) \bigg)\log^2 \eta_{ij}
+\frac13 \log^3 \eta_{ij}
\\
&{}
+ \bigg( \frac{\pi^2}6 
+ 4 \Lmax^2
- \Li_2 (1-\eta_{ij})
\bigg)\log \eta_{ij}
\\
{}&
+ 4 \Lmax \,  \Li_2 (1-\eta_{ij})
- \Li_3(1-\eta_{ij})
-3 \Li_3 (\eta_{ij} )
\bigg]
\\
{}&
+
\bigg[  \frac23 \, \Lmax
\bigg(\pi^2 
- 4 \Lmax^2
\bigg) - 7 \zeta_3\bigg] \sum_{\substack{i=1}}^{\Np} \ColT{i}^2 \; .
\end{split}
\ee

Moving to collinear limits, we define the hard-collinear operator as
\index{I!$\IColl^{(k)}$}
\begin{equation}
    \IColl^{(k)}(\epsilon) = \sum_{i=1}^{\Np} \frac{2}{k} \frac{\Gamma_{i,\fl{i}}^{(k)}}{\epsilon} \; , 
    \label{Eq:Appendix_Ic_k_general_definition}
\end{equation}
where $\Gamma_{i,\fl{i}}^{(k)}$ is given in Eq.~\eqref{Eq:Gamma_1_2_definition_k_general} if $i=1,2$ and in Eq.~\eqref{Eq:Gamma_i_3...N_definition_k_general} if $i \in [3,\Np]$. 
To treat the hard-collinear limits of the real-virtual matrix element we have introduced
\index{I!$\IColltilde$}
\begin{equation}
    \IColltilde(2\ep) = \sum_{i=1}^{\Np} \frac{\Gamma^{\rm 1L}_{i,f_i}}{2\ep} \; ,
\end{equation}
where $\Gamma^{\rm 1L}_{i,f_i}$ is given in Eq.~\eqref{Eq:Gamma_i_1L_1_2_definition} if $i=1,2$ and in Eq.~\eqref{Eq:Gamma_i_1L_i_3_...N_definition} if $i \in [3,\Np]$. We note that the following relations hold
\index{Collinear operators!Relations}
\begin{equation}
\begin{split}
    \IColltilde(2\epsilon) & = \IColl(2\epsilon) + \order{\ep} \; , \\
        \IColl^{(4)}(\ep) & = \IColl(2\epsilon) + \order{\ep^0} \;. 
\label{Eq:relation_between_IColltilde_and_IColl}
\end{split}
\end{equation}

Furthermore, we have used the $\ep$-finite operator $\ITot$ defined as
\index{I!$\ITot$}
\begin{equation}
    \ITot(\ep) = \IVirt(\ep) + \ISoft(\ep) + \IColl(\ep) \; ,
\end{equation}
to simplify the NLO and NNLO calculations. Its expansion in $\ep$ reads 
\be
\ITot(\ep) = \sum_{n=0}^\infty \ep^n \ITot^{(n)} \; ,
\ee
with expansion coefficients given by
\index{I!$\ITot$! Expansion}
\index{I!$\ITot^{(0)}$}
\index{I!$\ITot^{(1)}$}
\be
\begin{split}
    \ITot^{(0)} ={}& -\sum_{\inotj}^{\Np} \ColT{i} \cdot \ColT{j} \bigg[
    \left( 2\Lmax +\frac12 \log\eta_{ij} \right) \log \eta_{ij} -\frac12 L_{ij} \left(L_{ij} + \frac{2 \gamma_i}{\ColT{i}^2}\right) \\
    & +\Li_2(1-\eta_{ij}) + \frac{\pi^2}{2} \lambda_{ij} \bigg] \\
    & + \sum_{i=1}^{\Np} \ColT{i}^2 \bigg[2\Ltildei^2 - \frac{\pi^2}{6} - \frac{2\gamma_i}{\T_i^2} \, \Ltildei \, \bar{\theta}_{i2} + \left(\frac{67}{9} - \frac{11}{3} \Ltildei - \frac{2\pi^2}{3}\right) \theta_{i2} \bigg] \; ,\\
   \ITot^{(1)} ={}& \sum_{\inotj}^{\Np} \T_i \cdot \T_j \bigg[\frac{1}{6} \big(L_{ij}^3 + \log^3\eta_{ij}\big) + 2\Lmax^{2} \log\eta_{ij} -\frac{\pi^2}{2} \lambda_{ij} L_{ij} \\
   & + \left(\Lmax - \frac{1}{2} \log(1-\eta_{ij})\right)\log^2\eta_{ij} + 2\Lmax \, \Li_2(1-\eta_{ij}) - \Li_3(\eta_{ij}) \\
   & - \Li_3(1-\eta_{ij}) + \frac{\gamma_i}{2\T_i^2} \big(L_{ij}^2 - \pi^2 \lambda_{ij}\big) \bigg] \\
   & + \sum_{i=1}^{\Np} \T_i^2 \bigg[-\frac{4}{3} \Ltildei^3 + \frac{\pi^2}{3} \Ltildei - 3\zeta_3 + \left(2\Ltildei^2 - \frac{\pi^2}{6}\right)\gamma_i \, \bar{\theta}_{i2} + \bigg(\frac{808}{27} - \frac{134}{9} \Ltildei \\
   & + \frac{11}{3}\Ltildei^2 + \pi^2 \left(\frac{4}{3}\Ltildei - \frac{55}{36}\right) - 16\zeta_3\bigg) \theta_{i2} \bigg] \; ,
    \end{split}
    \label{eq:IT_fin}
\ee 
\index{T!$\theta_{i2}$}
\index{T!$\bar{\theta}_{i2}$}
where $\theta_{i2} = 1$ if $i > 2$ and $0$ otherwise, and $\bar{\theta}_{i2} = 1 - \theta_{i2}$.

While discussing the rearrangement of the single-unresolved terms (cf. Section~\ref{SubSec:X_fin}), we have introduced a variant of the virtual, soft and collinear operators, valid in the case of $N+1$ final-state partons. In particular, we have defined
\index{I!$\IVirtONLO$}
\be
\begin{split}
\IVirtONLO(\ep)
={}&
\ICatbar^{\Np+1}(\ep)
+
\Big(\ICatbar^{\Np+1}(\ep)\Big)^\dag \; , 
\end{split}
\ee
with $\ICatbar^{\Np+1}$ defined as in Eq.~\eqref{eq:Icatbar_app}, 
up to replacing $\Np \mapsto \Np+1$. Similarly, we have also used
\index{I!$\ISoftONLO$}
\be
\ISoftONLO(E_{\Fp})
=
-  \frac{(2E_{\Fp}/\mu)^{-2\eps}}{\ep^2}  \sum_{\inotj}^{\Np+1} \eta_{ij}^{-\eps} K_{ij} \left(\scprod{\ColT{i}}{\ColT{j}} \right) \; ,
\ee
and
\index{I!$\ICollONLO$}
\be
\ICollONLO(E_{\Fp})
= \sum_{i=1}^{\Np+1}
\frac{\Gamma_{i,f_i}}{\ep}\bigg|_{\Emax \mapsto E_\Fp} \; ,
\ee
where one needs to set $\Emax \mapsto E_\Fp$ in the definition of $\Gamma_{i,f_i}$, see Eqs.~(\ref{Eq:Gamma_i_Leg3_to_N_definition}, \ref{eq:gamma_expansion_is0}). 
 
\section{Partitions at NLO and NNLO for an arbitrary number of final-state particles}
\label{appendix:b}
\label{sect:partitions}

To treat the infrared singularities of a process with a large number of final state particles,  we require partitions that separate resolved
and potentially unresolved partons. To construct them, we consider a process that involves $N_p$ partons at leading order 
and an arbitrary  colorless final state
\be
f_1(p_1)+f_2(p_2) \to f_3(p_3) + ... + f_{N_p}(p_{N_p} ) +X \; .
\ee

At next-to-leading order, we need to add another particle to the final state to describe the real-emission process.  We denote the corresponding list of final-state partons
in this case as 
$\myset_{N+1} = \{f_3,f_4,... \, ,f_{N_p},f_{N_p+1}\}$, 
where $N = N_p -2$ is the number of final-state partons at leading order.

In principle, any of these final state partons can become unresolved. Suppose we want to describe a situation when this happens with  a parton $i$. We then write the set of $N+1$ partons as 
\be
\myset_{N+1} = \{i,\myset_{N}^{(i)}\} \; ,
\label{eq_myset_def}
\ee
where $\myset_N^{(i)} = \myset_{N+1}/\{i\}$
and introduce the function 
\index{D!$d^{(i)}$}
\be
d^{(i)} = \prod \limits_{k \in \myset_N^{(i)} } p_{k,\perp} \prod \limits_{\substack{l,m \in \myset_{N}^{(i)} \\  l < m }}  (1-\cos \theta_{lm}) \; ,
\ee
where $p_{k,\perp}$ is the transverse momentum of parton $k$.\footnote{We note that in the case of only one hard jet, $k$, $d^{(i)}$ reduces to $p_{k,\perp}$.}
These functions are used to construct the partitions 
\index{Damping factor!NLO!Definition}
\index{D!$\Delta^{(i)}$}
\be
\label{eq:Delta_NLO_def}
\Delta^{(i)} =  \frac{d^{(i)}}{\sum \limits_{j \in \myset_{N+1}} d^{(j)} } \; , 
\ee
where $i \in \myset_{N+1}$. It follows from their  definition that the functions $\Delta^{(i)}$ provide a partition of 
unity
\index{Damping factor!NLO!Properties}
\be
\sum \limits_{i \in \myset_{N+1}} \Delta^{(i)} = 1 \; .
\ee

It is straightforward to determine the action of soft and collinear operators on the partition functions. In the soft limit of parton $k$,  described by the operator $S_k$, we find
\index{Damping factor!NLO!Properties}
\be
S_k \Delta^{(i)} = \delta_{ki} \; .
\ee
In the limit where partons $l$ and $m$ become collinear, we have 
\index{Damping factor!NLO!Properties}
\be
C_{lm} \Delta^{(i)}
=
\begin{cases}
 0 \; ,  \qquad     & l, m \ne i \; , \\
 1 \; ,      & l=i,\; m \in \{1,2\} \; ,\\
 z_{i,m} \; , & l = i, m \in \myset_N \; ,
\end{cases}
 \ee
\index{Z!$z_{i,m}$}
where  $z_{i,m} = E_i/(E_i+E_m)$ and we assumed that partons $1$ and $2$ are in  the initial state. The limits 
 obtained by the interchange of $l$ and $m$ assignments follow naturally from the above formulas and  are not shown for this reason.

 A new element required for NNLO computations is the double-real emission process. To construct the 
 corresponding partition functions, we consider   an extended set of final-state 
 partons 
 \be
\myset_{N+2} =  \{f_3,f_4,...,f_{N_p+1},f_{N_p+2} \} \; .
 \ee
 Two of these final-state partons can become unresolved and we assume that 
 this happens with partons $i$ and $j$. 
 We then write $\myset_{N+2} = \{(i,j),\myset_{N}^{(ij)}\}$, define 
 functions $d^{(ij)}$  as follows
\index{D!$d^{(ij)}$}
\be
d^{(ij)}  = \prod \limits_{k \in \myset_N^{(ij)} } p_{k,\perp} \prod \limits_{\substack{l,m \in \myset_{N}^{(ij)} \\ l < m }}  (1-\cos \theta_{l m}) \; ,
\ee
and use them
to construct the NNLO partitions 
\index{D!$\Delta^{(ij)}$}
\be
\label{eq:Delta_NNLO_def}
\Delta^{(ij)} = \frac{ d^{(ij)}}{\sum \limits_{(lm) \in \myset_{N+2}}  d^{(lm)}} \; ,
\ee
Similar to the NLO case the functions 
$\Delta^{(ij)}$ provide partition of unity
\index{Damping factor!NNLO!Properties}
\be
\sum \limits_{(ij) \in \myset_{N+2}}^{} \Delta^{(ij)} = 1 \; ,
\ee
where the sum is over unordered pairs $(ij)$.

For the NNLO computation, we require 
the double-soft 
 ($\DoubSoft{lm}$), the single-soft  ($\Soft{l}$), the collinear  ($\Coll{lk})$  and the triple-collinear  ($\Coll{lk,m}$) limits of the partition 
functions $\Delta^{(ij)}$.
The double-soft limit reads
\index{Damping factor!NNLO!Properties}
\be
\DoubSoft{lm} \,
\Delta^{(ij)} = \delta_{(ij),(lm)} \; ,
\ee
where the Kronecker delta indicates that the unordered pair $(ij)$ should coincide with the unordered pair $(lm)$
for this limit to be different from zero.   The single-soft limit is
\index{Damping factor!NNLO!Properties}
\be
\Soft{l} \,  \Delta^{(ij)}
=
\begin{cases}
  0 \, ,  &  l \ne i, ~ l \ne j \; , \\
  \Delta^{(j)} \, , & l = i \; , \\
  \Delta^{(i)} \, , & l = j \; , 
\end{cases}
\label{eq:SoftonDeltaij}
\ee
where $\Delta^{{(i)}}$ and $\Delta^{(j)}$ in the above formulas are NLO partitions constructed 
for sets $\myset_{N+1} =\{j,\myset_{N}^{(ij)}\} $
and $\myset_{N+1} = \{i,\myset_{N}^{(ij)}\} $, 
respectively. 

  Next, we consider the collinear limits. We find
  \index{Damping factor!NNLO!Properties}
  \be
  \label{eqb.14}
  \Coll{lk} \, 
  \Delta^{(ij)}
  =
\begin{cases}
  0 \; , &  l \not= i,j, ~ k \not= i,j \; , \\
  \Delta^{(k')} \; , & l \in \{1,2\}, ~ k \in \{i,j\} \;  , \\
  \Delta^{([ij])} \; , & \{l,k\} = \{i,j\} \; , \\  
  z_{k,i} \Delta^{(j)} \; , & l = i, ~ k \neq j \; ,             
\end{cases}
\ee
where $k'=i$ if $k=j$ and $k'=j$ if $k=i$, and  $[ij]$ represents the ``clustered particle'' whose four-momentum is given 
by $p_{[ij]} = ( 1 +E_j/E_i) \,  p_i$  and 
the function $\Delta^{([ij])}$ is constructed from the set $\myset_{N+1} 
= \{[ij],\myset_{N}^{(ij)} \}$. In the final line of Eq.~\eqref{eqb.14}, the $\Delta$-function is constructed using  the transverse momentum of the clustered particle $[kl]$.

It is instructive to explain how 
the last formula in 
Eq.~(\ref{eqb.14}) is derived, since the other formulas 
in that equation can be computed  in a similar way. 
To describe the collinear  $i ||k$ limit, where $k$ is a final state particle, we write $\Delta^{(ij)}$ as follows
\index{Damping factor!NNLO!Definition}
\index{D!$\Delta^{(ij)}$}
\be
\Delta^{(ij)} = \frac{d^{(ij)}}{ d^{(ij)} + d^{(ik)} + \sum \limits_{m \ne k,j }^{} d^{(im)} +d^{(kj)}
+ \sum \limits_{m \ne i,j }^{} d^{(k m)} + \sum \limits_{m,n \ne i,k}^{} d^{(mn)} } \; .
\label{eqb.15}
\ee
We now study what happens to the 
various entries in the above formula 
when  the relevant  limit is taken. 
First, we note that 
the numerator $d^{(ij)}$ does not contain $i$ but contains $k$. We replace $p_{\perp,k}$ with $p_{\perp,[ik]}$ and write the resulting
expression as 
\index{Damping factor!NNLO!Properties}
\be
C_{ik} \, d^{(ij)} = \frac{E_{k}}{E_k + E_i} \; d^{(j)} = z_{k,i} \; d^{(j)} \; , 
\ee
where $d^{(j)} $ is constructed using 
the list $\{j,\myset_{N+2}^{(ij)}(k \to [ki])\}$.
The various entries in the denominator 
of Eq.~(\ref{eqb.15}) 
behave as follows 
\begin{equation}
  \begin{split}
    & 
    C_{ik} \,   d^{(ik)} = d^{([ik])} \; , \\
    &
    C_{ik} \sum \limits_{m \ne k,j }^{N} d^{(im)} = z_{k,i} \sum \limits_{m \ne k,j }^{N} d^{(m)}\; ,
  \end{split}
\qquad\quad
  \begin{split}
    &
    C_{ik} \, d^{(kj)} = z_{i,k} \, d^{(j)} \; ,
    \\
    &
    C_{ik}  \sum \limits_{m \ne i,j }^{N} d^{(k m)} = z_{i,k} \sum \limits_{m \ne i,j }^{N} d^{ (m)} \; .
  \end{split}
\end{equation}
Therefore
\be
C_{ik} \,  \Delta^{(ij)} = z_{k,i} \frac{d^{(j)} }{ d^{([ik])}  + d^{(j)} + \sum_{m \ne j,i,k} d^{(m)} } = z_{k,i} \, \Delta^{(j)} \; , 
\ee
with $\Delta^{(j)}$ being a NLO partition where partons $i$ and $k$ that appear in  the original list of partons are clustered
together. 

Finally, formulas for triple-collinear
limits can be derived in a similar way. 
We find that the only non-vanishing limits are  
  \be
  \label{eqb.19}
  \Coll{kij} \, 
  \Delta^{(ij)}
  =
\begin{cases}
  1 \, , & k \in \{1,2\}\;  , \\
  z_{k,ij} \Delta^{(ij)} \; , & k \in \{3,... \} \; ,   \\  
\end{cases}
\ee
\index{Z!$z_{k,ij}$}
where  $z_{k,ij} = E_k/(E_k  + E_i + E_j)$.\\

In addition to $\Delta$-partitions, which allow us to 
separate resolved and potentially unresolved partons,
we require angular partition functions $\omega$. 
These functions are supposed  to define   possible 
collinear  singular directions between unresolved partons.
Below we give an example of how such functions 
can be designed. 

We begin with the construction of these angular partition functions at NLO.   To this end, we consider
a situation where parton $\Fp$ is potentially unresolved, so that $\widetilde{\myset}_{\Np+1}=\{\Fp,\widetilde{\myset}_{\Np}^{(\Fp )}\}$, where $\widetilde{\myset}_{\Np}^{(\Fp )}$ extends the definition in Eq.~\eqref{eq_myset_def} so as to include also the initial-state partons. We  define the quantities
\be
g_{kl} = \rho_{kl}^{-1} \; , 
\ee
and use them to write the function $\omega^{\Fp i}$ as 
\index{O!$\ww{\Fp i}$}
\begin{align}
\ww{\Fp i} = \frac{g_{i\Fp}}{\sum \limits_{j \in \widetilde{\myset}_{\Np}^{(\Fp) }} \, g_{j\Fp}} \; , 
\qquad 
i \in \widetilde{\myset}_{\Np}^{(\Fp) } \; .
\label{eq:sector_NLO}\; 
\end{align}
Since
\index{Partition functions!Properties}
\begin{equation}
\sum \limits_{i \in \widetilde{\myset}_{\Np}^{(\Fp) }}
\omega^{\Fp i } =1 \; ,
\qquad C_{k \Fp} \; \omega^{\Fp i} = \delta_{ki} \; ,
\end{equation}
the functions 
$\omega^{\Fp i}$ possess  the required properties  
to be used as angular partitions in  NLO computations. 

We continue with the discussion of the NNLO case, 
where  
partons $\Fp$ and $\Sp$ are  potentially unresolved
and the remaining $N_{p}$ hard partons are  described 
by the set $\myset_{N}^{(\Fp\Sp)}$.  We proceed 
as follows. First, we employ the NLO partitions 
to construct a partition of unity in 
the following way 
\be
1 = \sum  \limits_{\substack{i,j=1 \\ i \ne j}}^{\Np} \omega^{\Fp i} \omega^{\Sp j} 
+ \sum  \limits_{\substack{i,j=1}}^{\Np}
\delta_{ij} \, 
\omega^{\Fp i} \omega^{\Sp j} \; .
\ee
The two sums  on the right-hand side are almost the right partitions for double-  and triple-collinear 
limits except  for the fact that the collinear 
$\Fp || \Sp$ singularity is present in both terms 
of this formula.  However,  we  
would like to move it into 
the triple-collinear partition.   To achieve this, 
we introduce 
yet another partition of unity which involves 
$\rho_{\Fp \Sp}$, $\rho_{i \Fp}$ and $\rho_{j \Sp}$ only
and write 
\be
1 = \frac{\rho_{\Fp \Sp}}{d_{\Fp \Sp i j}}
+
\frac{\rho_{i \Fp }+ \rho_{j \Sp}}{d_{\Fp \Sp i j}} \; ,
\ee
where
\be
\label{eq:d_Fp_Sp_ij}
d_{\Fp \Sp i j}= \rho_{\Fp \Sp}
+ \rho_{i \Fp}
+ \rho_{j \Sp} \; .
\ee
We now employ  these expressions to define the 
double-collinear partition
\index{O!$\ww{\Fp i ,\Sp j}$}
 \be
 \label{eq:DC_partitions}
\ww{\Fp i ,\Sp j}
=
\ww{\Fp i}
\ww{\Sp j}
\frac{\rho_{\Fp \Sp}}{d_{\Fp \Sp i j}} \; , 
\qquad i \neq j \; ,
\ee
and the triple-collinear partition 
\index{O!$\ww{\Fp i, \Sp i }$}
\be
\label{eq:TC_partitions}
\ww{\Fp i, \Sp i }
= \ww{\Fp i} \ww{\Sp i}
+ 
\ww{\Sp i}
\sum_{\substack{j=1 \\ j \ne i}}^{\Np}
\frac{\rho_{j\Fp} \, \ww{\Fp j}}{d_{\Fp \Sp j i}}
+ 
\ww{\Fp i}
\sum_{\substack{j=1 \\ j \ne i}}^{\Np}
\frac{\rho_{j \Sp} \,  \ww{\Sp j}}{d_{\Fp \Sp i j}} \; .
\ee
It is easy to check  that the following identity
holds
\be
1 = \sum_{\substack{i, j=1 \\ i \ne j}}^{\Np} \ww{\Fp i, \Sp j} 
+ \sum_{i=1}^{\Np}  \ww{\Fp i, \Sp i} \; 
.
\ee

\vspace*{0.3cm}
The partitions constructed in Eqs.~\eqref{eq:DC_partitions} and~\eqref{eq:TC_partitions}
satisfy all the properties that we need for NNLO QCD computations.
In particular, each partition  selects a minimal number of collinear singularities and satisfies the following 
 relations 
\begin{equation}
\label{eq:NNLO_splitting_prop_qq}
  \begin{split}
    &
    \Coll{i \Fp}\ww{\Fp i, \Sp j}
    =
    \ww{\Fp i, \Sp j}_{\Fp\parallel i} = \lim_{\rho_{i \Fp}\rightarrow 0}\ww{\Fp i, \Sp j} \; , 
    \\
    &
    \Coll{\Fp \Sp}\ww{\Fp i, \Sp j}
    = \delta_{ij} \,
    \ww{\Fp i, \Sp j}_{\Fp\parallel \Sp} = \lim_{\rho_{\Fp \Sp}\rightarrow 0}\ww{\Fp i, \Sp i} \; , 
    \\
    &
    \Coll{i \Fp }\, \Coll{\Fp \Sp}\ww{\Fp i, \Sp j}
    =
    \delta_{ij} \, \Coll{i \Fp}\, \Coll{\Fp \Sp} \; ,
    \\
    &
    \Coll{i\Fp} \, \Coll{j\Sp} \, \ww{\Fp k, \Sp l} = \delta_{ki} \, \delta_{lj} \, \Coll{\Fp i} \, \Coll{\Sp j}  \; , 
    \\
    &
    \Coll{\Fp\Sp,i}  \, \ww{\Fp j\Sp j}  = \delta_{ij} \, \Coll{\Fp \Sp,i} \; , 
  \end{split}
\qquad \quad
  \begin{split}
    &
    \Coll{j \Sp}\ww{\Fp i, \Sp j}
    =
    \ww{\Fp i, \Sp j}_{\Sp\parallel j} = \lim_{\rho_{j \Sp}\rightarrow 0}\ww{\Fp i, \Sp j} \; , 
    \\
    {}
    \\
    &
    \Coll{j \Sp }\, \Coll{\Fp \Sp}\, \ww{\Fp i, \Sp j}
    = \delta_{ij} \, \Coll{j \Sp }\, \Coll{\Fp \Sp} \; ,
    \\
    &
    \Coll{i\Fp} \, \Coll{i\Sp} \, \ww{\Fp j, \Sp j} = \delta_{ij} \,  \Coll{i\Fp} \, \Coll{i\Sp} 
    \; . 
    \\
    &
    {}
  \end{split}
\end{equation}
We note that these relations are important for 
 simplifying  the 
required subtraction terms. The partitions in Eqs~\eqref{eq:DC_partitions} and~\eqref{eq:TC_partitions} correspond to those defined in Eq.~(B.14) in Ref.~\cite{Caola:2017dug} when we restrict them to the case of color-singlet production, i.e.~$\Np=2$.

\section{Details of the NLO calculation}\label{sec:AppNLOdetails}

The goal of this appendix is to provide further details about the NLO  computation described in Section~\ref{sect:nlo}.  
In particular, we would like to show that 
the operator $\ITot(\ep)$ introduced in Eq.~(\ref{eq3.2}) does not contain  poles in $\ep$. 
According to Eq.~(\ref{eq3.2}), $\ITot(\ep)$ is given by a sum of three terms that describe   virtual, soft and 
hard-collinear contributions. 

We begin with the $\ep$-expansion of the operator $\ISoft$ defined in  Eq.~\eqref{eq:NLO_soft0}.
We report its definition here for convenience
\index{I!$\ISoft$}
\be
\ISoft (\eps)
=
      -  \frac{(2\Emax/\mu)^{-2\eps}}{\ep^2}  \sum_{\inotj}^{\Np}  \eta_{ij}^{-\eps} K_{ij} \left(\scprod{\ColT{i}}{\ColT{j}} \right)
     \; .
    \ee
The function  $K_{ij}$ is defined in Eqs.~\eqref{eq:Kij_defn0}. We note that
its  expansion  in $\ep$ reads
\begin{equation}
    K_{ij} = 1 + \Kijtwo \eps^2 + \order{\eps^3} \; ,
    \qquad
    \Kijtwo= \text{Li}_2(1-\eta_{ij}) - \frac{\pi^2}{6} \; .
\end{equation}
Although 
it is  straightforward  
 to construct the expansion of $\ISoft$, 
arranging it in a particular way is helpful for an efficient demonstration of 
the cancellations of infrared poles. 

We note that the  $I$-operators include  quantities  raised to  
$\ep$-dependent powers. For example, in 
the case of $\ISoft$, there are factors 
$(2 E_{\rm max}/\mu)^{-\ep}$
and $\eta_{ij}^{-\ep}$. 
The expansion of such 
quantities in $\ep$ starts with $1$ and 
it is convenient to make this explicit.  To this end, we 
introduce the function 
\begin{equation}
    f_k(x) = \frac{x^{-k \ep} - 1}{\ep} \; .
\end{equation}
such that $\fn{k}(x) \sim \order{\eps^0}$ as $\eps \to 0$. We then use this function to write 
$\ISoft$ as 
\index{I!$\ISoft$}
\be
\begin{split}
    \lint \ISoft (\eps)\colorprod \FLM \rint  ={}& - \frac{1}{\eps^2} \, 
    \Big\langle
    \Big[1+ \eps f_2(2\Emax/\mu) \Big]     \\
    & \times \sum_{\inotj}^{\Np} \Big[1 + \ep f_1(\eta_{ij}) + \eps^2 K_{ij}^{(2)}\Big] \left(\ColT{i} \cdot \ColT{j} \right)\colorprod \FLM \Big\rangle \; ,
    \end{split}
\ee
where ${\cal O}(\ep)$ 
terms have been neglected.
Since we only need terms through 
${\cal O}(\ep^0)$, we can simplify the above 
equation further. We find 
\index{Soft operators!Expansion}
\index{I!$\ISoft$}
\be
\begin{split}
    \lint \ISoft (\eps)\colorprod \FLM \rint = & ~ -\frac{1}{\ep^2} \sum_{\inotj}^{\Np} \llint  
    \Big[1+\ep \fn{2}(2\Emax/\mu) \Big] 
     \left( \ColT{i} \cdot \ColT{j} \right) \colorprod \FLM \rrint 
        \\
    & -  \sum_{\inotj}^{\Np} \llint \frac{\fn{1}(\eta_{ij})}{\eps}  \left( \ColT{i} \cdot \ColT{j} \right) \colorprod \FLM \rrint 
\\
    & -  \sum_{\inotj}^{\Np} \llint \left[\fn{1}(\eta_{ij}) \fn{2}(2\Emax/\mu) + K_{ij}^{(2)}\right] \left( \ColT{i} \cdot \ColT{j} \right) \colorprod \FLM \rrint \;  .
    \label{eq.a3}
\end{split}
\ee
Next, we  note that in the first term on the right-hand side in Eq.~(\ref{eq.a3}), 
only the 
color charge operators depend on the summation indices $i$ and $j$.
For this reason, the summation over one of the indices 
can be performed  using the  color conservation condition 
\index{Color space and algebra}
\be
\sum \limits_{k=1}^{\Np} \ColT{k} |{\cal M} \rangle_c= 0 \; .
\ee
It follows that 
\be
    \sum_{j \ne i}^{\Np} \lec \ampM{} \sepc \ColT{i} \cdot \ColT{j} \sepc \ampM{ } \ric=  - \ColT{i}^2 |\ampM{ }|^2 \; ,
    \label{eq:remove_col_corr}
\ee
and we obtain 
\index{I!$\ISoft$}
\be
\label{Eq:NLO_soft_real_corrections_final_result}
\begin{split}
    \lint \ISoft (\eps)\colorprod \FLM \rint = & \sum_{i=1}^{\Np}  
   \llint  \Big[1+\ep \fn{2}(2\Emax/\mu) \Big] \, \frac{\ColT{i}^2}{\eps^2} \, 
      \FLM 
    \rrint 
   \\
    & -  \sum_{\inotj}^{\Np} \llint \frac{\fn{1}(\eta_{ij})}{\eps}  \left( \ColT{i} \cdot \ColT{j} \right) \colorprod \FLM \rrint \\
    & -  \sum_{\inotj}^{\Np} \llint \left[\fn{1}(\eta_{ij}) \fn{2}(2\Emax/\mu) + K_{ij}^{(2)}\right] \left( \ColT{i} \cdot \ColT{j} \right) \colorprod \FLM \rrint \;  .
\end{split}
\ee
It is seen  from the above equation 
that the  residue of the $1/\eps^2$ pole is proportional to the sum of  the Casimir factors $\ColT{i}^2$. We recall that  the infrared poles of the one-loop amplitude 
described by Catani's function 
exhibit a similar feature. 
The $1/\eps$ pole in the second line 
of Eq.~(\ref{Eq:NLO_soft_real_corrections_final_result})
contains color correlations, while the terms in the third line are $\ep$-finite.\\

We turn to the virtual corrections.  We have introduced the operator  $\IVirt(\eps)$ in Eq.~\eqref{eq:IVirt_defn0}, and we display it  here  for convenience 
\index{I!$\IVirt$}
\index{I!$\ICatbar$}
\begin{equation}
\IVirt(\ep) = \ICatbar(\ep) + \ICatbar^\dagger(\ep)\, , 
\qquad 
    \ICatbar(\ep) = \frac{1}{2} \sum_{\inotj}^{\Np} \frac{\mathcal{V}_i^\text{sing}(\eps)}{\ColT{i}^2}   \left(\scprod{\ColT{i}}{\ColT{j}} \right)
    \left(\frac{\musq} {2p_i\cdot p_j}\right)^\epsilon e^{i \pi\lambda_{ij} \ep} \;. 
\end{equation}
The quantities  $\lambda_{ij}$ and $\mathcal{V}_i^\text{sing}(\eps)$ are defined in Eq.~\eqref{eq:I1Cat0}.    
Expanding in $\ep$, we find
\index{Virtual operators!Expansion}
\index{I!$\IVirt$}
\be
\begin{split}
   \lint \IVirt(\eps) \colorprod \FLM\rint ={} & \sum_{\inotj}^{\Np} \frac{\mathcal{V}_i^\text{sing}(\eps)}{\ColT{i}^2} 
   \bigg\langle \Big[ 1 + \ep \fn{1}(s_{ij}/\mu^2  ) - \frac{\pi^2}{2} \lambda_{ij}\eps^2 +\order{\eps^3}\Big] \\
   & \times (\ColT{i}\cdot \ColT{j}) \colorprod \FLM \bigg\rangle \; .
   \label{eq:ICatFLM}
  \end{split}
  \ee
In the first term on the right-hand side of Eq.~\eqref{eq:ICatFLM},  we can use  
color conservation to sum over the index $j$. Doing so allows us to   write the  virtual contributions as follows 
\begin{equation}
\begin{split}
  \lint \IVirt(\eps) \colorprod \FLM\rint = 
  & -  \sum_{i=1}^{\Np} \left[\frac{\ColT{i}^2}{\eps^2}   + \frac{\gamma_i}{\epsilon}\right] \lint \FLM\rint 
  %
  %
   +  \sum_{\inotj}^{\Np} \llint \frac{\fn{1}(s_{ij}/\mu^2)}{\epsilon}  (\ColT{i} \cdot \ColT{j}) \colorprod \FLM \rrint 
   \\
   & +   \doublesum{i,j=1}{j \not= i}{\Np}  \llint\left[\frac{\gamma_i}{\ColT{i}^2} \, \fn{1}( s_{ij}/\mu^2) - \frac{\pi^2}{2} \lambda_{ij}\right] (\ColT{i} \cdot \ColT{j}) \colorprod \FLM \rrint   \; , 
   \label{eq:NLO_virt_exp}
\end{split}    
\end{equation}
where we have dropped all terms 
beyond $\order{\eps^0}$.  
Since  $\fn{n}(x) \sim \mathcal{O}(\eps^0)$, 
poles in the color-correlated structures appear only  at $\mathcal{O}(\eps^{-1})$, while all terms in the last  line are finite. 

Comparing Eqs.~\eqref{Eq:NLO_soft_real_corrections_final_result} and~\eqref{eq:NLO_virt_exp},  we observe 
that the $\mathcal{O}(\eps^{-2})$ poles  cancel among these two contributions. 
Furthermore, we note that the function $f_1(s_{ij}/\mu^2)$ in Eq.~(\ref{eq:NLO_virt_exp}) can be written as 
\be    
\fn{1}(s_{ij}/\mu^2)  = 
\fn{1}(\eta_{ij}) +
\fn{1}(2E_i/\mu) + \fn{1}(2E_j/\mu) + \ep \, g_{ij} \; .
\ee
The first term on the right-hand side above  
is the function that appears in the 
soft contribution $\ISoft$, the next two 
terms depend  on one of the two indices 
$i$ or $j$, and the last term 
 \be
 g_{ij} =
 \fn{1}(2E_i/\mu) \fn{1}(2E_j/\mu) + \fn{1}(4 E_iE_j/\mu^2) \fn{1}(\eta_{ij}) \; , 
 \ee
 is  $\mathcal{O}(\eps^0)$. Thus we can further simplify the expression 
 for $\IVirt$ by making use of color conservation. We find 
 \begin{equation}
   \sum_{\inotj}^{N} \llint \frac{\fn{1}(2E_i/\mu) + \fn{1}(2E_j/\mu)}{\eps} (\ColT{i} \cdot \ColT{j}) \colorprod \FLM\rrint = - 2\sum_{i = 1}^{N}\frac{\ColT{i}^2}{\eps} \llint \fn{1}(2E_i/\mu) \, \FLM\rrint \; .
\end{equation}
Upon combining 
soft and virtual $I$-operators, 
we obtain the following result 
\be
\begin{split}
    & \llint \big[\IVirt(\eps)+\ISoft(\eps) \big]\colorprod \FLM \rrint \\ 
    = & \sum_{i=1}^{\Np}  \llint \left[ \frac{\ColT{i}^2}{\eps} \Big(\fn{2}(2\Emax/\mu) - 2\fn{1}(2E_i/\mu)\Big) - \frac{\gamma_i} {\epsilon} \right] \FLM \rrint + \order{\eps^0} \\
     =& - \sum_{i=1}^{\Np} \llint \left( 2 L_i \frac{\ColT{i}^2}{\eps} + \frac{\gamma_i}{\eps} \right) \FLM \rrint + \mathcal{O}(\eps^0) \; ,
\label{eq:soft_plus_virt}
\end{split}
\ee
where we  substituted the expansion 
of $f_{1,2}(x)$ in $\ep$ and 
used $L_i = \log(\Emax/E_i)$. The above equation  
implies that the $\ep$-divergences proportional to correlators 
of color charges cancel in the sum of the virtual and soft functions,   
$\IVirt$ and $\ISoft$.

To understand the cancellation of the  remaining poles, we  need  to combine the above result with 
the operator $\IColl(\eps)$
defined  in Eq.~\eqref{eq:IColl_definition0}.  We repeat its definition 
here for convenience
\index{I!$\IColl$}
\be
    \IColl(\eps) = \sum_{i=1}^{\Np}  \frac{\Gamma_{i,f_i}}{\epsilon} \; . 
    \ee
The generalized collinear anomalous 
 dimension 
 $\Gamma_{i,f_i}$ that appears in  the above equation 
  can be found in  
  Eq.~(\ref{eq:gamma_expansion_is0}). 
 Expanding it in powers of $\ep$, we find
\index{Generalized anomalous dimension!Expansion}
\be
    \Gamma_{i,f_i} = \gamma_i + 2 \ColT{i}^2 L_i + \mathcal{O}(\eps) \; , \qquad i=1, ...\, , \Np \; ,
\label{eq:gamma_expansion_fsis}
\ee
so that $\IColl(\eps)$ becomes 
\be
   \IColl(\eps) = \sum_{i=1}^{\Np}  
   \left (  2 L_i \frac{\ColT{i}^2}{\ep}  + \frac{\gamma_i}{\ep} 
    \right ) + {\cal O}(\ep^0) \; .
    \label{eq.a16}
\ee
Comparing this result 
with Eq.~(\ref{eq:soft_plus_virt}), 
we conclude that the following combination of $I$-operators
\index{I!$\ITot$}
\be
    \lint \ITot(\eps) \colorprod \FLM \rint 
    ={}  \llint \big[\IVirt(\eps) + \ISoft(\eps) + \IColl(\eps)\big] \colorprod  \FLM\rrint  \; ,
    \ee
is finite, as stated in the main text. 
Finally, we note that the cancellation between the initial-state collinear singularities and the PDFs renormalization has been discussed in
detail in Section~\ref{sec:nlo}.

\section{Partitions and sectors for the NNLO collinear limits} \label{sect:collinearseparation}
\par

In Section~\ref{sec:nnlo} we defined the soft-subtracted double-real contribution $\SigmaRR$, and we discussed the extraction of its collinear singularities. To do so, we first  split the angular phase space into partitions using the functions $\omega^{\Fp i, \Sp j}$ defined in Appendix~\ref{appendix:b}, and then further split the triple-collinear angular partitions into sectors using 
\index{T!$\theta^{(a)}$}
\index{T!$\theta^{(b)}$}
\index{T!$\theta^{(c)}$}
\index{T!$\theta^{(d)}$}
\be
\begin{split}
    \theta^{(a)} = & ~ \Theta\left(\eta_{i\Sp} < \frac{\eta_{i \Fp}}{2}\right) \; ,
    \;\;\;\;\;\;\;\;\;\;\;\;\;\;\;\theta^{(c)} = ~ \Theta\left(\eta_{i\Fp} < \frac{\eta_{i \Sp}}{2}\right) \; ,  \\
    \theta^{(b)} = & ~ \Theta\left(\frac{\eta_{i \Fp}}{2} < \eta_{i\Sp} < \eta_{i \Fp}\right) \; ,  
    \;\;\;\;\;\theta^{(d)} = ~ \Theta\left(\frac{\eta_{i \Sp}}{2} < \eta_{i\Fp} < \eta_{i \Sp}\right) \; .
    \label{eq:theta}
\end{split}
\ee
It follows that 
\begin{equation}
    \theta^{(a)} + \theta^{(b)} + \theta^{(c)} + \theta^{(d)} \equiv 1 \; . 
\end{equation}
A parametrization of the angular phase space that naturally achieves this sectoring is given in Ref.~\cite{Czakon:2010td} and is detailed in Appendix~\ref{sec:phase_space}. This procedure ensures that each partition and sector contains the minimal number of singular collinear limits. We then apply the appropriate collinear operators and write $\SigmaRR$ as the sum of four distinct contributions 
\begin{equation}
    \SigmaRR = \sum_{i=1}^{4} \SigmaRR^{(i)} 
    \equiv
\sum_{i=1}^{4} \lint \oS_{\Fp \Sp} \, \oS_{\Sp} \, \Omega_i \, \Delta^{(\Fp \Sp)} \, \THmn \FLM(\Fp,\Sp) \rint \; ,
\end{equation}
where the four quantities $\Omega_{i}$
provide  the partition of unity 
\begin{equation}
    \sum_{i=1}^{4} \Omega_i \equiv 1 \; .
\end{equation}
They read (cf. Refs.~\cite{Caola:2017dug,Caola:2019nzf,Caola:2019pfz})
\index{O!$\Omega_{1}$}
\index{O!$\Omega_{2}$}
\index{O!$\Omega_{3}$}
\index{O!$\Omega_{4}$}
\begin{align}
    \Omega_{1} = & ~ \sum_{\inotj}^{\Np} \oC_{i \Fp} \oC_{j \Sp} [\rmd p_\Fp] [\rmd p_\Sp] \, \omega^{\Fp i, \Sp j} \nonumber \allowdisplaybreaks\\
    & + \sum_{i=1}^{\Np} \Big[\oC_{i \Sp} \theta^{(a)} + \oC_{\Fp \Sp} \theta^{(b)} + \oC_{i \Fp} \theta^{(c)} + \oC_{\Fp \Sp} \theta^{(d)}\Big] [\rmd p_\Fp] [\rmd p_\Sp]  \, \oC_{\Fp \Sp, i}\; \omega^{\Fp i, \Sp i}  \; , \allowdisplaybreaks \label{eq:Omega1} \\
    \Omega_{2} = & ~ \sum_{i=1}^{\Np} \Big[\oC_{i \Sp} \theta^{(a)} + \oC_{\Fp \Sp} \theta^{(b)} + \oC_{i \Fp} \theta^{(c)} + \oC_{\Fp \Sp} \theta^{(d)}\Big] [\rmd p_\Fp] [\rmd p_\Sp]  \, C_{\Fp \Sp, i} \; \omega^{\Fp i, \Sp i} \;  , \allowdisplaybreaks \label{eq:Omega2}\\
    \Omega_{3} = & - \sum_{\inotj}^{\Np} C_{j \Sp} C_{i \Fp} [\rmd p_\Fp] [\rmd p_\Sp] \, \omega^{\Fp i, \Sp j} \; , \allowdisplaybreaks \label{eq:Omega3} \\
    \Omega_{4} = & ~ \sum_{\inotj}^{\Np} \Big[C_{i \Fp}[\rmd p_\Fp] + C_{j \Sp}[\rmd p_\Sp] \Big] \allowdisplaybreaks \, \omega^{\Fp i, \Sp j} \nonumber \allowdisplaybreaks  \\
    & + \sum_{i=1}^{\Np} \Big[C_{i \Sp} \theta^{(a)} + C_{\Fp \Sp} \theta^{(b)} + C_{i \Fp} \theta^{(c)} + C_{\Fp \Sp} \theta^{(d)}\Big] [\rmd p_\Fp] [\rmd p_\Sp] \, \omega^{\Fp i, \Sp i} \; , 
    \label{eq:Omega4}
\end{align}
where we have introduced the triple-collinear operator $C_{\Fp \Sp,i}$, which extracts the singular behavior in the limit $\rho_{i \Fp} \sim \rho_{i \Sp} \sim \rho_{\Fp \Sp} \to 0$. 
We note that in the above definitions of $\Omega_i$, $[\rmd p_{\Fp}]$ and $[\rmd p_{\Sp}]$
are phase-space elements for partons $\Fp$ and $\Sp$, and that they appear to the \emph{right} of the single collinear 
operators ($C_{i \Fp}$, $C_{\Fp \Sp}$, etc.) but to the \emph{left} of the triple-collinear operators $C_{\Fp \Sp,i}$. Therefore, the single-collinear operators  act on the phase-space elements, while the triple-collinear operators do not~\cite{Caola:2019nzf}. This allows us to use the results of Ref.~\cite{Delto:2019ewv} for $\Omega_2$.

\label{sec:collinear_contr}
\section{Phase-space parametrization and collinear limits} \label{sect:phasespace}

\subsection{Phase-space parametrizations for unresolved partons} 
\label{sec:phase_space}

\index{Phase-space parametrization}
In this subsection we describe   phase-space parametrizations 
for two unresolved partons that naturally achieve the angular sectoring required for NNLO computations \cite{Czakon:2010td}. 
We recall that there are two distinct kinematic  configurations that require different parametrizations. The first is a  triple-collinear configuration which requires
a
genuine $\NNLO$ parametrization   to describe strongly-ordered collinear limits.  The second is the case where the two partons are emitted by different hard legs and 
 can be described by two independent  $\NLO$-like parametrizations. 

In both cases, we begin by separating the energy and the angular parts of the phase space and write
\begin{align}
    [\rmd p_{\Fp}] [\rmd p_{\Sp}] = (\rmd E_\Fp \, E_\Fp^{1-2\ep}) \, (\rmd E_\Sp \, E_\Sp^{1-2\ep}) [\rmd \Omega_{\Fp \Sp}^{(d-1)}] \; ,
    \label{Eq:Appendix_dp_Fp_dp_Sp_parametrization}
\end{align}
where
\begin{equation}
    [\rmd\Omega_{\Fp \Sp}^{(d-1)}]  = [\rmd\Omega_{\Fp}^{(d-1)}] [\rmd\Omega_{\Sp}^{(d-1)}] \; , \qquad [\rmd\Omega_{i}^{(d-1)}] = \frac{\rmd\Omega_{i}^{(d-1)}}{2(2\pi)^{d-1}} \; .
\end{equation}

We first focus on the triple-collinear sectors and assume that the unresolved partons $\Fp$ and $\Sp$ are  emitted by a hard parton $i$, with $i \in [1,\Np]$.  It is convenient to choose 
 the momentum of parton $i$  as the reference direction. We then write
\begin{equation}
\begin{split}
    p_\Fp^\mu = & ~ E_\Fp \big(t^\mu +\cos \theta_{i \Fp} \, e_i^\mu +\sin \theta_{i \Fp} \, b^\mu \big) \; ,\\
    p_\Sp^\mu = & ~ E_\Sp \big(t^\mu +\cos \theta_{i \Sp} \, e_i^\mu + \sin \theta_{i \Sp} \, (\cos \phi_{\Fp\Sp} \, b^\mu + \sin \phi_{\Fp \Sp} \, a^\mu) \big) \; ,
    \label{Eq:p_Fp^mu_p_Sp^mu_parametrization}
\end{split}
\end{equation}
where
\begin{equation}
    t^{\mu} = (1,\vec{0}) \; , \qquad e_i^\mu = (0,\vec{n}_i) \; , \qquad p_i^\mu = E_i (t^\mu + e_i^\mu) \; .
    \label{eq:e4}
\end{equation}
Here $\vec{n}_i$ is a unit vector in $(d-1)$ spatial dimensions and  $a$ and $b$ are $d$-dimensional unit vectors such that 
\begin{align}
    t\cdot a = e_i \cdot a = t \cdot b = e_i \cdot b = a \cdot b = 0 \; .
\end{align}
We can use this parametrization to 
express the angular part of the phase space 
as 
\cite{Czakon:2010td}
\index{O!$\rmd\Omega_a^{(d-3)}$}
\begin{equation}
\begin{split}
	[\rmd\Omega_{\Fp \Sp}^{(d-1)}] 
    = &~ \frac{\rmd\Omega_b^{(d-2)} \rmd\Omega_a^{(d-3)}}{2^{6\epsilon} (2\pi)^{2d-2}} [\eta_{i\Fp}(1-\eta_{i\Fp})]^{-\epsilon} [\eta_{i\Sp}(1-\eta_{i\Sp})]^{-\epsilon} \\
    & \times  \frac{|\eta_{i\Fp} - \eta_{i\Sp}|^{1-2\epsilon}}{D^{1-2\epsilon}} \frac{\rmd \eta_{{i\Fp}} \, \rmd \eta_{{i\Sp}} \, \rmd \lambda}{[\lambda(1-\lambda)]^{\frac{1}{2} + \epsilon}} \; ,
\label{Eq:Appendix_dOmega5_dOmega6_general_writing}
\end{split}
\end{equation} 
where 
\begin{equation}
	D = \eta_{i \Fp} + \eta_{i \Sp} - 2\eta_{i \Fp} \, \eta_{i \Sp} + 2(2\lambda-1)\sqrt{\eta_{i \Fp} \, \eta_{i \Sp} (1-\eta_{i \Fp})(1-\eta_{i \Sp})} \; .
\end{equation}
The variable $\lambda$ parametrizes the dependence 
on the azimuthal angle $\phi_{\Fp \Sp}$ through the relation
\begin{equation}
	\sin^2\phi_{\Fp \Sp} = 4\lambda(1-\lambda) 
 \frac{|\eta_{i\Fp} - \eta_{i\Sp}|^2}{D^2} \; .
\end{equation}

The phase space can be split into four different sectors that we will refer to as $(a)$, $(b)$, $(c)$, $(d)$.
The following parametrizations are  chosen for each of the four sectors
\begin{align}
    a) & ~ \eta_{i \Fp} = ~ x_3 \; , &  
    \eta_{i \Sp} = & ~ x_3 x_4/2  \; ,
     \label{eq.e9}\\
    b) & ~ \eta_{i \Fp} = ~ x_3 \; , & 
    \eta_{i \Sp} = & ~ x_3 (1-x_4/2)  \; , \\
    c) & ~ \eta_{i \Fp} = ~ x_3 x_4/2 \; , & 
    \eta_{i \Sp} = & ~ x_3  \; ,
    \label{eq.e11} \\
    d) & ~ \eta_{i \Fp} = ~ x_3 (1-x_4/2)  \; , & 
    \eta_{i \Sp} = & ~ x_3  \; ,
\end{align}
with $0 \le x_{3,4} \le 1$.
We use them  to obtain explicit expressions for
\begin{equation}
    [\rmd\Omega_{\Fp \Sp}^{(i)}] = [\rmd\Omega_{\Fp \Sp}^{(d-1)}] \, \theta^{(i)} \; , \qquad i = ~ a, b, c, d \; ,
\end{equation}
with $\theta^{(i)}$ defined in Eq.~\eqref{eq:theta}. It turns out that the angular phase spaces for sectors $(a)$ and $(c)$ and for sectors $(b)$ and $(d)$ are identical.  For sectors $(a)$ and $(c)$ we find
\begin{equation}
\begin{split}
    [\rmd\Omega_{\Fp \Sp}^{(a,c)}] = &  \left[\frac{1}{8\pi^2} \frac{(4\pi)^\epsilon}{\Gamma(1-\epsilon)}\right]^2 \left[\frac{\Gamma^2(1-\epsilon)}{\Gamma(1-2\epsilon)}\right] \frac{[\rmd\Omega_b^{(d-2)}]}{[\Omega_b^{(d-2)}]} \frac{[\rmd\Omega_a^{(d-3)}]}{[\Omega_a^{(d-3)}]} \\
	& \times \frac{\rmd x_3}{x_3^{1+2\epsilon}} \frac{\rmd x_4}{x_4^{1+\epsilon}} \frac{\rmd \lambda}{\pi [\lambda(1-\lambda)]^{\frac{1}{2}+\epsilon}} \Big(256 F_\epsilon^{(a,c)}\Big)^{-\epsilon} 4F_0^{(a,c)} x_3^2 x_4 \; ,
	\label{Eq:Appendix_dOmega56^(a,c)_general_writing}
\end{split}
\end{equation}
where 
\begin{equation}
    F_\ep^{(a,c)} = \frac{(1-x_3) (1- x_3 x_4/2) (1-x_4/2)^2}{4 [N(x_3,x_4/2,\lambda)]^2} \; , \\
    \qquad
    F_0^{(a,c)} = \frac{1-x_4/2}{2 N(x_3, x_4/2,\lambda)} \; .
\end{equation}
For sectors $(b)$ and $(d)$ we obtain 
\begin{equation}
\begin{split}
    [\rmd\Omega_{\Fp \Sp}^{(b,d)}] = & \left[\frac{1}{8\pi^2} \frac{(4\pi)^\epsilon}{\Gamma(1-\epsilon)}\right]^2 \left[\frac{\Gamma^2(1-\epsilon)}{\Gamma(1-2\epsilon)}\right] \frac{[\rmd\Omega_b^{(d-2)}]}{[\Omega_b^{(d-2)}]} \frac{[\rmd\Omega_a^{(d-3)}]}{[\Omega_a^{(d-3)}]} \\
	& \times \frac{\rmd x_3}{x_3^{1+2\epsilon}} \frac{\rmd x_4}{x_4^{1+2\epsilon}} \frac{\rmd \lambda}{\pi [\lambda(1-\lambda)]^{\frac{1}{2}+\epsilon}} \Big(256 F_\epsilon^{(b,d)}\Big)^{-\epsilon} 4F_0^{(b,d)} x_3^2 x_4^2 \; ,
	\label{Eq:Appendix_dOmega^dc_general_writing}
\end{split}
\end{equation}
where 
\begin{equation}
    F_\ep^{(b,d)} = \frac{(1-x_3) (1-x_4/2) (1-x_3(1-x_4/2))}{4 [N(x_3,1-x_4/2,\lambda)]^2} \; , 
    \quad
    F_0^{(b,d)} = \frac{1}{4 N(x_3, 1-x_4/2,\lambda)} \; .
\end{equation}
The function 
$N(x_3, x_4, \lambda)$ introduced  in the above equations reads
\begin{equation}
    N(x_3, x_4, \lambda) = 1+x_4(1-2x_3) - 2(1-2\lambda) \sqrt{x_4(1-x_3)(1-x_3x_4)} \; .
\end{equation}

To simplify the subtraction terms, we need particular collinear  limits of the unresolved phase space.  To obtain those, 
we note that the following identities hold
\begin{equation}
\label{eq.e19}
  \begin{split}
    & 
    \lim_{x_4 \to 0} F_{\ep}^{(a,c)} = \frac{1-x_3}{2} \; ,
    \\
    & \lim_{x_4 \to 0} F_{\ep}^{(b,d)} = \frac{1}{64 \lambda^2} \; ,
  \end{split}
\qquad\quad
  \begin{split}
    &
    \lim_{x_4 \to 0} F_0^{(a,c)} = \frac{1}{2} \; ,
    \\
    &
    \lim_{x_4 \to 0} F_0^{(b,d)} = \frac{1}{16\lambda(1-x_3) } \; .
  \end{split}
\end{equation}
The $x_4 \to 0$ limit corresponds to the 
$\Sp || i$ and  $\Fp || i$ collinear limits in sectors $(a)$ and $(c)$, respectively, 
and to the $\Fp || \Sp$ limit in sectors $(b)$ and $(d)$.
The singular   quantities in  sectors $(a)$ and $(c)$  are $\eta_{i \Sp}$ and $\eta_{i \Fp}$, respectively, and they are given in Eqs~\eqref{eq.e9} and~\eqref{eq.e11}. For sectors $(b)$ and $(d)$, the limit of the corresponding singular variable is more complex.  It reads  
\be
\lim_{x_4 \to 0} \eta_{\Fp \Sp} =  \lim_{x_4 \to 0} \frac{x_3 x_4^2}{4N(x_3, 1-x_4, \lambda)} = \frac{x_3 x_4^2}{16 \lambda (1-x_3)} \equiv \bar{\eta}_{\Fp \Sp} \;.
\ee
\\

The phase-space parametrization is significantly simpler for the double-collinear partitions. Consider the case when parton $\Fp$ is collinear to parton $i$ and parton 
$\Sp$ to parton $j$, with $i \not=j$. We parametrize the momenta $p_\Fp$ and $p_\Sp$ using the momenta of partons $i$ and $j$ respectively, i.e.
\be
\begin{split}
    & p^\mu_{\Fp} = E_{\Fp}( t^\mu+ \cos \theta_{i \Fp} \, e_i^\mu + 
    \sin \theta_{i \Fp} \, b_{\Fp}^\mu) \; , \\
    & p^\mu_{\Sp} = E_{\Sp}( t^\mu+ \cos \theta_{j \Sp} \, e_j^\mu + 
    \sin \theta_{j \Fp} \, b_{\Sp}^\mu) \; ,
\end{split}    
\ee
and set 
\begin{equation}
     \eta_{i\Fp} = x_3 \; , \qquad
     \eta_{j\Sp} = x_4 \; . \label{Eq:Appendix_dc_x3_x_4_defs}
\end{equation}
We then write the angular phase space for  the double-collinear partition  $[\rmd\Omega_{\Fp \Sp}^{\rm dc}]$ as 
\begin{equation}
    [\rmd\Omega_{\Fp \Sp}^{\rm dc}] \equiv [\rmd\Omega_{\Fp}^{(d-1)}] [\rmd\Omega_{\Sp}^{(d-1)}] \; ,
\end{equation}
where 
\begin{equation}
\begin{split}
    [\rmd\Omega_{\Fp}^{(d-1)}]
    = & \left[\frac{1}{8\pi^2} \frac{(4\pi)^\epsilon}{\Gamma(1-\epsilon)}\right]  2^{4-4\ep} \frac{[\rmd \Omega_{\Fp}^{(d-2)}]}{[\Omega^{(d-2)}]} \frac{\rmd  x_3}{x_3^{1+\ep}} (1 - x_3)^{-\ep} x_3 \; , \\
    [\rmd\Omega_{\Sp}^{(d-1)}]
    = & \left[\frac{1}{8\pi^2} \frac{(4\pi)^\epsilon}{\Gamma(1-\epsilon)}\right]  2^{4-4\ep} \frac{[\rmd \Omega_\Sp^{(d-2)}]}{[\Omega^{(d-2)}]} \frac{\rmd  x_4}{x_4^{1+\ep}} (1 - x_4)^{-\ep} x_4 \; .
\end{split}
\end{equation}

\subsection{Action of the collinear operators on the phase space}\label{sect:pppandcl}

\index{Phase-space parametrization}
In our definitions of the angular terms $\Omega_{1,..,4}$ in  Eqs~\eqref{eq:Omega1}-(\ref{eq:Omega4}), the collinear operators act on the phase space of the two unresolved partons. 
As we have seen, it is useful to rewrite the subtraction terms in such a way that these operators do not act on the phase-space measure. 
We have quoted the results in the main text of the paper without deriving them, see e.g.~Eq.~\eqref{eq4.33}. The goal of this subsection is to provide the omitted details.

We begin by considering the double-collinear  partitioning with  a collinear operator $C_{i \Fp}$; an example can be found in the first term on the right-hand side of Eq.~\eqref{Eq:Sigma_RR1c^a}. 
Since in the double-collinear parametrization of the phase space  the collinear limit $i ||\Fp$ is controlled by the variable $x_3$ (see Eq.~\eqref{Eq:Appendix_dc_x3_x_4_defs}), we find 
\begin{equation}
\begin{split}
    \int C_{i\Fp} \frac{[\rmd \Omega_{\Fp \Sp}^{\rm dc}]}{\rho_{i\Fp}} \,  [...] 
    = & \left[\frac{1}{8\pi^2} \frac{(4\pi)^\epsilon}{\Gamma(1-\epsilon)}\right]  2^{3-4\ep} \int \frac{[\rmd \Omega_\Fp^{(d-2)}]}{[\Omega^{(d-2)}]} [\rmd \Omega_\Sp^{(d-1)}] \int \limits_{0}^{1} \frac{\rmd  x_3}{x_3^{1+\ep}} \,C_{i\Fp}  [...]  \\
    = & - \left[\frac{1}{8\pi^2} \frac{(4\pi)^\epsilon}{\Gamma(1-\epsilon)}\right]  \frac{2^{3-4\ep}}{\ep} \int \frac{[\rmd \Omega_\Fp^{(d-2)}]}{[\Omega^{(d-2)}]} [\rmd \Omega_\Sp^{(d-1)}] \, C_{i\Fp} [...]  \; ,
\label{Eq:Appendix_sec_ac_collinear_limit_swap_phase_space_intermediate_step}
\end{split}
\end{equation}
where $[...]$ stands for generic non-singular contributions whose exact form is not relevant for the following discussion. 
If we repeat the above steps without acting with $C_{i\Fp}$ on the phase space, we find
\begin{equation}
\begin{split}
   &  \int \frac{[\rmd \Omega_{\Fp \Sp}^{\rm dc}]}{\rho_{i\Fp}} \, C_{i\Fp}\, [...] 
    \\
&     =  \left[\frac{1}{8\pi^2} \frac{(4\pi)^\epsilon}{\Gamma(1-\epsilon)}\right]  2^{3-4\ep} \int \frac{[\rmd \Omega_\Fp^{(d-2)}]}{[\Omega^{(d-2)}]} [\rmd \Omega_\Sp^{(d-1)}]
     \int \limits_{0}^{1} \frac{\rmd  x_3}{x_3^{1+\ep}} (1 - x_3)^{-\ep} C_{i \Fp} \,  [...] \\
    & = - \frac{\Gamma^2(1-\ep)}{\Gamma(1-2\ep)} \left[\frac{1}{8\pi^2} \frac{(4\pi)^\epsilon}{\Gamma(1-\epsilon)}\right]  \frac{2^{3-4\ep}}{\ep} \int \frac{[\rmd \Omega_\Fp^{(d-2)}]}{[\Omega^{(d-2)}]} [\rmd \Omega_\Sp^{(d-1)}] \, C_{i \Fp} \, [...] \; .
    \label{Eq:Appendix_sec_ac_collinear_limit_swap_phase_space_intermediate_stepB}
\end{split}
\end{equation}
Comparing  the two formulas, we conclude that
\begin{equation}
   \int C_{i\Fp} \frac{[\rmd \Omega_{\Fp \Sp}^{\rm dc}]}{\rho_{i\Fp}} \, [...] 
   = \frac{\Gamma(1-2\ep)}{\Gamma^2(1-\ep)} \int \frac{[\rmd \Omega_{\Fp \Sp}^{\rm dc}]}{\rho_{i\Fp}} \, C_{i\Fp} \, [...]  \; .
   \label{Eq:Appendix_sec_ac_collinear_limit_swap_phase_space}
\end{equation}
We can use the above relation when rewriting Eq.~\eqref{eq4.16}
 as  Eq.~\eqref{Eq:Sigma_N_3}.
 Since in this 
case we have two collinear operators $C_{j \Sp} C_{i \Fp} $, we need 
to apply it twice, i.e.
\begin{equation}
    \int C_{j\Sp} C_{i\Fp} \frac{[\rmd \Omega_{\Fp \Sp}^{\rm dc}]}{\rho_{i\Fp} \, \rho_{j\Sp}} \,  [...] 
   = \left[\frac{\Gamma(1-2\ep)}{\Gamma^2(1-\ep)}\right]^2 \int \frac{[\rmd \Omega_{\Fp \Sp}^{\rm dc}]}{\rho_{i\Fp} \, \rho_{j\Sp}} \, C_{j\Sp} C_{i\Fp} \, [...]  \; ,
\end{equation}
so that Eq.~\eqref{eq4.16} becomes
\begin{equation}
\begin{split}
    \SigmaRRcc = & - \left[\frac{\Gamma(1-2\epsilon)}{\Gamma^2(1-\epsilon)}\right]^2 \sum_{\inotj}^{\Np} \lint \oS_{\Fp \Sp} \oS_{\Sp} C_{j \Sp} C_{i \Fp} \, \omega^{\Fp i, \Sp j} \Delta^{(\Fp \Sp)} \, \THmn \FLM(\Fp,\Sp) \rint \; . 
\end{split}
\end{equation}
We stress that the absence of the phase space $[\rmd p_\Fp] [\rmd p_\Sp]$ in the above equation indicates that the collinear operators  $C_{j\Sp} C_{i\Fp}$ no longer act on it.

 Similar formulas can also be derived for the triple-collinear partitions that involve sector $\theta^{(c)}$. As an example, we discuss the second term on the right-hand side of Eq.~\eqref{Eq:Sigma_RR1c^a}.
 In this case, the collinear limit $i || \Fp$
 corresponds to the $x_4 \to 0$ 
 limit in the phase space parametrization in 
 Eq.~\eqref{Eq:Appendix_dOmega56^(a,c)_general_writing}.
 We use Eq.~\eqref{eq.e19} to 
 compute this limit and find  
\begin{equation}
\begin{split}
    \int C_{i\Fp} \frac{[\rmd \Omega_{\Fp \Sp}^{(c)}]}{\rho_{i\Fp}}  \,  [...] 
    = \frac{\Gamma(1-2\ep)}{\Gamma^2(1-\ep)} \int \frac{[\rmd \Omega_{\Fp \Sp}^{(d-1)}]}{\rho_{i\Fp}} \, (\eta_{i\Sp}/2)^{-\ep} C_{i\Fp} \,  [...]  \; ,
    \label{Eq:Appendix_sec_ac_collinear_limit_swap_phase_space_tc_sector}
\end{split}
\end{equation}
where the integration over the angular variables of parton 
$\Fp$ on the right hand side of Eq.~(\ref{Eq:Appendix_sec_ac_collinear_limit_swap_phase_space_tc_sector}) is \emph{not restricted} to sector $(c)$ anymore.  It follows from the above discussion that Eq.~\eqref{Eq:Sigma_RR1c^a} can be rewritten as
\begin{equation}
\begin{split}
    \SigmaRRcACDC = & ~ \frac{\Gamma(1-2\epsilon)}{\Gamma^2(1-\epsilon)} \bigg\langle \mathcal{S}(\Fp,\Sp) \bigg[\sum_{\inotj}^{\Np} C_{i\Fp} \, \omega^{\Fp i,\Sp j} 
    \\
    & + \sum_{i =1}^{\Np} (\eta_{i\Sp}/2)^{-\ep} C_{i\Fp} \, \omega^{\Fp i,\Sp i}\bigg] \Delta^{(\Fp \Sp)} \FLM(\Fp,\Sp)\bigg\rangle \; .
    \end{split}
\end{equation}
This expression is the starting point to obtain Eqs.~\eqref{eq4.33} and \eqref{eq4.34a}.

Finally, we perform similar manipulations
for sector $(b)$ where the collinear 
limit of interest is $\Fp || \Sp$.  
This limit corresponds to $x_4 \to 0$ 
in the phase space parametrization 
given in Eq.~\eqref{Eq:Appendix_dOmega^dc_general_writing}. 
Using Eq.~\eqref{eq.e19} we find 
\begin{equation}
\begin{split}
   \int  C_{\Fp \Sp} \frac{[\rmd\Omega_{\Fp \Sp}^{(b,d)}]}{\rho_{\Fp \Sp}}  [...] 
    = & \left[\frac{1}{8 \pi^2} \, \frac{(4\pi)^\ep}{\Gamma(1-\ep)}\right] N_\ep^{(b,d)} \; \int [\rmd\Omega_{[\Fp \Sp]}^{(d-1)}] \, \\
    & \times \eta_{i[\Fp \Sp]}^{-\ep} \, \big(1-\eta_{i[\Fp \Sp]}\big)^\ep \, \rmd\Lambda \, \frac{\rmd\Omega_a^{(d-3)}}{[\Omega^{(d-3)}]} \, \frac{\rmd x_4}{x_4^{1+2\ep}} \; C_{\Fp \Sp}  [...]  \;.
\end{split}
\label{eq.e32}
\end{equation}
The normalization constants $N_\ep^{(b,d)}$ 
that appear in Eq.~(\ref{eq.e32})
can be found  in Eq.~\eqref{eq:normalisation}, while $[\rmd\Omega_{[\Fp \Sp]}^{(d-1)}]$ is the (exact) angular phase space of the clustered parton $[\Fp \Sp]$, whose momentum $p_{[\Fp \Sp]} = p_\Fp + p_\Sp$ must be computed in the strict collinear limit. Furthermore we have introduced a new variable $\Lambda$ such that
\index{L!$\Lambda$}
\be
    %
    \rmd \Lambda =  ~ \frac{\Gamma(1+\ep) \, \Gamma(1-\ep)}{\Gamma(1+2\ep) \, \Gamma(1-2\ep)} \, 
    \frac{\lambda^{-1/2+\ep}(1-\lambda)^{-1/2-\ep}}{\pi} \rmd \lambda \;. 
\ee

We note that the action of the operator 
$C_{\Fp \Sp}$ on the  matrix 
element squared  
is non-trivial  because it can lead to  integrands that  
depend on the parameter $\lambda$ and the transverse vector $a^\mu$. This phenomenon, known as spin correlations, is discussed in the  next appendix. Here we consider only those terms for which the action of $C_{\Fp \Sp}$  in Eq.~\eqref{Eq:Sigma_RR1c_b_definition}  does not lead to such terms. In this case  we can integrate over $x_4$, the directions of $a^\mu$, and the azimuthal variable $\Lambda$ using 
\be
\label{eq:Int_Lambda}
\int \rmd  \Lambda = 1 \; .
\ee
Comparing  the result with the one that is obtained when the collinear operator $C_{\Fp \Sp}$  
does not act on the phase space, we find 
\be
   \int  C_{\Fp \Sp} \frac{[\rmd\Omega_{\Fp \Sp}^{(b,d)}]}{\rho_{\Fp \Sp}}[...] 
 =  2^{2\ep - 1} \frac{\Gamma(1+2\ep) \Gamma(1-2\ep)}{\Gamma(1+\ep) \Gamma(1-\ep)}
\int  \frac{[\rmd\Omega_{\Fp \Sp}^{(d-1)}]}{\rho_{\Fp \Sp}} C_{\Fp \Sp}  [...]  \; , 
\label{eq.e34}
  \ee
where the integration over the angular 
variables of partons on the right-hand side is unrestricted. We use this relation in Eq.~\eqref{eq:4.46} and the analysis that follows.\\

As we just mentioned, the action of the collinear operator $C_{\Fp \Sp}$ on matrix elements may result in a limit that 
depends on  
$\lambda$ and $a^\mu$. In such cases 
Eq.~(\ref{eq.e34}) cannot be used.  
To understand how to proceed, we write  (see Appendix \ref{sec:splin_correlation})
\be
C_{\Fp \Sp} \, \FLM(\Fp, \Sp)  
=
\frac{\gsb^2 \,}{E_\Sp E_{\Fp} \, \rho_{\Fp\Sp}} \, 
\bigg[{}
- g^{\mu\nu} P_{gg}^{(0)}(z)
+
P_{gg}^\bot(z) \, \kappa_{\bot, (b)}^{\mu} \kappa_{\bot, (b)}^{\nu}
\bigg] \FLMmunu([\Fp \Sp]) \; , 
\label{eq:NNLO_C45_appE}
\ee
where vector $\kappa_{\bot, (b)}$ is a
unit space-like 
vector which is orthogonal to $p_{\Fp}$
\be
\kappa_{\bot, (b)} \cdot p_{\Fp } = 0 \; .
\ee
Using the phase space parametrization 
for sector $(b)$, we can write this vector 
as 
\index{K!$\kappa_{\bot, (b)}$}
\begin{align}
    \kappa_{\bot, (b)}^{\mu}  = a^\mu \sqrt{1-\lambda} + r_{i, (b)}^\mu \, \sqrt{\lambda} \; , 
\label{eq:kappa_decomp}
\end{align}
where vectors $a$ and $b$ were introduced in Eq.~\eqref{Eq:p_Fp^mu_p_Sp^mu_parametrization} and $r_{i,(b)}$ is the auxiliary spacelike vector ($r_{i,(b)} \cdot r_{i,(b)} = -1$) defined as 
\index{R!$r^{\mu}_{i,(b)}$}
\begin{align}
    r^{\mu}_{i,(b)} = & ~ \sin \theta_{i\Fp} \, e_i^{\mu} - \cos \theta_{i\Fp} \, b^\mu  \; .
    \label{Eq:r_i_b_definition}
\end{align}
The momentum 
of the clustered parton $[\Fp \Sp]$ is aligned with 
the momentum $p_\Fp$, which does not 
depend on $\lambda$ and $a^\mu$. Since  $\FLM([\Fp \Sp])$ is independent of $\lambda$ and $a^\mu$, we can  integrate 
over $\rmd\Omega_a^{(d-3)}$ and $\rmd\Lambda$.
Specifically, we  need to calculate 
\begin{align}
    \lint \kappa_{\bot,(b)}^\mu \, \kappa_{\bot,(b)}^\nu \rint = & ~ \int \rmd\Lambda \; \frac{\rmd\Omega^{(d-3)}_a}{\Omega^{(d-3)}} \,\; \kappa_{\bot,(b)}^\mu \, \kappa_{\bot,(b)}^\nu \; .
\end{align}
To compute this integral, we use Eq.~\eqref{eq:Int_Lambda} together with
\begin{equation}
  \begin{split}
    & 
    \int \rmd\Lambda \, \frac{\rmd\Omega^{(d-3)}_a}{\Omega^{(d-3)}} \, a^\mu = 0 \; ,
    \\
    &
    \int \rmd\Lambda \, \lambda = \frac{1+2\ep}{2} \; ,
  \end{split}
\qquad
  \begin{split}
    &
    \int \rmd\Lambda \, \frac{\rmd\Omega^{(d-3)}_a}{\Omega^{(d-3)}} \; a^\mu \, a^\nu= - \frac{g^{\mu \nu}_{\bot, (d-3)_a }}{d-3}\;, 
    \\
    &
    \int \rmd\Lambda \, (1-\lambda) = \frac{1-2\ep}{2} \; ,
  \end{split}
\end{equation}
and find 
\be
\begin{split}
    \lint \kappa_{\bot,(b)}^\mu \, \kappa_{\bot,(b)}^\nu \rint = & - \frac{g_{\perp,(d-3)}^{\mu\nu}}{2} + \frac{1+2\epsilon}{2} r_{i, (b)}^\mu r_{i, (b)}^\nu \\
	= & ~ \frac{1}{2}\bigg[g_{\perp,(d-3)}^{\mu\nu} + r_{i, (b)}^\mu r_{i, (b)}^\nu\bigg] + \ep \, r_{i, (b)}^\mu r_{i, (b)}^\nu \\
    \equiv & - \frac{g^{\mu \nu}_{\bot, (d-2)}}{2} + \ep \, r_{i, (b)}^\mu \, r_{i, (b)}^\nu \; .
\end{split}
\ee
We then obtain 
 \begin{align} \label{eq:kappa_and_r}
    \lint \kappa_{\bot,(b)}^\mu \, \kappa_{\bot,(b)}^\nu \rint \, \FLMmunu = \frac12 \FLM
    + \ep \, r_{i, (b)}^\mu \, r_{i, (b)}^\nu \, \FLMmunu \; ,
\end{align}
where  we used 
\begin{align}
    - g^{\mu \nu}_{\bot, (d-2)} \FLMmunu =  - g^{\mu \nu} \, \FLMmunu = \FLM \; ,
\end{align}
as allowed by the transversality of scattering amplitudes.

\section{Spin correlations}\label{sec:splin_correlation}

\index{Phase-space parametrization}
In this appendix we  discuss  the double-real contributions where  the so-called spin correlations appear. 
These effects arise  in sectors  $(b)$ and $(d)$ in the  limits when gluons $\Fp$ and $\Sp$ become
collinear to each other. 
To make this appendix self-contained, we start by considering the $\Fp || \Sp$  limit, which is described by the following expression (see also Eq.~\eqref{eq:NNLO_C45_appE})
\index{C!$C_{\Fp \Sp}$}
\be
\begin{split}
& C_{\Fp \Sp} \, \FLM(\Fp, \Sp)  
=
  \frac{\gsb^2 \, }{E_{\Fp} E_{\Sp} \, \rho_{\Fp \Sp}} \, 
 \Pgg^{\mu \nu} (z) \, 
 \FLMmunu([\Fp \Sp])
 \\
= &{}
\frac{\gsb^2 \,}{E_\Sp E_{\Fp} \, \rho_{\Fp\Sp}} \, 
\bigg[
\Pgg^{(0)}(z) \FLM([\Fp \Sp])
+
\Pgg^\bot(z) \, \kappa_{\bot, (b)}^{\mu} \kappa_{\bot, (b)}^{\nu} \, 
\FLMmunu([\Fp \Sp])
\bigg] \; , 
\label{eq:NNLO_C45}
\end{split}
\ee
 where the splitting functions were introduced in Appendix~\ref{subsec:split_tree_level}, $z = E_{\Fp}/(E_{\Fp} +E_{\Sp})$, and $\kappa_{\bot, (b)}$ is defined in Eq.~\eqref{eq:kappa_decomp}. The four-momentum of the clustered parton $[\Fp \Sp]$
  is equal to
  \begin{align}
p_{\Fp\Sp}^\mu = (E_\Fp+E_\Sp) \, n_\Fp^\mu = \frac{E_\Fp+E_\Sp}{E_\Fp} \, p_\Fp^\mu \; ,
\end{align}
where the vector $n_{\Fp}$ is a light-like vector defined as $n_\Fp = p_{\Fp}/E_{\Fp}$.
To proceed further, we assume that the collinear limit $\Fp || \Sp$ occurs in a particular triple-collinear
  partitioning, characterized by the
  partition
  function $\omega^{\Fp i, \Sp i}$, 
  and to restrict our analysis to sector $(b)$. The contribution that we are interested in reads (see Eq.~\eqref{Eq:Sigma_RR1c_b_definition})
  \be
\begin{split}
\label{eq:NNLO_Cmn_b}
\SigmaRRcBi
= &~
 \lint
\oS_{\Fp \Sp} \, 
\oS_{\Sp} \,  
\THmn\,
\Coll{\Fp\Sp} \,
\theta^{(b)} [\rmd p_{\Fp}][\rmd p_{\Sp}] \,
\omega^{\Fp i, \Sp i} \, 
\Delta^{(\Fp\Sp)} \,   
\FLM(\Fp,\Sp)
\rint
\\
= &
-\frac{[\alpha_{s}]}{2\ep } \, 
N_\ep^{(b,d)}  \; 
\bigg\langle
\oS_{\Fp \Sp} \, 
\oS_{\Sp}
\int \limits_0^{E_{\rm max}}
\frac{\rmd E_{\Fp}}{E_\Fp^{2\ep-1}} \frac{\rmd E_{\Sp}}{E_\Sp^{2\ep-1}} \, 
\THmn \, 
 \int
 [\rmd\Omega_{[\Fp\Sp]}] \, 
 \sigma_{i[\Fp \Sp]}^{-\ep} \,
\Delta^{([\Fp \Sp])}
 \\
& \times
  \omega^{\Fp i,  \Sp i}_{\Fp\parallel \Sp} \, 
\frac{1}{E_\Fp E_\Sp} \, 
\Big[
P_{gg}(z,\ep) \, \FLM([\Fp\Sp])
+ \ep \, P_{gg}^\bot(z) \, 
r_{i, (b)}^\mu \, r_{i, (b)}^\nu \, 
\FLMmunu([\Fp\Sp])
\Big]
\bigg\rangle
 \; ,
\end{split}
\ee
where we recall that $\sigma_{ij} = \eta_{ij}/(1-\eta_{ij})$.
To  derive Eq.~\eqref{eq:NNLO_Cmn_b}
we  
exploited the parametrization presented in Appendix~\ref{sect:phasespace}, integrated over the angles of  parton $\Sp$ and used the relation displayed in Eq.~\eqref{eq:kappa_and_r}. All the splitting functions that appear in 
Eq.~(\ref{eq:NNLO_Cmn_b}) can be found in 
Appendix~\ref{sec:Splitting}.

We note that the hard matrix element squared  appears in Eq.~\eqref{eq:NNLO_Cmn_b} in two distinct ways: once as
$\FLM([\Fp \Sp])$ and once  as $\FLMmunu([\Fp \Sp])$, where the open spin indices refer to the clustered parton.
In fact, the relation between the two contributions reads
\index{F!$\FLMmunu$}
\begin{align}
\FLM([\Fp \Sp]) 
=
\sum \limits_{\lambda_{[\Fp \Sp]}} \varepsilon^{\lambda_{[\Fp \Sp]}}_{\mu}
\varepsilon^{\lambda_{[\Fp \Sp]},*}_\nu \FLMmunu\big([\Fp \Sp]\big) 
= 
-g_{\mu \nu} \, 
\FLMmunu([\Fp \Sp])
\; ,
\end{align}
where the sum runs over the physical polarizations of the clustered parton $[\Fp \Sp]$ and the last step follows from the transversality
of $\FLMmunu$. 

In Eq.~\eqref{eq:NNLO_Cmn_b} the only term that requires further discussion is the one proportional to $\FLMmunu([\Fp \Sp])$. In fact, we find it convenient
to split these terms in such a  way that the coefficient
of $\FLM([\Fp \Sp])$ in  Eq.~\eqref{eq:NNLO_Cmn_b} is the spin-averaged $g \to gg$ splitting function $ \Pgg$ (c.f. Eq.~\eqref{def:Pgg_av})
and the soft subtraction term associated with it. We will refer to all other contributions that appear in the expression
for $\SigmaRRcBi$ as ``spin-correlated''.
Hence, we write 
\begin{align}
& \SigmaRRcBi
=
-\frac{\asbr}{2\ep } \, 
N_\ep^{(b,d)}  \; 
\bigg\langle 
\int \limits_0^{E_{\rm max}}
\frac{\rmd E_{\Fp}}{E_\Fp^{2\ep-1}} \frac{\rmd E_{\Sp}}{E_\Sp^{2\ep-1}} \, 
\Theta_{\Fp \Sp} 
\int
[\rmd\Omega_{[\Fp\Sp]}] \, 
\sigma_{i[\Fp \Sp]}^{-\ep} \,
\omega^{\Fp i , \Sp i}_{\Fp\parallel \Sp}
\nonumber 
 \\
& \times
\bigg\{
 \frac{1}{E_\Fp\, E_\Sp} \, 
P_{gg}(z)  \,\oS_{\Fp \Sp} \, \Delta^{([\Fp\Sp])} \,  
\FLM\big([\Fp\Sp]\big)
- \frac{2 C_A}{E_\Sp^2} \, \oS_{\Fp}\, \Delta^{(\Fp)} \,  \FLM(\Fp)
\label{eq:NNLO_Cmn_b_v2}
 \\
& + \frac{\ep}{E_\Fp E_\Sp} \, 
\Big[P_{gg}^\bot(z) \, 
\big(  r_{i, (b)}^\mu \, r_{i, (b)}^\nu+g^{\mu \nu} \big) 
- P_{gg}^{\bot, r}(z) \, g^{\mu  \nu} 
\Big] 
\oS_{\Fp \Sp}\, \Delta^{([\Fp \Sp])} \, 
\FLMmunu([\Fp \Sp])
\bigg\}
\bigg\rangle 
\nonumber
\; ,
\end{align}
where we have used the relation 
\index{Splitting functions!Relations}
\begin{align} 
\Big[P_{gg}(z, \ep) 
+
 \ep \, P_{gg}^\bot(z) 
 \Big] \FLM([\Fp\Sp])
 =
 P_{gg}(z) \, \FLM([\Fp\Sp])
-  
\ep \, 
g^{\mu \nu} \,  
P_{gg}^{\bot, r}(z) \FLMmunu([\Fp\Sp])
 \; ,
\end{align}
with $\Pgg^{\bot, r}$ defined in Eq.~\eqref{def:Pgg_bot_r}.
 The second line in  Eq.~\eqref{eq:NNLO_Cmn_b_v2} contains ``spin-averaged'' and the third line ``spin-correlated'' contributions. They read
\be
\begin{split}
\label{eq:NNLO_Cmn_b_sa}
\SigmaRRcBiSA
= &
-\frac{\asbr}{2\ep } \, 
N_\ep^{(b,d)} 
\bigg\langle 
\int \limits_0^{E_{\rm max}}
\frac{\rmd E_{\Fp}}{E_\Fp^{2\ep-1}} \frac{\rmd E_{\Sp}}{E_\Sp^{2\ep-1}} \, 
\Theta_{\Fp \Sp} 
\int
[\rmd\Omega_{[\Fp\Sp]}] \, 
\sigma_{i[\Fp \Sp]}^{-\ep} \,
 \omega^{\Fp i , \Sp i}_{\Fp\parallel \Sp}
 \\
&
\times
\left[
 \frac{1}{E_\Fp\, E_\Sp} \, 
P_{gg}(z)  \,\oS_{\Fp \Sp} \, \Delta^{([\Fp\Sp])} \,  
\FLM([\Fp\Sp])
- \frac{2 C_A}{E_\Sp^2} \, \oS_{\Fp}\, \Delta^{(\Fp)} \,  \FLM(\Fp)
\right]
\bigg\rangle \; ,
\end{split}
\ee
and 
\be
\begin{split}
\label{eq:NNLO_Cmn_b_sc}
& \SigmaRRcBiSC
=
-\frac{\asbr}{2} \, 
N_\ep^{(b,d)}  \; 
\bigg\langle 
\int \limits_0^{E_{\rm max}}
\frac{\rmd E_{\Fp}}{E_\Fp^{2\ep-1}} \frac{\rmd E_{\Sp}}{E_\Sp^{2\ep-1}} \, 
\Theta_{\Fp \Sp} 
\int
[\rmd\Omega_{[\Fp\Sp]}] \, 
 \sigma_{i[\Fp \Sp]}^{-\ep} \,
 \omega^{\Fp i , \Sp i}_{\Fp\parallel \Sp}\\
& \times \frac{1}{E_\Fp E_\Sp} 
\Big[P_{gg}^\bot(z) \, 
\big(  r_{i, (b)}^\mu \, r_{i, (b)}^\nu+g^{\mu \nu} \big) 
- P_{gg}^{\bot, r}(z) \, g^{\mu  \nu} 
\Big] 
\oS_{\Fp \Sp}\, \Delta^{([\Fp \Sp])} \, 
\FLMmunu([\Fp \Sp])
\bigg\rangle 
\; .
\end{split}
\ee

We continue with the 
discussion of the spin-correlated collinear 
limits. We find that 
after adding the contribution of sector $(d)$
to $\SigmaRRcBiSC$,
the energy-ordering constraint 
disappears and we obtain 
the following expression for the full
spin-correlated part
\be
\begin{split}
\label{eq:NNLO_C45_b}
& \SigmaRRcBDiSC
=
-\frac{\asbr}{2 } \, 
N_\ep^{(b,d)} 
\bigg\langle 
\int \limits_0^{E_{\rm max}}
\frac{\rmd E_{\Fp}}{E_\Fp^{2\ep-1}} \frac{\rmd E_{\Sp}}{E_\Sp^{2\ep-1}} \, 
\int
[\rmd\Omega_{[\Fp\Sp]}] \, 
\sigma_{i[\Fp \Sp]}^{-\ep} \,
 \omega^{\Fp i , \Sp i}_{\Fp\parallel \Sp}
\\
&
\times \frac{1}{E_\Fp E_\Sp} 
\Big [ P_{gg}^\bot(z) \, 
\big(  r_{i}^\mu \, r_{i}^\nu+g^{\mu \nu} \big) 
- P_{gg}^{\bot, r}(z) \, g^{\mu  \nu} 
\Big] 
\oS_{\Fp \Sp}\, \Delta^{([\Fp \Sp])} \, 
\FLMmunu([\Fp \Sp])
\bigg\rangle 
\; ,
\end{split}
\ee
where we have relabelled $r_{i, (b)}$
as $r_{i}$ for brevity. 
We note that the energy integration for each of the two particles $\Fp$ and $\Sp$   extends to $\Emax$. As we discussed in Section~\ref{Sec:Extracting_singularities_from_Sigma_RR_1c}, this leads
to a possible contribution of the unphysical region $E_{[\Fp \Sp]} >  \Emax$. Since $\Emax$ is chosen  in such
a way that $\FLMmunu([\Fp \Sp])$ has no support for $E_{[\Fp \Sp]} > \Emax$, only the soft subtraction term contributes
in this case. Hence,  in the above formula 
we can write 
\index{S!$\oS$}
\begin{align}
\oS_{\Fp \Sp} = \Theta_{[\Fp \Sp],{\rm max}} \; \oS_{[\Fp \Sp]}
- \Theta_{{\rm max},[\Fp \Sp]}  \Soft{[\Fp \Sp]} \; ,
\end{align}
where in the first (second) term 
on  the right hand side 
the energy of the clustered particle is restricted to be smaller (larger) than $E_{\rm max}$, respectively. 
The integration over the energies of partons $\Fp$ and $\Sp$
can be rearranged  to conform with the above 
splitting of the soft operator 
\begin{align}
\int \limits_0^{\Emax} \rmd  E_{\Fp} \int \limits_0^{\Emax} \rmd E_{\Sp} 
=
\int \limits_0^{\Emax} \rmd E_{[\Fp \Sp]} \, E_{[\Fp \Sp]}
\int \limits_0^1
\rmd z
+
\int \limits_{\Emax}^{2\Emax} \rmd  E_{[\Fp \Sp]} \, E_{[\Fp \Sp]}
\int \limits_{1-\frac{\Emax}{E_{[\Fp\Sp]}}}^{\frac{\Emax}{E_{[\Fp \Sp]}}} 
\rmd z \; .
\end{align}
Following this rearrangement, we have (cf. Eq.~\eqref{eq:4.52A})
\be
  \SigmaRRcBDiSC = \SigmaRRcBDiSCI + \SigmaRRcBDiSCII \; ,
  \label{eq:sc_phys_unphys}
  \ee
  where
\be
\begin{split} 
    \SigmaRRcBDiSCI 
    ={} & -\frac{\asbr}{2 } \, N_\ep^{(b,d)}  \; 
\bigg\langle 
\int [\rmd  p_{[\Fp \Sp]}] \, 
E_{[\Fp \Sp]}^{-2\ep}  \,
\Theta_{{\rm max},  [\Fp \Sp]}\,  
\sigma_{i[\Fp \Sp]}^{-\ep} \, 
\omega^{\Fp i , \Sp i}_{\Fp\parallel \Sp}
\\
& \times
\int \limits_{0}^{1} \frac{\rmd z}{\big[z (1-z)\big]^{2\ep}}
\; \Big [ P_{gg}^\bot(z) \, 
\big(  r_{i}^\mu \, r_{i}^\nu+g^{\mu \nu} \big) 
- P_{gg}^{\bot, r}(z) \, g^{\mu  \nu} 
\Big] \\
& \times \oS_{[\Fp \Sp]}\, \Delta^{([\Fp \Sp])} \,
\FLMmunu([\Fp \Sp])
\bigg\rangle
\;,
\end{split}
\ee
and
  \be
\begin{split} 
  \SigmaRRcBDiSCII 
  ={} &
-\frac{\asbr}{2 } \, 
N_\ep^{(b,d)}  \; 
\bigg\langle 
\int [\rmd p_{[\Fp \Sp]}]  \;
E_{[\Fp \Sp]}^{-2\ep} \;
\Theta_{{\rm max},[\Fp \Sp] } \; 
\sigma_{i[\Fp \Sp]}^{-\ep} \, 
\omega^{\Fp i , \Sp i}_{\Fp\parallel \Sp} \; 
\\
& \times \!\!\!\!
\int \limits_{1-\frac{\Emax}{E_{[\Fp\Sp]}}}^{\frac{\Emax}{E_{[\Fp \Sp]}}} 
\frac{\rmd z}{ 
\big[z (1-z)\big]^{2\ep}}
\; 
\Big [ P_{gg}^\bot(z) \, 
\big(  r_{i}^\mu \, r_{i}^\nu+g^{\mu \nu} \big) 
- P_{gg}^{\bot, r}(z) \, g^{\mu  \nu} 
\Big] \\
& \times S_{[\Fp \Sp]}\, \Delta^{([\Fp \Sp])} \, 
\FLMmunu([\Fp \Sp])
\bigg\rangle
\; ,
\end{split}
\ee
where $[\rmd p_{[\Fp \Sp]}]$ identifies the phase space of the clustered parton $[\Fp \Sp]$. We first discuss  $\SigmaRRcBDiSCI$, where the integration over $z$ decouples
from the rest and can be easily performed. We find
  \be
\begin{split} 
 \SigmaRRcBDiSCI 
  = &~
\frac{\asbr}{2} \, 
N_\ep^{(b,d)}  \; 
\bigg\langle  
\int [\rmd  p_{[\Fp \Sp]}] \,
E_{[\Fp \Sp]}^{-2\ep}\,
\Theta_{{\rm max}, [\Fp \Sp]} \;
\sigma_{i[\Fp \Sp]}^{-\ep} \, \omega^{\Fp i , \Sp i}_{\Fp\parallel \Sp} \; 
\\
& \hspace{-10mm} \times
\Big [ {\gamma}_{\bot, g\rightarrow gg}^{22} \, 
\big(  r_{i}^\mu \, r_{i}^\nu+g^{\mu \nu} \big) 
- {\gamma}_{\bot, g\rightarrow gg}^{22, r} \, g^{\mu  \nu} 
\Big] 
\oS_{[\Fp \Sp]}\, \Delta^{([\Sp \Sp])} \, 
\FLMmunu([\Fp \Sp])
\bigg\rangle
\;,
\end{split}
\ee
where the functions ${\gamma}_{\bot, g\rightarrow gg}^{22}$ and 
${\gamma}_{\bot, g\rightarrow gg}^{22, r}$ are given in Eq.~\eqref{eq:gamma_tilde}.
Since this contribution is soft-regulated, the
only singularity left there is $[\Fp \Sp] || i$. To regularize and extract  this 
collinear divergence, we insert $1 = \oC_{i[\Fp \Sp]}
+ C_{i[\Fp \Sp]}$ into the above formula and obtain
\be
\SigmaRRcBDiSCI 
=
\SigmaRRcBDiSCIone 
+
\SigmaRRcBDiSCItwo \; ,
\ee
where
\be
\begin{split} 
\SigmaRRcBDiSCIone
 = &~
\frac{\asbr}{2} \, 
N_\ep^{(b,d)}  \; 
\bigg\langle 
\int [\rmd p_{\Fp}] \,  E_{\Fp}^{-2\ep}   \, \Theta_{{\rm max}, \Fp} \;
C_{i\Fp} \,
\sigma_{i \Fp}^{-\ep} \, 
\omega^{\Fp i , \Sp i}_{\Fp\parallel \Sp} \; 
\\
&  \hspace{-10mm} \times \Big [ {\gamma}_{\bot, g\rightarrow gg}^{22} \, 
\big(  r_{i}^\mu \, r_{i}^\nu+g^{\mu \nu} \big) 
- {\gamma}_{\bot, g\rightarrow gg}^{22, r} \, g^{\mu  \nu} 
\Big] 
\oS_{\Fp}\, \Delta^{(\Fp)} \, 
\FLMmunu(\Fp)
\bigg\rangle
\;,
\label{eq:Ibdsc}
\end{split}
\ee
and
\be
\begin{split} 
\SigmaRRcBDiSCItwo
= &~
\frac{\asbr}{2} \, 
N_\ep^{(b,d)}  \; 
\Big\langle 
\mathcal{O}_{\rm NLO}^{(i)}
E_{\Fp}^{-2\ep}\,
\theta_{{\rm max} , \Fp} \;
\sigma_{i \Fp}^{-\ep} \, 
\omega^{\Fp i , \Sp i}_{\Fp\parallel \Sp}
\\
& \hspace{-10mm} \times \Big [ {\gamma}_{\bot, g\rightarrow gg}^{22} \, 
\big( r_{i}^\mu \, r_{i}^\nu+g^{\mu \nu} \big) 
- {\gamma}_{\bot, g\rightarrow gg}^{22, r} \, g^{\mu  \nu} 
\Big] 
\Delta^{(\Fp)} \, 
\FLMmunu(\Fp)
\Big\rangle
\;.
\end{split}
\label{eq.h21}
\ee
We note that  we have relabelled $[\Fp \Sp] \mapsto \Fp$ when writing the above 
equations. 
The function $\SigmaRRcBDiSCItwo$ is a fully-regulated single-unresolved contribution which is finite in the limit $\ep \to 0$ and can be numerically integrated
in four space-time dimensions. 

On the other hand, the quantity  $\SigmaRRcBDiSCIone$ will include a $1/\ep$ pole once we integrate over the unresolved parton $\Fp$. To do this, we need to evaluate the soft and collinear limits of $r_i^\mu r_i^\nu \FLMmunu([\Fp \Sp])$, which we have not encountered before. Doing so requires us to revisit the construction of the vectors $r_i$. 
We recall that, following Eq.~\eqref{Eq:p_Fp^mu_p_Sp^mu_parametrization}, the angular parametrization employs the direction of parton $i$ as a reference axis, so that (cf. Eq.~\eqref{eq:e4})
\be
p_i^\mu = E_i (t^\mu +  e_i^\mu ) \; ,
\ee
where $t$ is a time-like vector with $t^2 = 1$ and $e_i$ is a space-like vector with $e_i^2 = -1$. 
The momentum of the clustered particle $[\Fp \Sp]$
is defined as
\be
p_{[\Fp \Sp]}^\mu = E_{[\Fp \Sp]} \left ( t^\mu + \cos \theta_{[\Fp \Sp] i} \, e_i^\mu + \sin \theta_{[\Fp \Sp] i} \, b_i^\mu \right ) \; ,
\ee
with 
\be
b_i^\mu \,  e_{i,\mu} = 0 \; .
\ee
The vector $r_{i}$ reads 
\be
r_i^{\mu} = \sin \theta_{[\Fp \Sp]i} \, 
e_{i}^\mu 
- \cos \theta_{[\Fp \Sp]i} \,  b_{i}^\mu
\; ,
\ee
from which it follows that
\be
p_{[\Fp \Sp]}^{\mu} \, r_{i,\mu} = 0 \; .
\ee
This implies that $r_{i}$ is a valid polarization vector for the clustered gluon ${[\Fp \Sp]}$. 
Armed with this understanding, it is   straightforward to write a general expression for the soft limits $S_{[\Fp \Sp]}$ of 
spin-correlated amplitudes-squared. We find
\be
r_i^\mu r_i^\nu \,  
S_{[\Fp \Sp]} \, 
\FLMmunu([\Fp \Sp])
= - \gsb^2 \sum \limits_{k, l=1}^{\Np} \frac{ (p_k \cdot r_i) \, (p_l \cdot r_i)}
{(p_k \cdot p_{[\Fp \Sp]}) \, 
(p_l \cdot p_{[\Fp \Sp]})} \;  
\left(\scprod{\ColT{k}}{\ColT{l}} \right)\colorprod \FLM 
\; .
\label{eq:d21}
\ee
One also needs to consider the limit $C_{i [\Fp \Sp]}$ of this expression, which develops singularities arising from two contributions in the sum: first from $k = i, l=i$, and second from $k =i, l \ne i$ or $k \ne i, l = i$.

We begin with the first one and write 
\be
\frac{ (p_i \cdot   r_i) (p_i \cdot r_i)}
{(p_i \cdot p_{[\Fp \Sp]}) \,  
(p_i \cdot p_{[\Fp \Sp]})} 
= 
\frac{1}{E_{[\Fp \Sp]}^2}\frac{\sin^2\theta_{[\Fp \Sp] i}}
{\big(1-\cos \theta_{[\Fp \Sp]i}\big)^2} =  \frac{1}{E_{[\Fp \Sp]}^2}
\frac{\big(2 - \rho_{[\Fp \Sp]i}\big)}
{\rho_{[\Fp \Sp]i}} \; ,
\label{eq:d22}
\ee
where we used the explicit parametrization of momenta $p_i$ and $p_{[\Fp \Sp]}$ and the vector $r_i$.   The collinear limit
of the term in Eq.~\eqref{eq:d21} with $k=i, l=i$ therefore reads 
\be
-\gsb^2  \, 
C_{i[\Fp \Sp]} \; \frac{  (p_i \cdot r_i) \,  (p_i\cdot r_i)}
{(p_i \cdot p_{[\Fp \Sp]}) (p_i \cdot p_{[\Fp \Sp]})} \; 
\left(\scprod{\ColT{i}}{\ColT{i}} \right)\colorprod \FLM 
=
 \frac{ -2 \gsb^2 } {E_{[\Fp \Sp ]}^2 \; \rho_{[\Fp \Sp]i} } \;  \ColT{i}^2  \; \FLM \; .
 \label{eq:F.26A}
\ee
Next, we consider terms with $k=i$ and $l \ne i$
\be
-\gsb^2 \sum \limits_{l \ne i}
  \frac{ (p_i \cdot r_i) \,  (p_l \cdot r_i)}{(p_i \cdot p_{[\Fp \Sp]}) \,  (p_l \cdot p_{[\Fp \Sp]} )} \;  
  \left(\scprod{\ColT{i}}{\ColT{l}} \right)\colorprod \FLM \; .
 \ee
Since $p_i \cdot  r_i \sim  \sin\theta_{[\Fp \Sp]i}$ 
and $p_i \cdot p_{[\Fp \Sp]} \sim (1-\cos \theta_{[\Fp\Sp]i})$,
and all other factors in the above expression are regular in the limit  $\theta_{[\Fp \Sp]i} \to 0$, we conclude
that this contribution is actually integrable in the collinear limit $[\Fp \Sp]\, || \, i$. The same conclusion holds for the symmetric $k \ne i$ and $l = i$ 
terms. Hence, we find the following result
\be
C_{i[\Fp \Sp]} \, S_{[\Fp \Sp]} \, 
r_i^\mu r_i^\nu \,  
\FLMmunu([\Fp \Sp]) =
 -\frac{2 \gsb^2} {E_{[\Fp \Sp ]}^2 \; \rho_{[\Fp \Sp]i} } \; 
 \ColT{i}^2  \; \FLM \; .
\ee
This coincides with the limit without spin correlations, $C_{i [\Fp \Sp]} \,  S_{[\Fp \Sp]}  F_{\rm LM}([\Fp \Sp])$, 
so that
\be
C_{i[\Fp \Sp]} \; S_{[\Fp \Sp]} (g^{\mu \nu} + r_i^ \mu r_i^\nu) \,  \FLMmunu([\Fp \Sp]) = 0 \; .
\ee
We can use this cancellation to simplify $\SigmaRRcBDiSCIone$ in Eq.~\eqref{eq:Ibdsc}. We write 
\be
\begin{split} 
\SigmaRRcBDiSCIone
= &~
\frac{\asbr}{2 } \, 
N_\ep^{(b,d)}  \; 
\bigg\langle 
\int [\rmd p_{[\Fp\Sp] }] \,
E_{[\Fp \Sp]}^{-2\ep}   \, 
\Theta_{{\rm max} ,  [\Fp\Sp]} \,
\eta_{i[\Fp\Sp]}^{-\ep} 
\\
& \times
\; \Big [ {\gamma}_{\bot, g\rightarrow gg}^{22} \, 
 C_{i[\Fp\Sp]} \big(  r_{i}^\mu \, r_{i}^\nu+g^{\mu \nu} \big) 
- {\gamma}_{\bot, g \rightarrow gg}^{22, r} \, g^{\mu  \nu}   C_{i[\Fp\Sp]} \, \oS_{[\Fp\Sp]}\, 
\Big]  \\
&\times \Delta^{([\Fp\Sp])} \, 
\FLMmunu([\Fp \Sp])
\bigg\rangle 
\; .
\label{eq:d28}
\end{split}
\ee
The only limit in the above 
equation that we have not yet encountered is 
\be
C_{i[\Fp\Sp]} \,
r_{i}^\mu \, 
r_{i}^\nu \, 
\FLMmunu([\Fp\Sp]) \; .
\ee
As we will show later, it evaluates
to 
\be
\label{eq:coll_spin_corr}
C_{i[\Fp\Sp]} \,
r_{i}^\mu \, 
r_{i}^\nu \, 
\FLMmunu([\Fp\Sp])
=
\frac{\gsb^2}{p_i\cdot p_{[\Fp\Sp]}}
\cdot 
\begin{cases}
 P^{\rm spin}_{f_i f_{[i\Fp]}}(z) \otimes \FLM^{(i)}(z\cdot [\Fp \Sp i]) \; , 
 \qquad i \le 2 \; , \\[5pt]
P^{\rm spin}_{f_i f_{[i\Fp]}}(z) \, \FLM \; , 
\qquad \quad \qquad \quad \, \quad
i > 2
\; ,
\end{cases}
\ee
where  the splitting functions are given by 
the following equations 
\index{P!$P^{\rm spin}_{ij}$}
\index{F!$\FLM^{(i)}$}
\be
\label{eq:P_spin}
P^{\rm spin}_{f_i f_{[i \Fp ]}}(z) 
= 
\begin{cases}
\frac12  \; C_F \frac{(1+z)^2}{1-z} \; , \hspace{36mm}
f_i=f_{[i \Fp ]}=\{q, \bar{q} \} \; , 
\\[8pt]
2\, \Ca \, 
\left[
\frac{z}{1-z} 
+ 
\frac{(1-z)/z +z(1-z)}{2(1-\ep)} 
\right] \; ,
\qquad
f_i=f_{[i \Fp ]}=g \; ,
\end{cases}
\ee
and we have adopted the convention that $\FLM^{(1)} (z \cdot 1 ) \equiv \FLM(z\cdot 1_a,2_b, ...)/z$ and
$\FLM^{(2)} (z \cdot 2 ) \equiv \FLM(1_a,z\cdot 2_b, ...)/z$.

We are now in the position to evaluate the limits of Eq.~\eqref{eq:d28} and to integrate over the angular phase space.
For the final-state emissions ($i > 2$) we find 
  \be
\begin{split} 
\SigmaRRcBDiSCIone\Big|_{i>2}
  = &~
\frac{\asbr}{2 } \, 
N_\ep^{(b,d)}  \; 
\bigg\langle 
\int [\rmd  p_i]  [\rmd  p_{[\Fp \Sp]}]  \, 
E_{[\Fp \Sp]}^{-2\ep} \,
\Theta_{{\rm max},  [\Fp \Sp]}\,
\eta_{i[\Fp\Sp]}^{-\ep} 
\\
& \times
\Big [{\gamma}_{\bot, g\rightarrow gg}^{22} \,
 \frac{\gsb^2}{E_{[\Fp \Sp]} E_i \rho_{i[\Fp \Sp]}}
\left ( P_{gg}^{\rm spin}(z) - P_{gg}(z) \right )
z \,   \FLM ([\Fp \Sp i])
  \\
  & + {\gamma}_{\bot, g\rightarrow gg}^{22, r} \, \Coll{i[\Fp \Sp]} 
 \oS_{[\Fp \Sp]}\, 
z \,  \FLM\big([\Fp \Sp]) 
\Big]
\bigg\rangle
\;,
\label{eq:d12}
\end{split}
\ee
while for  $i \le 2$ we find 
  \be
\begin{split} 
  \SigmaRRcBDiSCIone\Big|_{i \le 2}
  = &~
\frac{\asbr}{2 } \, 
N_\ep^{(b,d)}  \; 
\bigg\langle 
\int 
[\rmd  p_{[\Fp \Sp]}]  \, 
E_{[\Fp \Sp]}^{-2\ep} \,
\Theta_{{\rm max},  [\Fp \Sp]}\,
\eta_{i[\Fp\Sp]}^{-\ep} 
\\
& \times
\; \Big [ 
{\gamma}_{\bot, g\rightarrow gg}^{22} \,
\frac{\gsb^2}{E_{[\Fp \Sp]} E_i \rho_{i[\Fp \Sp]}}
\left ( P_{qq}^{\rm spin}(z) - P_{qq}(z) \right )
\otimes
\FLM^{(i)} (z \cdot [\Fp \Sp i])
  \\ 
  & + {\gamma}_{\bot, g\rightarrow gg}^{22, r} \,
 \Coll{i[\Fp \Sp]} 
 \oS_{[\Fp \Sp]}\, 
 \FLM([\Fp \Sp]) 
\Big]
\bigg\rangle
\;.
\label{eq:I_Omega_4_Ia}
\end{split}
\ee
We can then integrate over the remaining energy and angular variables using the formulas in Appendix~\ref{app:useful_integral} and obtain
  \be
\begin{split} 
\SigmaRRcBDiSCIone\Big|_{i > 2}
   = &~
\frac{\asbr^2}{4\ep} \, 
\frac{2^{2\ep}\,\Gamma(1-\ep) \, \Gamma(1-2\ep)}{\Gamma(1-3\ep)} 
N_\ep^{(b,d)}  \;  
\bigg\langle
\left(\frac{2 E_i}{\mu}\right)^{-4\ep} 
\\
& \hspace{-10mm} \times 
\bigg[
- {\gamma}_{\bot, g\rightarrow gg}^{22} \, 
 \Big[
  \gamma_{z, g \rightarrow gg}^{24}  
  -
 \gamma_{z, g \rightarrow gg}^{24, \, \rm spin}  
  \Big]
  +
{\gamma}_{\bot, g\rightarrow gg}^{22, r} \, 
\gamma^{24}_{z,g \rightarrow gg}(\ep, L_i) 
\bigg] \, \FLM
\bigg\rangle
\;,
\\ \\
  \SigmaRRcBDiSCIone\Big|_{i \le 2}
  = &~
\frac{\asbr^2}{4\ep} \, 
\frac{2^{2\ep}\,\Gamma(1-\ep) \, \Gamma(1-2\ep)}{\Gamma(1-3\ep)} 
N_\ep^{(b,d)}  \;  
\bigg\langle
\left(\frac{2 E_i}{\mu}\right)^{-4\ep} \; 
\\
& \times 
\bigg\{
 {\gamma}_{\bot, g\rightarrow gg}^{22} \, 
\int_0^1 \rmd z \, 
(1-z)^{-4\ep} \Big[
P_{qq} (z) - P_{qq}^{\rm spin} (z)
  \Big]  \otimes \FLM^{(i)}(z \cdot i)
  \\
& - 
{\gamma}_{\bot, g\rightarrow gg}^{22, r} \,
\int_0^1 \rmd z \, 
\CalP_{qq}^{(4)}(z,L_i)
 \otimes 
\FLM^{(i)}(z\cdot i)
\bigg \}
\bigg\rangle
\;,
\label{eq:I_Omega_4_Ia_v2}
\end{split}
\ee
where we have defined
\index{G!$\gamma_{z, g \rightarrow gg}^{24, \, \rm spin}$}
\be
 \gamma_{z, g \rightarrow gg}^{24, \, \rm spin}   = 
 - \int_0^1 dz \Big[\;  
 z^{-2\ep} \, (1-z)^{-4\ep} \, 
 z \, P_{gg}^{\rm spin}(z)  
   -
   2 \, \Ca \,  
   (1-z)^{-1-4\ep}  \Big] \; .
\ee
Finally, combining emissions off different legs,  we write $\SigmaRRcBDSCIone$ as
\be
\label{eq.h45}
\begin{split}
& 
\SigmaRRcBDSCIone
=
\frac{\asbr^2}{4\ep} \, 
\frac{2^{2\ep}\,\Gamma(1-\ep) \, \Gamma(1-2\ep)}{\Gamma(1-3\ep)} 
N_\ep^{(b,d)}  \;
\\
& \times
\bigg\{-
\sum_{i=3}^{\Np}
\bigg\langle
\bigg(\frac{2 E_i}{\mu}\bigg)^{-4\ep} \;
{\gamma}_{\bot, g\rightarrow gg}^{22} \, 
 \Big[
  \gamma_{z, g \rightarrow gg}^{24}  
  -
 \gamma_{z, g \rightarrow gg}^{24, \, \rm spin}  
  \Big] \FLM
\bigg\rangle
\\
& +
\sum_{i=1}^2  
\bigg\langle
\bigg(\frac{2 E_i}{\mu}\bigg)^{-4\ep} 
{\gamma}_{\bot, g\rightarrow gg}^{22} \, 
\int \limits_0^1 \rmd z \, 
(1-z)^{-4\ep} \Big[
P_{qq} (z) - P_{qq}^{\rm spin} (z)
  \Big]  \otimes 
\FLM^{(i)}(z\cdot i) \bigg\rangle
\bigg\}
  \\
  & + 
 \frac{\asbr^2}{4\ep}
  N_{sc}^{(b,d)}  \, 
  {\gamma}_{\bot, g\rightarrow gg}^{22, r}
 \sum_{i=1}^2
 \bigg\langle
\int \limits_0^1 \rmd z \, 
\CalP_{qq}^{(4), \rm gen}(z)
 \otimes 
\FLM^{(i)}(z \cdot i)
\bigg\rangle
\\
& +\frac{\asbr^2}{2} \, 
N_{sc}^{(b,d)} \, 
{\gamma}_{\bot, g\rightarrow gg}^{22, r} 
\lint
\IColl^{(4)}(\ep) \cdot  \FLM
\rint \; , 
\end{split}
\ee
where we have introduced
\be
\begin{split}
N_{\rm sc}^{(b,d)}
=
\frac{2^{2\ep}\,  \Gamma^3(1-2\ep)}{\Gamma(1-3\ep) \Gamma^3(1-\ep)} 
N_\ep^{(b,d)} \; .
\end{split}
\ee

We return to the ``unphysical" contribution $\SigmaRRcBDiSCII$ of Eq.~\eqref{eq:sc_phys_unphys}. Using  Eq.~\eqref{eq:d21}, we can immediately obtain the soft limit $S_{[\Fp \Sp]} \FLMmunu[\Fp \Sp]$. Integrating over $E_{[\Fp \Sp]}$ and $z$, we find 
\be
\begin{split} 
\SigmaRRcBDiSCII
= &~
\asbr \, \gsb^2 \, 
\left(\frac{E_{\rm max}}{\mu}\right)^{-4\ep} \, 
\bigg\langle 
\int 
\rmd  \Omega_{[\Fp \Sp]}\,
\sigma_{i[\Fp\Sp]}^{-\ep} \, 
\omega^{\Fp i , \Sp i}_{\Fp\parallel \Sp}  \, \Big [ \delta_{g}^{\bot} (r_i^\mu r_i^\nu + g^{\mu \nu})
\\
& - \delta_{g}^{\bot, r} \, g^{\mu  \nu} 
  \Big] \sum \limits_{k,l=1}^{\Np} \frac{ n_{k,\mu} \,  n_{l,\nu}}{\big(n_k  \cdot n_{[\Fp \Sp]}\big)  \, \big(n_l \cdot  n_{[\Fp \Sp]} \big)} \; 
\left(\scprod{\ColT{k}}{\ColT{l}} \right)\colorprod \FLM 
  \bigg \rangle \; ,
\end{split}
\label{eq:I_Omega4_bd_sc_II}
\ee
where $\delta_g^{\perp}$ and  $\delta_g^{\perp,r}$ are given in  Eq.~\eqref{Eq:delta_g_def}.
At this point, we introduce the 
functions 
\index{W!${\cal W}_r^{(i)}$}
\index{W!$\Wac{i}$}
\be
\label{eq:part_dep_contr}
\begin{split}
\lint {\cal W}_r^{(i)}  \tensprod \FLM\rint
\equiv &
\int \frac{[\rmd\Omega_{[\Fp\Sp]}^{(d-1)}]}{[\Omega^{(d-2)}]}  
\bigg\langle\sigma_{i[\Fp\Sp]}^{-\ep} \, 
\omega^{\Fp i , \Sp i}_{\Fp \parallel \Sp} \,  
\\
& \times
\big(  r_{i}^\mu \, r_{i}^\nu
+
g^{\mu \nu} \big)
\sum_{\substack{k,l=1}}^{\Np} \frac{ n_{k,\mu} \,  n_{l,\nu}}{\big(n_k  \cdot n_{[\Fp \Sp]}\big)  \big(n_l \cdot  n_{[\Fp \Sp]} \big)} \,  \FLM^{(kl)} 
\bigg\rangle\; , 
\\
\lint \Wac{i} \tensprod \FLM\rint 
\equiv &
-  \ep \, 2^{2\ep} 
\int \frac{[\rmd\Omega_{[\Fp \Sp]}^{(d-1)}]}{[\Omega^{(d-2)}]} \bigg\langle
\sigma_{i[\Fp\Sp]}^{-\ep} \, 
\omega^{\Fp i , \Sp i}_{\Fp \parallel \Sp} \, 
\\
&
\times 
\sum_{\substack{k,l=1 \\ k \neq l}}^{\Np} \frac{ n_k \cdot n_l}{\big(n_k  \cdot n_{[\Fp \Sp]}\big)   \big(n_l \cdot  n_{[\Fp \Sp]} \big)} \,  \FLM^{(kl)} 
\bigg\rangle\; ,
\end{split}
\ee
where we have used the shorthand notation
$
\FLM^{(ij)} = \left(\scprod{\ColT{i}}{ \ColT{j}} \right) \colorprod \FLM
$, and write Eq.~\eqref{eq:I_Omega4_bd_sc_II} as
\be
\begin{split} 
\label{eq:I_Omega4_bd_sc_IInew}
\SigmaRRcBDiSCII ={}&
\asbr^2 \, 
\left(\frac{E_{\rm max}}{\mu}\right)^{-4\ep} \, 
\Big\langle 
\delta_{g}^{\bot} \, 
{\cal W}_r^{(i)}  \tensprod \FLM 
+ \delta_{g}^{\bot, r} \,
\frac{2^{-2\ep}}{\ep}
\Wac{i} \tensprod \FLM 
\Big \rangle \; .
\end{split}
\ee
The function ${\cal W}_r^{(i)}$ is finite in $\ep$ because the pole  arising from the term  proportional to $r_i^{\mu} r_i^{\nu}$ cancels with that arising from the $g^{\mu \nu}$ term. 
This can be understood as follows: the most singular contribution affecting the
term proportional to $r_i^{\mu} r_i^{\nu}$ stems from the combination
$k=l=i$, since the partition functions damp all other potential collinear configurations. In this case, the singularity is proportional to $2C_{f_i} /\rho_{[\Fp \Sp]i}$, as we already saw in  Eq.~\eqref{eq:F.26A}. On the other hand, the singularity proportional to $g^{\mu\nu}$ can only arise when $k=i, l\neq i$ and $k\neq i, l=1$, given that $n_l^2 = n_k^2 =0$. We can then isolate the divergent ratio $1/(n_i\cdot n_{[\Fp \Sp]})$ and sum over colors, obtaining precisely $-2C_{f_i} /\rho_{[\Fp \Sp]i}$. We conclude that ${\cal W}_r^{(i)}$ does not contribute to the pole content of $\SigmaRRcBDiSCII$. 

By contrast, the term in Eq.~\eqref{eq:I_Omega4_bd_sc_IInew} containing the function $\Wac{i}$ does contain singularities of $\order{\ep^{-1}}$, which could (in principle) be dependent on the partitions $\omega^{\Fp i , \Sp i}_{\Fp \parallel \Sp}$. This would imply that the pole structure of $\SigmaRRcBDiSCII$ would depend on the choice of partition functions. However, in Appendix~\ref{sec:part_dep} we will show that the sum over all the external legs of $\Wac{i}$ can be written as 
\index{W!$\Wac{i}$}
\begin{equation}
\begin{split}
    \sum_{i=1}^{\Np} \lint \Wac{i} \tensprod \FLM\rint
    ={} & 
    - \epsilon \, 2^{2\epsilon} \sum_{i=1}^{\Np} \sum_{\substack{k,l=1\\ k\neq l}}^{\Np} 
    \int \frac{[\rmd \Omega_\Fp^{(d-1)}]}{[\Omega^{(d-2)}]} \bigg\langle \frac{\rho_{kl}}{\rho_{k\Fp} \, \rho_{l\Fp}} \, 
    \sigma_{i\Fp}^{-\ep} \, \omega_{\Fp \parallel \Sp}^{\Fp i, \Sp i} \, \FLM^{(kl)} \bigg\rangle
    \\
    ={}&
    2 \sum_{\substack{i,j=1 \\i\neq j}}^{\Np} 
    \llint
    \eta_{ij}^{-\ep} K_{ij} \FLM^{(ij)}
    \rrint
    \\
    {}&
    + \sum_{i=1}^{\Np}    
    \Big[
    N_c(\epsilon)  \, \T_i^2
    \llint
    \FLM
    \rrint
    + 
    \ep^2 \, 
    \llint
    \Wacfin{i} \tensprod \FLM
    \rrint
    \Big]
    \; ,
\label{Eq:AppendixF_partition_dependent_general_def_bd_sectors}
\end{split}
\end{equation}
where we have relabelled $[\Fp \Sp] \mapsto \Fp$. It is clear from the above equation that the poles of $\SigmaRRcBDiSCII$  are in fact \emph{independent} of the partition functions, whose explicit form only affects the finite remainder $\Wacfin{i}$ given in Eq.~\eqref{Eq:Appendix_Wacfin_i_def}.\footnote{Changing the form of the partition functions would also change the value obtained from numerical integration for the fully-regulated term $\SigmaFullyRes$ (cf.~Eq.~\eqref{Eq:Sigma_N+2^fin_def}). These changes would compensate each other such that the physical cross section does not depend on the explicit expression for the partition functions.}
Summing over emissions from all legs, we find
\be
\begin{split} 
\label{eq:SigmabdscII_AppF}
\Sigma_{RR,1c}^{(b,d),{\rm sc},II} ={}&
~ 2 [\alpha_s]^2 \delta_g^{\perp,r}(\epsilon) \left(\frac{\Emax}{\mu} \right)^{-2\epsilon} \\
    &{} \times \bigg[-\lint \ISoft(\epsilon) \colorprod \FLM \rint 
     + \frac{(2\Emax/\mu)^{-2\epsilon}}{2\epsilon^2} N_c(\epsilon) \sum_{i=1}^{N} \T_i^2 \; \lint\FLM\rint \bigg] \\
     &{}+ \asbr^2 \, 2^{-2\ep} \delta_g^{\perp,r}(\ep) \, 
    \left(\frac{\Emax}{\mu}\right)^{-4\ep} \sum_{i=1}^{\Np} \lint \Wacfin{i}\colorprod \FLM \rint \\ 
     & {} +[\alpha_s]^2 \delta_{g}^{\bot}\left(\frac{\Emax}{\mu} \right)^{-4\eps}    
\sum_{i=1}^{\Np} \lint {\cal W}_r^{(i)}  \tensprod \FLM  \rint\Biggr] \; .
\end{split}
\ee

The complete result for spin-correlated contributions 
is obtained upon combining Eqs.~\eqref{eq.h21}, \eqref{eq.h45} and \eqref{eq:SigmabdscII_AppF}. 
\\

It remains to prove the results for spin-correlated
splitting functions introduced  in  Eq.~\eqref{eq:coll_spin_corr}.
To this end, we  consider the cases where $i$ is the initial-state or the final-state parton separately. 
We begin with the discussion of the final-state splitting, in which case $i$ is a gluon. Since
$r_i$ can be considered to be  the polarization vector of the clustered gluon, the calculation 
of the collinear limit in Eq.~(\ref{eq:coll_spin_corr}) is 
equivalent  to the 
computation of a $g \to gg$ splitting for \emph{polarized}
gluons. The corresponding results can be found in Ref.~\cite{Ellis:1996mzs}. To understand how they can be used,
we note that Ref.~\cite{Ellis:1996mzs} defines polarization vectors relative to the decay plane formed by the momenta of the final state particles, there called  $b$ and  $c$. Their momenta define a two-dimensional plane in   $(d-1)$-dimensional space (we discard the temporal
 component for obvious reasons). We need
 $(d-2)$ polarization vectors to fully describe the quantum state of a gluon. Hence,
 for each of the gluons, we choose one polarization vector to lie in the plane defined by the momenta and $(d-3)$ to be orthogonal to that plane.
 It is clear that we can choose the ``out-of-the-plane'' polarization vectors to be the same for the three gluons
 $a,b,c$. 

\begin{table*}[t]
\centering
\begin{tabular}{|c|c|c|c|}
  \hline
  $a$ & $b$ & $c$ & $F_{3g}(z,\varepsilon_a, \varepsilon_b, \varepsilon_c) $ \\
  in & in & in &  $(1-z)/z + z/(1-z) + z(1-z) $ \\
  in & out & out  &  $ z(1-z) $ \\
  out & in & out &  $z/(1-z)$  \\
  out  & out  & in & $(1-z)/z $ \\
  \hline
\end{tabular}
\label{table:d1}
\caption{The table from Ref.~\cite{Ellis:1996mzs}, page 160. Note that we use $z = 1-E_b/E_a$ at variance
with Ref.~\cite{Ellis:1996mzs}. }
\end{table*}

The dependence of the $g \to gg $ 
splitting on the polarization of the partons is characterized by the function $F_{3g}(z)$ shown in Table~\ref{table:d1}.  One can use this function
to write the collinear limit of the scattering amplitude as follows~\cite{Ellis:1996mzs}
\be
|\mathcal{M}_{n+1}(\varepsilon_b, \varepsilon_c)|^2 \sim \frac{4 \gsb^2 \, \Ca}{(p_{[\Fp\Sp]}+p_i)^2} \, F_{3g}(z;\varepsilon_a,\varepsilon_b,\varepsilon_c)
  |\mathcal{M}_{n}(\varepsilon_a)|^2 \; .
\ee
As explained in Ref.~\cite{Ellis:1996mzs}, 
this formula implies that the polarizations of the parent and daughter partons are kept fixed.  For our purposes, we identify  parton $[\Fp \Sp]$ with parton $b$ and
parton $i$ with $c$. Therefore we need
to sum over the polarizations of partons $a$ and $c$ and keep the polarization of the gluon $b$ fixed and equal to $r_i$.
Note that, since
this polarization is composed of vectors $e_i$ and $b_i$, it is ``in-plane'', according to the language of Ref.~\cite{Ellis:1996mzs}.
Hence, for our purposes we require 
 \be
\begin{split} 
& C_{i[\Fp \Sp]} r^\mu_i  
r^\nu_i 
\FLMmunu
( [\Fp \Sp])
=  \frac{4 \gsb^2 \Ca^2}{(p_{[\Fp \Sp]} + p_i)^2} \FLMmunu([\Fp \Sp i]) 
 \\
 & \times \bigg\{
 \varepsilon_a^\mu({\rm in}) \, \varepsilon_a^\nu ({\rm in})
F_{3g}(a_{\rm in}, b_{\rm in}, c_{\rm in} ) 
+ \sum \limits_{\rm out} \varepsilon_{a}^{\mu}({\rm out}) \, \varepsilon_{a}^{\nu}({\rm out}) F_{3g}(a_{\rm out}, b_{\rm in}, c_{\rm out} )
\bigg \} \; .
\end{split} 
\ee
The ``in-plane'' polarization for the gluon $a$ in the collinear limit is $b$. It remains to write the sum for the ``out-of-plane'' polarizations, which reads
\be
\sum \limits_{\rm out} \varepsilon_{a}^{\mu}({\rm out}) \, 
\varepsilon_{a}^{\nu}({\rm out})=
-g^{\mu \nu} + t^\mu t^\nu + \frac{e_i^\mu e_i^\nu}{e_i^2} + \frac{ b_i^\mu b_i^\nu}{b_i^2} \; .
\ee
Thanks to the transversality of $\FLMmunu([\Fp \Sp i]) $ w.r.t.~$p_{[\Fp \Sp i]}$, we find
\be
t^\mu \FLMmunu([\Fp \Sp i]) = - e_i^\mu \FLMmunu([\Fp \Sp i]) \; .
\ee
This implies that
\be
\left ( t^\mu t^\nu + \frac{e_i^\mu e_i^\nu}{e_i^2} \right ) \FLMmunu([\Fp \Sp i]\big) = 0 \; .
\ee
Hence, we obtain 
 \be
\begin{split} 
C_{i [\Fp \Sp]} \,
r^\mu_i  
r^\nu_i \,   \FLMmunu([\Fp \Sp]\big)
= &~ 
\frac{2 \gsb^2 \, \Ca}{E_{[\Fp \Sp]} E_i \rho_{i [\Fp\Sp]}} \FLMmunu([\Fp \Sp i]) \bigg  \{ b^\mu b^\nu F_{3g}(a_{\rm in}, b_{\rm in}, c_{\rm in} )   
 \\
 & + (-g^{\mu \nu} + b^\mu b^\nu) F_{3g}(a_{\rm out}, b_{\rm in}, c_{\rm out} )
\bigg \} \\
= &~ \frac{ \gsb^2}{ E_{[\Fp \Sp]} E_i \rho_{i [\Fp \Sp]} } \FLMmunu([\Fp \Sp i]) \;P_{gg}^{\, r, \, \mu \nu}(z) \; , 
\end{split}
\ee
where $z = E_i/(E_i + E_{[\Fp \Sp]})$ and
\index{P!$P_{gg}^{\, r, \, \mu \nu}$}
\begin{align}
P_{gg}^{\, r, \, \mu \nu}(z) = 2 \Ca
\bigg[
- \frac{z}{1-z} \, g^{\mu \nu}
+ \left( \frac{1-z}{z}+z(1-z)\right) b^\mu b^\nu\bigg] \; .
\end{align}
\begin{table*}[t]
\centering
\begin{tabular}{|c|c|c|c|}
  \hline
  $\lambda_a$ & $\varepsilon_b$ & $\lambda_c$ & $F_{qqg}(z,\lambda_a, \varepsilon_b, \lambda_c) $ \\
  $\pm$  & in & $\pm$  &  $(1+z)^2/(1-z) $ \\
  $\pm$  & out & $\pm$  &  $ (1-z) $ \\
  \hline
\end{tabular}
\caption{  The table from Ref.~\cite{Ellis:1996mzs}, page  160, 
that can be used to compute 
$q \to qg$ splittings.}
\label{table:d2}
\end{table*}
Since we will have to use this result in Eq.~\eqref{eq:d12}, where the integration over directions of $b$ decouples from the rest, 
we will only require the  spin-averaged version of $P_{gg}^{\ r,\,  \mu \nu}$, that is
\begin{align}
\label{eq:NNLO_C45_leg3_spin_aver}
\lint P_{gg}^{\, r, \,  \mu \nu}(z)  \rint = (-g^{\mu \nu}) P_{gg}^{\rm spin}(z) \, ,
\end{align}
where (c.f.~Eq.~\eqref{eq:P_spin})
\index{P!$P_{gg}^{\rm spin}$}
\be
P_{gg}^{\rm spin}(z) =
2 \Ca
\bigg[
\frac{z}{1-z} 
+ 
\frac{(1-z)/z +z(1-z)}{2(1-\ep)} 
\bigg] \; ,
\ee
Since the spin-averaging also applies to the standard collinear limit $C_{i[\Fp \Sp]} \FLM([\Fp\Sp])$,
we obtain
\be
\begin{split} 
& C_{i[\Fp \Sp]} ( r_i^\mu r_i^\nu + g^{\mu \nu} ) \FLMmunu([\Fp \Sp])
 =
 \frac{\gsb^2}{E_{[\Fp \Sp]} E_i \rho_{i[\Fp \Sp]}}
\left ( P_{gg}^{\rm spin}(z) - P_{gg}(z) \right ) \FLM([\Fp \Sp i])
\\
& =
- \frac{1-2\ep}{1-\ep} \; \frac{\gsb^2 \, \Ca}{E_{[\Fp\Sp]} E_i \rho_{i[\Fp \Sp] }} \; \frac{(1-z)(1+z^2)}{z}
 \FLM ([\Fp \Sp i]) \; .
\end{split} 
 \ee

 \vspace*{0.5cm}
To describe the initial-state splitting, we require the $q \to q^* g$ splitting.  To compute it,
we start from the final state $q^* \to q g$ and then perform the parton crossing. Similar to the gluon case, we need
to keep the gluon polarized.  The polarization-dependent
splitting functions can again be found in Ref.~\cite{Ellis:1996mzs}; they are reproduced in Table~\ref{table:d2}. 
We only need to consider the ``in plane'' polarization of the gluon and sum over quark polarizations. 
Performing the crossing, we find 
\be
\begin{split} 
  & C_{[\Fp \Sp] i} r^\mu_i  r^\nu_i \,
  \FLMmunu([\Fp \Sp])
=  
\frac{4 \gsb^2 \, }{E_{ [\Fp \Sp]} E_{i} \; \rho_{i [\Fp \Sp]}} \;  \Pqq^{\rm spin}(z) \; \frac{ \FLM(z \cdot i, ... )}{z},
\end{split} 
\ee
where $z = 1- E_{[\Fp \Sp]}/E_i$, $i = 1,2$, and $\Pqq^{\rm spin}$ is given in Eq.~\eqref{eq:P_spin}.

\section{Partition-dependent contribution} \label{sec:part_dep}
In this appendix, we discuss two 
contributions that appear in the 
computation of double-unresolved 
limits. They are required to obtain terms in the final result in 
the second line of Eq.~\eqref{eq:finite_reminder_SB}
and in the third line of Eq.~\eqref{sigma_elastic_def}, respectively.
They read  
\index{W!$\Wbd{i}$}
\index{W!$\Wac{i}$}
\begin{equation}
\begin{split}
    \sum_{k=1}^{\Np} \langle \Wbd{k} \tensprod \FLM\rangle \equiv & \sum_{k=1}^{\Np}\sum_{\inotj}^{\Np} \langle \mathcal{W}_k^{\Fp \parallel \Sp, (ij)} \FLM^{(ij)} \rangle \\
    = & - \epsilon \, 2^{2\epsilon} \sum_{k=1}^{\Np} \sum_{\inotj}^{\Np} \int \frac{[\rmd \Omega_\Fp^{(d-1)}]}{[\Omega^{(d-2)}]} \llint 
    \sigma_{k\Fp}^{-\ep} \, 
    \frac{\rho_{ij}}{\rho_{i\Fp} \, \rho_{j\Fp}} \, \omega_{\Fp \parallel \Sp}^{\Fp k, \Sp k} \, \FLM^{(ij)}\rrint \; ,
    \label{Eq:Appendix_partition_dependent_general_def_bd_sectors}
\end{split}
\end{equation}
and 
\begin{equation}
    \langle \Wac{k} \tensprod \FLM\rangle = - \epsilon \, 2^{2\epsilon} \sum_{\inotj}^{\Np} \int \frac{[\rmd \Omega_\Fp^{(d-1)}]}{[\Omega^{(d-2)}]} \llint \big[(\eta_{k \Fp}/2)^{-\epsilon} - 1\big] \frac{\rho_{ij}}{\rho_{i\Fp} \, \rho_{j\Fp}} \, \omega^{\Fp k,\Sp k}_{k \parallel \Sp} \FLM^{(ij)}\rrint \; ,
    \label{Eq:Appendix_partition_dependent_general_def_ac_sectors}
\end{equation}
where we have used the shorthand notation $\FLM^{(ij)} = \left(\scprod{\ColT{i}}{ \ColT{j}} \right) \colorprod \FLM$, 
which will appear in this appendix.

\subsection*{Extracting singularities from \texorpdfstring{$\Wbd{k}$}{}} 
\label{subsec:W}
\par 
We first investigate Eq.~\eqref{Eq:Appendix_partition_dependent_general_def_bd_sectors}. We note that  the contribution of 
$\lint\Wbd{k} \tensprod \FLM\rint$ 
to cross sections will be multiplied 
by  $1/\ep^2$ which 
originates from the integration over gluon energies.  For this reason, 
we require the expansion of 
Eq.~(\ref{Eq:Appendix_partition_dependent_general_def_bd_sectors}) through 
${\cal O}(\epsilon^2)$. 
We also note  that, thanks to  the partition functions $\omega_{\Fp \parallel \Sp}^{\Fp k, \Sp k}$ that appear in Eq.~\eqref{Eq:Appendix_partition_dependent_general_def_bd_sectors}, the only 
allowed collinear 
singularities correspond to the kinematic 
configurations where 
 $\Fp || k   $. 
 To isolate such divergences, we 
 write 
\begin{equation}\label{eq.i4}
\begin{split}
    \sum_{k=1}^{\Np} \lint \Wbd{k} \tensprod \FLM\rint 
    = & - \ep \, 2^{2\ep} \sum \limits_{k=1}^{\Np} \sum \limits_{\inotj}^{\Np} \; \int \frac{[\rmd \Omega_\Fp^{(d-1)}]}{[\Omega^{(d-2)}]} \bigg\langle \Big[(\iden - C_{k \Fp}) \left (\sigma_{k \Fp}^{-\ep}-1 \right) \\
    & + 1 - C_{k \Fp} \left ( 1 - \sigma_{k \Fp}^{-\ep} \right) \Big] \frac{\rho_{ij}}{\rho_{i \Fp} \, \rho_{j \Fp}} \, \omega_{\Fp \parallel \Sp}^{\Fp k, \Sp k} \, \FLM^{(ij)} \bigg\rangle \; .
\end{split}
\end{equation}
Next, we note that the first term in the above equation is ${\cal O}(\ep^2)$ already.  The second term allows us to sum over index $k$ using the relation 
\begin{equation}
    \sum_{k = 1}^{\Np} \omega_{\Fp \parallel \Sp}^{\Fp k, \Sp k} = 1 \; ,
\end{equation}
and the last one can be simplified since the collinear $C_{k \Fp}$ limit selects particular contributions from the sum.  \\
We now consider the second and the third term in more detail.  The former reads 
\begin{equation}
    - \ep \, 2^{2\ep} \sum \limits_{k=1}^{\Np} \sum \limits_{\inotj}^{\Np} \int \frac{[\rmd \Omega_\Fp^{(d-1)}]}{[\Omega^{(d-2)}]} \llint \frac{\rho_{ij}}{\rho_{i \Fp } \, \rho_{j \Fp}} \, \omega_{\Fp \parallel \Sp}^{\Fp k, \Sp k} \FLM^{(ij)} \rrint
    = 2 \sum_{\inotj}^{\Np} \lint \eta_{ij}^{-\ep} K_{ij}\; \FLM^{(ij)} \rint \; .
\end{equation}
To compute the contribution of the third term, we note 
that 
\begin{equation}
    C_{k \Fp} \left ( 1 - \sigma_{k \Fp}^{-\ep} \right ) \frac{\rho_{ij}}{\rho_{i \Fp} \, \rho_{j \Fp}} \, \omega_{\Fp \parallel \Sp}^{\Fp k, \Sp k} 
    =  (1 - \eta_{k \Fp}^{-\ep} )  \frac{1}{\rho_{k \Fp}} \left ( \delta_{ik} + \delta_{jk} \right ) \; .
\end{equation}
Using this expression  in Eq.~(\ref{eq.i4}), it becomes possible 
to sum either over $j$ or $i$ using 
the color conservation condition.
We obtain 
\begin{equation}
    \ep \, 2^{2\ep} \sum \limits_{k=1}^{N} \sum \limits_{\inotj}^{\Np} \; \int \frac{[\rmd \Omega_\Fp^{(d-1)}]}{[\Omega^{(d-2)}]} \llint C_{k \Fp} \left (1 - \sigma_{k \Fp}^{-\ep} \right ) \frac{\rho_{ij}}{\rho_{i \Fp} \, \rho_{j \Fp}} \omega_{\Fp \parallel \Sp}^{\Fp k, \Sp k}  \FLM^{(ij)}\rrint
    = N_c(\ep) \sum_{i=1}^{\Np} \T_i^2\;  \langle\FLM\rangle \; ,
\end{equation}
where 
\index{N!$N_c$}
\begin{equation}
    N_c(\epsilon) = \frac{2\Gamma^2(1-\epsilon)}{\Gamma(1-2\epsilon)} - \frac{\Gamma(1-\epsilon)  \Gamma(1-2\epsilon)}{\Gamma(1-3\epsilon)} = 1 + \mathcal{O}(\epsilon^3) \;  .
\end{equation}
Combining all the relevant terms, we find 
\begin{equation}
\begin{split}
    \sum_{k=1}^{\Np} \lint \Wbd{k} \colorprod \FLM\rint = & ~ 2 \sum_{\inotj}^{\Np} \lint \eta_{ij}^{-\ep} K_{ij} \FLM^{(ij)}\rint + N_c(\ep)\sum_{i=1}^{\Np} \T_i^2 \langle\FLM\rangle \\
    & + \ep^2 \, \sum_{k=1}^{\Np} \lint \Wbdfin{k} \colorprod \FLM\rint \; ,
\end{split}
\end{equation}
where
\index{W!$\Wbdfin{i}$}
\begin{equation} \label{Eq:Wbd_fin_def}
    \lint \Wbdfin{k} \colorprod \FLM\rint
    = \sum_{\inotj}^{\Np} \int \frac{\rmd \Omega_\Fp^{(3)}}{2\pi} \llint (\iden - C_{k \Fp}) \log \left(
    \sigma_{k \Fp}
    \right) \, \frac{\rho_{ij}}{\rho_{i\Fp} \, \rho_{j\Fp}} \,  \omega_{\Fp \parallel \Sp}^{\Fp k, \Sp k} \, \FLM^{(ij)}\rrint \; .
\end{equation}
Notice that $\Wbdfin{k}$ is finite in $\ep$, thus we evaluate it in $d=4$ dimensions.

\subsection*{Extracting singularities from \texorpdfstring{$\Wac{k}$}{}}
We can compute the second contribution $\langle \Wac{k} \tensprod \FLM\rangle$ shown in  Eq.~\eqref{Eq:Appendix_partition_dependent_general_def_ac_sectors} in the same way. As in the previous case, we introduce collinear subtraction operators as
\index{W!$\Wac{i}$}
\begin{equation}
\begin{split}
    \langle \Wac{k} \tensprod \FLM\rangle = &
    - \epsilon \, 2^{2\epsilon} \sum_{\inotj}^{\Np} \int \frac{[\rmd \Omega_\Fp^{(d-1)}]}{[\Omega^{(d-2)}]} \\
    & \times \llint (\iden - C_{k\Fp} + C_{k\Fp}) \big[(\eta_{k \Fp}/2)^{-\epsilon} - 1\big] \frac{\rho_{ij}}{\rho_{i\Fp} \, \rho_{j\Fp}} \,  \omega^{\Fp k,\Sp k}_{k \parallel \Sp} \, \FLM^{(ij)}\rrint \; .
\end{split}
\end{equation}
The term with $(\iden - C_{k \Fp})$ leads 
to an ${\cal O}(\ep^2)$ contribution that  we express  through  
\index{W!$\Wacfin{i}$}
\begin{equation}
    \lint \Wacfin{k} \tensprod \FLM\rint 
    = \sum_{\inotj}^{\Np} \int \frac{\rmd \Omega_\Fp^{(3)}}{2\pi} \llint (\iden - C_{k \Fp}) \log \left (\frac{\eta_{k\Fp}}{2}\right) \frac{\rho_{ij}}{\rho_{i\Fp} \, \rho_{j\Fp}} \, \omega^{\Fp k,\Sp k}_{k \parallel \Sp} \, \FLM^{(ij)} \rrint \; .
    \label{Eq:Appendix_Wacfin_i_def}
\end{equation}
\color{black}

The calculation of the term with $C_{k \Fp}$ proceeds exactly as already explained in the previous subsection.
We use 
\begin{equation}
    C_{k \Fp} \, \omega^{\Fp k,\Sp k}_{k \parallel \Sp} \frac{\rho_{ij}}{\rho_{i\Fp} \, \rho_{j\Fp}} = \frac{1}{\rho_{k \Fp}} \left( \delta_{i k}+\delta_{jk} \right) \; ,
\end{equation}
sum over one of the color indices and employ the following integral
\begin{equation}
\begin{split}
    \int \frac{[\rmd \Omega_\Fp^{(d-1)}]}{[\Omega^{(d-2)}]} \frac{2}{\rho_{k\Fp}} \left[\Big(\frac{\eta_{k\Fp}}2\Big)^{-\epsilon} - 1\right] 
    = & ~ \frac{2^{-2\ep}}{\ep} \left[\frac{2 \, \Gamma^2(1-\epsilon)}{\Gamma(1-2\epsilon)} - \frac{2^\ep \, \Gamma(1-\epsilon) \Gamma(1-\epsilon)}{\Gamma(1-3\epsilon)}\right] \\
    \equiv & ~ \frac{2^{-2\ep}}{\ep} N_k(\ep) \; ,
\end{split}
\end{equation}
to find the  final result
\begin{equation}
\label{eq:W_a//n}
    \langle\Wac{k} \tensprod \FLM\rangle = \epsilon^2 \langle\Wacfin{k} \colorprod \FLM\rangle - N_k(\ep) \, \T_k^2 \langle\FLM\rangle \; .
\end{equation}

\section{Triple color-correlated  contributions to real-virtual corrections}
\label{sec:trip_color_corr_RV}
In this appendix we discuss the computation of the triple color-correlated component arising from the integrated soft limit of the real-virtual contribution. The relevant factorization formula in the soft limit is given in Eq.~\eqref{eq:SoftLim_RV}, and we are interested in the final term
\index{S!$S_{\Fp}^{\rm tri}$}
\be
\Soft{\Fp}^{\rm tri} \, \FLRV({\Fp}) = - \asbr \, 
 \frac{4\pi \,\Gamma(1+\ep) \Gamma^3(1-\ep)}{\ep \, \Gamma(1-2\ep)}  \sum_{
 (ijk) }^{\Np} \kappa_{ij} \, 
 S_{ki}(p_{\Fp}) \,
\Big(
S_{ij}(p_{\Fp})\Big)^\ep \FLM^{(kij)}
\; ,
\label{eq:Soft_RV_tri_defn}
\ee
where $(ijk)$  labels triplets with different $i$, $j$ and $k$ and 
we have used the notation 
\index{F!$\FLM^{(kij)}$}
\index{Color space and algebra}
\be
\label{def:tripole}
\FLM^{(kij)} =  \lint \amp_{0} \big|  f_{abc} \,  T_k^a \, T_i^b \, T_j ^c \, 
\big|  \amp_{0}  \rint \; ,
\ee
to indicate the triple color-correlated matrix element. The phase factor $\kappa_{ij}$ is reported in Eq.~\eqref{eq:phase_factors}, and the eikonal factor $S_{ij}$ in Eq.~\eqref{eq:defn_eikonal}. Here we just recall that $\kappa_{ij}$ is 
completely symmetric under the exchange $i \leftrightarrow j$ and (obviously) is independent of $k$.

We begin by pointing out that the triple color-correlated matrix element gives a non-zero contribution only when there are at least four colored  particles in the Born-level process. Indeed, with three colored particles one can use color conservation to obtain the following identity 
\index{Color space and algebra}
\begin{align}
f_{abc} \,  T_1^a \, T_2^b \, T_3^c \, 
\big|  \amp_{0}  \rint
= - 
f_{abc} \,  T_1^a \,  T_2^b \, \big( T_1^c + T_2^c \big) \, 
\big|  \amp_{0}  \rint
= 0 \; .
\end{align}

Our goal is to integrate Eq.~\eqref{eq:Soft_RV_tri_defn} over the phase space of the soft gluon with momentum $p_{\Fp}$. We begin by integrating over the energy $E_{\Fp}$ and obtain 
\be
\begin{split}
\lint \Soft{\Fp}^{\rm tri} \, \FLRV \rint
={}&
 -\asbr^2 \, 
 \frac{4\pi^{3-\ep} \, 2^\ep\,\Gamma(1+\ep) \Gamma^4(1-\ep)}{\ep^2 \, \Gamma(1-2\ep)}\left(\frac{4\Emax^{2}}{\mu^2}\right)^{-2\ep} \; \\
 &
 \times \; \sum_{(ijk)}
 \llint
  \kappa_{ij} \; G^{kij} \;\FLM^{(kij)}
  \rrint\; ,
  \label{eq:h10}
  \end{split}
  \ee
   In Eq.~\eqref{eq:h10} we have defined 
\index{G!$G^{kij}$}
  \be
G^{kij} =   
\int \frac{\rmd\Omega^{(d-1)}_{\Fp}}{2(2\pi)^{d-1}}
 \frac{\rho_{ki}}{\rho_{k \Fp} \rho_{i \Fp}}
 \left (\frac{\rho_{ij}}{\rho_{i \Fp} \rho_{j \Fp}}  \right)^{\ep} \; ,
\ee
which is a function of the angular variables $\rho_{ij}, \rho_{ik}$ and $\rho_{jk}$.
We note that since $\kappa_{ij}$ 
is a symmetric tensor  
and $\FLM^{(kij)}$ is fully anti-symmetric,  only the anti-symmetric contribution $G^{kij} - G^{kji}$ can contribute to the sum, whereas the 
symmetric part drops out.  We will use 
this result when writing intermediate expressions for $\lint \Soft{\Fp}^{\rm tri} \, \FLRV({\Fp}) \rint$.

To perform the remaining integration over the soft-gluon angle, 
we employ the Mellin-Barnes representation of $d$-dimensional angular integrals presented in Ref.~\cite{Somogyi:2011ir}, and write the integral as
\be
\begin{split}
\label{eq:MBrep}
G^{kij} = & \int \frac{\rmd\Omega^{(d-1)}_{\Fp}}{2(2\pi)^{d-1}}
\frac{\rho_{ki}}{\rho_{k \Fp} \rho_{i \Fp}}
 \left (\frac{\rho_{ij}}{\rho_{i \Fp} \rho_{j \Fp}}  \right)^{\ep} 
 \\
= &~ \rho_{ki}\,\rho_{ij}^\ep\int \limits_{-i\infty}^{+i\infty}\frac{dz_{ij}\,dz_{jk}\,dz_{ki}}{(2\pi i)^3} \, \frac{\pi^{-2+\ep}}{2^{4+2\ep}}\,
\Gamma(-z_{ij})\Gamma(-z_{ki})\Gamma(-z_{jk})  \\
& \times \Gamma(1+\ep+z_{ij}+z_{ki})\Gamma(-1-3\ep-z_{ij}-z_{ki}-z_{jk})\Gamma(\ep+z_{ij}+z_{jk})\, \\
& \times \Gamma(1+z_{ki}+z_{jk})\frac{1}{\Gamma(-4\ep)\Gamma(\ep)\Gamma(1+\ep)}
\,\eta_{ij}^{z_{ij}}\,\eta_{ki}^{z_{ki}}\,\eta_{jk}^{z_{jk}} \; .
\end{split}
\ee
In the above equation  we have  introduced the three complex Mellin-Barnes variables $z_{ij},z_{ki},z_{jk}$, and $\eta_{ij}=\rho_{ij}/2$.
The integration contour has to be chosen in such a way that the poles of $\Gamma(... + x)$ are separated from the poles of $\Gamma(... - x)$, with $x$ being a generic integration variable. In order to resolve the singularity structure in $\ep$ we employ the packages {\tt MBresolve}~\cite{Smirnov:2009up} and {\tt MB}~\cite{Czakon:2005rk}, which allow us to express our original integral as a linear combination of integrals that can be safely expanded in $\ep$ under the integration sign, and whose integration contours are straight vertical lines in the complex plane. 
Upon applying this procedure we find that it is possible to express the function $G^{kij}$ up to 
${\cal O}(\ep^0)$ in terms of  classical and generalized polylogarithms (GPLs)~\cite{Goncharov:1998kja,Goncharov2001} up to weight three. 
It is convenient to write the final result for the angular integral as follows
\be
\begin{split}
G^{kij} = &\int \frac{\rmd\Omega^{(d-1)}_{\Fp}}{2(2\pi)^{d-1}}
\frac{\rho_{ki}}{\rho_{k \Fp} \rho_{i \Fp}} 
\left (\frac{\rho_{ij}}{\rho_{i \Fp} \rho_{j \Fp}}  \right)^{\ep}
 = -\frac{\ep^2}{4\pi^2}\bigg[\frac{2^{-\ep}\pi^{\ep}\,\Gamma(1-\ep)}{\Gamma(1-4\ep)\Gamma^2(1+\ep)}\bigg] \, 
{\overline G}^{\, kij} \; ,
\end{split}
\label{eq:Gbardef}
\ee
where 
\index{G!${\overline G}^{\, kij}$}
\be
\begin{split}
\label{eq:RVGpoles}
{\overline G}^{\, kij}  =
&~ \frac{3}{4 \epsilon ^3}  + \frac{1}{2 \ep^2}\Big[
\log \left(\eta _{ij}\right)-3 \log \left(\eta _{ik}\right)- \log \left(\eta_{jk}\right) \Big] 
+\frac{1}{\ep }\bigg[\frac{1}{2} \log ^2\left(\eta _{ij}\right) 
\\
   & + \log \left(\eta _{ij}\right) \left(-\log \left(\eta _{ik}\right)-\log
   \left(\eta _{jk}\right)\right)
   +\log
   \left(\eta _{ik}\right) \left(\log \left(\eta _{jk}\right)-2\log \left(1-\eta_{ik}\right)\right)
   \\
   &-2\Li_2\left(\eta _{ik}\right)+\frac{3}{2} \log ^2\left(\eta
   _{ik}\right)
   +\frac{1}{2} \log ^2\left(\eta _{jk}\right)+\pi ^2\bigg] + \mathcal{O}(\ep^0) \; .
\end{split}
\ee
Note that the $\epsilon$-dependent prefactor in Eq.~\eqref{eq:Gbardef} starts at $\mathcal{O}(\epsilon^2)$, so that the whole angular integral is effectively $\order{\ep^{-1}}$, as expected.

Inserting the result for $G^{kij}$ into Eq.~\eqref{eq:h10} we get the following final result for the triple color-correlated contribution to the real-virtual counterterm
 \be
 \label{eq:tripolefin}
 \lint \Soft{\Fp}^{\rm tri} \, \FLRV \rint
=
\asbr^2 \,
\frac{\pi \Gamma^5(1-\ep)}{\Gamma(1-2\ep)\Gamma(1-4\ep)\Gamma(1+\ep)}\,\left(\frac{4\Emax^2}{\mu^2}\right)^{-2\ep} 
\sum_{(ijk)}^{}
\llint \kappa_{ij}\, \overline{G}^{\, kij}\, \FLM^{(kij)} \rrint \; .
\ee
To proceed further it is convenient to 
split the function ${\overline G}^{\, kij}$ 
into  contributions using its symmetry  properties  under  the  $ i \leftrightarrow j$ permutations. We write 
\be
\overline{G}^{\, kij} = \overline{G}^{\, kij}_s+\overline{G}^{\, kij}_r\; ,
\ee
with 
\index{G!$\overline{G}^{\, kij}_s$}
\index{G!$\overline{G}_r^{\, kij}$}
\begin{align}
\label{eq:int_tripole_A}
\overline{G}^{\, kij}_s = &\frac34 \frac{1}{\ep^3} + \frac{1}{2\ep^2}
\log \left(\frac{\eta_{ij}}{\eta_{jk} \eta_{ik}} \right)+ 
\frac{1}{\ep} \bigg[ \frac{2\pi^2}{3} +\frac12 \log^2 \left(\frac{\eta_{ij}}{\eta_{jk} \eta_{ik}} \right)\bigg] + \mathcal{O}(\ep^0) \; ,
\end{align}
and
\begin{align}
\label{eq:int_tripole_B}
\overline{G}_r^{\, kij} = - \frac{\log \eta_{ik}}{\ep^2}  
+\frac{1}{\ep} \Big[   \log^2 \eta_{ik}  
+2\Li_2 (1-\eta_{ik}) \Big] + \mathcal{O}(\ep^0) \; .
\end{align}

 The function $\overline{G}_s^{\, kij}$ is symmetric under  $i \leftrightarrow j$ permutations; hence, it does not contribute to $\lint \Soft{\Fp}^{\rm tri} \, \FLRV(\Fp) \rint$ and can be dropped.  Note also that the function 
$\overline{G}_{r}^{\, kij}$, up to $\order{\ep^{-1}}$, is symmetric under 
the $i \leftrightarrow k$ permutation.  It follows that $\lint \Soft{\Fp}^{\rm tri} \, \FLRV(\Fp) \rint$ is free of poles for processes with a color-singlet initial state, as in this case we have $\kappa_{ij}= -1$.

For a hadron collider process with two incoming and any number of outgoing partons, the function 
$\kappa_{ij}$ reads 
\index{K!$\kappa_{ij}$}
\be
\kappa_{ij} = -1 + 2 \delta_{i 1} \delta_{j 2}
+ 2 \delta_{i 2} \delta_{j 1} \; ,
\ee
from which it follows that
\be
\begin{split}
\sum_{(ijk)}
\llint \kappa_{ij}\, 
\overline{G}^{\, kij}\, \FLM^{(kij)} \rrint  
& =
\sum_{(ijk)}
\llint \kappa_{ij}\, \overline{G}_r^{\, kij}\, \FLM^{(kij)}
\rrint 
\\
& =
2 \sum_{k \ne 1,2 }
\llint \left ( 
\overline{G}_r^{\, k12}-\overline{G}_r^{\, k21} \right ) \, \FLM^{(k12)} \rrint \; . 
\end{split}
\ee
Using this result together with  Eq.~\eqref{eq:int_tripole_B} and  
Eq.~\eqref{eq:tripolefin}, we obtain the final formula for the poles in the  triple color-correlated contribution to the soft limit of the real-virtual corrections. It reads 
\be
\begin{split}
\label{eq:trip_RV_final}
& \lint \Soft{\Fp}^{\rm tri} \, \FLRV \rint
=
\asbr^2
\sum_{k \ne 1,2}^{} 
\bigg\langle  \FLM^{(k 12)} 
\bigg\{
\frac{2 \pi}{\ep^2}
\log \frac{\eta_{2k}}{\eta_{1k}}
+ 
\frac{2 \pi}{\ep}
\bigg[
\log^2 \eta_{1k} -\log^2 \eta_{2 k}
\\
& \qquad 
+
2 \log\left(\frac{4 \Emax^2}{\mu^2}\right)
\log \left( \frac{\eta_{1 k}}{\eta_{2 k}}\right)
+ 2 \Li_2(1-\eta_{1k})
- 2 \Li_2(1-\eta_{2k})
\bigg]
+ \mathcal{O}(\ep^0)
\bigg\}\bigg\rangle \; 
.
\end{split}
\ee
We now present the formula for the $\mathcal{O}(\ep^0)$ terms of Eq.~\eqref{eq:RVGpoles}.
We exploit once again the symmetry properties of the triple color-correlated  contribution under the exchange of $i \leftrightarrow j$ indices and therefore only present results for the antisymmetric part. The result reads 
\index{G!$\overline{G}_{r,\rm fin}^{\, kij}$}
\begin{align}
\label{eq:fin_part_trip}
\overline{G}_{r,\rm fin}^{\, kij} = &~ \Li_2(\eta_{ij}) \log
   \left(\frac{\eta_{ik}}{\eta_{jk}}\right)-\Li_2(\eta_{ik})
   \log \left(\frac{\eta_{jk}}{\eta_{ij}
   \eta_{ik}}\right)+\Li_2(\eta_{jk}) \log
   \left(\frac{\eta_{ik}}{\eta_{ij} \eta_{jk}}\right) 
   \allowdisplaybreaks \notag \\
   & +\log (\eta_{ik})
   \Li_2\left(-\frac{\eta_{ik}-\eta_{jk}}{1-\eta_{ik}}\right)+\log (\eta_{ik})
   \Li_2\left(-\frac{\eta_{ik}-\eta_{jk}}{\eta_{jk}}\right) 
   +3
   \Li_3(1-\eta_{ik})
   \notag \\
   &-3 \Li_3(1-\eta_{jk})
   +\Li_2\left(\frac{1-\eta_{jk}}{1-\eta_{ik}}\right) \log (\eta_{ik}
   \eta_{jk})+\Li_2\left(\frac{\eta_{ik}}{\eta_{jk}}\right) \log (\eta_{ik} \, \eta_{jk}) 
   \allowdisplaybreaks \notag \\
   & -\log (\eta_{jk})
   \Li_2\left(-\frac{\eta_{jk}-\eta_{ik}}{\eta_{ik}}\right)-\log
   (\eta_{jk})
   \Li_2\left(-\frac{\eta_{jk}-\eta_{ik}}{1-\eta_{jk}}\right)
   +\Li_3(\eta_{ik})
   \notag \\
   & -\Li_3(\eta_{jk})
   +\log ^2(\eta_{ik})
   \bigg[\frac{1}{2} \log
   \left(\frac{1-\eta_{jk}}{\eta_{ij}}\right)+\log
   \left(\frac{\eta_{jk}-\eta_{ik}}{\eta_{jk}}\right)\bigg]
   \allowdisplaybreaks \notag \\
   & +\log
   (\eta_{ik}) \bigg[-\frac{1}{2} \log ^2(\eta_{ij})+\log
   (1-\eta_{ij}) \log (\eta_{ij})+\frac{1}{2} \log
   ^2\left(\frac{1-\eta_{jk}}{1-\eta_{ik}}\right)
   \allowdisplaybreaks \notag \\
   & +\frac{1}{2} \log
   ^2\left(\frac{\eta_{ik}}{\eta_{jk}}\right)+\log (1-\eta_{jk}) \log
   (\eta_{jk} (\eta_{jk}-\eta_{ik}))+\log ^2(\eta_{jk})-\frac{13 \pi^2}{6}\bigg]
   \allowdisplaybreaks \notag \\
   & +\log (1-\eta_{jk}) \Big[-\log (\eta_{jk}) \log
   \left(\frac{\eta_{ij}}{\eta_{jk}-\eta_{ik}}\right)-\log
   ^2(\eta_{jk})-\frac{\pi ^2}{6}\Big]
   \allowdisplaybreaks \notag \\
   & +\log (1-\eta_{ik}) \bigg[\log(\eta_{ik}) \Big[\log
   \Big(\frac{\eta_{ij}}{\eta_{jk}-\eta_{ik}}\Big)-\log
   (1-\eta_{jk})\Big] +\log ^2(\eta_{ik})
   \allowdisplaybreaks  \\
   & -\log (\eta_{jk}) \log
   (\eta_{jk}-\eta_{ik})-\frac{\log
   ^2(\eta_{jk})}{2}+\frac{\pi ^2}{6}\bigg]+\frac{1}{2} \log
   (\eta_{ij}) \log ^2(\eta_{jk})
   \allowdisplaybreaks \notag \\
   & +\frac{1}{2} \log ^2(\eta_{ij}) \log
   (\eta_{jk})-\log (1-\eta_{ij}) \log (\eta_{ij}) \log
   (\eta_{jk}) -\frac{1}{3} \log^3 \left(\frac{\eta_{ik}}{\eta_{jk}}\right)
   \allowdisplaybreaks \notag \\
   &  -\frac{1}{2} \log
   \left(\frac{1-\eta_{jk}}{1-\eta_{ik}}\right) \log
   ^2\left(\frac{\eta_{ik}}{\eta_{jk}}\right) -\frac{1}{2} \log
   ^2\left(\frac{1-\eta_{jk}}{1-\eta_{ik}}\right) \log
   \left(\frac{\eta_{ik}}{\eta_{jk}}\right)
   \allowdisplaybreaks \notag \\
   &  -\log ^2(\eta_{jk}) \log
   (\eta_{jk}-\eta_{ik}) +\frac{2}{3} \pi ^2 \log
   \left(\frac{\eta_{ik}}{\eta_{jk}}\right)+\log
   \left(\frac{\eta_{ik}}{1-\eta_{ik}}\right) 
   \allowdisplaybreaks \notag \\
   & \times \Big[\frac{\pi
   ^2}{6} -\log (1-\eta_{jk}) \log (\eta_{jk})\Big]  -\frac{1}{2} \log
   ^3(\eta_{ik})+\log ^2(1-\eta_{ik}) \log (\eta_{ik})
   \allowdisplaybreaks \notag \\
   & +\frac{\log
   ^3(\eta_{jk})}{2} -\log ^2(1-\eta_{jk}) \log (\eta_{jk})+\frac{3}{2}
   \pi ^2 \log (\eta_{jk})-\frac{1}{6} \pi ^2 \log
   \left(\frac{\eta_{jk}}{1-\eta_{jk}}\right)
   \allowdisplaybreaks \notag \\
   & + \log (\eta_{ij})
   \Big[G(\tilde{\eta}_{ik},w^+,1)-G(\tilde{\eta}_{jk},w^+,1)+G(\tilde{\eta}_{ik},w^-,1)-G(\tilde{\eta}_{jk},w^{-},1)\Big]
   \allowdisplaybreaks \notag \\
   & -G(\tilde{\eta}_{ik},w^+,\tilde{\eta}_{jk},1)+G(\tilde{\eta}_{jk},w^+,\tilde{\eta}_{ik},1)
   +G(\tilde{\eta}_{ik},w^+,1,1)-G(\tilde{\eta}_{ik},w^+
   ,\tilde{\eta}_{ik},1)
   \allowdisplaybreaks \notag \\
   & -G(\tilde{\eta}_{jk},w^+,1,1)+G(\tilde{\eta}_{jk}
   ,w^+,\tilde{\eta}_{jk},1)-G(\tilde{\eta}_{ik},w^-,\tilde{\eta}_{jk},1) +G(\tilde{\eta}_{jk},w^-,\tilde{\eta}_{ik},1)
   \allowdisplaybreaks \notag \\
   & +G(\tilde{\eta}_{ik},w^-,1,1)
   -G(\tilde{\eta}_{ik},w^-,\tilde{\eta}_{ik},1)-G(\tilde{\eta}_{jk},w^-,1,1)+G(\tilde{\eta}_{jk},w^-,\tilde{\eta}_{jk}
   ,1) \notag \; ,
\end{align}
where we defined 
\be
\begin{split}
w^{\pm} = &\frac{
2 -\eta_{ij}-\eta_{ik}-\eta
   _{jk}
\pm \sqrt{\left(\eta_{ij}-\eta_{ik}-\eta
   _{jk}\right){}^2-4 \eta_{ik}\eta_{jk} (1-\eta_{ij} )}}{2 \left(\eta_{ik} \eta_{jk}-\eta
   _{ik}-\eta_{jk}+1\right)} \; ,
\end{split}
\ee
and $\tilde{\eta}_{ab} = 1/(1-\eta_{ab})$. 

The expression in Eq.~\eqref{eq:fin_part_trip} is well defined in the region $\eta_{ik}<\eta_{jk}$.
However, this is sufficient  to cover the entire phase space since the other region can be obtained by swapping indices $i$ and $j$.  Thanks to the antisymmetry of the result under such an  exchange, this  only amounts to an overall sign change.

\label{sec:tripole_RV}


\section{Collection of functions used in the  final result}
\label{app:final_res_fun}

In this appendix we collect  all the functions that are necessary to write the final result 
for the NNLO QCD contribution to the  
partonic cross section of the process 
$q \bar q \to X+Ng$ given in Section~\ref{sec:final_res}.
For the reader's convenience, we attempt to make this appendix as self-contained as possible. 

  We use the following notations
\index{Z!$\bar{z}$}
\index{D!$\D_n$}
\index{L!$\Ltildei$}
\index{L!$L_i$}
\index{L!$\Lmax$}
  \begin{equation}
 \begin{split}
    \bar{z} = 1-z \; ,
\qquad
    \D_n(z) = \left[\frac{\log^n(1-z)}{1-z}\right]_+ \; ,
    \end{split}
\end{equation}
 \begin{equation}
 \begin{split}
\Ltildei = \log \left(\frac{2 E_i}{\mu}\right) \; , 
\quad 
L_i = \log \left(\frac{\Emax}{E_i} \right) \; , 
\quad 
\Lmax = \log \left(\frac{2\Emax}{\mu} \right) \; .
\end{split}
\end{equation}
\\

To present the double-boosted contribution in Eq.~\eqref{eq:finite_double_boosted} we have used
the following splitting function   
\index{P!$\PNLO_{qq}$}
\be
\label{eq:PNLO_app_final_res}
\begin{split}
    \PNLO_{qq}(z, E_i) 
    = \Cf \Big[\bar{z} + 4\D_1(z) + \big[4\D_0(z) + 3\delta(\bar{z})\big] \Ltildei - 2 (1+z)\big[\Ltildei + \log(\bar{z})\big]\Big] \; .
\end{split}
\ee
The single-boosted contribution in Eq.~\eqref{eq:finite_reminder_SB} depends on the function $\Wacfin{i}$, defined in Eq.~\eqref{Eq:Appendix_Wacfin_i_def}, and an operator 
 $\ITot^{(0)}$,  reported in Eq.~\eqref{eq:IT_fin}.  We have also introduced  the function
\index{P!$\CalP_{qq}^{\cal W}$}
\begin{equation}
\begin{split}
    \CalP_{qq}^{\cal W} (z, E_i)
    = & - \frac{1}{2\epsilon} \frac{\Gamma^2(1-\epsilon)}{\Gamma(1-2\epsilon)} \left(\frac{2E_i}{\mu}\right)^{-4\epsilon} \Big[\CalP_{qq}^{(4)}(z,L_i) - e^{-2\epsilon L_i} \CalP_{qq}^{(2)}(z,L_i)\Big] \\
    = & ~ \Cf \Big[\big[1 + z - 2\D_0(z) \big] L_i + 2\D_1(z) + \delta(\bar{z}) L_i^2 - (1+z) \log(\bar{z})\Big] \; , 
    \label{Eq:Sec5_Pqq_calW_def}
\end{split}
\end{equation}
where in the second line 
we have taken the $\ep \to 0$ limit. Furthermore, we use
\index{P!$\PNNLO_{qq}$}
\be
\PNNLO_{qq}(z, E_i)
    =
\Cf^2 P^{\rm NNLO,a}_{qq}(z, E_i) 
+ \Cf \Ca P^{\rm NNLO,na}_{qq}(z, E_i) \; ,
\label{eq:PNNLO_app}
    \ee
with
\begin{align}
    & P^{\rm NNLO,a}_{qq}(z, E_i)
    = 2 \Ltildei^2 \bigg [8 \mathcal{D}_1(z)+6 \mathcal{D}_0(z)-\frac{ \left(3 z^2+1\right) \log (z)}{\bar{z}}-4(z+1) \log(\bar{z}) \notag \allowdisplaybreaks \\
    & - z-5\bigg ] + \Ltildei \bigg[ 24 \mathcal{D}_2(z)
    +12 \mathcal{D}_1(z)
    -\frac{8 \pi^2}{3}  \mathcal{D}_0(z)
    +\frac{8
       \Li_2(\bar{z})}{\bar{z}}-\left(1+3 z^2\right) \frac{\log ^2(z)}{\bar{z}} \notag \allowdisplaybreaks \\
    & +4 \left(1+z+z^2\right) \frac{\log (z)}{\bar{z}}- \log (\bar{z}) \left(\frac{8 z^2 \log(z)}{\bar{z}}+ 2 (5+z)\right) -12 (1+z) \log ^2(\bar{z}) \notag \allowdisplaybreaks \\
    & +\frac{4 \pi^2}{3} (z+1)+9-7 z\bigg ] +
    8 \mathcal{D}_3(z) -\frac{8 \pi^2}{3}  \mathcal{D}_1(z) + 16 \zeta_3  \mathcal{D}_0(z) 
    -2 \left(5+3z^2\right) \frac{\Li_3(z)}{\bar{z}} \notag \allowdisplaybreaks \\
    & -\left(5-3 z^2\right) \frac{\Li_3(\bar{z})}{\bar{z}} +\frac{\log (\bar{z})}{\bar{z}} \bigg [\left(7-z^2\right) \Li_2(\bar{z})-6 \left(1+z^2\right) \log^2(z)
    +\frac{4 \pi^2}{3} (1-z^2) \label{Eq:Appendix_Pqq_NNLO_def} \\
    & +\left(7 - 2z +7 z^2\right) \log (z) +\bar{z} \left (6-\frac{9}{2} z\right ) -4 (z+1)\bar{z} \log ^2(\bar{z}) \bigg] + \frac{\log (z)}{\bar{z}} \bigg[\left ( \frac{5}{2}-\frac{9 z}{2} \right )  \bar{z}
    \notag \allowdisplaybreaks \\
    & - 2 \left(1+z^2\right) \left ( \Li_2(\bar{z})-\frac{5 \pi^2}{6} \right )   \bigg]+3 \bar{z} \left ( \Li_2(\bar{z}) +\frac{2 \pi^2}{9} \right ) +\left(\frac{5}{4}+\frac{13}{12} z^2 \right) \frac{\log ^3(z)}{\bar{z}} \notag \allowdisplaybreaks \\
    & +\frac{2 \zeta_3(1+7z^2) }{\bar{z}}-\log ^2(\bar{z}) \bigg[2\bar{z} -\left ( \frac{5}{2}-\frac{3z^2}{2}\right) \frac{\log (z)}{\bar{z}}\bigg]  + 8 z+\frac{z}{2} \log^2(z)-9  \notag \allowdisplaybreaks \\
    &+ \delta(\bar{z}) \bigg[\left(\frac{9}{2}-\frac{4 \pi ^2}{3}\right) \Ltildei^2 +\left(16 \zeta_3 + \frac{9}{2} \log2\right) \Ltildei + \frac{3 \pi ^2}{16}-\frac{\pi ^4}{45}-\frac{9}{16} \log^22\bigg] \; \notag,
\end{align}
and
\be
\begin{split}
    & P^{\rm NNLO,na}_{qq}(z, E_i) =
    -\frac{11}{3} \Ltildei^2  \Big[ 
    2 \mathcal{D}_0(z) - 1-z 
    \Big] 
    +  
    \Ltildei
    \bigg  [
    \left ( \frac{134}{9}-\frac{2 \pi ^2}{3}\right) \mathcal{D}_0(z)  -\frac{44}{3}\mathcal{D}_1(z)
    \\
    & + 2 \left(1+ z^2\right) \frac{\Li_2(z)}{\bar{z}}+\left[\frac{2}{3} 
       + \frac{11}{3} z^2+2(1+z^2)\log(\bar{z})\right] \frac{\log
       (z)}{\bar{z}}
       + \frac{22}{3} (z+1) \log (\bar{z})
       \\
    & 
       +\frac{2\pi^2}{3}  
        -\frac{52}{9}  
        -  \frac{91 z}{9}
    \bigg ] 
    -\frac{22}{3}\mathcal{D}_2(z)
    +\left(\frac{134}{9}-\frac{2 \pi ^2}{3}\right) \mathcal{D}_1(z)
    +
    \bigg[9 \zeta_3-\frac{208}{27}+\frac{11 \pi ^2}{6} 
      \\
    &
     -\frac{2 \log
       2}{3} \bigg] \mathcal{D}_0(z)
        -\frac{\left ( 1+ 6z + 19z^2\right)}{6} \frac{\Li_2(z)}{
       \bar{z}}+\frac{2 \log 2}{3}
       +\frac{\left(1+z^2\right)}{4 \bar{z}} 
       \Big[  2 \Li_3(z) - 8
       \Li_3(\bar{z})
        \\
    & 
       -2 \big( \log(z)-2 \log(\bar{z})\big)\Li_2(z) 
       - \log^2(z) \log(\bar{z})
       + 4 \log(z) \log^2(\bar{z})
       \Big]
       + \frac{2+11z^2}{8 \bar{z}}  \log^2(z)
       \\
       &
       +\frac{11}3(1+z) \log^2(\bar{z})
       +\frac{20-57z-49z^2}{36\bar{z}} \log(z)
       - \log \bar{z} \bigg[\frac{52}{9}   
       + \frac{173}{18} z
       -\frac12 (1-z)\log (z)
    \\
    &
       - \frac{\pi ^2}{6}  \frac{1+ 3z^2}{\bar{z}}\bigg]
      -\frac{5-4 z^2}{\bar{z}} \zeta_3
    +\frac{\pi^2}{36 } \frac{12z+49z^2-35}{\bar{z}}
    +\frac{563}{108}    
    +\frac{197}{108}z
    + \delta(\bar{z}) \bigg\{\Lmax^2 \bigg[\frac{64}{9}
    \\
    &   
    -\frac{\pi ^2}{3}+\frac{22 \log 2}{3}\bigg]
    + \Ltildei^2 \left(\frac{\pi ^2}{3}
    -\frac{227}{18}-\frac{22}{3} \log 2\right)
    + \Lmax \bigg[\frac{11 \zeta_3}{2}
    -\frac{22 \pi ^2}{9}+\frac{383}{54}
    \\
    & 
    -\frac{77}{3} \log ^22
    -\frac{125 \log 2}{9}\bigg]
    + \Ltildei \bigg[\frac{263}{6} -7 \zeta_3 -\frac{7 \pi ^2}{9}
    +\frac{11 \log ^22}{3}
    +\left(\frac{224}{9}-\frac{4 \pi ^2}{3}\right) \log 2\bigg]
    \\
    &     
     - 2 \Li_4(1/2) +\frac{22 \log ^32}{9}
    +\zeta_3 \left(\frac{217}{8}+\frac{25 \log 2}{4}\right)
    +\frac{211 \pi ^4}{1440}
    -\frac{1561}{36}
    -\frac{103 \pi ^2}{432}
    -\frac{\log ^42}{12}
    \\
    &
      +\left(\frac{15 \pi ^2}{4}-\frac{284}{9}\right) \log 2 
    +\left(\frac{5 \pi ^2}{12} -\frac{415}{36}\right) \log ^22\bigg\} \; .
\end{split}
\ee

It remains to discuss functions 
that contribute to  ${\rm d} {\hat \sigma}_{\rm el}^{\rm NNLO}$, see Eq.~\eqref{sigma_elastic_def}.
The quantity $\Iccfin$ collects color-correlated contributions and reads
\index{I!$\Iccfin$}
\begin{equation}
\begin{split}
    \Iccfin = &~ 
    \frac{1}{2} \big(\ITot^{(0)}\big)^2
    + K \ITot^{(0)} 
    + \Ca \bigg[\frac{11}{6} \left(\ITot^{(1)} + \ISofttilde^{(1)} - 2\ISoft^{(1)} + \frac{\pi^2}{24}\IVirt^{(-1)}\right)\\
    & + \ISoft^{(-1)} \bigg[\left(\frac{2 \pi ^2}{3}-\frac{131}{18}+\frac{22}{3} \log 2\right) \Lmax - \frac{17 \zeta_3}{4} +\frac{1975}{108} -\frac{11}{12} \pi ^2\\
    & -11 \log ^22-\frac{2}{3} \pi ^2 \log 2\bigg] 
    + \ISoft^{(0)} \left(\frac{\pi ^2}{3}-\frac{131}{36}+\frac{11 \log 2}{3}\right)\bigg] \; ,
    \label{eq:Iccfin_defn}
\end{split}
\end{equation}
where $K$ is a constant given in Eq.~\eqref{eq:constants_VV} and 
 $\ISoft^{(n)}$, $\ISofttilde^{(n)}$, $\IVirt^{(n)}$, $\ITot^{(n)}$ are the coefficients of the $n$-th power in the $\ep$-expansion of the corresponding operators reported in 
Appendix~\ref{sec:subsec_operator_defn}.
The finite part of the triple color-correlated operator  is given by
\index{I!$I_{\rm tri}^{\rm fin}$}
\index{I!$I_{\rm tri}^{\rm (cc),fin}$}
\index{G!$\overline{G}_{r,\rm fin}^{\, kij}$}
\be
    I_{\rm tri}^{\rm fin} = I_{\rm tri}^{\rm (cc),fin} 
   + \sum_{(ijk)}^{\Np}
 \kappa_{ij}\, 
\overline{G}_{r,\rm fin}^{\, kij}\, F^{(kij)} \; ,
    \label{eq:Itrifin_defn}
\ee
where $F^{(kij)} = f_{abc} \,  T_k^a \, T_i^b \, T_j ^c$. We note that $I_{\rm tri}^{\rm (cc),fin}$ corresponds to the $\order{\ep^0}$ contributions of the operator $\Itricc$ in Eq.~\eqref{eq:tripole_VVplusRVa} and reads
\index{I!$I_{\rm tri}^{\rm (cc),fin}$}
\begin{equation}
\begin{split}
    I_{\rm tri}^{\rm (cc),fin} 
    =
    \frac{\pi}{2} \sum_{(ijk)}^{\Np} F^{(kij)} & \bigg[\frac{1}{2} \big(\delta^-_{ij} +\delta^-_{ji}\big) \big(\delta^+_{jk} + \delta^+_{kj} + 4\phi_{jk} \big) + L_{jk} \big(\xi^-_{ij} + \xi^-_{ji} \big) \\
    & - \lambda_{ij} \big(\xi^+_{jk} + \xi^+_{kj} + 4\psi_{jk} \big)\bigg] \;,
\end{split}
\end{equation}
where
\index{D!$\delta^+_{ij}$}
\index{D!$\delta^-_{ij}$}
\index{F!$\phi_{ij}$}
\begin{equation}
\begin{split}
    \delta^+_{ij} = &~ \frac{1}{2} L_{ij}^2+\frac{\gamma_i}{\ColT{i}^2} \, L_{ij} -\frac{1}{2} \pi^2 \lambda_{ij}^2  \; , \\
    \delta^-_{ij} = &~ \frac{\gamma_i}{\ColT{i}^2} \, \lambda_{ij}  + L_{ij}\,  \lambda_{ij} \; , \\
    \phi_{ij} = & -2 \Lmax \log(\eta_{ij}) - \frac{1}{2} \log^2(\eta_{ij}) - \Li_2(1-\eta_{ij}) \; , \\
    \xi^+_{ij} = &\; \frac{1}{6} L_{ij}^3 + \frac{\gamma_i}{2\T_i^2} L_{ij}^2 - \frac{\pi^2}{2} \bigg(L_{ij} + \frac{\gamma_i}{\T_i^2} \bigg) \lambda_{ij}^2 \;, \\
    \xi^-_{ij} = &\; \frac{1}{2} L_{ij}^2  \lambda_{ij} + \frac{\gamma_i}{\T_i^2} \lambda_{ij} L_{ij} - \frac{\pi^2}{6} \lambda_{ij}^3 \;, \\
    \psi_{ij} = &\; 2 \Lmax^2 \log{\eta_{ij}} + \bigg(\Lmax - \frac12 \log{(1-\eta_{ij})}\bigg) \log^2{\eta_{ij}} + \frac{1}{6} \log^3{\eta_{ij}} \\
    & + 2\Lmax \Li_2(1-\eta_{ij}) - \Li_3(1-\eta_{ij}) - \Li_3(\eta_{ij}) \;.
\end{split}
\end{equation}
Furthermore, the term $\overline{G}_{r,\rm fin}^{\, kij}$  can be found  in Eq.~\eqref{eq:fin_part_trip}. 
\\

The operator $I_{\rm unc}^{\rm fin}$ in 
Eq.~\eqref{sigma_elastic_def} collects color-uncorrelated contributions. It reads
\index{I!$I_{\rm unc}^{\rm fin}$}
\begin{equation}
\begin{split}
    I_{\rm unc}^{\rm fin} = & \sum_{i=1}^{\Np} D_{\rm c}(E_i) 
    + \ISoft^{(-2)} \Ca \bigg\{\bigg[\frac{2 \pi ^2}{3}-\frac{131}{18}+\frac{22 \log 2}{3}\bigg] \Lmax^2 -\frac{935 \zeta_3}{72}
     +\frac{9607}{324} \\
    &+\bigg[-8 \zeta_3-\frac{11 \pi ^2}{6}+\frac{1433}{108}\bigg] \log 2 - \frac{\pi ^2 \left(945+199 \pi ^2\right)}{1440}  -\frac{11}{3} \log^32\\
    & + \left(\frac{143}{36}-\frac{\pi ^2}{3}\right) \log ^22\bigg\} \; ,
\label{eq:cunc_fin_app}
\end{split}
\end{equation}
where we have introduced
\begin{align}
    D_{\rm c}(E_i) 
    = &\; \Ca \Cf \bigg\{L_i \bigg[\frac{2}{9} \Big(3 \pi ^2-64-66 \log 2\Big) \Ltildei - 16 \zeta_3 \notag \\
    & +\frac{1}{27} \Big(802-36 \pi ^2 \log 2+ 3 (131+33 \log 2) \log 2 \Big)\bigg] \notag \\
    & +\frac{1}{9} \Big(3 \pi ^2-64-66 \log 2\Big) L_i^2 + \frac{1}{6} \Big(9 \pi ^2-64-66 \log 2\Big) \Ltildei \\
    & -12 \zeta_3+\frac{1}{36} \Big(802-36 \pi ^2 \log 2+3 (131+33 \log 2) \log 2 \Big)\bigg\} \notag \\
    & + \Cf^2 \left(-\frac{3}{16} \left(\pi^2 - 3 \log ^22\right)-\frac{9}{2} \Ltildei \log 2\right) \; , \notag 
\end{align}
if $i=1,2$, and 
\begin{align}
    D_{\rm c}(E_i) 
    = &\; \Ca^2 \bigg\{\bigg[-\frac{15 \zeta_3}{2}+\frac{1010}{27}-22 \log ^22-\frac{1}{6} \pi ^2 (11+ 8\log 2)\bigg] \Ltildei \notag \\
    & + \left[-\frac{21 \zeta_3}{2}+\frac{1987}{54}-22 \log ^22+\frac{2 \log 2}{3}-\frac{2}{9} \pi ^2 (11+6 \log 2)\right] L_i \notag \allowdisplaybreaks \\
    & - 2 \Li_4(1/2) + \zeta_3 \left(\frac{583}{24}+\frac{25 \log 2}{4}\right)+\frac{47 \pi ^4}{160}-\frac{40201}{648} \allowdisplaybreaks\\
    & - \frac{\log 2}{36}  \Big[713+\log 2 \Big(316+\log 2 \big(3\log 2-88\big)\Big)\Big] \notag \allowdisplaybreaks\\
    & +\frac{\pi^2}{432} \Big[259+36 \log 2 \big(33+5 \log 2 \big)\Big]\bigg\} \; , \notag
\end{align}
if $i \in [3,\Np]$.

The function $\gamma^{\cal W}$ 
in Eq.~\eqref{sigma_elastic_def}
is a combination of anomalous dimensions.  
It is given by 
\index{G!$\gamma^{\mathcal{W}}$}
\begin{equation}
\begin{split}
    \gamma^{\mathcal{W}}(L_i)
    = &~ \frac{1}{2\epsilon} \frac{\Gamma^2(1-\epsilon)}{\Gamma(1-2\epsilon)} \left(\frac{2E_i}{\mu}\right)^{-4\epsilon} \Big[\gamma_{z,g \to gg}^{24}(\epsilon,L_i) - e^{-2\epsilon L_i} \gamma_{z,g \to gg}^{22}(\epsilon,L_i)\Big]  \\
    = &~\Ca   \left[\frac{203}{72} +  L_i \left(\frac{11}{6} + L_i\right)\right]  \; ,
    \label{Eq:Sec5_gamma_calW_def}
\end{split}
\end{equation}
where in the second line we have taken $\ep \to 0$.
Furthermore, the functions $\Wbdfin{i}$, $\Wacfin{i}$ 
 and ${\cal W}_r^{(i)}$ are given 
in Eqs~\eqref{Eq:Wbd_fin_def}, \eqref{Eq:Appendix_Wacfin_i_def}  and~\eqref{eq:part_dep_contr}, respectively.
The quantities $\delta_g^{(0)}$ and $\delta_g^{\perp}$ correspond to
\index{D!$\delta_g^{(0)}$}
\index{D!$\delta_g^{\perp}$}
\begin{equation}
    \delta_g^{(0)} 
    = \Ca \bigg( -\frac{131}{72} + \frac{\pi^2}6 + \frac{11}6 \log 2\bigg) \; ,
    \qquad 
    \delta_g^{\perp} = \Ca \bigg( \frac{13}{36} -\frac{\log2}3 \bigg)
    \; .
\label{eq:delta_fin_app}
\end{equation}

The finite remainder of the double-soft 
integrated subtraction term is given by 
\index{S!$\SmnFin$}
\begin{align}
    & \SmnFin = \notag \\
    & = \left[\amu\right]^2 \sum_{\inotj}^{\Np} \Ca \bigg\langle \bigg\{-\frac{\text{Si}_2(2 \delta_{ij})}{6 \tan(\delta_{ij})} -\frac{11}{3} \text{Ci}_3 (2 \delta_{ij}) - 2G_{-1,0,0,1}(\eta_{ij}) \notag\\
    & +\frac{7}{2} G_{0,1,0,1}(\eta_{ij}) -\frac{5}{24} \log ^4(\eta_{ij}) -\frac{1}{12} \log^4(1+\eta_{ij}) + \frac{1}{2} \log (1-\eta_{ij}) \log^3(\eta_{ij}) - \bigg[\frac{5 \pi ^2}{12} \notag \allowdisplaybreaks \\
    & + \frac{11}{12} \log (1-\eta_{ij}) + \frac{7}{4} \log^2(1-\eta_{ij})\bigg] \log^2(\eta_{ij}) + \frac{\pi^2}{12} \log^2(1+\eta_{ij})-\frac{7}{4} \Li_2(\eta_{ij}){}^2 \notag \allowdisplaybreaks \\
    & + 3 \Li_4(\eta_{ij}) - 5 \Li_4\left(1-\frac{1}{\eta_{ij}}\right) - 5 \Li_4(1-\eta_{ij}) -2 \Li_4\left(\frac{1}{1+\eta_{ij}}\right)+\Li_4\left(\frac{1-\eta_{ij}}{1+\eta_{ij}}\right) \notag \allowdisplaybreaks \\
    & - \Li_4\left(-\frac{1-\eta_{ij}}{1+\eta_{ij}}\right) - \frac{1}{2} \Li_4\left(1-\eta_{ij}^2\right) - \Li_2(\eta_{ij}) \bigg[\log ^2(\eta_{ij}) + \frac{11}{6} \log (\eta_{ij}) + \frac{1+2 \pi ^2}{12}  \label{eq_SmnFin} \allowdisplaybreaks \\
    & + \frac{11 \log 2}{3}\bigg] + \Li_2(-\eta_{ij}) \left[2 \log (1-\eta_{ij}) \log (\eta_{ij})+2 \Li_2(1-\eta_{ij})-\frac{\pi ^2}{3}\right] \notag \allowdisplaybreaks \\
    & - 2 \log(1-\eta_{ij}) \Li_3(-\eta_{ij}) + 2\Li_3(1-\eta_{ij}) \Big(\log(1+\eta_{ij}) - \log (\eta_{ij})\Big) + \Li_3(\eta_{ij}) \bigg[\frac{11}{6} \notag \allowdisplaybreaks \\
    & + 2 \log (\eta_{ij}) - 2 \log (\eta_{ij}+1) - 7 \log(1-\eta_{ij})\bigg] + \log 2 \bigg[-\frac{11}{3} \log (1-\eta_{ij}) \log (\eta_{ij}) \notag \allowdisplaybreaks \\
    & +\frac{21}{4}\zeta_3 + \frac{33 \pi^2 - 868}{108}\bigg] +\frac{11}{2} \zeta_3 \log (1-\eta_{ij})-\frac{7}{4} \zeta_3 \log(1+\eta_{ij}) + 6 \Li_4\left(\frac{1}{2}\right) \notag \allowdisplaybreaks \\
    & + \log (\eta_{ij}) \left[\left(\pi ^2-\frac{1}{12}\right) \log (1-\eta_{ij}) + 2 \zeta_3 - \frac{1}{6}\right] -\frac{11}{24}\zeta_3 + \frac{137 \pi ^2}{432} -\frac{17 \pi ^4}{160} + \frac{649}{162} \notag \allowdisplaybreaks \\
    & +\frac{\log ^42}{4}-\frac{11 \log ^32}{9} - \frac{137 + 9 \pi^2}{36} \log^2 2\bigg\} \, (\T_i \cdot \T_j) \colorprod \FLM \bigg\rangle \; . \notag
\end{align}
In the above equation we used 
$\delta_{ij} = \theta_{ij}/2$, where $\theta_{ij}$
is the opening angle between momenta of partons $i$ and $j$. The Clausen functions are defined as
\begin{equation}
    \text{Ci}_n(z) = \frac{\Li_n(e^{iz}) + \Li_n(e^{-iz})}{2} \; , 
    \qquad
    \text{Si}_n(z) = \frac{\Li_n(e^{iz}) - \Li_n(e^{-iz})}{2i} \; ,
\end{equation}
and $G_{a_1, a_2, ... , a_m}(x)$ are the standard Goncharov polylogarithms.

The last two functions in Eq.~\eqref{sigma_elastic_def}
are $\FLVfinsq$ and $\FLVVfin$, which refer to the  infrared-finite components of the one-loop squared amplitude and the two-loop amplitude interfered with tree level, respectively.



\bibliographystyle{JHEP}
\bibliography{biblio.bib}


\printindex

\end{document}